\documentclass[prb,notitlepage,twocolumn,superscriptaddress]{revtex4}
\usepackage{amsmath,amssymb,color,graphicx,hyperref,amsthm,bm}
\usepackage{epstopdf}

\definecolor{TW-color}{RGB}{0,130,128}
\definecolor{MS-color}{RGB}{128,0,128}
\definecolor{EB-color}{RGB}{111,0,111}
\definecolor{Error-color}{RGB}{250,50,50}

\graphicspath{{./images/}}

\begin{document}

\title{Non-London electrodynamics in a multiband London model: anisotropy-induced non-localities and multiple magnetic field penetration lengths. }
\author{Mihail  Silaev}
 \affiliation{Department of
Physics and Nanoscience Center, University of Jyv\"askyl\"a, P.O.
Box 35 (YFL), FI-40014 University of Jyv\"askyl\"a, Finland}
\author{Thomas Winyard}
\affiliation{Department of Physics, KTH-Royal Institute of Technology, Stockholm, SE-10691 Sweden}
\affiliation{School of Mathematics, University of Leeds, Leeds LS2 9JT, United Kingdom}
\author{Egor Babaev}
\affiliation{Department of Physics, KTH-Royal Institute of Technology, Stockholm, SE-10691 Sweden}

\begin{abstract}
The London model describes strongly type-2 superconductors as massive vector field theories, where the magnetic field decays exponentially at the length scale of the London penetration length. This also holds for isotropic multi-band extensions, where the presence of multiple bands merely renormalises the London penetration length. We show that, by contrast, the magnetic properties of anisotropic multi-band London models are not this simple, and the anisotropy leads to the inter-band phase differences becoming coupled to the magnetic field. This results in the magnetic field in such systems having $N+1$ penetration lengths, where $N$ is the number of field components or bands. That is, in a given direction, the magnetic field decay is described by $N+1$ modes with different amplitudes and different decay length scales. 
For certain anisotropies we obtain magnetic modes with {\sl complex masses.  That means that
magnetic field decay is not described by a monotonic exponential increment set by a real
penetration length but instead is oscillating.  
Some of the penetration lengths are shown to diverge away from the superconducting phase transition when the 
mass of the phase-difference mode vanishes. }
Finally the anisotropy-driven hybridization of the London mode with the Leggett modes can provide an effectively non-local magnetic response in the nominally local London model. Focusing on the two-component model, we discuss the magnetic field inversion that results from the effective non-locality, both near the surface of the superconductor and around vortices. In the regime where the magnetic field decay becomes non-monotonic, the multiband London superconductor is shown to form weakly-bound states of vortices.

\end{abstract}

\maketitle
\section{Introduction}

The work by F. London  and H. London that formulated hydro-magnetostatic theory of superconductivity is one of the most influential works in condensed matter physics.
It demonstrated that the magnetic field in a superconductor is described by a massive vector field theory\cite{London1935}
\begin{equation}
\nabla^2 {\bm B}=\frac{1}{\Lambda^2}{\bm B}.
\label{1}
\end{equation}
One of the direct consequences is that an externally applied magnetic field decays in a superconductor exponentially at the length scale of the London magnetic field penetration length $\Lambda$: ${\bm B}\propto {\bm B}_0e^{-r/\Lambda}$ . The superconducting current  $\bm J$ varies at the same length scale due to its relation to the magnetic field ${\bm J} = \boldsymbol{\nabla}\times {\bm B}$. The work also paved the way to Anderson's demonstration of the Higgs mechanism \cite{Anderson1963}.

Multi-band superconducting materials are currently of central interest. For these systems the pairing of electrons is supposed to take place in several Fermi surfaces, formed due to the overlapping of electronic bands
 \cite{Mazin2003,Damascelli2000,Kamihara2008,Mazin2008,Kuroki2008,Chubukov2008,Boeker2017}.

The range of validity of London electrodynamics is well understood in single component or single band superconductors and should not, in general, be applicable at the length scale $\xi_0$ associated with the superconducting carrier, namely the estimated Cooper pair size. In a weak-coupled BCS superconductor $\xi_0$ can exceed the magnetic field penetration length $\Lambda$, leading to markedly different electrodynamics. Such a state cannot be described by the local London model \cite{Bardeen1957,Bardeen1957a} 
 and is termed Pippard electrodynamics \cite{Pippard1953}.
 In contrast, the limiting case when $\xi_0 \ll \Lambda$ is described by local London electrodynamics, which is typically applicable to extreme type-2 superconductors. 

In this paper we will only be interested in multiband anisotropic materials. Furthermore we focus on the case where $\xi_0^{(\alpha)} \ll \Lambda$, where $\alpha$ is a band index
 and therefore where one can neglect the effects that lead to Pippard electrodynamics.
 We will  study how the crystal anisotropy of superconducting materials affects their magnetic properties even at the level of London's model. 
 In a single component case, the anisotropy effects in London electrodynamics are well studied \cite{Kogan1981,Balatskii,Buzdin}. 
 In the single component case the anisotropy leads, in general, to an electrodynamic kernel that is characterized by two real-valued penetration lengths, corresponding to the different polarizations of the magnetic field.
We show that the situation is principally different in the multiband London model due to the presence of additional massive modes, associated with the variations of the phase differences between order parameter components, known as the Leggett modes \cite{Leggett1966}.
In the isotropic case the magnetic and Leggett modes are de-coupled, however we show that, in general, this coupling appears in the London model with the introduction of crystal anisotropy. On the qualitative level, this coupling arises due to differing anisotropy in the superconducting bands which enables the gradients of the inter-band phase differences to produce non-zero transverse charge currents, which generate magnetic field. 
The London model then exhibits $N+1$ magnetic modes, where $N$ is the number of superconducting components. Namely, the two independent components of the magnetic field and $N-1$ interband phase differences, yielding in total $N+1$ massive scalar fields \cite{smiseth2005field}
 \footnote{ 
  A more general formalism is required to discuss  
 multi-band systems with frustrated interband interaction where
 the Leggett mode is mixed with density modes\cite{Carlstroem2011a,Garaud2017a},
 yielding additional length scales and thus additional magnetic field penetration lengths when generalized to 
 the anisotropic case.
 }
 We analyse this behaviour in detail using the minimal model of the two-component superconductor.
  The results can be straightforwardly generalized to the larger number of bands, e.g. the three-component model with phase frustrations yielding the intrinsically complex $s+is$ state. This opens interesting possibilities to study the behaviour of  magnetic signatures of the broken-time reversal symmetry phase transition to the $s+is$ characterized by the soft 
  Leggett mode which becomes massless at the transition point
  \cite{Carlstroem2011a, Weston2013}.
One of the  interesting properties is that the masses and correspondingly the relaxation lengths can become complex under certain conditions. The unconventional behaviour of magnetic modes will be shown to result in magnetic field reversal near the boundaries and vortices leading to the formation of vortex bound states.

 \section{The Model}
 It is illustrative to view the London model as a constant density limit of the Ginzburg-Landau model for complex fields 
 { (although indeed the model is much more general and is valid at low temperatures)}. The simplest Ginzburg-Landau free energy
 density for an $N$-component anisotropic system is given by  
 \begin{equation} 
 F =  \sum^{N}_{\alpha = 1} \left( \lambda^{-1}_{ij\alpha} D_j\psi_{\alpha}\right)
 \left(\lambda^{-1}_{ik\alpha}  \overline{D_k\psi_{\alpha}}\right) + \boldsymbol{B}^2 + F_p ,
 \label{GL}
 \end{equation}
 \noindent where $\boldsymbol{D} = i \boldsymbol{\nabla} + 2\pi e \boldsymbol{A}/c$ is the covariant derivative,
 $e$ and $c$ are the electron charge and light velocity respectively 
 {(hereafter we use the units with $\hbar =1$)}, 
  the fields $\psi_\alpha = \left|\psi_\alpha\right|e^{i\theta_\alpha}$ represent the different superconducting components. 
 Greek indices will always be  used to denote superconducting components and Latin indices will be spatial with the summation 
 principle applied for repeated Latin indices. 

 The inverse mass tensors $\lambda^{-1}_{ij\alpha}$ represents a 3 dimensional diagonal matrix for each component,
 \begin{equation}
 \lambda^{-1}_{ij\alpha} = \left(\begin{array}{ccc} \lambda^{-1}_{x \alpha} & & 
 \\ & \lambda^{-1}_{y \alpha} & 
 \\ & & \lambda^{-1}_{z \alpha} \end{array} \right).
 \end{equation}
 \noindent In Eq.(\ref{GL}) $F_p$ collects together the potential (non-gradient) terms which can be any from a large range. 
 The simplest example is the standard single band potential terms and the Josephson inter-band coupling term
 \begin{eqnarray} 
 F_p &=& \sum^N_{\alpha = 1}\frac{\gamma_{\alpha}}{4}\left({\psi^0_\alpha}^2 - \left|\psi_\alpha\right|^2\right)^2  
 \nonumber 
 \\ 
 && - \sum^N_{\alpha  = 1}\sum_{\beta < \alpha}\eta_{\alpha \beta} \left|\psi_\alpha\right|\left|\psi_\beta\right| 
 \cos{\left(\theta_{\alpha\beta}\right)},
 \end{eqnarray}
where ${\psi^0_\alpha}$, $\gamma_{\alpha}$ and $\eta_{\alpha\beta}$ are positive real constants. The second term above is the Josephson inter-band coupling, where $\theta_{\alpha\beta} = \theta_\alpha - \theta_\beta$ is the inter-band phase difference between components $\alpha$ and $\beta$. The simplest multi-component system is the isotropic case $\lambda_\alpha = I_3$ without inter-band coupling $\eta_{\alpha \beta} = 0$, which has the maximal symmetry $U(1)\times U(1)... = U(1)^N$. However the introduction of the simplest non-frustrated Josephson terms, that in the ground state locks all phase differences to zero, breaks this symmetry to a $U(1)$ symmetry. 
 In the  isotropic case, the phase differences are neutral massive modes. In the absence of a coupling to the gauge field the  phase sum  gives rise to the massless Goldstone mode. In the presence of a gauge field coupling that mode is the massive London mode.  

 As we are interested in strongly type 2 systems we neglect density variations, but, following Leggett, we retain the massive degree of freedom associated with the phase difference mode \cite{Leggett1966} 
 \footnote{The rationale behind not discarding a massive mode is that even in single-component London hydrodynamics the massless phase variable combines with the gauge field yielding a massive mode. Hence, other modes cannot in principle be discarded solely on the grounds that they are massive. However in practice, for isotropic theories all other massive modes are decoupled in the linear equations. We demonstrate that in the anisotropic case, such decoupling in general does not take place. The reason we do not retain massive modes associated with density fields is that they do decouple in the linear equations that we consider here.}  
 We consider the limit $\left|\psi_\alpha\right|^2 \approx const$, leading to the following free energy (for brevity below we refer to the approximation that neglects density variations as the London limit).
 \begin{equation}\label{Eq:LondonEnergy}
 F = \frac{ 16\pi^2}{c^2} \sum^N_{\alpha = 1} (\hat \lambda_\alpha^2 {\bm j_\alpha}\cdot{\bm j_\alpha} )
 - \sum^N_{\alpha = 1}\sum_{\beta < \alpha} J_{\alpha\beta} \cos\theta_{\alpha\beta} + {\bm B}^2 .
 \end{equation}
 \noindent Here $\bm{j}_\alpha$ are the partial superconducting currents
 \begin{equation}\label{Eq:Currents}
 \frac{4\pi}{c} {\bm j}_{\alpha}=\hat\lambda_{\alpha}^{-2}
 \left( \frac{\Phi_0}{2\pi} \boldsymbol{\nabla}\theta_{\alpha}-{\bm A} \right),
 \end{equation}
 where $\Phi_0=\pi c/e$ is the flux quantum, $\hat\lambda_{k}$ are coefficients characterizing the contribution of each band to the Meissner screening, ${\bm A}$ is the vector potential and $J_{\alpha\beta}$ the Josephson coupling. 
 Parameters of the London model (\ref{Eq:LondonEnergy}) are related to that of the Ginzburg-Landau functional  
 Eq.(\ref{GL}) by  $J_{\alpha\beta} = \eta_{\alpha\beta} |\psi_\alpha||\psi_\beta|$ and $ \hat\lambda_{\alpha}^{-2} = (2e|\psi_\alpha|/c )^2 \lambda_\alpha^{-2} $, where the matrix indices are suppressed for $\hat \lambda$.
   
 Considering the total current, which is a sum of partial current contributions from each band 
 ${\bm j} = \sum_\alpha{\bm j}_\alpha$ we get an expression for the magnetic field, which provides an 
 extension of the London
 theory in multicomponent systems \cite{Silaev2015c,Garaud2016}
 \begin{align}\label{Eq:MagneticField}
 & {\bm B} = - \frac{4\pi}{c}\nabla\times \left( \hat\lambda_L^2 {\bm j} \right)
 + \\ \nonumber
 & \frac{\Phi_0}{2\pi N}\sum_{\alpha>\beta}\nabla\times
 \left[ \hat\lambda_L^2 \left(\hat\lambda_{\alpha}^{-2} - \hat\lambda_{\beta}^{-2} \right) \nabla {\theta}_{\alpha\beta}
 \right],
 \end{align}
 where $N$ is the number of components,
 $ \hat\lambda_L^2 = (\sum_\alpha \hat\lambda^{-2}_{\alpha})^{-1}$
 and $\theta_{\alpha\beta} =\theta_\alpha-\theta_\beta$ are the relative inter-band phases. 
 Expression (\ref{Eq:MagneticField}) shows that the phase difference gradients in the second term can generate 
 magnetic field in anisotropic materials. 
  We will demonstrate that the coupling between magnetic field and interband phase differences leads to
  the non-local magnetic response. 
  Importantly the non-locality here has nothing to do with the Pippards non-locality 
  \cite{Pippard1953} associated with the Cooper pair dimension \cite{Bardeen1957,Bardeen1957a}.
  Rather it can be obtained already within the the standard anisotropic London model, 
  which has only ``local" terms. 
  In result of that, the magnetic response of  such a system is characterized by the  
  multiple magnetic modes with different penetration lengths which depend on the degree 
  of anisotropy in different bands and the strength of inter-band pairing interaction. 
  In next sections of the paper we will consider the influence of these anisotropic effects on the magnetic response 
 of multi-band superconductors and demonstrate that they many  observable physical consequences 
 both for the Meissner and vortex states. 
 
 To connect the general model given by the Eqs.(\ref{Eq:LondonEnergy},\ref{Eq:Currents}, \ref{Eq:MagneticField}) 
 with  real materials we can use  
  the multiband weak-pairing pairing models and express  the Josephson couplings $J_{\alpha\beta}$
 through the  microscopic parameters, such as the pairing coefficients  
 $V_{ij}$, Fermi velocities in different bands and the lengths $\hat\lambda_{k}$ characterising magnetic 
 responses of each band  in Eq.(\ref{Eq:Currents}).
 For the generic case of a two-band superconductor there is only one Josephson energy scale $E_J = J_{12}$, which determines the Leggett mode frequency \cite{Leggett1966}. 
 Since we are dealing with a static but spatially inhomogeneous problem and aim to 
 describe the coupling between the phase difference and magnetic field, it is natural to 
 construct the inverse characteristic length scale as follows $k_0 = [ 2\pi^2 E_J/ \Phi_0^2 ]^{1/4}$.
 This parameter does not depend on the anisotropy or condensate stiffness. Therefore
 it can be expressed through the weak-coupling pairing coefficients
 \begin{equation} \label{Eq:k0}
  k_0^4 = -\frac{32 \pi^3}{\Phi^2_0}  \nu_1 (\hat V^{-1})_{12}  \Delta_1\Delta_2,
\end{equation}  
where $\Delta_{i}$ is the superconducting gap amplitude for the $i$th band.   
The density of states in the $i$-th band $\nu_i$ can be found using the magnetic 
response length $(\hat\lambda^{-2}_i)_{\alpha\beta} = 
\nu_i\langle v_{F\alpha} v_{F\beta} \rangle_{FS} (2\pi)^3/\Phi_0^2 $, where $v_{F\alpha}$ is the $\alpha$-th
spatial component of the Fermi velocity, and $\langle ... \rangle_{FS}$ denotes the Fermi surface average.
For the parameters \cite{Mazin2002,Moshchalkov2009} 
  of the uniaxial anisotropic two-band superconductor MgB$_2$ we get $k_0\lambda_{1\perp} =1.3$ for the  $\sigma$-band with larger gap and $k_0\lambda_{2\perp} = 0.9$ for the $\pi$-band with smaller gap. Although the London length anisotropy in MgB$_2$ is quite weak  \cite{Pal2006}, 
  it has  a rather pronounced anisotropy of the partial magnetic responses in the 
  almost cylindrical $\sigma$-band. According to the expression above it is determined by 
  the Fermi velocity anisotropy $v_{F\perp}/v_{Fz} \approx 8.6 $ \cite{Brinkman2002}. 
  Therefore we can estimate $\lambda_{2z}\approx \lambda_{2\perp}$  and $\lambda_{1z}\approx 8.6 \lambda_{1\perp}$.
  
  The other example of multiband anisotropic superconductor is Sr$_2$RuO$_4$.
  The nature of the superconducting state for this material is still highly debated.
 In order to make one more  estimate, we consider a
  weak-coupling three-band model  from Ref. \cite{Huang2016}.
  The suggested coupling matrix $\hat V$ for Sr$_2$RuO$_4$ contains two bands with strong interband interactions and the 
  third band which has much weaker interband pairing, such that the elements $V_{13}$ and $V_{23}$ are 
  much smaller than the others. Therefore one can use an effective two-band model to describe 
  the low-energy and large-scale variations of the interband phase differences 
  $\theta_{13}=\theta_{23}$ and assuming $\theta_{12}=0$.  
  Additionally, we use the same values of gaps, and lengths $\lambda_{i\perp}$ in all bands  \cite{Mackenzie2003} to obtain an estimation of $k_0\lambda_{\perp i} =2.8$. 
   The overall London length anisotropy in Sr$_2$RuO$_4$ is about $\lambda_{Lz}/\lambda_{L\perp} \approx 20 $, although 
   the anisotropy of each band contribution is not known.
Therefore, in realistic multiband compounds
     $k_0$ can be of the order of the 
     penetration length    
     and the anisotropy of $\hat\lambda_k$ in each band can change in the wide limits.
 
\section{Electromagnetic response}
In this section we consider the system in the absence of vortices and demonstrate that the anisotropy qualitatively changes the electromagnetic modes in multi-band superconductors. To obtain the equations of motion we rewrite the condensate phases in Eq.(\ref{Eq:LondonEnergy}) as $\theta_\alpha = \sum_\beta \theta_{\alpha\beta} /N + \theta_\Sigma$ where $\theta_\Sigma = \sum_\alpha \theta_\alpha /N$.
  Let us assume for simplicity the same strength of Josepson coupling between all bands $J_{\alpha\beta}= E_J$. Then varying the free energy (\ref{Eq:LondonEnergy}) by $\theta_{\alpha\beta}$ and ${\bm A}$ we obtain the system of coupled equations for the phase differences and magnetic field
  \begin{align} \label{Eq:Teta12}
  & \nabla\cdot [ (\hat\lambda_\alpha^{-2} + \hat\lambda_\beta^{-2}) \nabla\theta_{\alpha\beta} + N (\hat\lambda_\alpha^{-2} - 
  \hat\lambda_  \beta^{-2}) {\bm p}_s ] = 
  \\ \nonumber
  & N^2 k_0^4 \sin\theta_{\alpha\beta} , 
  \\
  \label{Eq:Q0}
  & \frac{2\pi}{\Phi_0} \nabla\times {\bm B} =
  \frac{1}{N}  \sum_{\beta > \alpha}  (\hat\lambda_\alpha^{-2} -
  \hat\lambda_\beta^{-2} ) \nabla\theta_{\alpha\beta} + \hat\lambda_L^{-2} {\bm p_s} ,
  \end{align}  
where we have introduced the gauge invariant term ${\bm p}_s = \nabla \theta_\Sigma - 2\pi {\bm A}/\Phi_0$.

In the absence of phase singularities we can choose the gauge so that the common phase is constant $\sum_\alpha \nabla \theta_\alpha = 0$ and therefore ${\bm p}_s = - 2\pi {\bm A}/\Phi_0$. For isotropic superconductors where the total current is ${\bm j} \propto {\bm p}_s $  this choice corresponds to the London gauge. In the anisotropic case the relation between ${\bm p}_s$ and the current is more complicated so that the gauge in general depends on the specific choice of the anisotropy parameters.

To study the linear electromagnetic response we linearise the above equations of motion and switch to the momentum representation. Then we get the algebraic relation between the current and vector potential
 \begin{equation}\label{Eq:LinearResponce}
 - \frac{4\pi}{c} j_i  = Q_{ij} ({\bm k}) A_j ,
 \end{equation}
 where $Q$ is known as the polarisation operator and tells us how changes to the gauge field relate to the current. It is given to be,
  \begin{equation}\label{Eq:KernelGen}
  Q_{ij} ({\bm k}) = -\hat\lambda_{Lij}^{-2}  +
  \sum_{\beta > \alpha} \frac{ (\hat\lambda_{\alpha ik}^{-2} - \hat\lambda_{\beta ik}^{-2}) 
  (\hat\lambda_{\alpha jl}^{-2} - 
  \hat\lambda_{\beta jl}^{-2}) k_kk_l }
 {N^2 k_0^4 + ({\bm k}\cdot  [ \hat\lambda_\alpha^{-2} + \hat\lambda_\beta^{-2} ] {\bm k}) } .
 \end{equation} 
During the derivation we used the commutator $\hat \lambda_\alpha^{-2} \hat \lambda_\beta^{-2} =  \hat \lambda_\beta^{-2} \hat \lambda_\alpha^{-2}$.    
 
  In the isotropic limit the kernel becomes local due to the second term in Eq.(\ref{Eq:Kernel}) becoming zero, through the London gauge choice ${\bm k}\cdot{\bm A} = 0$. In general one can see that the non-locality scale of $Q (\boldsymbol{k})$ is determined by the inter-band Josephson length $k^{-1}_{0}$ which is not related to the BCS 
  non-locality scale determined by the Cooper pair size.
    In the absence of coupling between $\theta_{\alpha\beta}$ and ${\bm B}$ we obtain the usual local response, similar to the single-component superconductors determined by the constant tensor $\hat\lambda_L$.     
 
  The complicated structure of the second term in $Q ({\bm k})$ points to some unusual magnetic properties of anisotropic multi-band superconductors. In particular, that it produces multiple magnetic modes in which the magnetic field and the inter-band phase difference are coupled. This is what has lead us to calling them magnetic phase difference modes. 
 
 \section{Magnetic modes and field screening in the two-band superconductor}
    
\begin{figure}[htb!]
 \centerline{\includegraphics[width=1.0\linewidth]{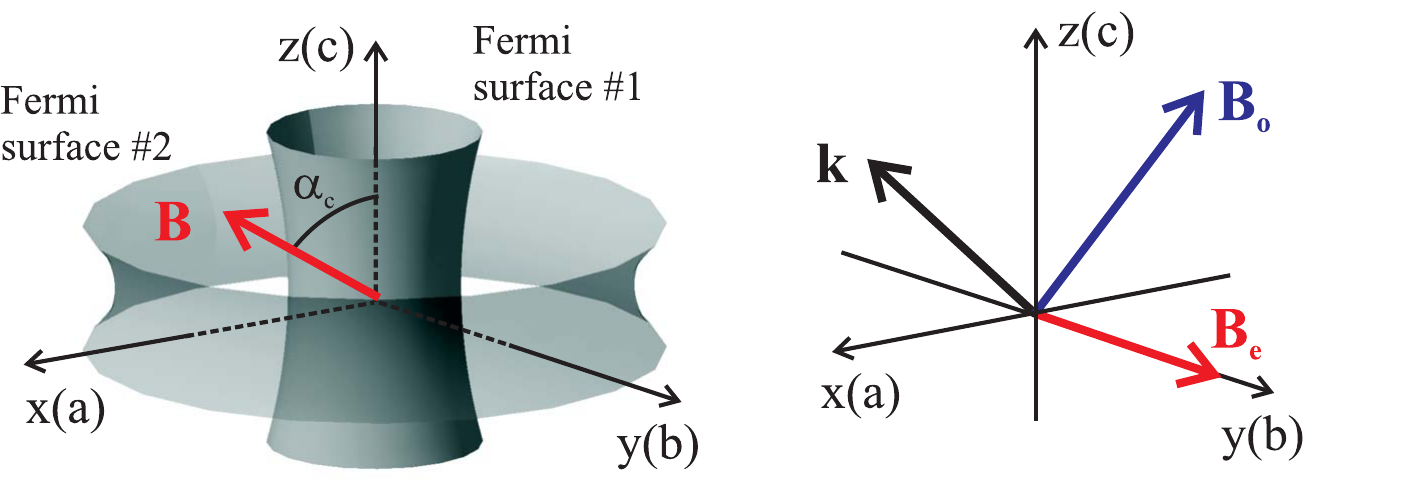}}
 \caption{\label{Fig:FS} (Left) Two Fermi surfaces in an anisotropic multi-band
 superconductor, qualitatively similar to that of the uni-axial compound
 MgB$_2$. The crystal structure anisotropy axis is $c$. The
 magnetic field ${\bm B}$ is declined with respect to the $c$ axis.
 (Right) Polarizations of one ordinary ${\bm B}_0$ and two extraordinary ${\bm B}_{e1,2}$
 magnetic modes in a multi-band superconductor with uni-axial
 anisotropy. The ordinary mode is decoupled from inter-band phase
 difference $\theta_{12}$ while the extraordinary modes give a coupled
 magnetic field and $\theta_{12}$.}
 \end{figure}

Let us consider the magnetic response of a two-band anisotropic superconductor. A good  example of such a kind of material is the MgB$_2$ compound \cite{Kortus2001} 
which has two superconducting Fermi surfaces with the structure qualitatively similar to the one shown in Fig.(\ref{Fig:FS})a. The two-band polarization operator (\ref{Eq:KernelGen}) with $N=2$ becomes
  \begin{equation}\label{Eq:Kernel}
  Q_{ij} ({\bm k}) = -\lambda_{Lij}^{-2}  +
 \frac{ (\hat\lambda_{1ik}^{-2} - \hat\lambda_{2ik}^{-2}) (\hat\lambda_{1jl}^{-2} - \hat\lambda_{2jl}^{-2}) k_kk_l }
 {4k_0^4 + ({\bm k}\cdot \hat\lambda_L^{-2}{\bm k}) } .
 \end{equation}
  
 \begin{figure}[htb!]
 \centerline{\includegraphics[width=1.0\linewidth]{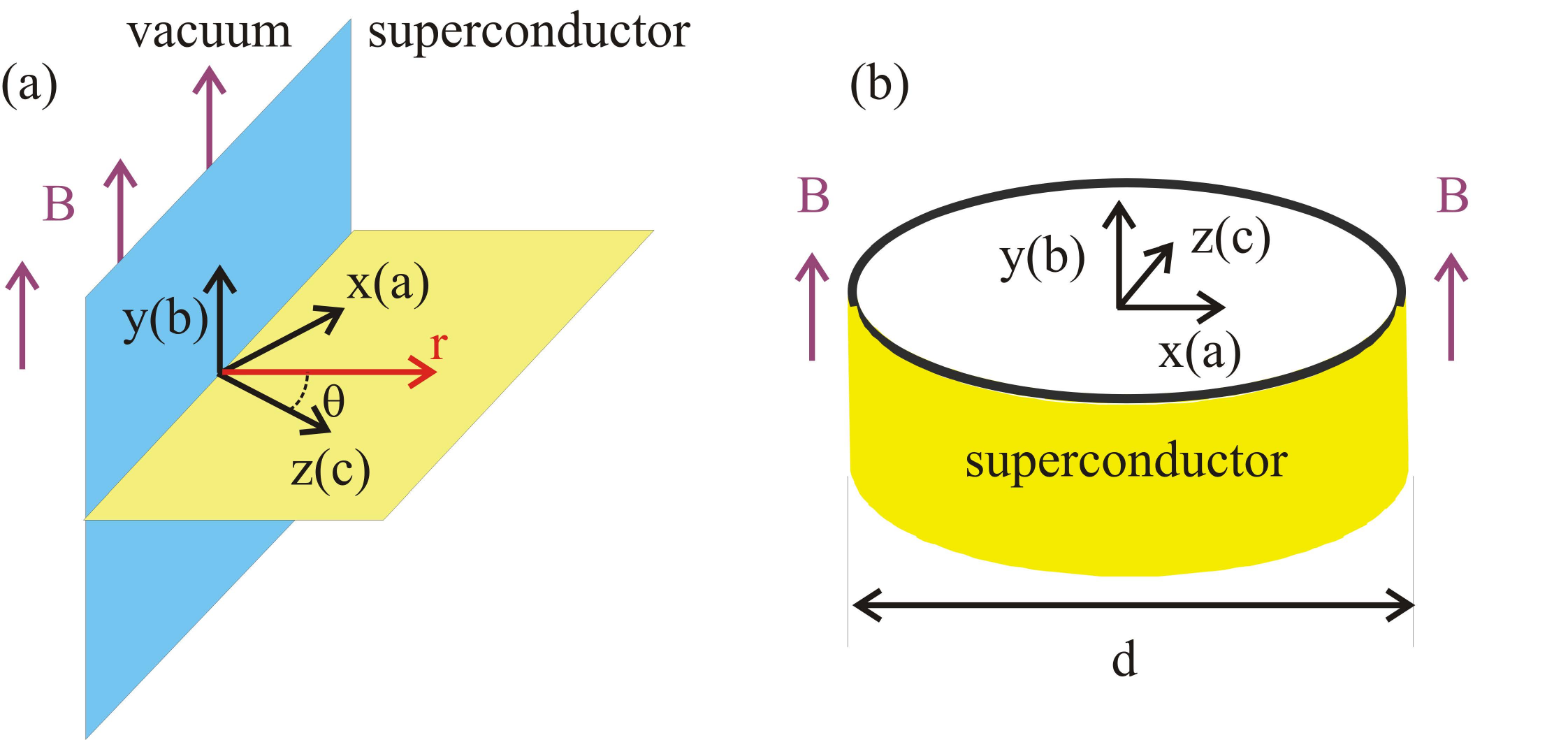}}
 \caption{\label{Fig:Meissner} 
 (a) Orientation of the superconductor boundary with respect to the crystal axes.
  The anisotropy axis $z(c)$ lies in the plane perpendicular to the boundary and makes the angle 
   $\theta$ to the normal direction. The field $\bm B$ applied parallel to the boundary depends on the coordinate $r$ 
   along the surface normal. 
   (b) The cylinder of anisotropic superconductor subjected to the external magnetic field applied along the cylinder axis $\bm y$.  }
 \end{figure}
  
In the limiting cases of large and small Josephson interaction (i) $k_0^4 \gg ({\bm k}\cdot \hat\lambda_L^{-2}{\bm k})$ 
and (ii) $k_0^4 \ll ({\bm k}\cdot \hat\lambda_L^{-2}{\bm k})$ the response becomes local but with
 very different magnetic field penetration lengths. From Eq. (\ref{Eq:Kernel}) we get 
 (i) $Q = -\hat \lambda_{L}^{-2} $ and
 (ii) $Q = -4\hat \lambda_{L}^{2} \hat \lambda_{1}^{-2} \hat \lambda_{2}^{-2}$. 
 In the first case the amplitude of $\theta_{12}$ is small so that the condensate phases are effectively ``glued together". 
 In the second case   $\theta_{12}$ can be large due to negligible inter-band Josephson coupling.

 The coupling between $\theta_{12}$  and magnetic field also leads to  especially important consequences 
 when $k_0^4 \sim  ({\bm k}\cdot \hat\lambda_L^{-2}{\bm k})$. In this case
 the magnetic response has multiple length scales. It means that the magnetic
 field penetration into such a superconductor is determined by the
 superposition of several fundamental modes, which can be found from Eq. \ref{Eq:Teta12}, by linearising the 
 Josephson term near the ground state value $\theta_{12}=0$ and searching for solutions in the form of plain waves 
 ${\bm B}, \theta_{12}\sim   e^{\bm{ ik\cdot r}}$. Thus we obtain the linear system
 \begin{align}\label{Eq:London2BandK}
 & {\bm k}\times ( \hat\lambda_L^2 {\bm k} \times{\bm h} ) - {\bm h} =
 {\bm k}\times ( \hat\lambda^{-2}_1 \hat\lambda_L^2  {\bm k} )\theta_{12} 
 \\ \label{Eq:London2BandK1}
 & {\bm k}\cdot ( \hat\lambda^{-2}_1\hat\lambda^{-2}_2\hat\lambda^{2}_L {\bm k})\theta_{12}
 + k_0^4 \theta_{12} =   {\bm k}\cdot ( \hat\lambda_1^{-2} \hat\lambda_L^{2} {\bm k}\times {\bm h} ),
 \end{align}
 where we denote ${\bm h}=2\pi{\bm B}/\Phi_0$.
 This system is of the sixth order, since magnetic field has only two independent components $\bm {k\cdot h} =0$. Hence in general for each direction of ${\bm k}$ there exists three different solutions with $Im k >0$. Therefore the system  (\ref{Eq:London2BandK}) can not be solved analytically. However as we will see below, we can ascertain properties and even analytical solutions for certain symmetries, including the most realistic model of uni-axial anisotropy.
  
  Assume that the $x$ and $y$ axes are equivalent $\lambda_{\alpha x}=\lambda_{\alpha y}$ and $z$ is the anisotropy axis. 
  In this case the general system (\ref{Eq:London2BandK}, \ref{Eq:London2BandK1}) splits into the second order and the fourth order equations which determine the usual and unconventional magnetic modes respectively discussed below. 
 Let us consider first the magnetic field with polarization ${\bm B}= \boldsymbol{B}_o$
 coplanar with the anisotropy axis and the wave vector, $\hat{\bm z}$ and ${\bm k}$ as shown in Fig.(\ref{Fig:FS}). 
 In this case we get $k= \pm i \lambda_{Lx}^{-1}$ and the magnetic field is decoupled  
 from the phase $\theta_{12}=0$. We call this magnetic mode the {\it ordinary} one. Alternately, let us consider the 
 {\it extraordinary} modes ${\bm B}= {\bm B}_e$, which are coupled to the inter-band phase. The magnetic field is perpendicular to 
 both ${\bm k}$ and $\hat{\bm z}$ as shown in Fig.(\ref{Fig:FS}). In this case 
 Eqs.(\ref{Eq:London2BandK},\ref{Eq:London2BandK1}) yield a finite coupling between the magnetic field and the inter-band phase difference. 
  
\begin{figure}[htb!]
 \centerline{\includegraphics[width=1.0\linewidth]{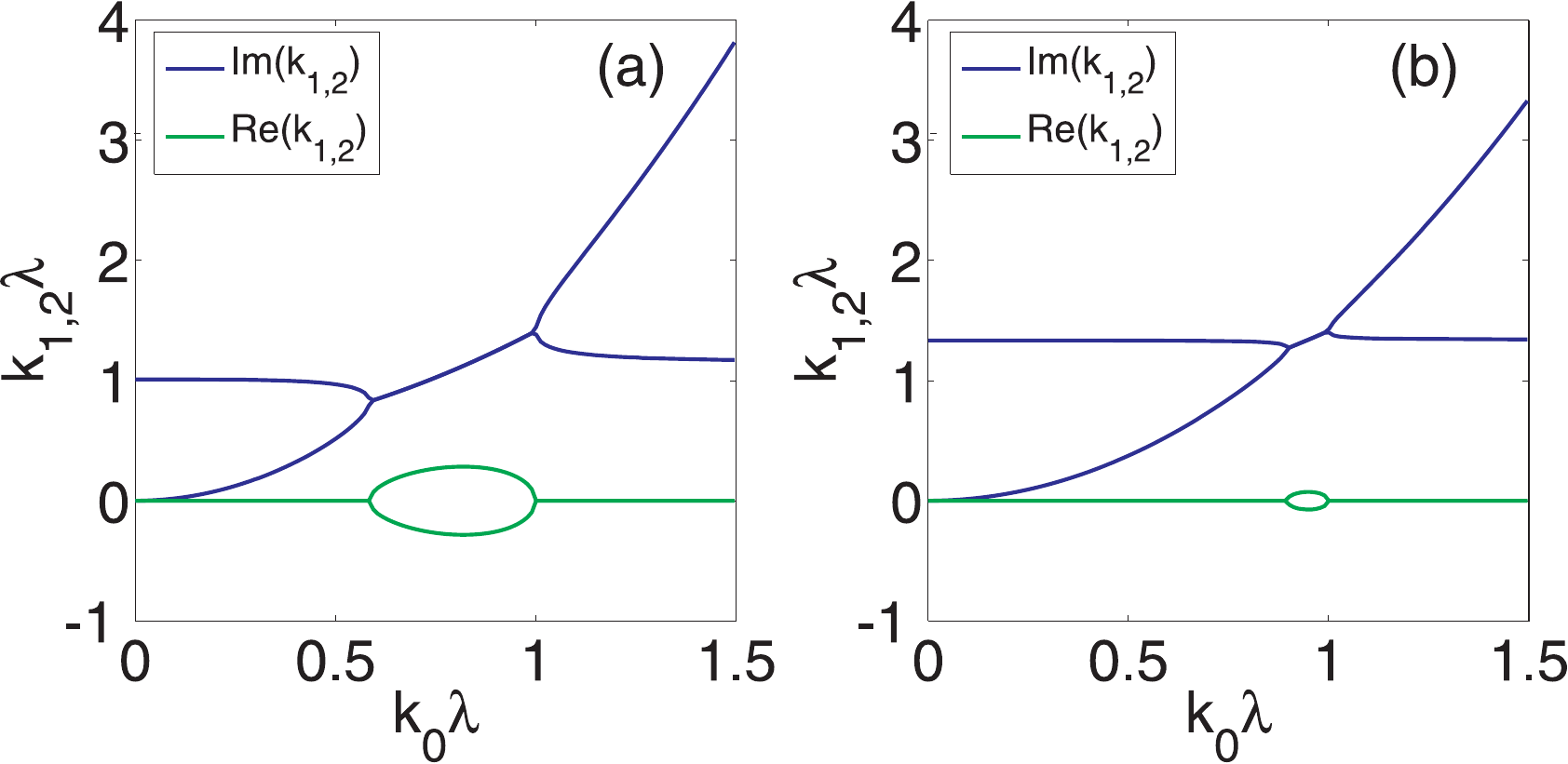}}
 \caption{\label{Fig:k12} Wavenumbers of the extraordinary magnetic modes as functions of the inter-band
 Josephson coupling $k_{1,2}=k_{1,2} (k_0)$. 
 (a) Strong anisotropy $\lambda_{2z} = 0.1 \lambda_{2\perp} = 0.1\lambda$
 (b) Weak anisotropy $\lambda_{2z} = 0.8 \lambda_{2\perp} = 0.8\lambda$.}
   \end{figure}

  To analyse the extraordinary mode in detail we parametrize the wave vector components as 
 $k_z = k \sin\theta$, $k_x= k\cos\theta\sin\varphi $ and
 $k_y= k\cos\theta\sin\varphi $  where $k_\perp = k \cos \theta$. The direction $\varphi$ drops out from the equations
  due to the rotational symmetry in $xy$ plane.
 Then we get the following relation between the interband phase and magnetic field amplitudes 
   \begin{equation}\label{Eq:hytheta12}
   h = \theta_{12}\frac{k_\perp k_z (\lambda^{-2}_{1\perp} \lambda^{2}_{L\perp} - \lambda^{-2}_{1z} \lambda^{2}_{Lz} )}
  {k_\perp^2 \lambda^{2}_{Lz} + k_z^2 \lambda^{2}_{L\perp} + 1 }.
  \end{equation}
 Note that in the isotropic case with 
  $\lambda_{1\perp}= \lambda_{1z}$ and  
  $\lambda_{L\perp}= \lambda_{Lz}$ 
  Eq.(\ref{Eq:hytheta12}) yields $h=0$ so that this mode becomes the non-magnetic pure phased-difference excitation.
  The wavenumber of the extraordinary mode is then given by the following 
  bi-quadratic equation
 \begin{align}\label{Eq:k}
 & (k_\perp^2 \lambda^{2}_{Lz} + k_z^2 \lambda^{2}_{L\perp} + 1) \nonumber 
 \\ \nonumber
 &(k_\perp^2 \lambda^{-2}_{1\perp} \lambda^{-2}_{2\perp}\lambda^{2}_{L\perp} + k_z^2 \lambda^{-2}_{1z}\lambda^{-2}_{2z}
 \lambda^{2}_{Lz} + k_0^4)+
 \nonumber \\
  & k_\perp^2 k_z^2 (\lambda^{-2}_{1\perp} \lambda^{2}_{L\perp} - \lambda^{-2}_{1z} \lambda^{2}_{Lz}  )^2  =0.
 \end{align}
 It has two complex solutions with ${\rm Im} k>0$ yielding two {\it extraordinary} magnetic modes with 
 the polarization shown schematically in Fig.(\ref{Fig:FS})b.  The wavenumber of these modes can have non-zero real parts 
 ${\rm Re} k\neq 0$ at the intermediate values of $k_0$, yielding the oscillating behaviour of the magnetic field. 
   Example solutions of Eq.(\ref{Eq:k}) corresponding to both modes  
 $k_{1,2}=k_{1,2}(k_0)$ are shown in Fig.(\ref{Fig:k12}) as functions of the interband pairing strength $k_0$. Here we assume that the first band is isotropic $\lambda_{1\perp} =  \lambda_{1z} = \lambda$. The 
 second band has either a weak anisotropy with 
 $\lambda_{2z} = 0.8  \lambda_{2\perp}$ (Fig.\ref{Fig:k12}a) or a strong 
 one with
  $\lambda_{2z} = 0.1 \lambda_{2\perp}$ (Fig.\ref{Fig:k12}b) and 
   $\lambda_{2\perp} = \lambda$.
  As shown in the Fig.(\ref{Fig:k12}) 
 for large and small Josephson couplings, the wavenumbers of the two modes are quite different. One of them is proportional to 
 $k_0^2$ and hence either diverges or goes to zero at $k_0\rightarrow \infty$ and $k_0\rightarrow 0$ respectively. 
 At the same time the other one tends to the constant values corresponding to the local response approximation discussed above.
    
The general solution of Eq.(\ref{Eq:k}) reads
\begin{equation}\label{Eq:kSolution}
k_{1,2}^2 = \frac{-b \pm \sqrt{b^2 - 4ac}}{2a} ,
\end{equation}
where,
 \begin{align}
 & a = \left(\lambda_{1\perp}^{-2} \cos^2 \theta + \lambda_{1z}^{-2} \sin^2 \theta\right)\left(\lambda_{2\perp}^{-2} \cos^2 \theta + 
 \lambda_{2z} ^{-2} \sin^2 \theta\right) ,
  \\
 & b = \lambda_{1\perp}^{-2}\lambda_{2\perp}^{-2}\lambda_{Lz}^{-2} \cos^2 \theta + 
 \lambda_{1z}^{-2}\lambda_{2z}^{-2}\lambda_{L\perp}^{-2} \sin^2  \theta 
 \nonumber \\
 & + k_0^4\left(\lambda_{L\perp}^{-2} \cos^2 \theta + \lambda_{Lz}^{-2} \sin^2 \theta\right),\\
 & c = k_0^4\lambda_{L\perp}^{-2}\lambda_{Lz}^{-2}.
 \end{align}
There are a few interesting limiting cases for $k_{1,2}$.
First, the strong inter-band coupling $k_0 \gg \lambda_{i\perp}^{-1},\lambda_{iz}^{-1}$ leads to $b^2 \gg 4ac$ and hence gives purely imaginary solutions,
\begin{gather}
k_1 = \frac{i k_0^2\sqrt{\lambda_{L\perp}^{-2}\cos^2 \theta + \lambda_{Lz}^{-2}\sin^2 \theta}}{\sqrt{\left(\lambda_{1\perp}^{-2}\cos^2 \theta + \lambda_{1z}^{-2}\sin^2 \theta\right)\left(\lambda_{2\perp}^{-2}\cos^2 \theta + \lambda_{2z}^{-2}\sin^2 \theta\right)}} , \nonumber \\
k_2 = \frac{i\lambda_{L\perp}^{-1}\lambda_{Lz}^{-1} }{\sqrt{\lambda_{L\perp}^{-2}\cos^2 \theta + \lambda_{Lz}^{-2}\sin^2 \theta}}
\end{gather}
The weak inter-band coupling $k_0 \ll \lambda_{i\perp}^{-1},\lambda_{iz}^{-1}$ leads again to $b^2 \gg 4ac$ giving slightly different purely imaginary solutions,
\begin{gather}
k_1 = \frac{i\sqrt{ \lambda_{1\perp}^{-2}\lambda_{2\perp}^{-2}\lambda_{Lz}^{-2}\cos^2 \theta +  \lambda_{1z}^{-2}\lambda_{2z}^{-2}\lambda_{L\perp}^{-2}\sin^2 \theta }}{\sqrt{\left(\lambda_{1\perp}^{-2}\cos^2 \theta + \lambda_{1z}^{-2}\sin^2 \theta\right)\left(\lambda_{2\perp}^{-2}\cos^2 \theta + \lambda_{2z}^{-2}\sin^2 \theta\right)}} ,  \nonumber \\
k_2 = \frac{i k_0^2 \lambda_{L\perp}^{-1}\lambda_{Lz}^{-1} }{\sqrt{\lambda_{1\perp}^{-2}\lambda_{2\perp}^{-2}\lambda_{Lz}^{-2}\cos^2 \theta + \lambda_{1z}^{-2}\lambda_{2z}^{-2}\lambda_{L\perp}^{-2}\sin^2 \theta}} .
\end{gather}

 The imaginary wavenumbers obtained in the limits considered above 
 correspond to the real-valued masses or the inverse decay length-scales
 of the magnetic modes. A completely different regime is possible when the 
 masses of magnetic modes become complex, resulting in damped oscillating behaviour of the magnetic field.
 Indeed in the range of parameters when $b^2 < 4ac$ the solutions (\ref{Eq:kSolution}) have finite real parts, the examples of such solutions are shown in Fig.(\ref{Fig:k12}) where ${\rm Re} k_{1,2} \neq 0$ for an interval of $k_0$ which expands with increasing degree of anisotropy.
   Generically, this regime can be realised when a strong anisotropy is applied in each band in different directions,
   $\lambda^{-1}_{1\perp} \gg \lambda^{-1}_{1z} $ and 
   $\lambda^{-1}_{2z} \gg \lambda^{-1}_{2\perp}$. 
 In this case we have $\lambda^{-2}_{L\perp}\approx\lambda^{-2}_{1\perp}$ and $\lambda^{-2}_{Lz}\approx\lambda^{-2}_{1z}$.
 Then in the wide rang of Josephson couplings 
 $\lambda^{-1}_{1\perp}\lambda^{-1}_{2z} \gg k_0^2 \gg 
 \lambda^{-1}_{1\perp}\lambda^{-1}_{2\perp}, \lambda^{-1}_{1z}\lambda^{-1}_{2z} \nonumber $
  we obtain  the wavenumber $k_{1,2} = (i\pm 1) k_0/\sqrt{2}$ which has the
  amplitudes of real and imaginary parts.

 \subsection{ Flux expulsion in the Anisotropic two-band model}
 
We now consider the problem of magnetic field screening at the surface of an anisotropic multiband superconductor. Let us consider cylindrical geometry with magnetic field applied in the $y$ direction  $\boldsymbol{B} = H_0 \hat{\boldsymbol{y}}$ parallel to the boundary of the superconducting sample as shown in Fig.(\ref{Fig:Meissner})a. 
Unlike the usual isotropic superconductors, the boundary orientation with respect to the crystal axes is important and is considered by the angle $\theta$ in polar coordinates, introduced previously in considering the normal modes. 
The wave vectors of excited magnetic modes are directed perpendicular to the surface $\bm k = k(\cos\theta, 0, \sin\theta )$.
We use $r$ to indicate the coordinate orthogonal to the boundary, such that in the presence of two magnetic field penetration lengths, 
the magnetic field decay and phase difference follow a double-exponential law,
\begin{eqnarray}
B &=& B_1 e^{ -r/\Lambda_1} + B_2 e^{ -r/\Lambda_2} \label{Eq:LondonsolB} \\
\theta_{12} &=& \theta^{(1)}_{12} e^{ -r/\Lambda_1} + \theta^{(2)}_{12} e^{ - r/\Lambda_2}
\label{Eq:Londonsoltheta}
\end{eqnarray}
\noindent 
where $\Lambda_1=-i \, k_1^{-1}$ and $\Lambda_2=-i \, k_2^{-1}$ are two magnetic field penetration lengths. 
We have previously made the assumption that $\theta_{12}$ is small and hence linear in nature. This does not necessarily have to be the case and indeed removing this assumption could lead to more unconventional physics e.g. oscillation of the inter-band phase difference parallel to the boundary. To discover if this has an effect one would have to consider the full equations numerically. For our 1-dimensional equation, the boundary conditions are given by the value of the magnetic field on the boundary due to the external field $B_1 + B_2 = H_0$ and the requirement for the normal current to vanish at the boundary $ \bm n\cdot\boldsymbol{j} = 0$. The contribution from each of the magnetic modes (\ref{Eq:LondonsolB}) 
 to the normal current can be found as 
 $\boldsymbol{n}\cdot \boldsymbol{j}_{\alpha} = 
 -i\boldsymbol{\nabla} \cdot \boldsymbol{j}_\alpha /k_\alpha = -ie E_J \theta_{12}/k_\alpha $. 
 The relation between phase difference and magnetic field in each of the mode is given by the Eq.(\ref{Eq:hytheta12}).
 
Then we get the following solution for the amplitudes $B_{1,2}$ in Eq.(\ref{Eq:LondonsolB}):

\begin{widetext}
\begin{eqnarray}
B_1 = \left(\frac{H_0 k_1^3}{k_1 - k_2}\right)\frac{\left(\lambda_{L\perp}^{-2}\cos^2\theta + \lambda_{Lz}^{-2}\sin^2\theta\right)k_2^2 + \lambda_{L\perp}^{-2}\lambda_{Lz}^{-2}}{k_1^2k_2^2\left(\lambda_{L\perp}^{-2}\cos^2\theta + \lambda_{Lz}^{-2}\sin^2\theta\right)+\lambda_{L\perp}^{-2}\lambda_{Lz}^{-2}\left(k_1^2 + k_1k_2 + k_2^2\right)}
 \\
B_2 = \left(\frac{H_0 k_2^3}{k_2 - k_1}\right)\frac{\left(\lambda_{L\perp}^{-2}\cos^2\theta + \lambda_{Lz}^{-2}\sin^2\theta\right)k_1^2 + \lambda_{L\perp}^{-2}\lambda_{Lz}^{-2}}{k_1^2k_2^2\left(\lambda_{L\perp}^{-2}\cos^2\theta + \lambda_{Lz}^{-2}\sin^2\theta\right) +
\lambda_{L\perp}^{-2}\lambda_{Lz}^{-2}\left(k_1^2 + k_1k_2 + k_2^2\right)}.
\end{eqnarray}
\end{widetext}

\begin{figure*}
\centerline{\includegraphics[width=0.326\linewidth]{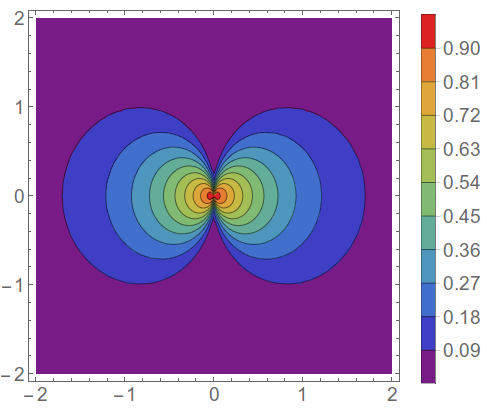}\includegraphics[width=0.368\linewidth]{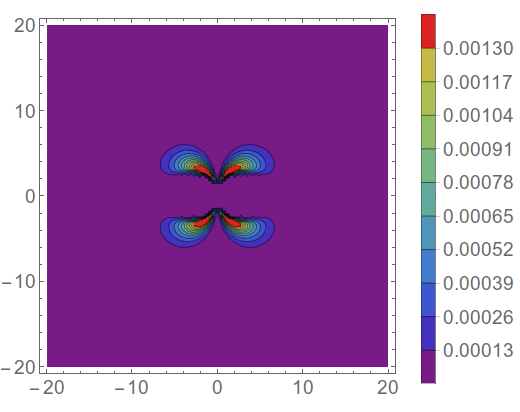}\includegraphics[width=0.326\linewidth]{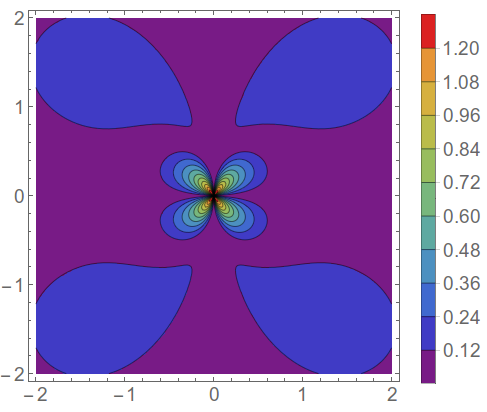}} \centerline{\includegraphics[width=0.326\linewidth]{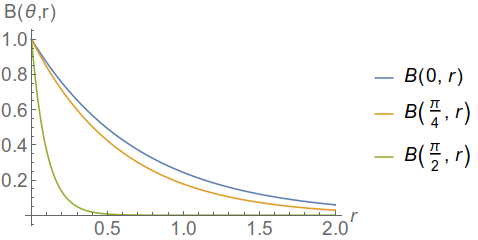}\includegraphics[width=0.368\linewidth]{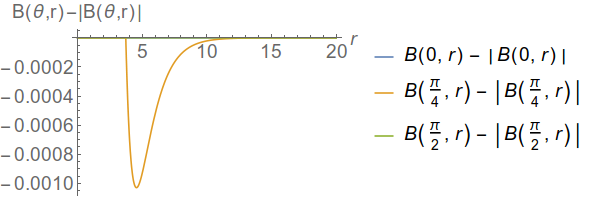}\includegraphics[width=0.326\linewidth]{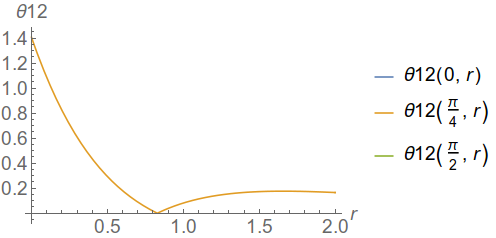}}
 \caption{
 A contour plot and a radial slice at various angles of the magnetic field, negative magnetic field and phase difference for the 1d boundary problem solution with strong anisotropy in a single component $k_0 = 0.7$, $\lambda_{2x}=\lambda_{2y}=\lambda_{1y}=1$ and $\lambda_{1x}=0.1\lambda_{1y}$. Vertical and horizontal axes on upper panels correspond to $y$ and $x$ directions respectively.
 A radial curve from the centre of the plot represents the field orthogonal to a 1d boundary crossing the origin in the x-y plane. This way every possible direction (or $\theta$) is plotted for equations \ref{Eq:LondonsolB} and \ref{Eq:Londonsoltheta} with the radial distance representing $r$. The plot quantities are: magnetic field $B_z$ (left), negative magnetic field $|B_z|-B_z$ (centre) and phase difference $\theta_{12}$ (right).}
\label{Fig:LondonBoundarySame}
\end{figure*}

\begin{figure*}
\centerline{\includegraphics[width=0.326\linewidth]{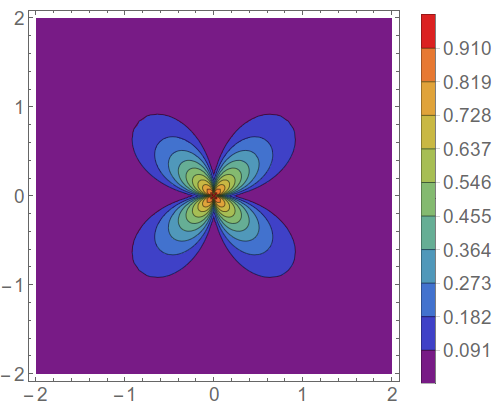}\includegraphics[width=0.368\linewidth]{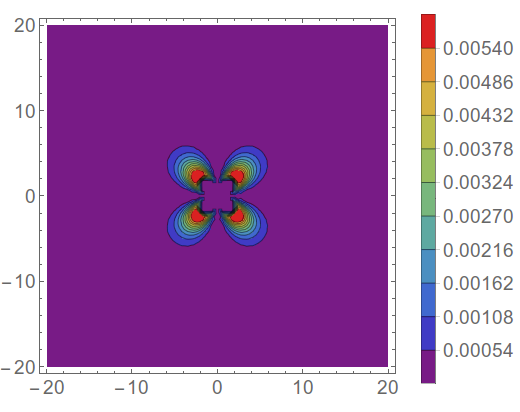}\includegraphics[width=0.326\linewidth]{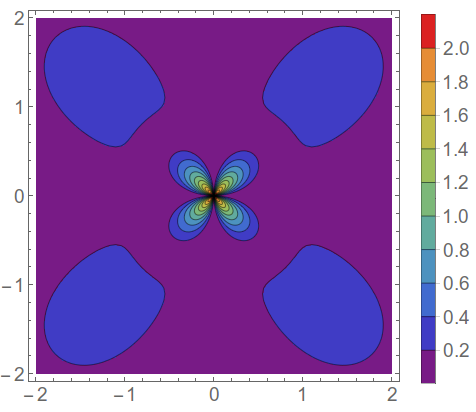}} \centerline{\includegraphics[width=0.326\linewidth]{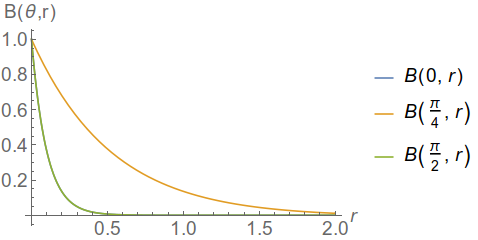}\includegraphics[width=0.368\linewidth]{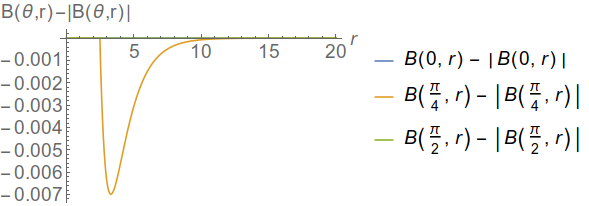}\includegraphics[width=0.326\linewidth]{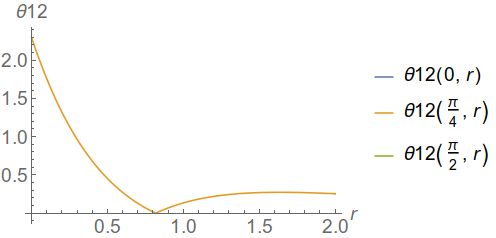}}
 \caption{ 
 A contour plot and a radial slice at various angles for the 1d boundary problem solution with strong anisotropy in opposite directions in each component $k_0 = 0.7$, $\lambda_{2x}=\lambda_{1y}=1$ and $\lambda_{1x}=\lambda_{2y}=0.1$. Vertical and horizontal axes on upper panels correspond to $y$ and $x$ directions respectively. A radial curve from the centre of the plot represents the field orthogonal to a 1d boundary crossing the origin in the x-y plane. This way every possible direction (or $\theta$) is plotted for equations \ref{Eq:LondonsolB} and \ref{Eq:Londonsoltheta} with the radial distance representing $r$. The plot quantities are Magnetic field $B_z$ (left), negative magnetic field $|B_z|-B_z$ (centre) and phase difference $\theta_{12}$ (right).}
\label{Fig:LondonBoundaryOpp}
\end{figure*}

Example solutions for various parameters are plotted in Figs. \ref{Fig:LondonBoundarySame} and \ref{Fig:LondonBoundaryOpp}. The two key features that differ from the isotropic case is the double-exponential decay of the magnetic field and a self-induced gradient of the phase difference between the superconducting components. The origin of both effects is the hybridization of the Leggett mode with the magnetic mode. Also note that the solution  matches the four-fold symmetry of the free energy, which is to be expected due to the anisotropy. Additionally, the limiting cases of $k_0 \rightarrow \infty$ and the isotropic case decouples the phase difference and magnetic field and we are left with a single penetration length as expected.

An interesting limit to consider is where the Leggett mode becomes massless, 
for example the zero Josephson coupling limit ($k_0 \rightarrow 0$). As the Josephson coupling becomes smaller one of the magnetic field penetration lengths diverges $\Lambda_1 \to \infty$. The physical consequence of a diverging magnetic field penetration length in our-two-scale system is markedly different from the divergence of magnetic field penetration length at $T_c$ in a single-component system; namely in our case it doesn't imply absence of magnetic screening. This is due to the amplitude of the mode vanishing as the penetration length diverges ($B_1 \to 0$). Simultaneously the other mode becomes the isotropic case deformed by the anisotropy, as one would expect if the anisotropy in the different components matched and could hence be rescaled. 

This implies that in such a limit, most of the magnetic field's amplitude decays at the length scale $\Lambda_2$ along with an increasingly long-range penetration of a small ``tail" of magnetic field.
The situation should occur for example in multi-band systems close to $s+is$ transitions where the Leggett mode becomes massless \cite{Lin2012,Carlstroem2011a,Weston2013,Silaev2015c}.

Finally, and rather interestingly, the multi-mode magnetic response implies that the magnetic field is not necessarily monotonic. This non monotonic behaviour can lead to field inversion for a range of parameters. This means interactions between vortices and boundaries and also inter-vortex interactions will be non-trivial. It is likely that should the external field be increased such that vortices enter into the sample, their position will be affected by the negative magnetic field which would attract the vortices. Again it should be emphasised that the field inversion here is not related to oscillatory behaviour of the magnetic field in the non-local Pippard's model.

This field inversion can be seen most clearly in the plots of the solutions for various parameters below. In Fig. \ref{Fig:LondonBoundarySame} we have looked at strong anisotropy in one of the components and in Fig. \ref{Fig:LondonBoundarySameAll} in the appendix we have plotted anisotropy in a single component for increasing strength. It is clear that the strength of the field inversion increases as the anisotropy is increased. Additionally the generated phase difference is also more pronounced. In Fig.\ref{Fig:LondonBoundaryOpp} we have looked at strong anisotropy in different directions in each component and in Fig.\ref{Fig:LondonBoundaryOppAll} in the appendix we have plotted anisotropy in opposite directions for increasing strength. Again as in the other plot the field inversion and phase difference are increased in magnitude at the anisotropy amount increases. Due to the choice of parameters however, we observe a far more symmetrical dihedral solution.

\begin{figure}
\centerline{\includegraphics[width=1.0\linewidth]{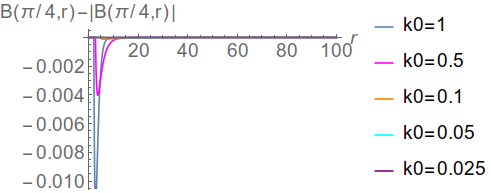}}\centerline{\includegraphics[width=1.0\linewidth]{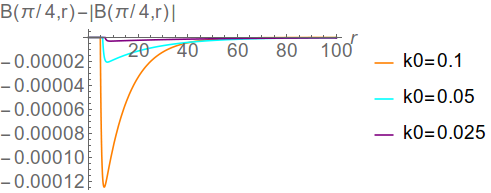}}
 \caption{A plot of the negative magnetic field at $\theta = \pi / 4$ for strong anisotropy $\lambda_{2x}=\lambda_{1y}=1$ and $\lambda_{1x}=\lambda_{2y}=0.1$ for various values of $k_0$.}
\label{Fig:negmagk0}
\end{figure}

Finally we have considered the negative magnetic field at $\theta = \frac{\pi}{4}$ for various strengths of $k_0$, shown in Fig.\ref{Fig:negmagk0}. We can observe here the effect of decreasing the Josephson coupling strength as discussed above. This leads to one of the modes becoming weaker (and also the strength of the negative magnetic field becoming weaker) 
but also long range.

 \subsection{Numerical solution for the boundary problem}
\begin{figure}[tb!]
\centerline{\includegraphics[width=0.35\linewidth,trim={0.0cm 2.4cm 0.25cm 0.0cm},clip]{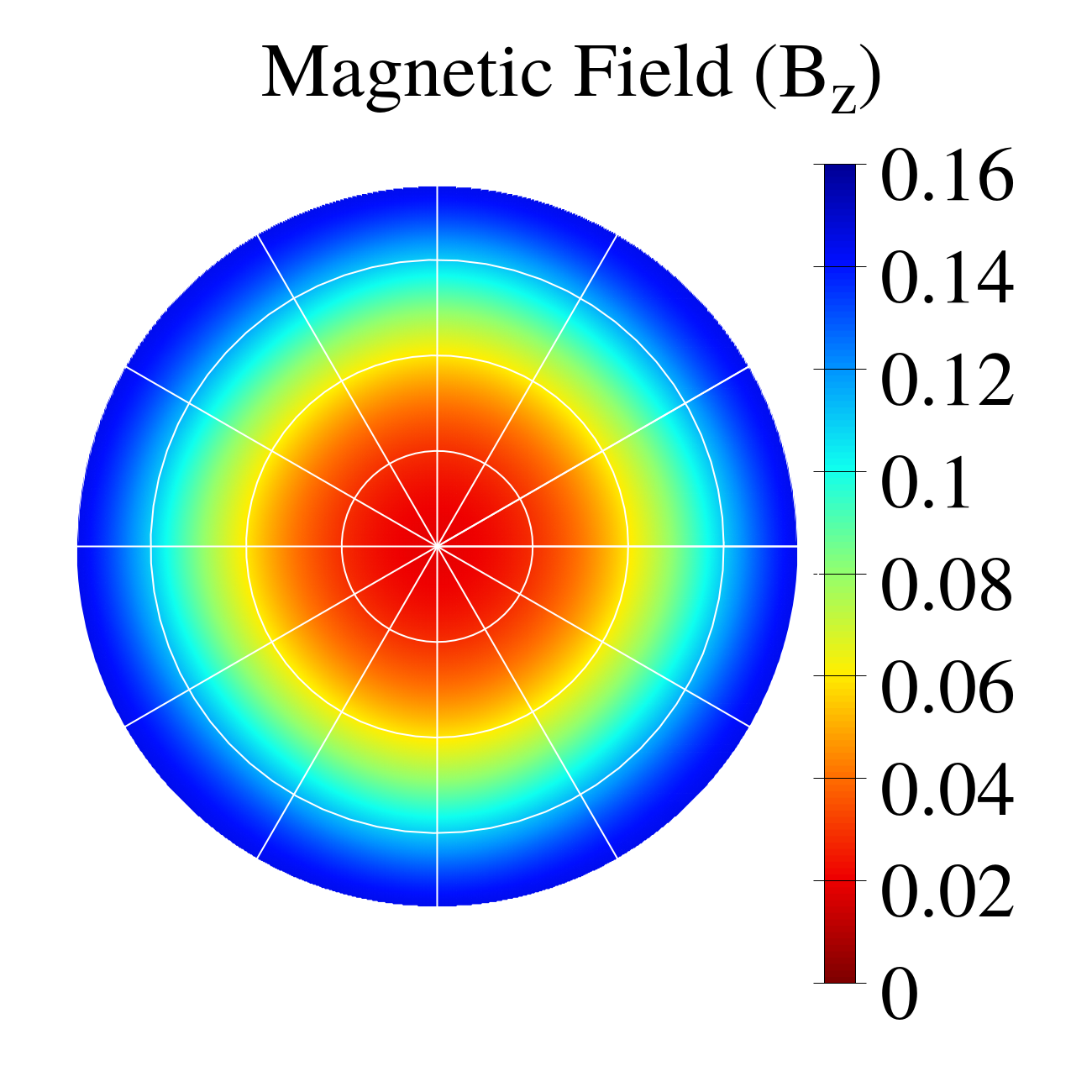}\includegraphics[width=0.35\linewidth,trim={0.0cm 2.4cm 0.25cm 0.0cm},clip]{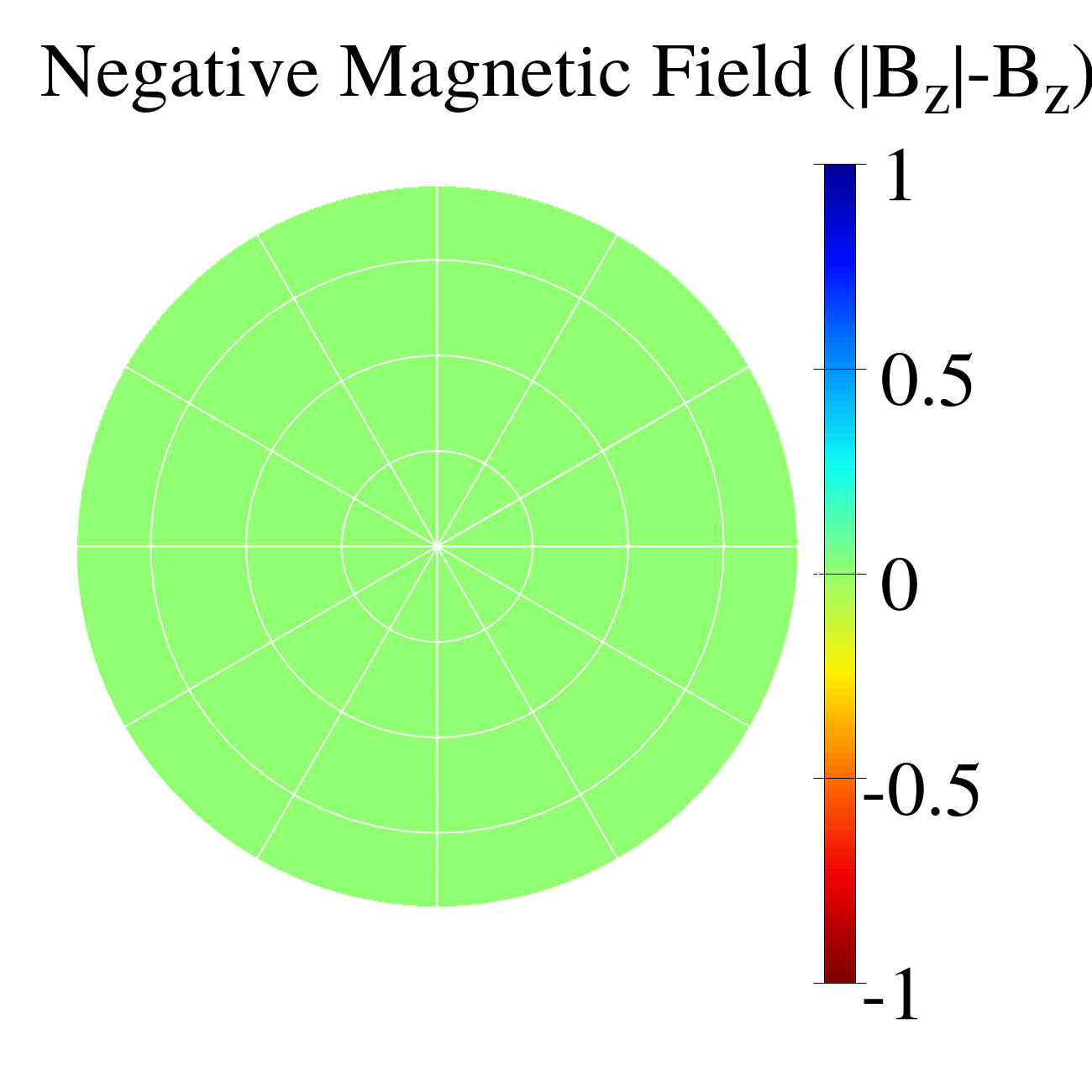}\includegraphics[width=0.35\linewidth,trim={0.0cm 2.4cm 0.25cm 0.0cm},clip]{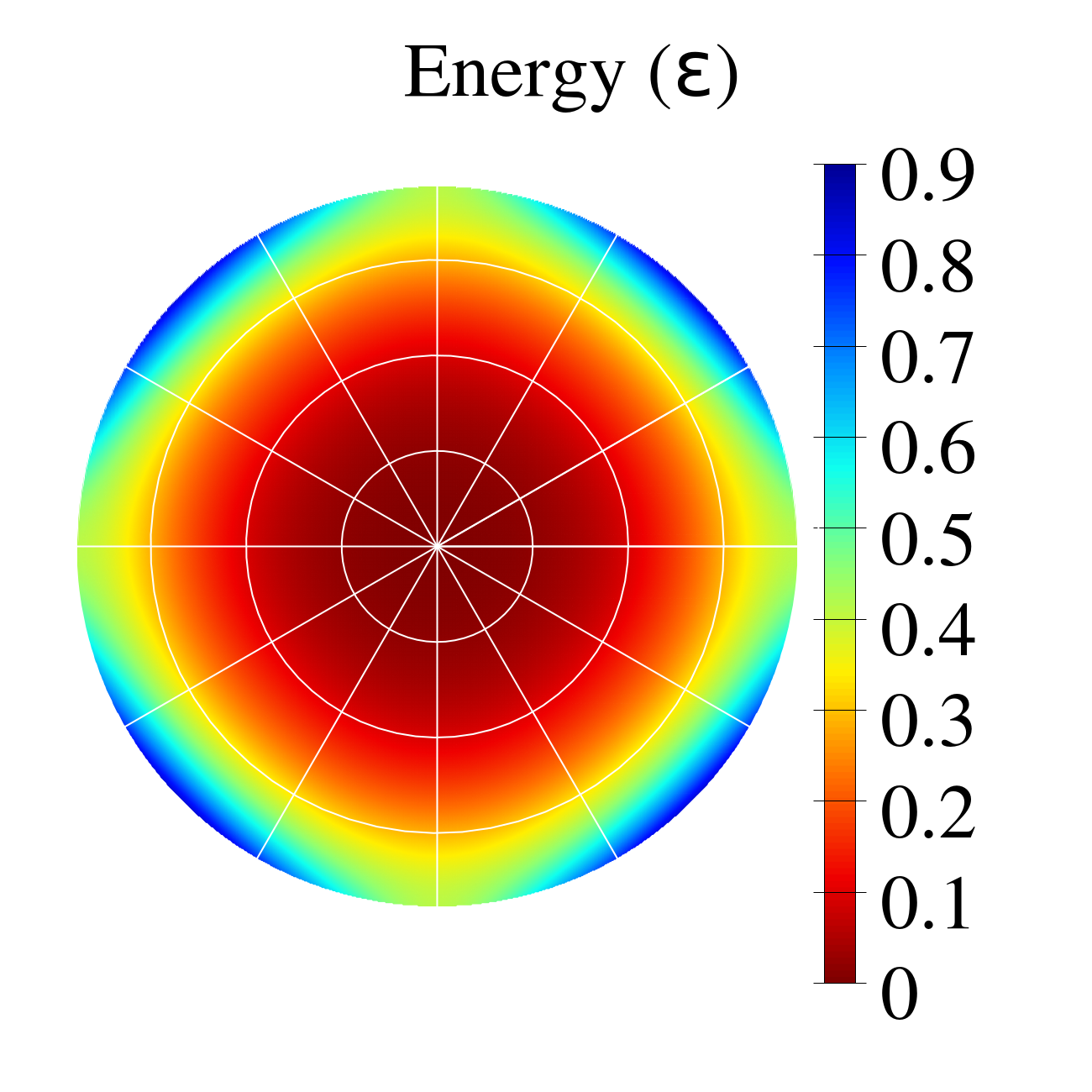}}
 \centerline{\includegraphics[width=0.35\linewidth,trim={1.4cm 2.8cm 0.0cm 1.7cm},clip]{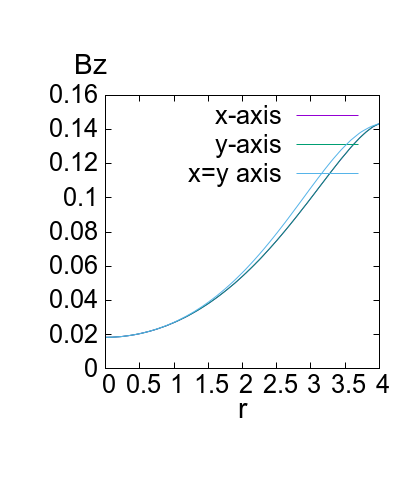}\includegraphics[width=0.35\linewidth,trim={1.4cm 2.8cm 0.0cm 1.7cm},clip]{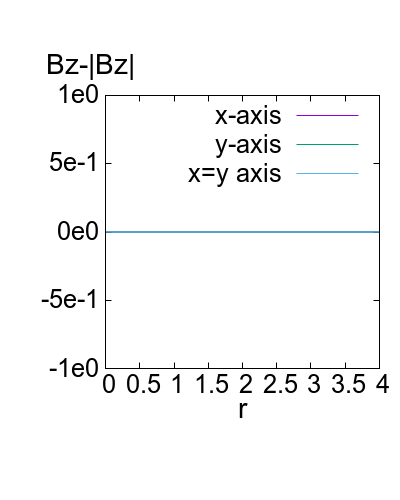}\includegraphics[width=0.35\linewidth,trim={1.4cm 2.8cm 0.0cm 1.7cm},clip]{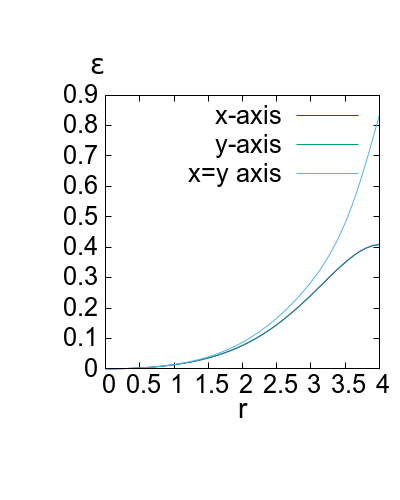}}
  \centerline{\quad(a) \enskip \quad \quad \quad \quad \quad \quad \quad (b)  \quad \quad \quad \quad \quad \quad \quad \quad (c)}
 \centerline{\includegraphics[width=0.35\linewidth,trim={0.0cm 2.4cm 0.25cm 0.0cm},clip]{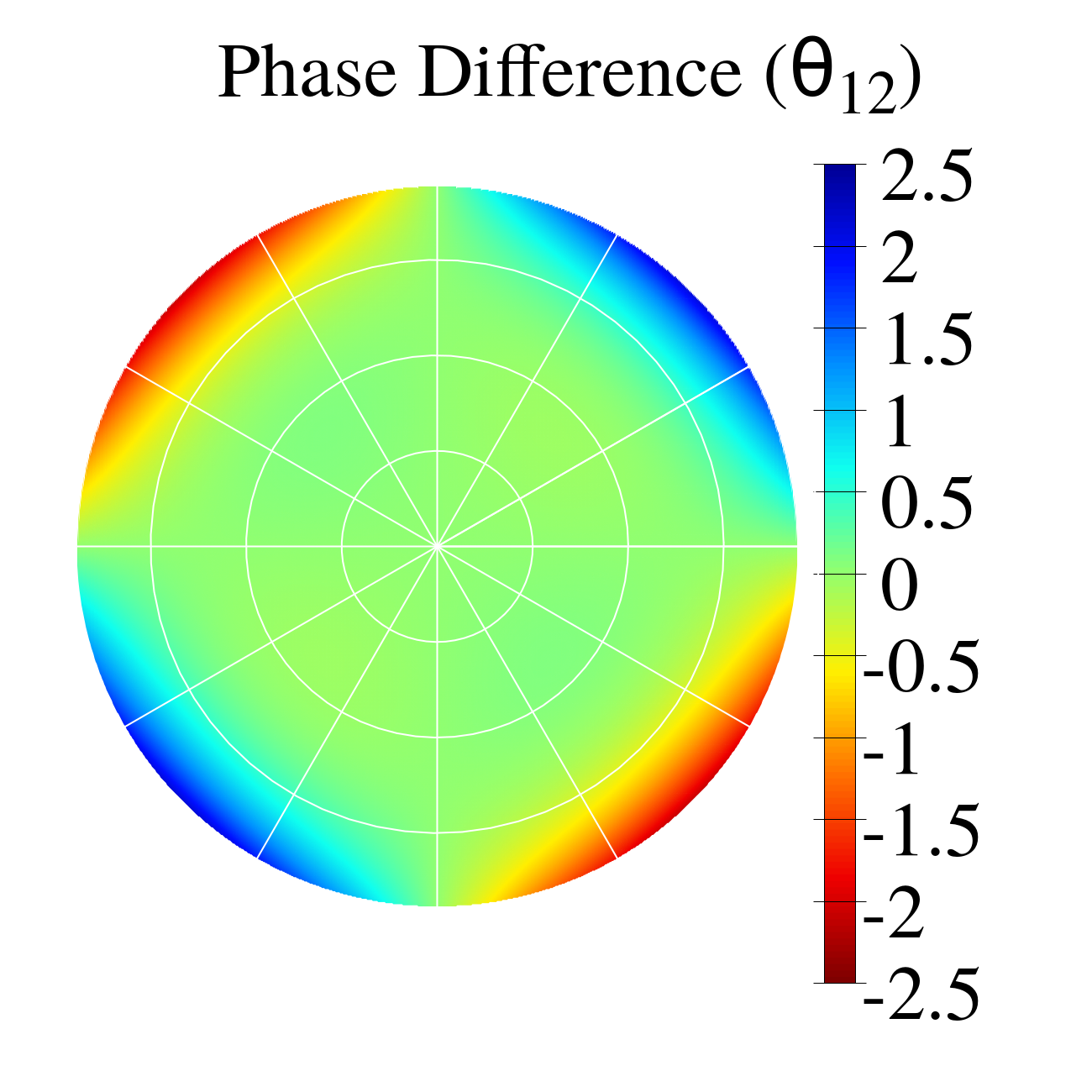}\includegraphics[width=0.35\linewidth,trim={0.0cm 2.4cm 0.25cm 0.0cm},clip]{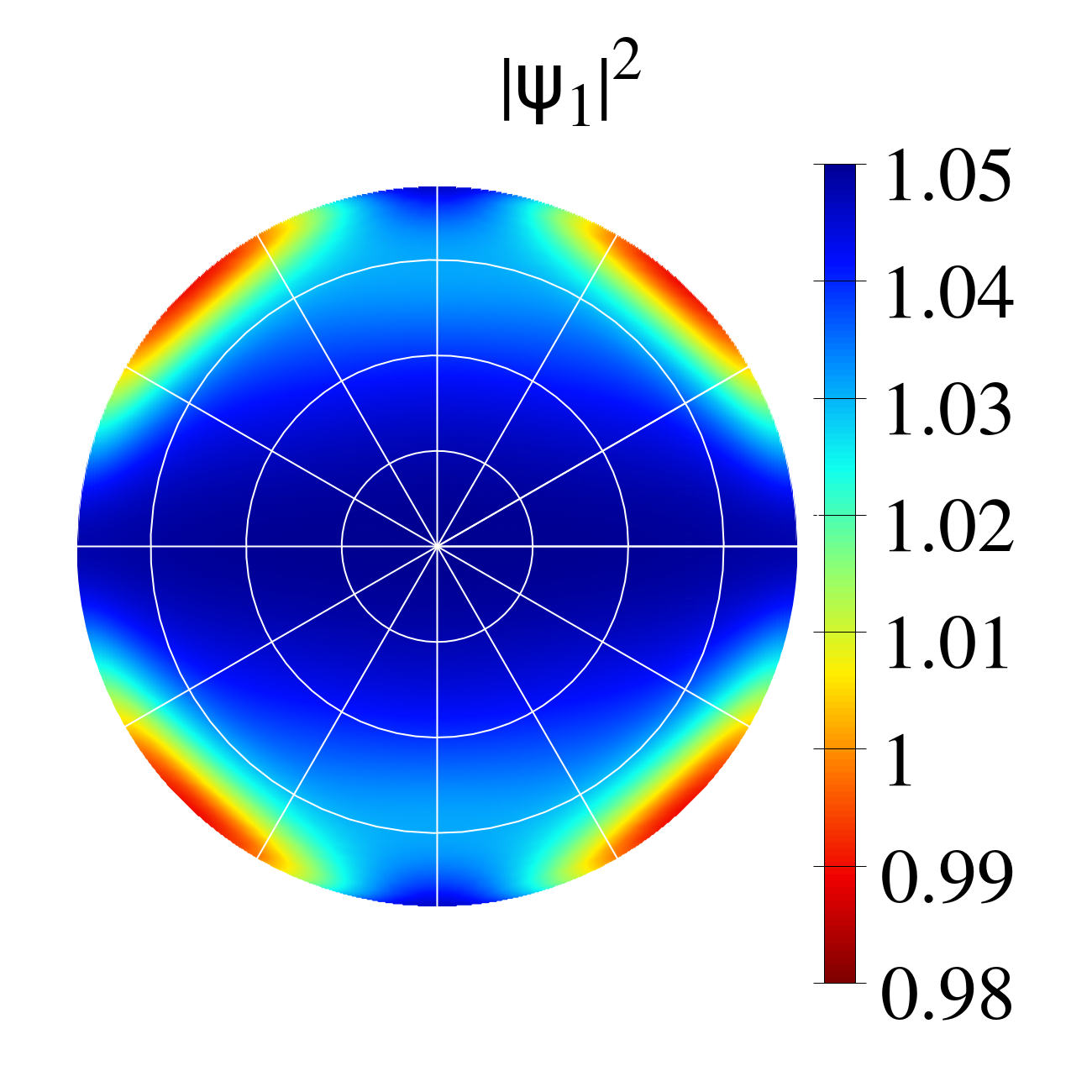}\includegraphics[width=0.35\linewidth,trim={0.0cm 2.4cm 0.25cm 0.0cm},clip]{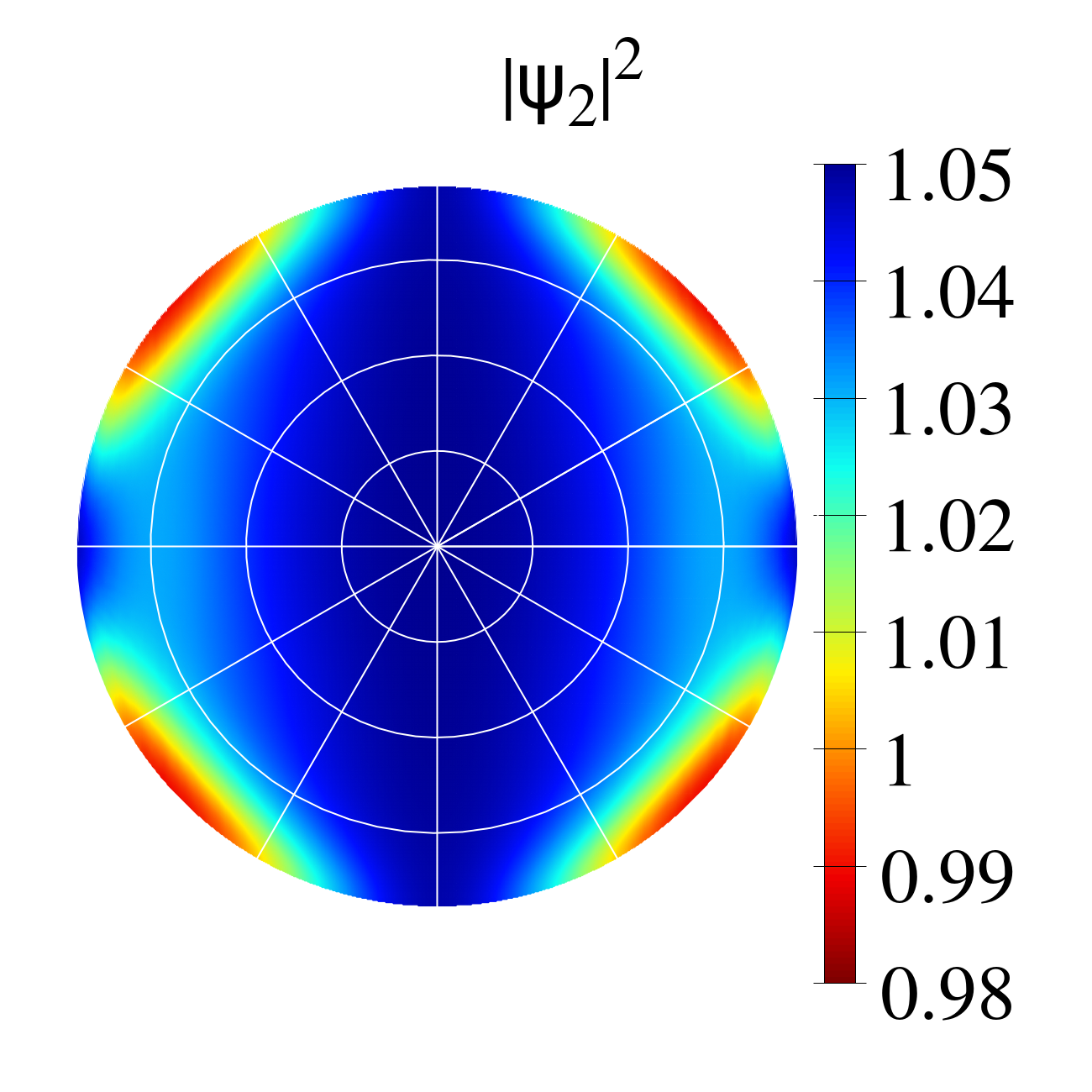}}
 \centerline{\includegraphics[width=0.35\linewidth,trim={1.4cm 2.8cm 0.0cm 1.7cm},clip]{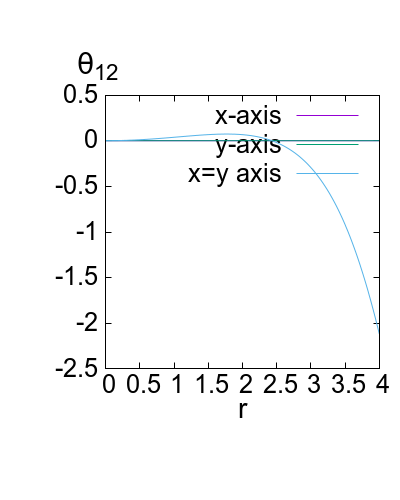}\includegraphics[width=0.35\linewidth,trim={1.4cm 2.8cm 0.0cm 1.7cm},clip]{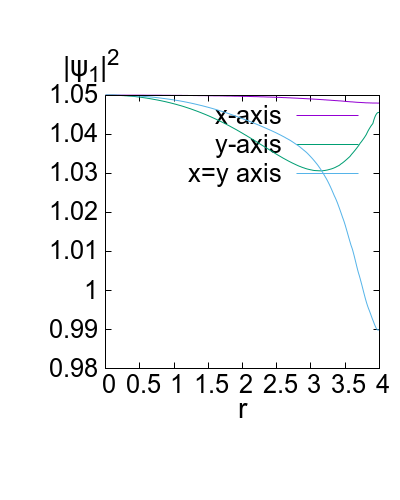}\includegraphics[width=0.35\linewidth,trim={1.4cm 2.8cm 0.0cm 1.7cm},clip]{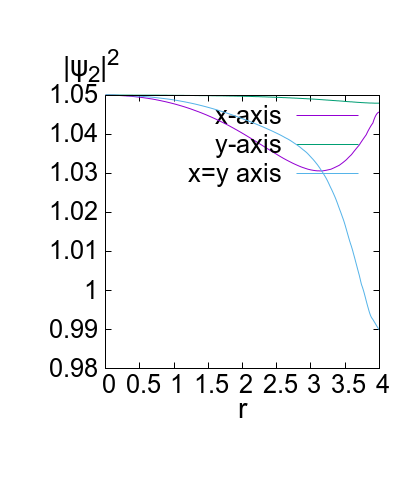}}
   \centerline{\quad(d) \enskip \quad \quad \quad \quad \quad \quad \quad (e)  \quad \quad \quad \quad \quad \quad \quad \quad (f)}
 \caption{\label{Fig:meisner1} Meissner state numerical solution for strong anisotropy in opposite directions on a disc of radius 5 $\lambda_{x1}^{-1}=\lambda_{y2}^{-1}=1$, $\lambda_{x2}^{-1} = \lambda_{y1}^{-1} = 0.1$, $\eta_{12} = 0.5$ and $\gamma_{1}=\gamma_{2}=10$. The full 2d plots are accompanied by 1d slices at $\theta = 0,\frac{\pi}{4}, \frac{\pi}{2}$ below. The quantities plotted are (a) $B_z$ magnetic field (b) $B_z-|B_z|$ negative magnetic field (c) $\mathcal{E}$ energy density (d) $\left|\phi_1\right|^2$ (e)$\left|\phi_2\right|^2$ (f)$\theta_{12}$ phase difference.}
 \end{figure}

\begin{figure}[tb!]
 \centerline{\includegraphics[width=0.35\linewidth,trim={1.85cm 2.4cm 0.0cm 0.0cm},clip]{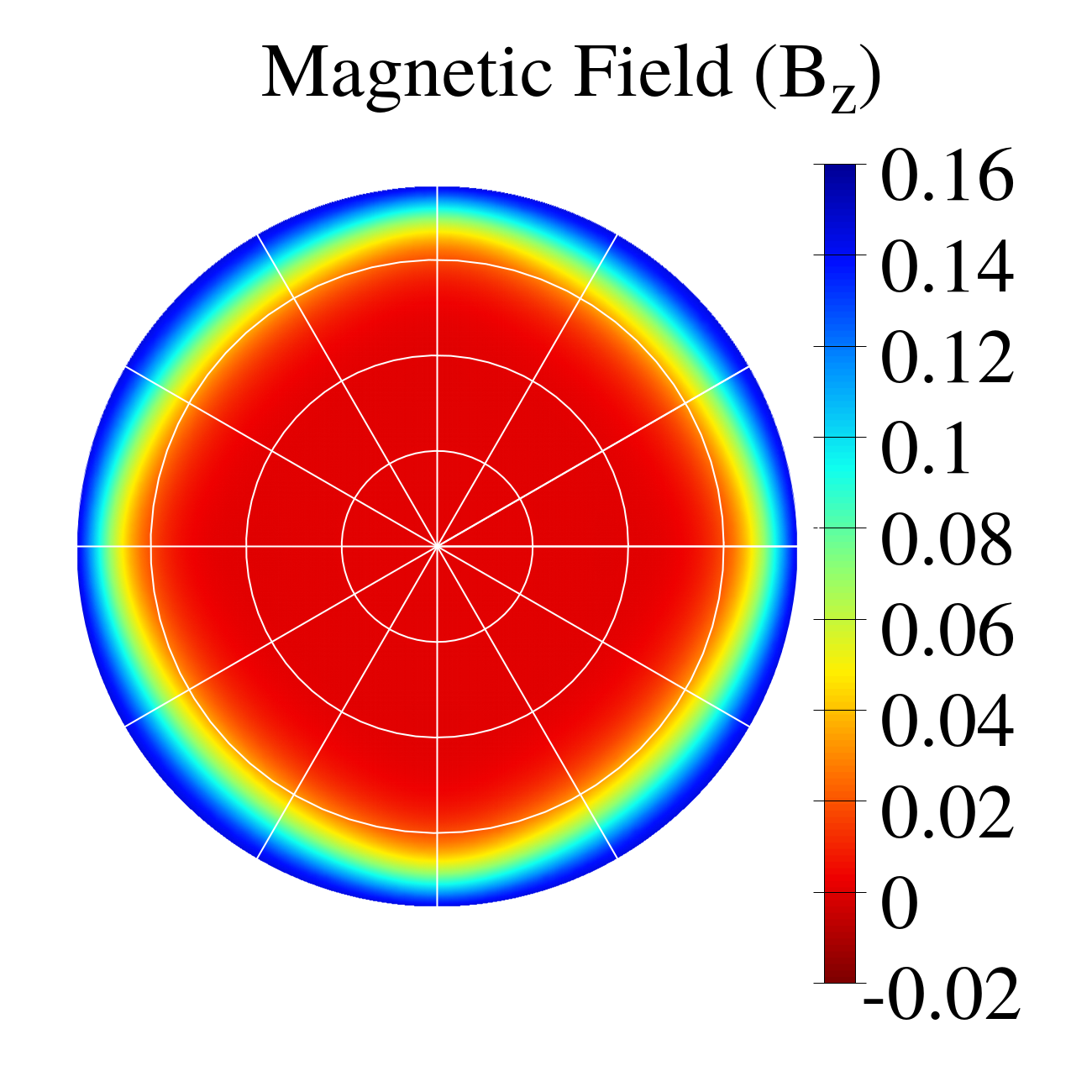}\includegraphics[width=0.35\linewidth,trim={1.85cm 2.4cm 0.0cm 0.0cm},clip]{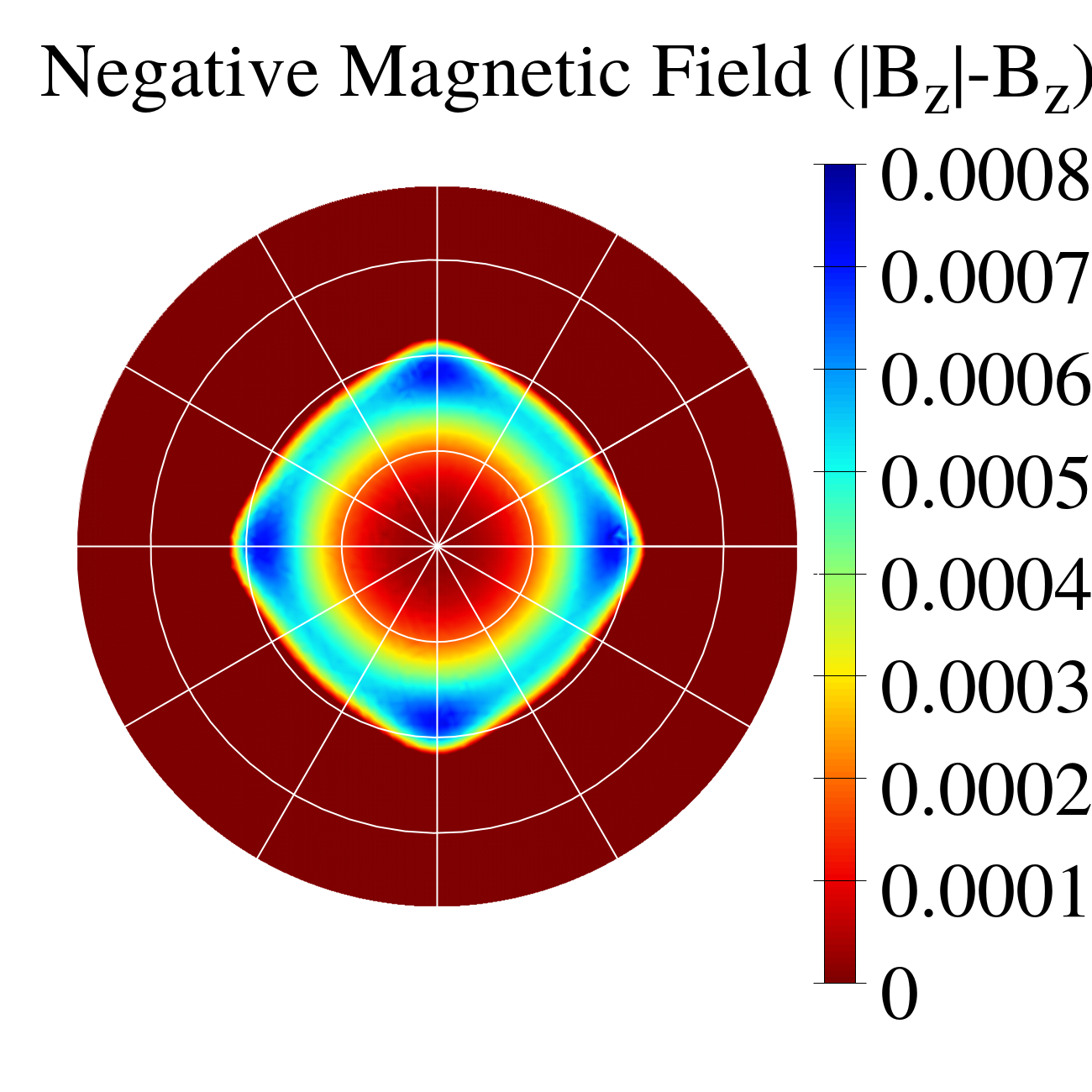}\includegraphics[width=0.35\linewidth,trim={1.85cm 2.4cm 0.0cm 0.0cm},clip]{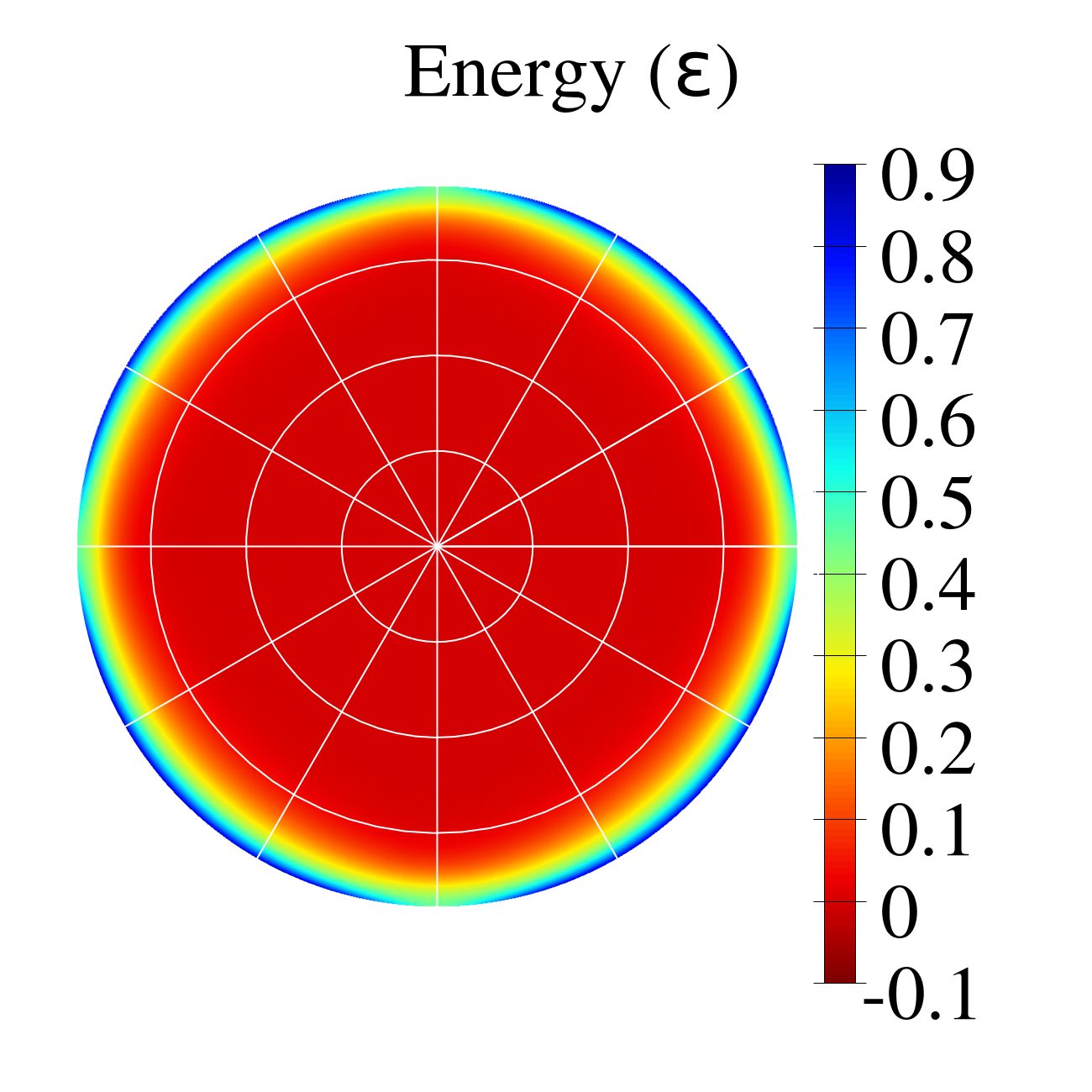}}
  \centerline{\includegraphics[width=0.35\linewidth,trim={1.4cm 2.8cm 0.0cm 1.7cm},clip]{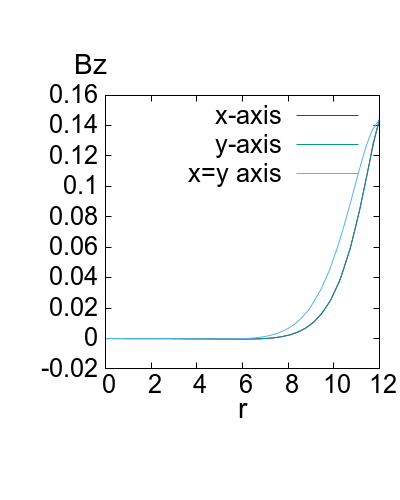}\includegraphics[width=0.35\linewidth,trim={1.4cm 2.8cm 0.0cm 1.7cm},clip]{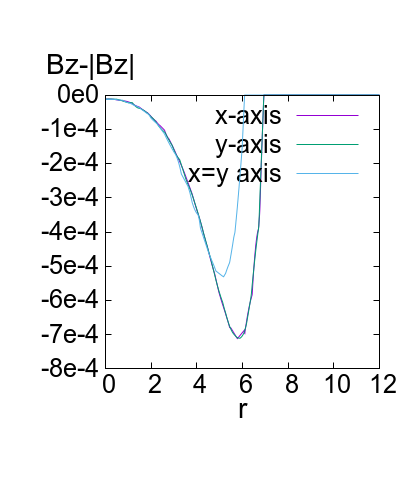}\includegraphics[width=0.35\linewidth,trim={1.4cm 2.8cm 0.0cm 1.7cm},clip]{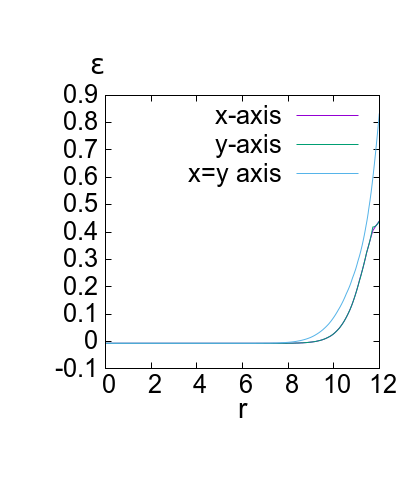}}
   \centerline{\quad(a) \enskip \quad \quad \quad \quad \quad \quad \quad (b)  \quad \quad \quad \quad \quad \quad \quad \quad (c)}
 \centerline{\includegraphics[width=0.35\linewidth,trim={1.9cm 2.4cm 0.0cm 0.0cm},clip]{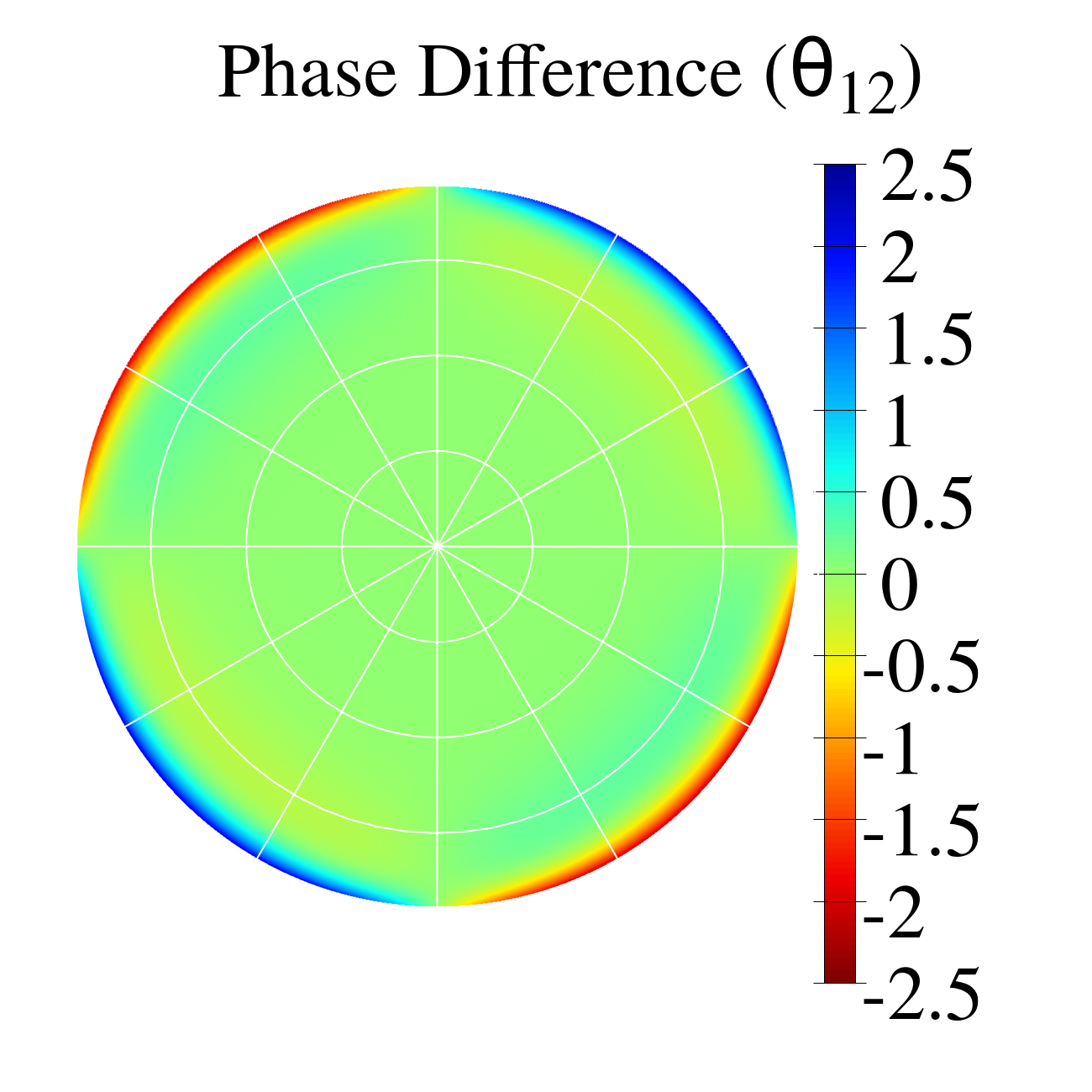}\includegraphics[width=0.35\linewidth,trim={1.9cm 2.4cm 0.0cm 0.0cm},clip]{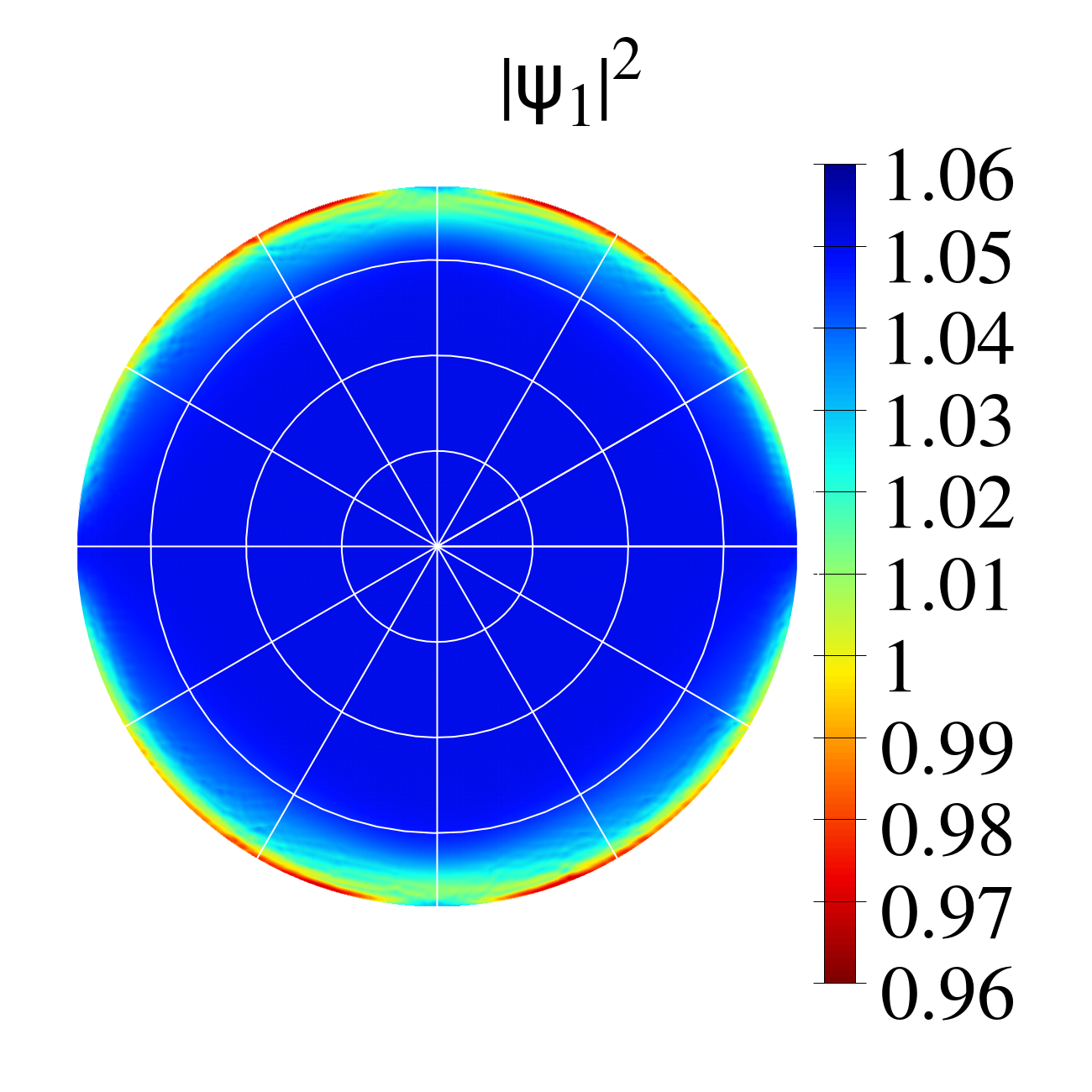}\includegraphics[width=0.35\linewidth,trim={1.9cm 2.4cm 0.0cm 0.0cm},clip]{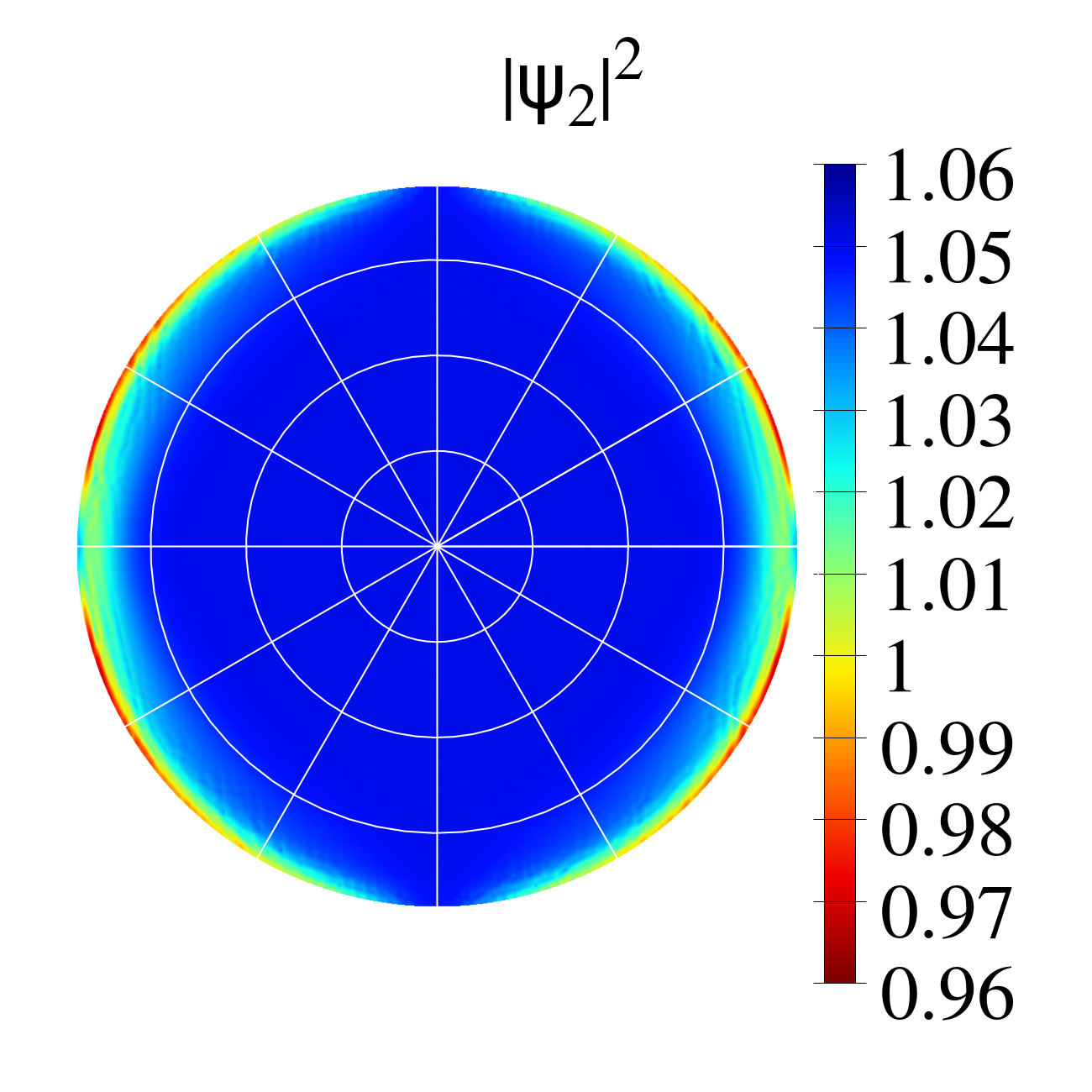}}
  \centerline{\includegraphics[width=0.35\linewidth,trim={1.4cm 2.8cm 0.0cm 1.7cm},clip]{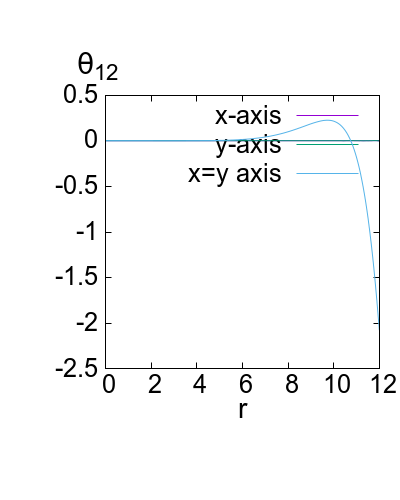}\includegraphics[width=0.35\linewidth,trim={1.4cm 2.8cm 0.0cm 1.7cm},clip]{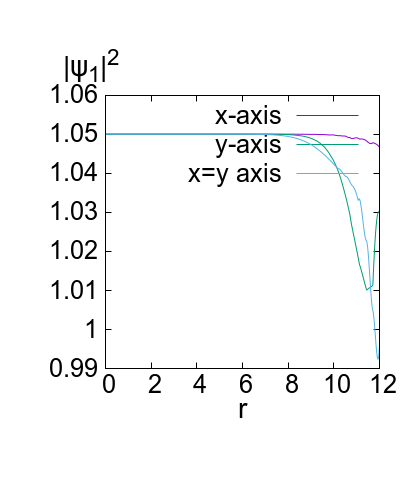}\includegraphics[width=0.35\linewidth,trim={1.4cm 2.8cm 0.0cm 1.7cm},clip]{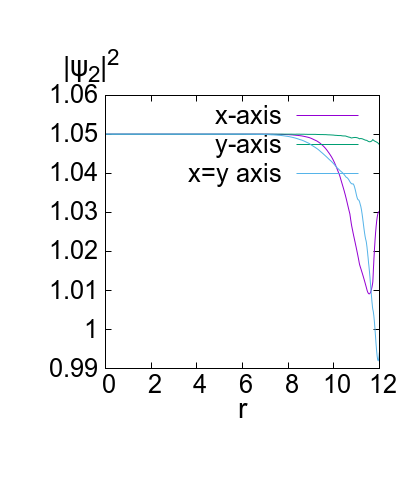}}
   \centerline{\quad(d) \enskip \quad \quad \quad \quad \quad \quad \quad (e)  \quad \quad \quad \quad \quad \quad \quad \quad (f)}
 \caption{\label{Fig:meisner2} Messiner state numerical solution for strong anisotropy in opposite directions on a disc of radius 12 $\lambda_{x1}^{-1}=\lambda_{y2}^{-1}=1$, $\lambda_{x2}^{-1} = \lambda_{y1}^{-1} = 0.1$, $\eta_{12} = 0.5$ and $\gamma_{1}=\gamma_{2}=10$. The full 2d plots are accompanied by 1d slices at $\theta = 0,\frac{\pi}{4}, \frac{\pi}{2}$ below. The quantities plotted are (a) $B_z$ magnetic field (b) $B_z-|B_z|$ negative magnetic field (c) $\mathcal{E}$ energy density (d) $\left|\phi_1\right|^2$ (e)$\left|\phi_2\right|^2$ (f)$\theta_{12}$ phase difference.} 
 \end{figure}

\begin{figure}[tb!]
 \centerline{\includegraphics[width=0.35\linewidth,trim={0.0cm 2.4cm 0.25cm 0.0cm},clip]{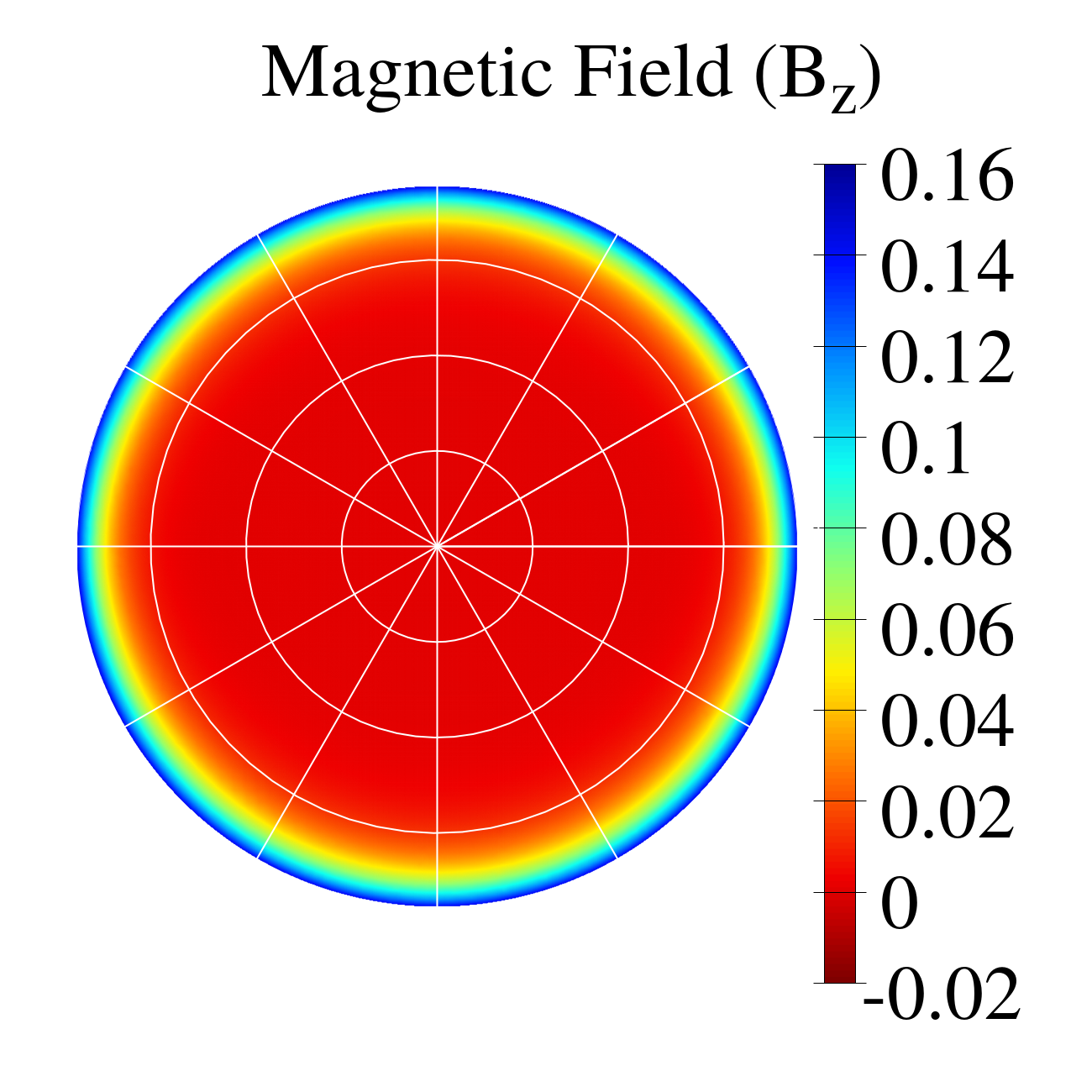}\includegraphics[width=0.35\linewidth,trim={0.0cm 2.4cm 0.25cm 0.0cm},clip]{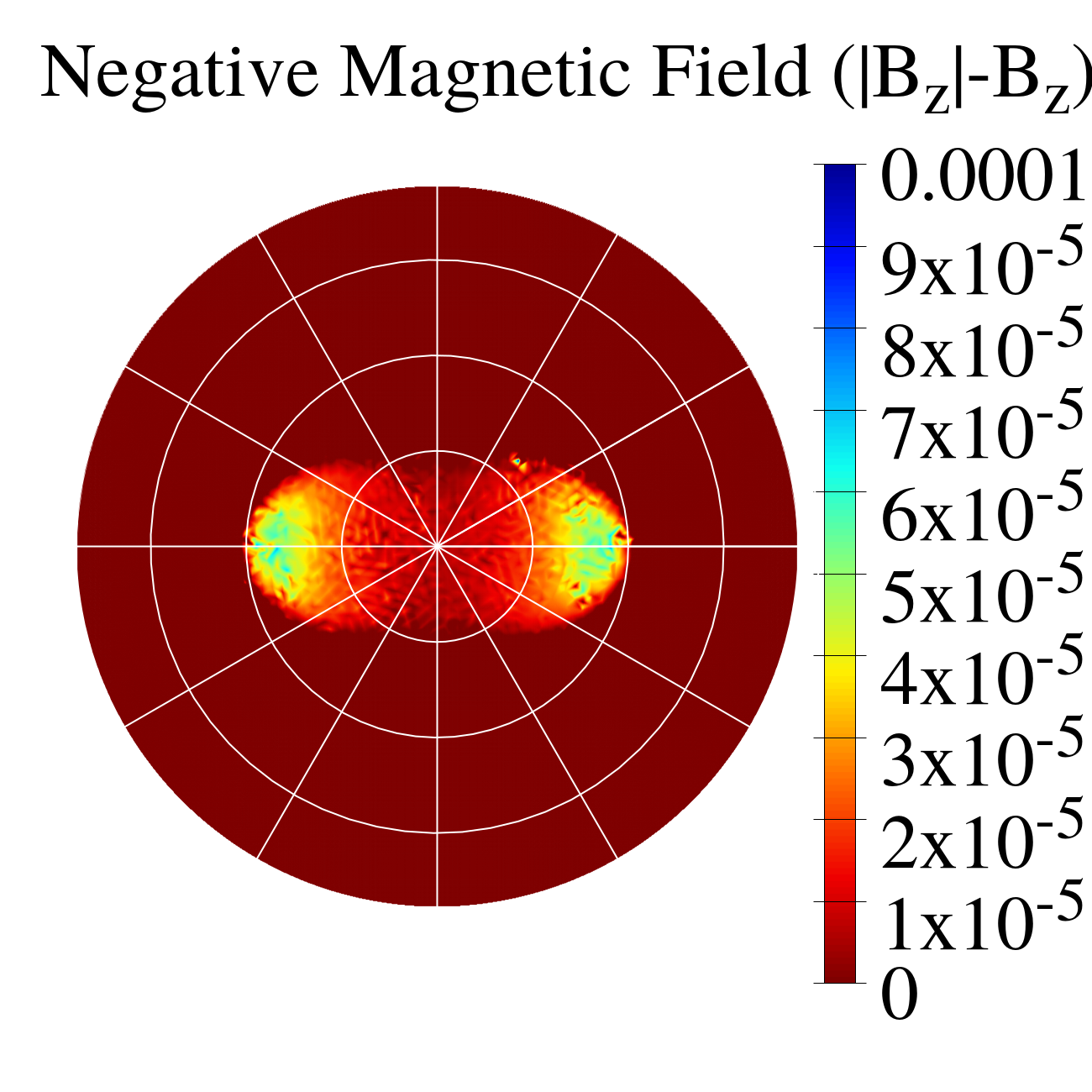}\includegraphics[width=0.35\linewidth,trim={0.0cm 2.4cm 0.25cm 0.0cm},clip]{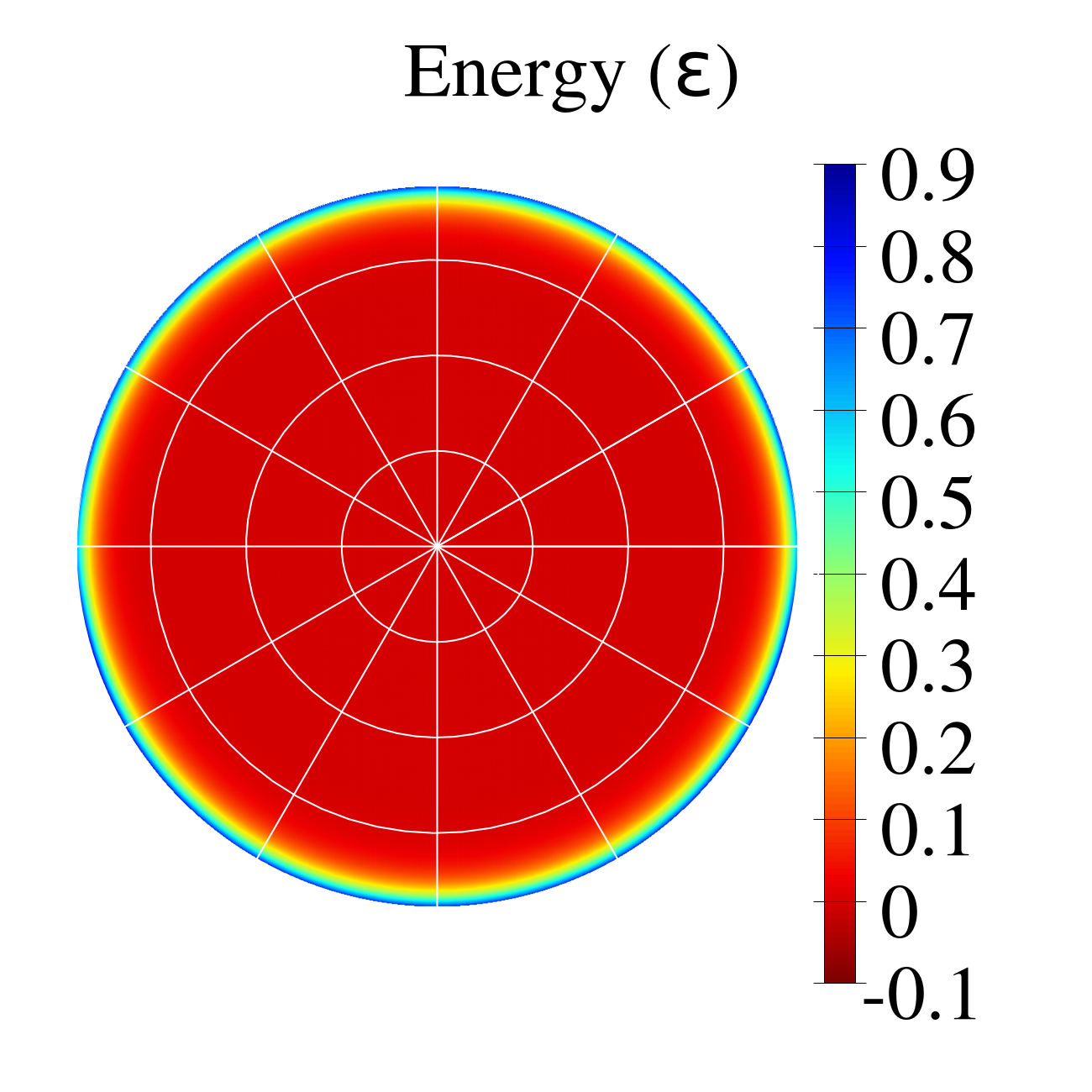}}
  \centerline{\includegraphics[width=0.35\linewidth,trim={1.4cm 2.8cm 0.0cm 1.7cm},clip]{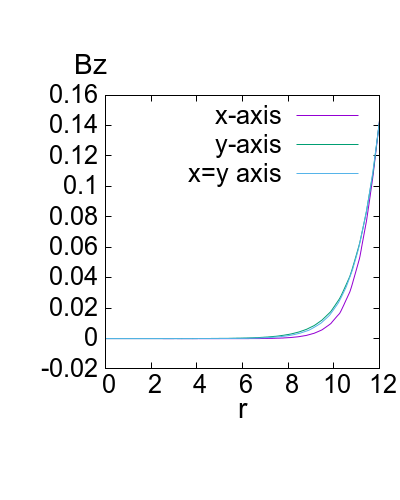}\includegraphics[width=0.35\linewidth,trim={1.4cm 2.8cm 0.0cm 1.7cm},clip]{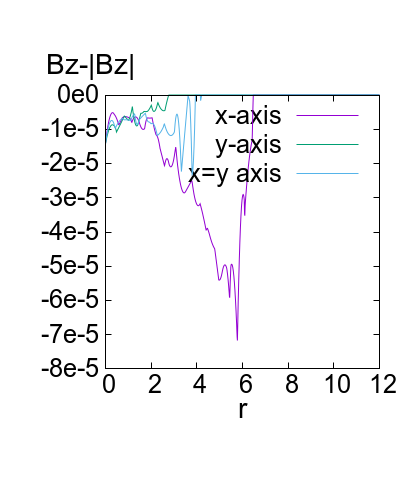}\includegraphics[width=0.35\linewidth,trim={1.4cm 2.8cm 0.0cm 1.7cm},clip]{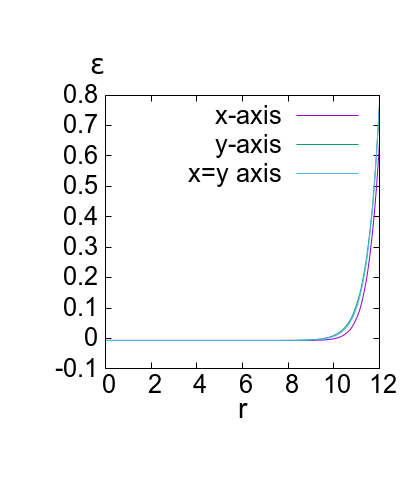}}
    \centerline{\quad(a) \enskip \quad \quad \quad \quad \quad \quad \quad (b)  \quad \quad \quad \quad \quad \quad \quad \quad (c)}
 \centerline{\includegraphics[width=0.35\linewidth,trim={0.0cm 2.4cm 0.25cm 0.0cm},clip]{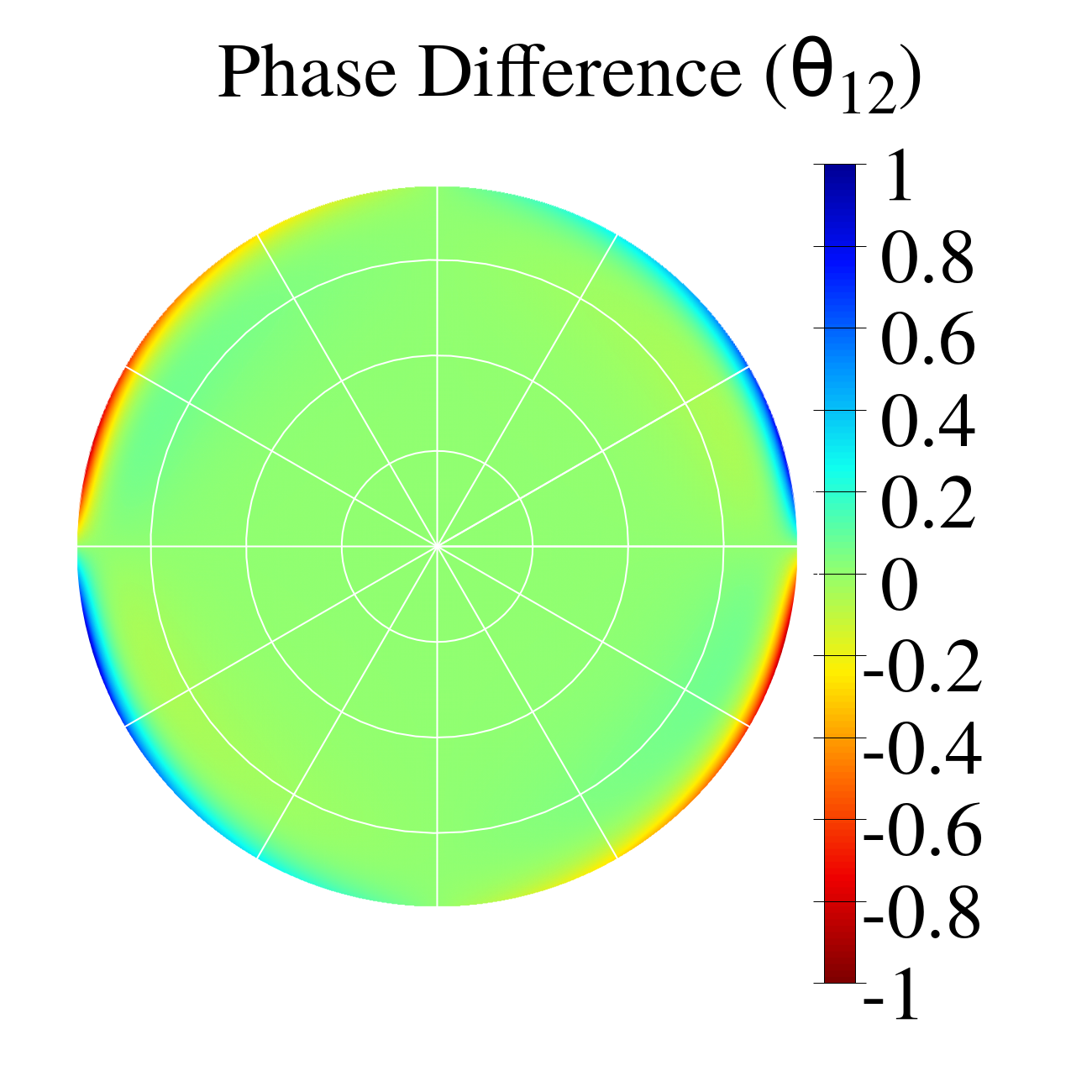}\includegraphics[width=0.35\linewidth,trim={0.0cm 2.4cm 0.25cm 0.0cm},clip]{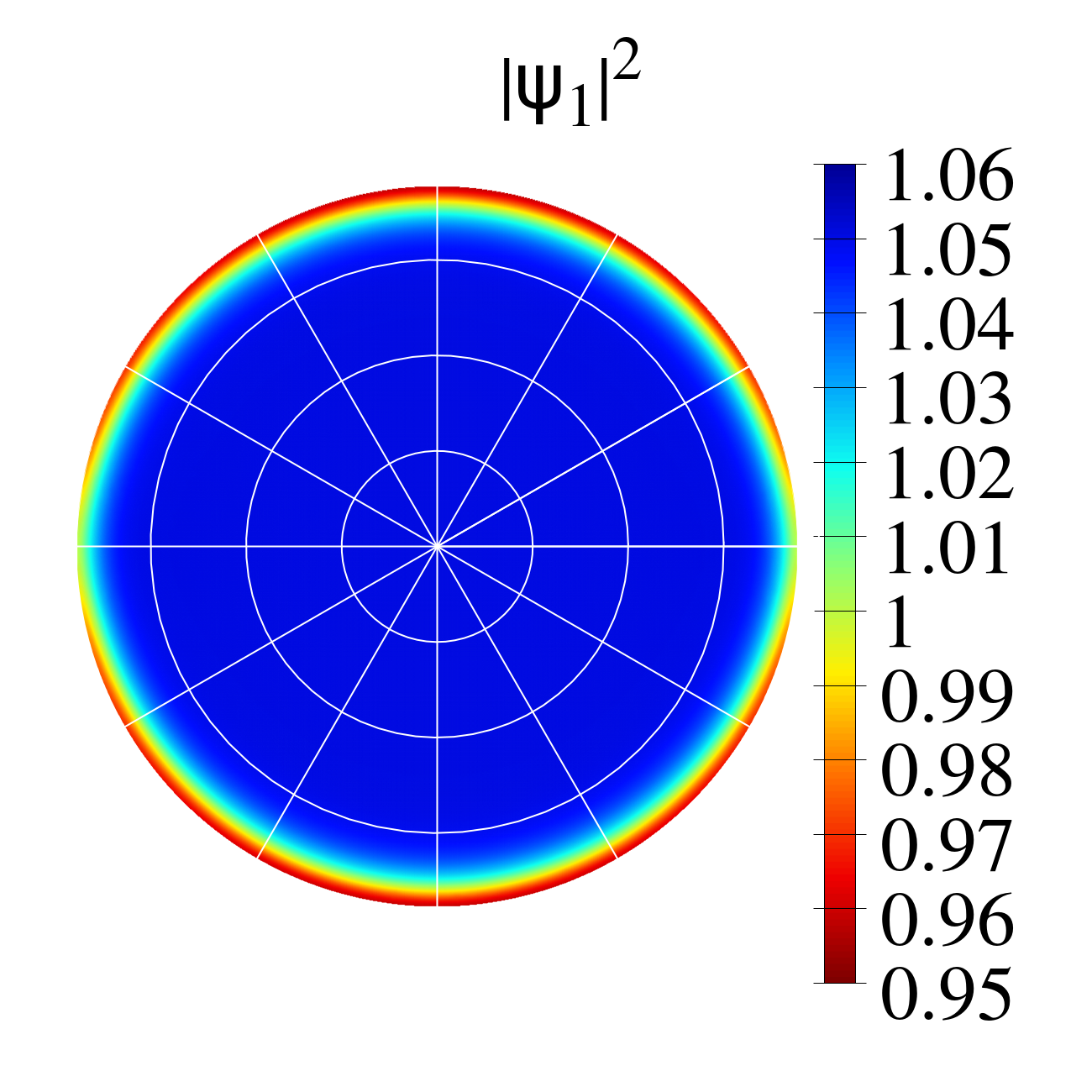}\includegraphics[width=0.35\linewidth,trim={0.0cm 2.4cm 0.25cm 0.0cm},clip]{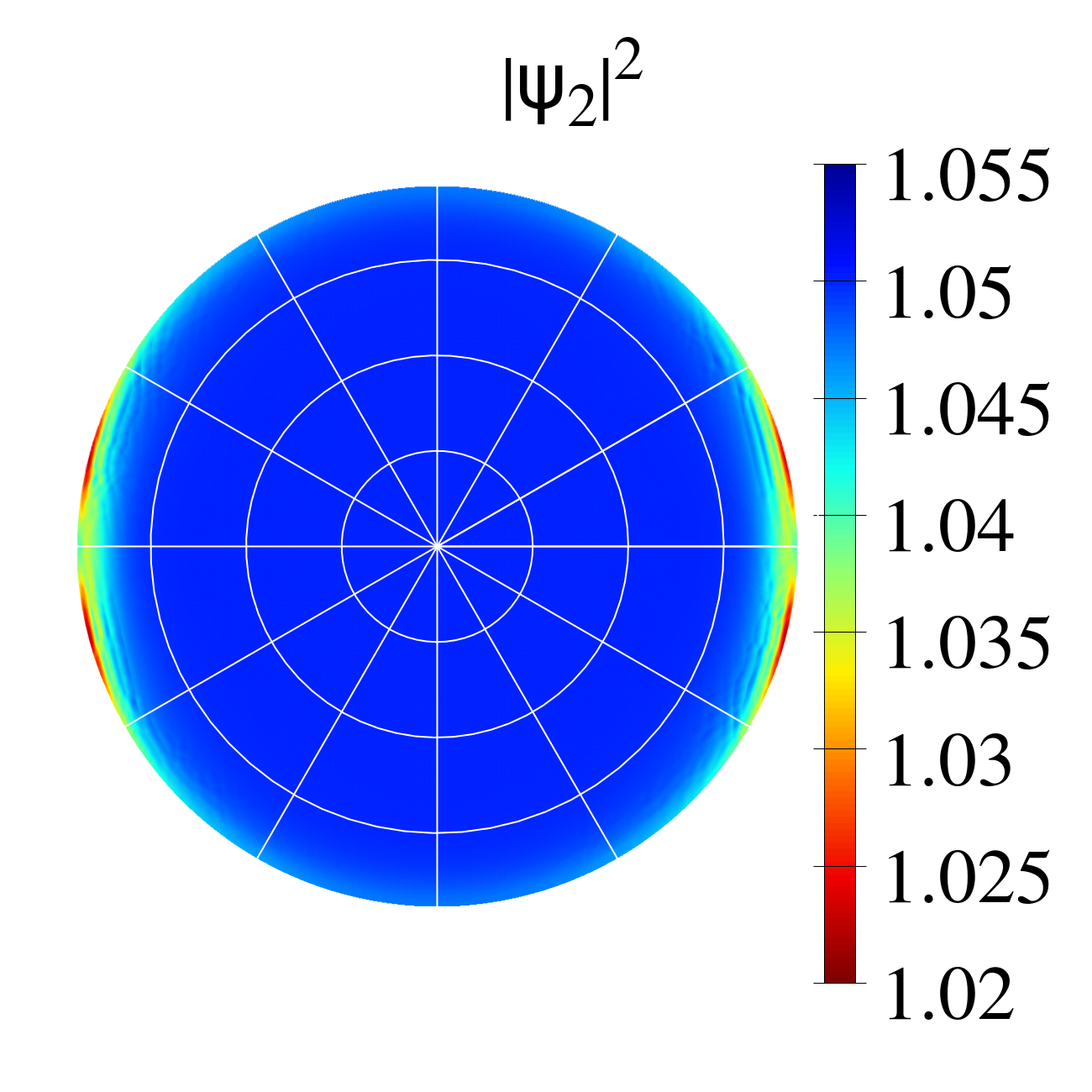}}
 \centerline{\includegraphics[width=0.35\linewidth,trim={1.4cm 2.8cm 0.0cm 1.7cm},clip]{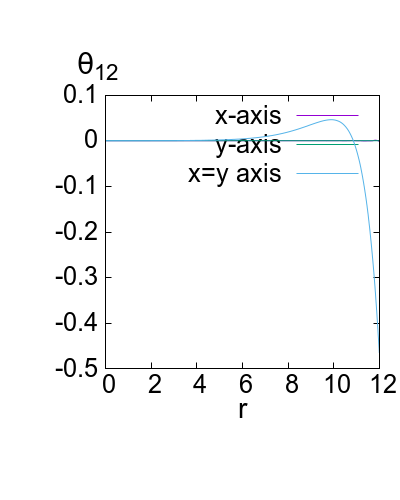}\includegraphics[width=0.35\linewidth,trim={1.4cm 2.8cm 0.0cm 1.7cm},clip]{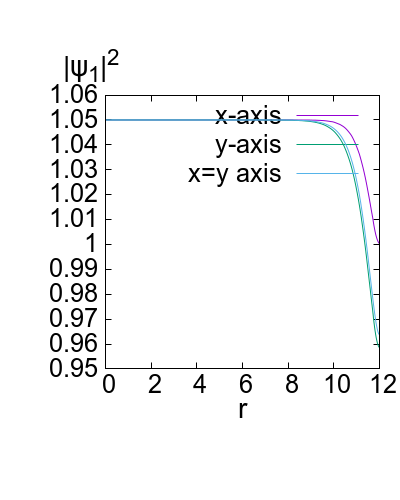}\includegraphics[width=0.35\linewidth,trim={1.4cm 2.8cm 0.0cm 1.7cm},clip]{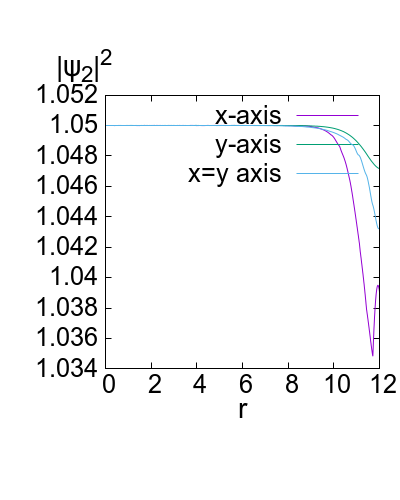}}
   \centerline{\quad(d) \enskip \quad \quad \quad \quad \quad \quad \quad (e)  \quad \quad \quad \quad \quad \quad \quad \quad (f)}
 \caption{\label{Fig:meisner3} 
  Messiner state numerical solution for strong anisotropy in similar directions on a disc of radius 12 $\lambda_{x1}^{-1}=\lambda_{y1}^{-1}=\lambda_{y2}^{-1}=1$, $\lambda_{x2}^{-1} = 0.1$, $\eta_{12} = 0.5$ and $\gamma_{1}=\gamma_{2}=10$. The full 2d plots are accompanied by 1d slices at $\theta = 0,\frac{\pi}{4}, \frac{\pi}{2}$ below. The quantities plotted are (a) $B_z$ magnetic field (b) $B_z-|B_z|$ negative magnetic field (c) $\mathcal{E}$ energy density (d) $\left|\phi_1\right|^2$ (e)$\left|\phi_2\right|^2$ (f)$\theta_{12}$ phase difference.} 
 \end{figure}

We now perform a numerical simulation of the Meissner state of the two-band anisotropic 
full Ginzburg-Landau model (\ref{GL}). 
This was performed using the FreeFem++ numerical library\cite{MR3043640,Hecht2007}, 
which utilises a finite element space, over which conjugate gradient flow is performed. The simulations were performed on a disc as the problem is directionally dependent. We minimise the Gibbs free energy $G = \int_{\mathbb{R}^3}F - \int_{\mathbb{R}^3}\boldsymbol{B}\cdot\boldsymbol{H} + \int_{\mathbb{R}^2} F_{surface}$, where $\boldsymbol{H}=H_z\boldsymbol{e}_z$ in an external field, applied orthogonal to our 2D system, where we have chosen a finite domain with the boundary conditions $\nabla \times \boldsymbol{A} = \boldsymbol{H}$ and for the current to vanish $\boldsymbol{n}\cdot\boldsymbol{j}=0$. We then slowly increase the external field strength $\left|H_z\right|$ in steps of $10^{-3}$, through the various Meissner states. 
When the external field is below the 1st critical value we get the Meissner state solutions shown in Fig. \ref{Fig:meisner1}, \ref{Fig:meisner2} and \ref{Fig:meisner3}. Note all simulations were run with ${\psi^0_\alpha} = e = \hbar = c = 1$ and $\gamma_\alpha = 10$, so in the strong type 2 regime, hence our results should be comparable with the London model above. 


In Fig. \ref{Fig:meisner1} we see a small sample in the Meissner state, with parameters 
$\lambda_{x1} = \lambda_{y2} = 1$, $\lambda_{y1} = \lambda_{x2}= 10$ and 
$\eta_{\alpha\beta}=0.5$. The key effect to note is the oscillation of the phase difference away from the axis: a consequence of the anisotropy driven hybridization of the Leggett mode and magnetic mode discussed above. When the angle of the boundary is away from $\theta = n\pi/2$. 
When looking at the magnetic field however we don't see the expected oscillation. 
This is due to the effect being long range, so if we increase the radius of the disc for similar parameters, as shown in Fig. \ref{Fig:meisner2}, we can see the inversion of magnetic field. 

We also have depicted the Meissner state for anisotropy in only one band in Fig. \ref{Fig:meisner3} for the parameters $\lambda_{x1} = \lambda_{y2} = \lambda_{x2} = 1$, $\lambda_{y1} = 10$ and $\eta_{\alpha\beta}=0.5$. This leads to a similar effect on the shape of the various plots, matching the symmetry of the anisotropy. 
However the negative component of the magnetic field is far smaller, which is as we predicted when we considered the limiting cases in the London model. Finally switching off the Josephson term removes the magnetic field inversion as was discussed in the London approximations. So the results from the full field numerics seem to qualitatively support the London model calculations and all the predictions that came from them.

\section{Vortex Structure}

Let us now consider how the multiple magnetic modes in anisotropic multi-band superconductors modify the vortex states in the London model. 
We begin by studying vortex solutions carrying a single flux quantum, where both components have $2\pi$ phase winding around the core. We later consider a solution for fractional flux vortex i.e. the vortex that has phase winding only in one phase and carries a fraction of flux quantum (for details of flux fractionalization and energetical preference of fractional and composite vortices in multiband systems see the discussions for isotropic 
 systems  \cite{Babaev2002a,Silaev2011b}
).
 
 \subsection{Integer-flux vortex}
The field distribution around a single vortex line can be found in the form of a Fourier transform,
 \begin{equation}\label{Eq:FourierTransform}
 {\bm B}({\bm r}) = \frac{\Phi_0}{2\pi} \int {\bm h} ({\bm k}) e^{i{\bm k}\cdot{\bm
 r}} d^2{\bm k},
 \end{equation}
 where ${\bm r}$ is a coordinate vector in the plane perpendicular to the vortex line and the 2D integration is done by the corresponding momentum space cross section.

 The components ${\bm h} ({\bm k})$ are now determined by the non-homogeneous system
   \begin{align}\label{Eq:London2BandKVort}
  & {\bm h} - {\bm k}\times ( \hat\lambda_L^2 {\bm k} \times{\bm h} ) +
    {\bm k}\times ( \hat\lambda^{-2}_1 \hat\lambda_L^2  {\bm k} )\theta_{12} = 2\pi {\bm n}_v 
    \\
  & {\bm k}\cdot ( \hat\lambda^{-2}_1\hat\lambda^{-2}_2\hat\lambda^{2}_L {\bm k})\theta_{12} +
   k_0^4 \theta_{12} - {\bm k}\cdot ( \hat\lambda_1^{-2} \hat\lambda_L^{2} {\bm k}\times {\bm h} ) =0
  \end{align}
where ${\bm n}_v$ is the direction of the vortex line.

The anisotropy in the plane perpendicular to the vortex line can be caused by two reasons. 
\textbf{(i)} When the magnetic field is directed along the $c$-axis, there can be 
  anisotropy in the $ab$ plane either if the crystal is biaxial or as a result of the strain-induced distortions. 
\textbf{(ii)} The effective anisotropy can be caused by a misalignment between the external field and the anisotropy $c$ axis. 
 The distribution of the magnetic field around vortices will be different in these two cases since in \textbf{(i)} only two of 
 the magnetic modes are excited and in 
 \textbf{(ii)} the amplitudes of all three magnetic modes are non-zero.
 
  In this paper we consider in detail only the case \textbf{(i)} when the magnetic field around the vortex is 
  ${\bm h} = h_z {\bm z}$,  where
   \begin{align}
   \label{Eq:Hz}
   & h_z = 2\pi \frac{ k^4( \lambda_{1x}^{-2}\lambda_{2x}^{-2}\lambda_{Lx}^{2}k_x^2 + 
   \lambda_{1y}^{-2}\lambda_{2y}^{-2}\lambda_{Ly}^{2}k_y^2 +
   k_0^4)}{a (k^2+k_1^2) (k^2+k_2^2)} 
   \\
  & a = ( \lambda_{1x}^{-2}k_x^2 + \lambda_{1y}^{-2}k_y^2) 
  ( \lambda_{2x}^{-2}k_x^2 + \lambda_{2y}^{-2}k_y^2).
   \end{align}
   Here the poles $k_{1,2}$ are given by the Eq.(\ref{Eq:kSolution}).

 Using expressions (\ref{Eq:Hz},\ref{Eq:FourierTransform}) one can consider the 
 asymptotics of the field far from the vortex center. We introduce the polar coordinates ${\bm k} = k(\cos\theta,\sin\theta,0)$ and integrate first by $k$ taking into account the symmetry  $h_{z}(k)= h_{z}(-k)$ which allows the integration to be extended to the domain $k<0$.
 Using Eq.(\ref{Eq:Hz}) we get the magnetic field distribution in the real space polar coordinates $(r,\varphi)$
 with the origin at the vortex center
 \begin{align}\label{Eq:BZFin}
 & B_z (r,\varphi) =  \Phi_0\left( h_1(\varphi) \frac{e^{-k_1 r}}{\sqrt{k_1r}} - h_2(\varphi) \frac{e^{-k_2 r}}{\sqrt{k_2r}}
 \right), \\ \label{Eq:hj}
 & h_j (\varphi) =  \frac{ k_j^2 (\lambda_{1x}^{-2}\lambda_{2x}^{-2}\lambda_{Lx}^{2}\cos^2\varphi +
 \lambda_{1z}^{-2}\lambda_{2z}^{-2}\lambda_{Lz}^{2}\sin^2\varphi )- k_0^4}{a(k_1^2 - k_2^2)},
 \end{align}
 where $h_j(\varphi)={\rm Res} (h_z(k,\varphi), ik_j(\varphi)) $
 is a residue of the function $h_z(k,\varphi)$ at the pole
 $k=ik_j(\varphi)$.  For the angle
 integration we used an approximation $k_jr\cos(\theta-\varphi) \approx k_j r ( 1- (\theta-\varphi)^2/2)$
 which is valid provided $k_j r \gg 1$.

 In general due to the $ab$-plane anisotropy Eqs.(\ref{Eq:BZFin},\ref{Eq:hj}) yield the expected fourfold magnetic
 field profile around a vortex. But the most important point is the non-monotonic field behaviour with field inversion at some distance from the vortex center, as with the boundary problems. For this it is necessary and sufficient to satisfy two conditions: $h_{2}<0$ and $k_1>k_2$, so that the mode with negative amplitude can become dominating at some distance from the vortex.

\begin{figure}[tb!]
 \centerline{\includegraphics[width=1.0\linewidth]{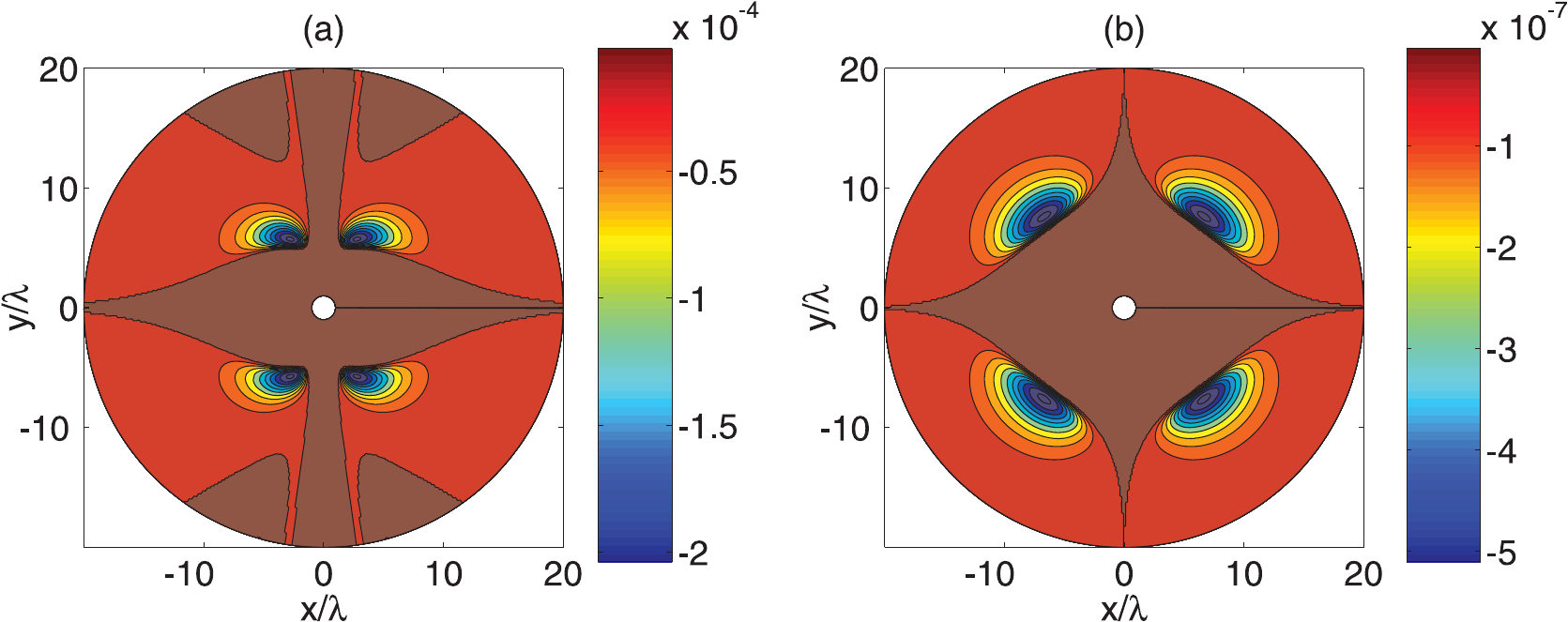}}
 \caption{\label{Fig:Bz} (Colour online)
 Negative part of the magnetic field $\tilde{B}_z ={B}_z - |{B}_z| $ distribution around a single vortex, normalized by $\Phi_0/\lambda^2$.'
  The parameters correspond to either
 (a) Strong anisotropy or (b) weak anisotropy.}
 \end{figure}

To demonstrate the possibility of field inversion we consider the parameters 
$\lambda_{1x}=0.7\lambda$, $\lambda_{1y}=0.8\lambda$, $\lambda_{2x}=\lambda_{2y}=\lambda$ and $k_0= 0.7/\lambda$ which 
results in the correct four-fold magnetic field profile with field inversion far from the vortex core. 
The negative parts of field distribution $B_z- |B_z|$ are shown in Fig.(\ref{Fig:Bz}). 
Field inversion leads to a non-trivial four-fold non-monotonic interaction between vortices and thus bound states. 
This will be considered in more detail below in Sec.\ref{Sec:VortexBoundStates}.

Finally the above modifications to vortex solutions point towards a new mechanism  for vortex stripe formation in multi-band superconductors.

 \subsection{Fractional vortex solution}
For fractional vortices we have a singularity appearing in just one of the components (chosen to be component 1 without loss of generality). It follows immediately that this will lead to a singularity appearing in $\theta_{12}$ itself unlike in the composite case. This means the previous approach of linearising the equations and approximating $\sin{\theta_{12}} \approx \theta_{12}$ is no longer valid. This is unsurprising as it is due to the presence of Josephson strings when the coupling is switched on. Here it is illustrative to consider the fractional vortex in the $U(1)\times U(1)$ model. 

We follow a similar procedure by Fourier transforming the equations of motion, however the singularity now exists in $\nabla\theta_{1}$, allowing us to utilise the fact that $ {\bm p}_s - \nabla\theta_{12}/2$ is singularity-free. The equations of motion can be rearranged to isolate the singularity and yield in the Fourier representation the following,
 \begin{align} \label{Eq:LondonFrac}
   & {\bm k}\times ( \hat\lambda_1^2 {\bm k} \times{\bm h} ) - {\bm h} = \\ \nonumber
  &  i{\bm k}\times [ \hat\lambda^{-2}_2 \hat\lambda_1^2  ({\bm p}_s + \frac{i}{2}\bm k \theta_{12}) ] + 2\pi {\bm n_v} ,
  \\
  & {\bm k}\cdot \lambda_2^{-2} \left({\bm p}_s + \frac{i}{2}\bm k \theta_{12}\right) = 0
  \end{align}
 
If the magnetic field lies in the $z$-direction only ${\bm h}=h_z\hat{\boldsymbol{z}}$ the solution of this system is given by,
 \begin{align}
  & h_z ({\bm k}) = \frac{2\pi }{\left(\lambda_{1y}^2 \cos^2 \theta + \lambda_{1x}^2 \sin^2 \theta\right) \left(k^2 + k_1^2\right)} \\
 & k_1^2 = \frac{\left( 1 + \lambda_{1y}^{2} \lambda_{2y}^{-2}\right) \lambda_{2y}^{2} \cos^2 \theta + \left( 1 + \lambda_{1x}^{2}\lambda_{2x}^{-2} \right) \lambda_{2x}^{2} \sin^2 \theta} 
{\left(\lambda_{1y}^2\cos^2\theta + \lambda_{1x}^2 \sin^2\theta\right)
\left(\lambda_{2y}^{2}\cos^2\theta + \lambda_{2x}^{2}\sin^2\theta\right)} .
 \end{align}
 where we have used $\bm k = k (\cos\theta, \sin\theta, 0)$ again.

\begin{figure}[htb!]
 \centerline{\includegraphics[width=0.5\linewidth]{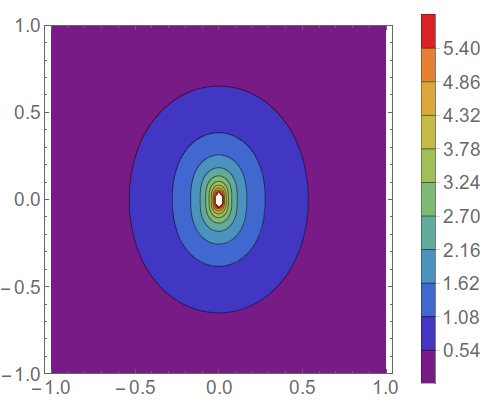}
  \put (-75,105) {(a)} \put (-130,60) {\rotatebox{90}{$y/\lambda$}}
 \includegraphics[width=0.5\linewidth]{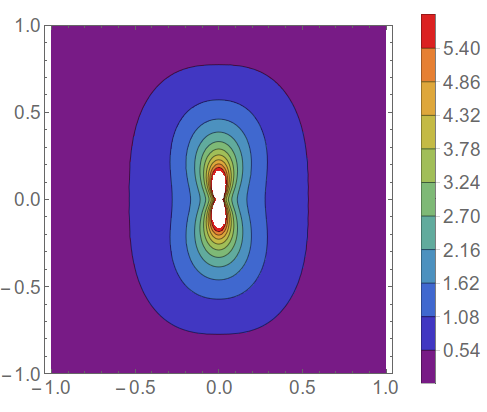}
 \put (-75,105) {(b)} 
 }
 \centerline{
 \includegraphics[width=0.5\linewidth]{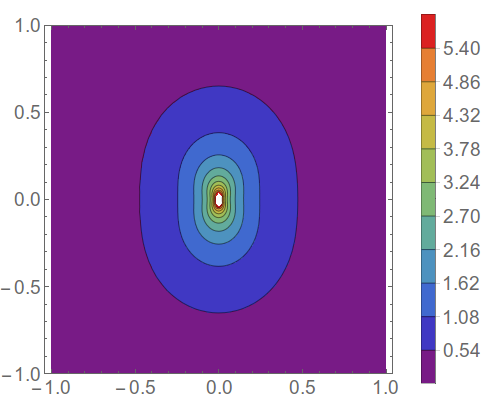}
 \put (-75,105) {(c)} \put (-75,-5) {$x/\lambda$} 
 \put (-130,60) {\rotatebox{90}{$y/\lambda$}}
 \includegraphics[width=0.5\linewidth]{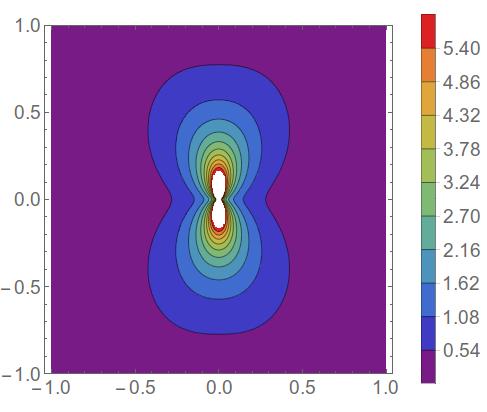}
 \put (-75,105) {(d)} \put (-75,-5) {$x/\lambda$} 
 } 
 \caption{\label{Fig:frac}
 Magnetic field $B_z$ for a fractional vortex in $U(1)\times U(1)$model (i.e. $k_0 = 0$), with various types of anisotropy (a) $\lambda_{1x}=0.8\lambda_{1y}$, $\lambda_{2x}=\lambda_{2y}=\lambda_{1y}$  (b) $\lambda_{1x}=0.4\lambda_{1y}$, $\lambda_{2x}=\lambda_{2y}=\lambda_{1y}$ (c) $\lambda_{1x}=\lambda_{2y}=0.8\lambda_{1y}$ (d) $\lambda_{1x}=\lambda_{2y}=0.4\lambda_{1y}$, $\lambda_{2x} = \lambda_{1y}$. The field is in the units of $\Phi_0/\lambda_{1y}^2$.}
   \end{figure}
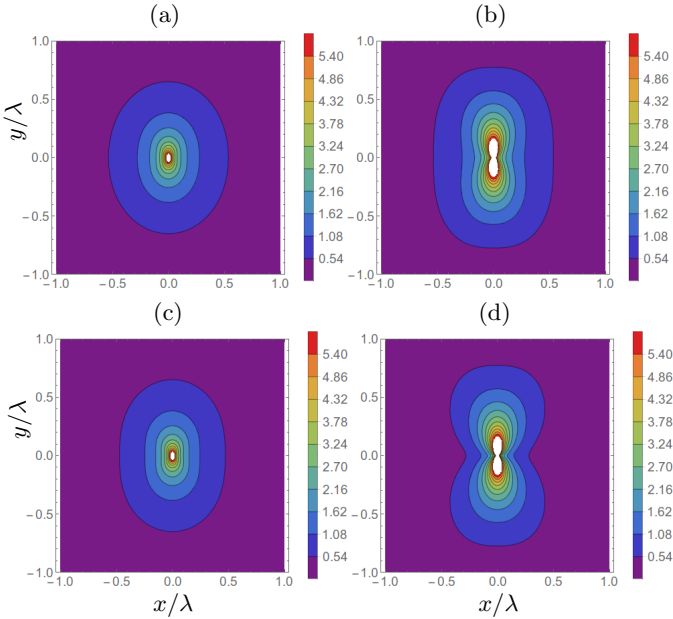

Using a similar method to before we can find the magnetic field to be,
\begin{equation}
B_z\left(r,\psi\right) = \frac{\Phi_0 e^{-k_1 r}}{\left(\lambda_{1y}^2 \cos^2 \theta + \lambda_{1x}^2 \sin^2 \theta\right)\sqrt{k_1 r}}.
\end{equation}
This illustrates that when interband Josephson coupling is zero $k_0 = 0$, the Leggett mode becomes non-magnetic 
and does not contribute to the  magnetic response despite gradients of the phase difference.
 \footnote{Should Josephson coupling be switched back on the second mode should appear.  A more complex situation for fractional vortices appears beyond the London model: there will be field inversion due to the presence of coexistent phase difference and relative density gradients   
 \cite{Babaev2009}. 
 } 
The solution is plotted for a number of values in Fig. \ref{Fig:frac}. 

 \subsection{Numerical Vortex Solutions}
We now consider the numerical solutions for type 2 vortices in the full anisotropic Ginzburg-Landau equations. 
All numerical simulations were performed using the FreeFem++ numerical library
\cite{Hecht2007,MR3043640}, which utilises a finite element space over which conjugate gradient flow is performed. We take expression (\ref{GL}) to be our energy functional, where all the solutions were found in a type 2 superconductor with parameters $\gamma_{\alpha} = 2$ and ${\psi^0_\alpha} = e = \hbar = c = 1$ for all components. We also restrict to the x-y 2-dimensional plane, where we can now refer to different solutions in terms of the number of flux quanta,
\begin{equation}
N = \frac{\Phi}{\Phi_0} = \frac{1}{2\pi}\int_{\mathbb{R}^2} B_z d^2 x.
\end{equation}

\begin{figure}[tb!]
  \centerline{
  \includegraphics[width=0.35\linewidth,trim={5.1cm 2.5cm 3.5cm 1.6cm},clip]
  {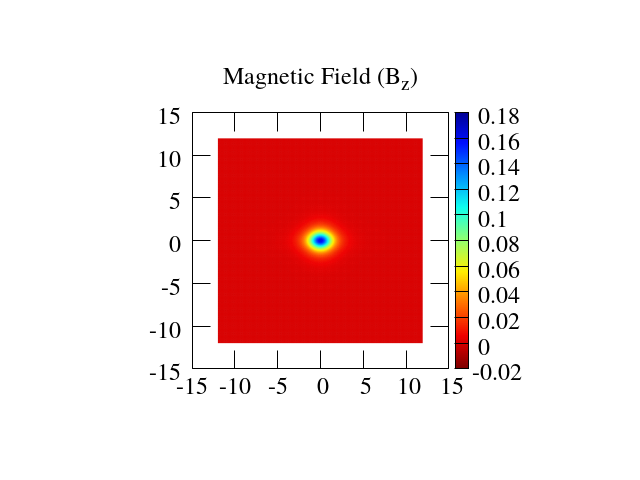} 
  \includegraphics[width=0.35\linewidth,trim={5.1cm 2.5cm 3.5cm 1.6cm},clip]
  {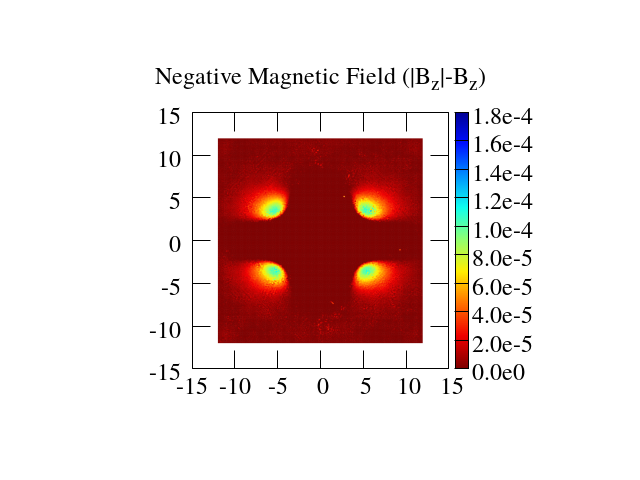}
  \includegraphics[width=0.35\linewidth,trim={5.1cm 2.5cm 3.5cm 1.6cm},clip]
  {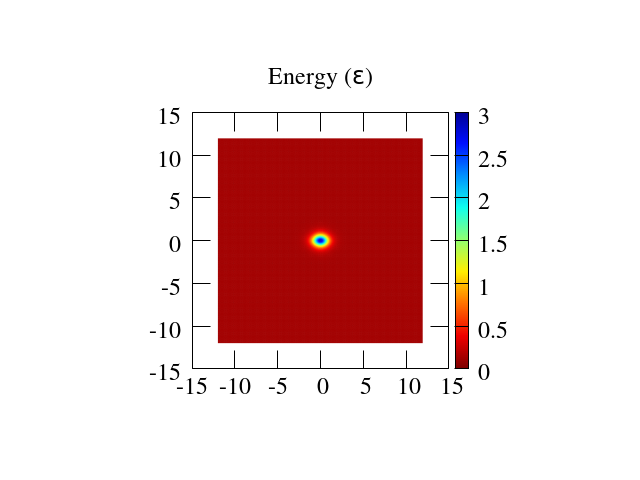}}
  \centerline{
  \includegraphics[width=0.35\linewidth,trim={5.1cm 2.5cm 3.5cm 1.6cm},clip]
  {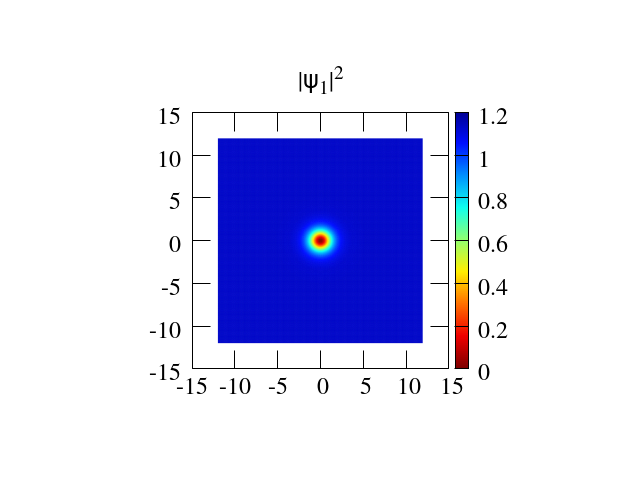}
  \includegraphics[width=0.35\linewidth,trim={5.1cm 2.5cm 3.5cm 1.6cm},clip]
  {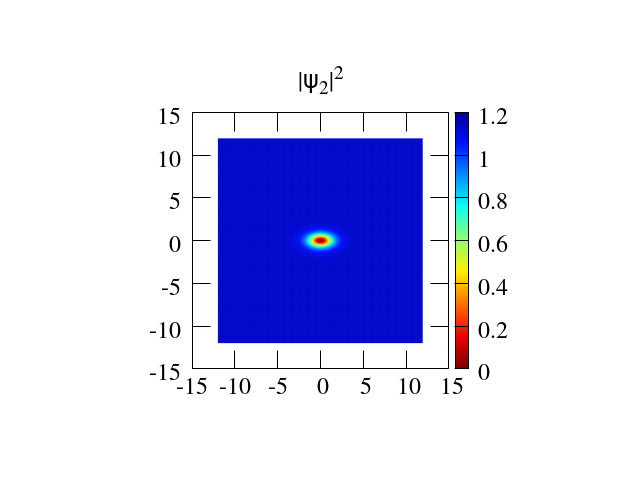}
  \includegraphics[width=0.35\linewidth,trim={5.1cm 2.5cm 3.5cm 1.6cm},clip]
  {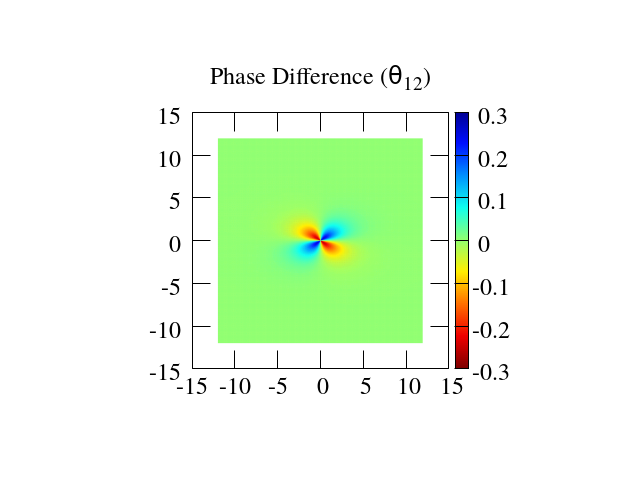}}
 \caption{\label{Fig:charge1oneweak} (Colour online)
 $N=1$ single quanta numerical solution for weak anisotropy in one band 
 $\lambda_{x1}=\lambda_{y1}=\lambda_{x2}=1$, 
 $\lambda_{y2} = 0.5$, $\eta_{12} = 0.5$ and $\gamma_{1}=\gamma_{2}=2$. The contour plots are (a) $B_z$ magnetic field 
 (b) $\left|B_z\right| - B_z$ negative magnetic field (c) $\mathcal{E}$ energy density (d) 
 $\left|\phi_1\right|^2$ 
 (e)$\left|\phi_2\right|^2$ (f)$\theta_{12}$ phase difference. }
 \end{figure}

\begin{figure}[tb!]
  \centerline{
  \includegraphics[width=0.35\linewidth,trim={5.1cm 2.5cm 3.5cm 1.6cm},clip]{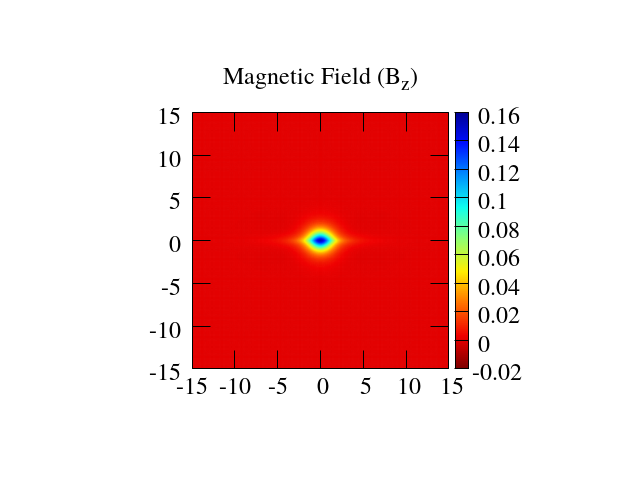}
  \includegraphics[width=0.35\linewidth,trim={5.1cm 2.5cm 3.5cm 1.6cm},clip]{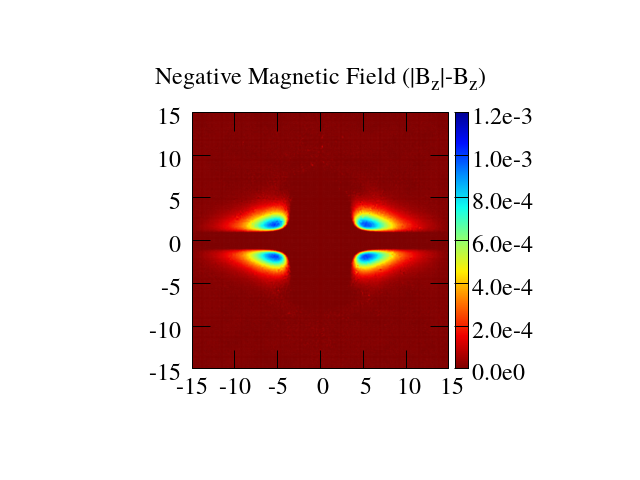}
  \includegraphics[width=0.35\linewidth,trim={5.1cm 2.5cm 3.5cm 1.6cm},clip]{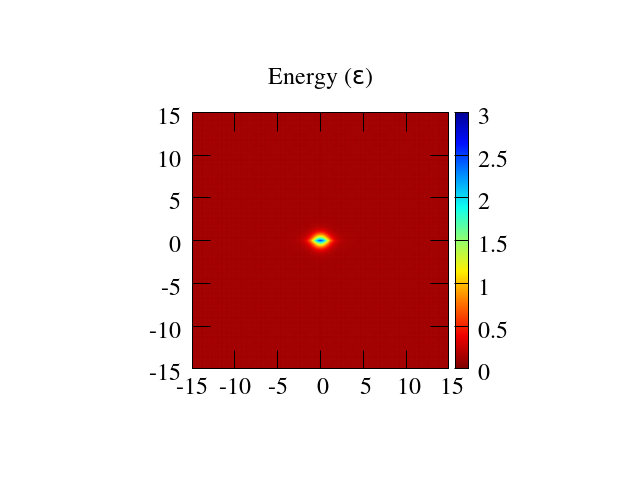}}
  \centerline{
  \includegraphics[width=0.35\linewidth,trim={5.1cm 2.5cm 3.5cm 1.6cm},clip]{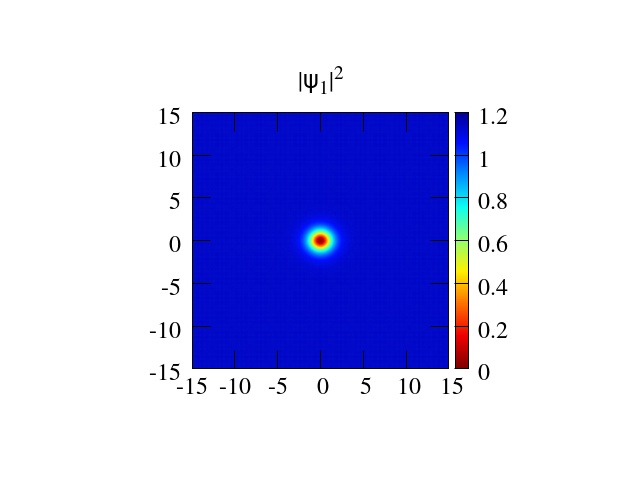}
  \includegraphics[width=0.35\linewidth,trim={5.1cm 2.5cm 3.5cm 1.6cm},clip]{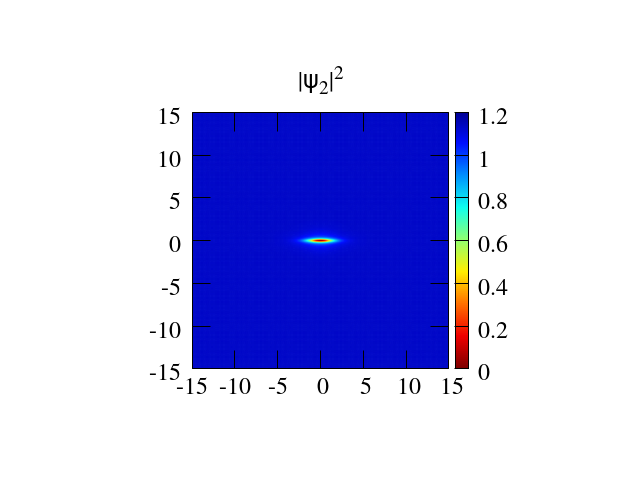}
  \includegraphics[width=0.35\linewidth,trim={5.1cm 2.5cm 3.5cm 1.6cm},clip]{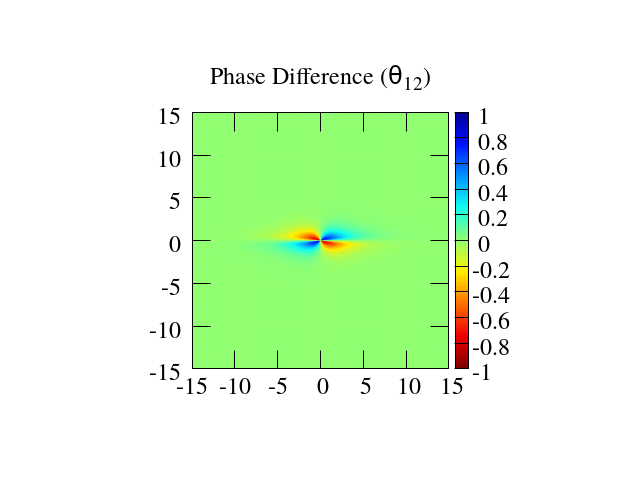}}
 \caption{\label{Fig:charge1one} (Colour online)
 $N=1$ single quanta numerical solution for strong anisotropy in one band $\lambda_{x1}=\lambda_{y1}=\lambda_{x2}=1$, 
 $\lambda_{y2} = 0.1$, $\eta_{12} = 0.5$ and $\gamma_{1}=\gamma_{2}=2$. The contour plots are (a) $B_z$ magnetic field 
 (b) $\left|B_z\right| - B_z$ negative magnetic field (c) $\mathcal{E}$ energy density 
 (d) $\left|\phi_1\right|^2$ (e)$\left|\phi_2\right|^2$ (f)$\theta_{12}$ phase difference. }
 \end{figure}

We first consider the single-quantum vortex solution with weak anisotropy in one band 
 $\lambda_{1x}=\lambda_{1y}=\lambda_{2x}$, 
 $\lambda_{2y} = 2\lambda$ shown in Fig. \ref{Fig:charge1oneweak}, 
 and strong anisotropy in one band 
 $\lambda_{1x}=\lambda_{1y}=\lambda_{2x}$, 
 $\lambda_{2y} = 10\lambda$ 
 shown in Fig. \ref{Fig:charge1one}. The salient point is the confirmation of magnetic field inversion for both sets of parameters, increasing in strength as the anisotropy is increased. This along with the self induced phase difference gradients confirm that the results from the above London model calculations, do transition into the full Ginzburg-Landau model as expected. Another thing to note is that the symmetry of the solutions match the broken symmetry of the energy functional as with the London model. The isotropic component appears to retain it's axial symmetry from an unbroken model, while the anisotropic components symmetry is broken to the expected 4 fold symmetry, or squashed as one might expect of a single component system.

\begin{figure}[tb!]
  \centerline{
  \includegraphics[width=0.35\linewidth,trim={5.1cm 2.5cm 3.5cm 1.6cm},clip]{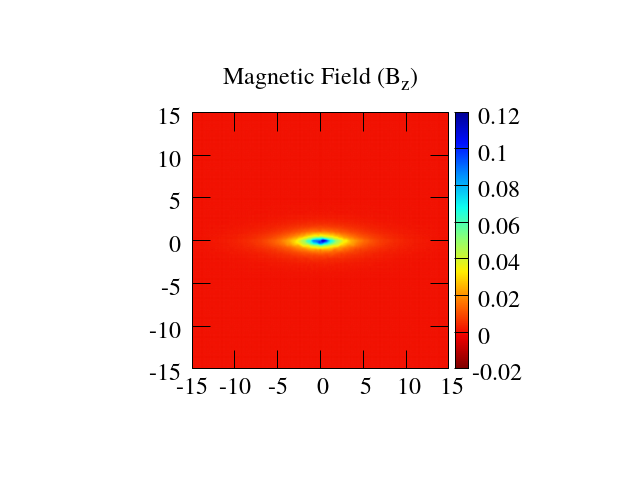}
  \includegraphics[width=0.35\linewidth,trim={5.1cm 2.5cm 3.5cm 1.6cm},clip]{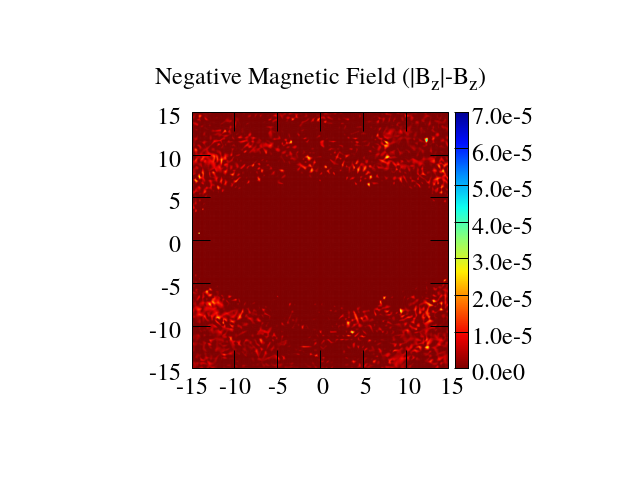}
  \includegraphics[width=0.35\linewidth,trim={5.1cm 2.5cm 3.5cm 1.6cm},clip]{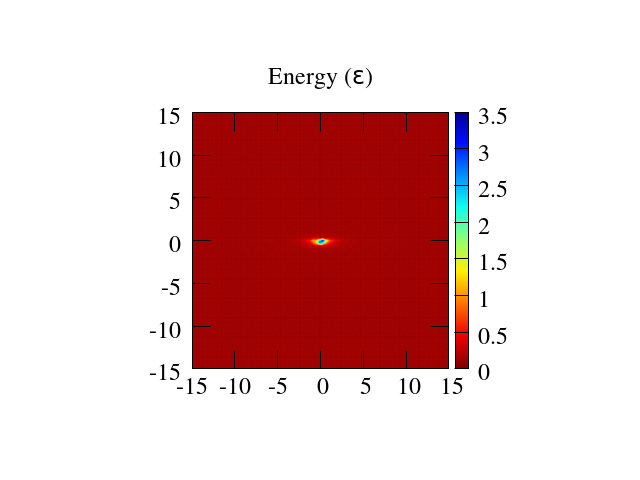}} 
  \centerline{
  \includegraphics[width=0.35\linewidth,trim={5.1cm 2.5cm 3.5cm 1.6cm},clip]{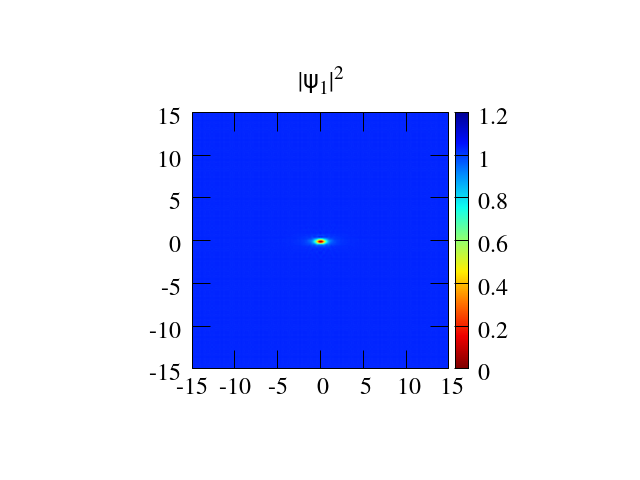}
  \includegraphics[width=0.35\linewidth,trim={5.1cm 2.5cm 3.5cm 1.6cm},clip]{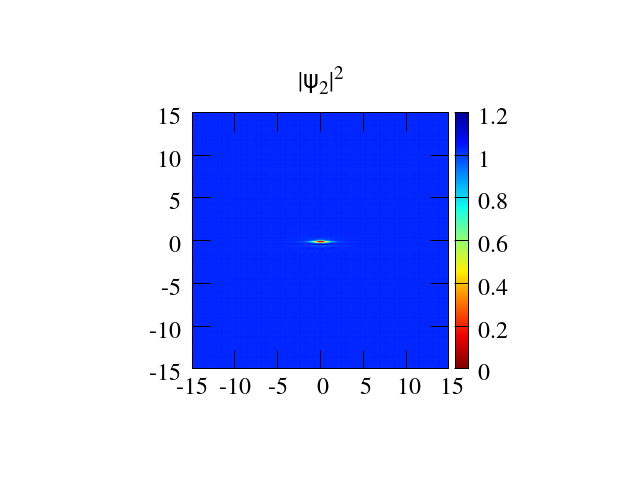}
  \includegraphics[width=0.35\linewidth,trim={5.1cm 2.5cm 3.5cm 1.6cm},clip]{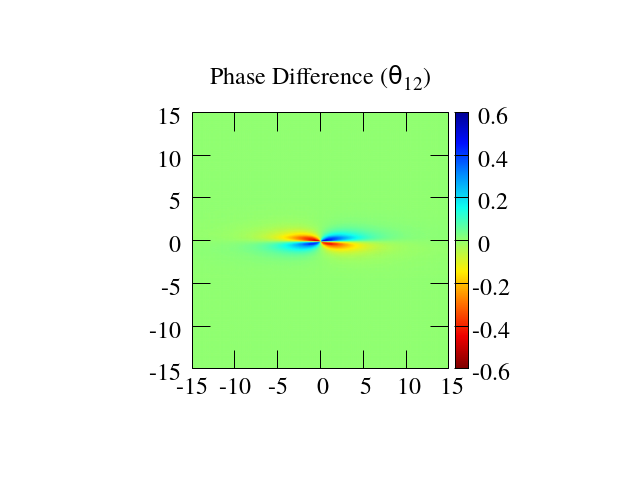}}
 \caption{\label{Fig:charge1same} (Colour online)
 $N=1$ single quanta numerical solution for anisotropy in equivalent directions in both bands $\lambda^{-1}_{1x} = 0.7$, $\lambda^{-1}_{1y} = 0.4$, $\lambda_{2x} = 1$ and $\lambda^{-1}_{2y} = 0.1$, $\eta_{12} = 0.5$ and $\gamma_{1}=\gamma_{2}=2$. The contour plots are (a) $B_z$ magnetic field (b) $\left|B_z\right| - B_z$ negative magnetic field (c) $\mathcal{E}$ energy density (d) $\left|\phi_1\right|^2$ (e)$\left|\phi_2\right|^2$ (f)$\theta_{12}$ phase difference. }
 \end{figure}

We now consider the effect of having anisotropy exhibited in both bands in a similar direction as shown in Fig. \ref{Fig:charge1same}. As the anisotropies in the two band approach each other the field inversion gets weaker and once they are the same the model is in many respects analogous to a single component anisotropic system. The closer the anisotropies are to each other in each of the bands the more diminished the exotic behaviour we have observed becomes and eventually vanishes as the anisotropies coincide. 

\begin{figure}[tb!]
  \centerline{
  \includegraphics[width=0.35\linewidth,trim={5.1cm 2.5cm 3.5cm 1.6cm},clip]{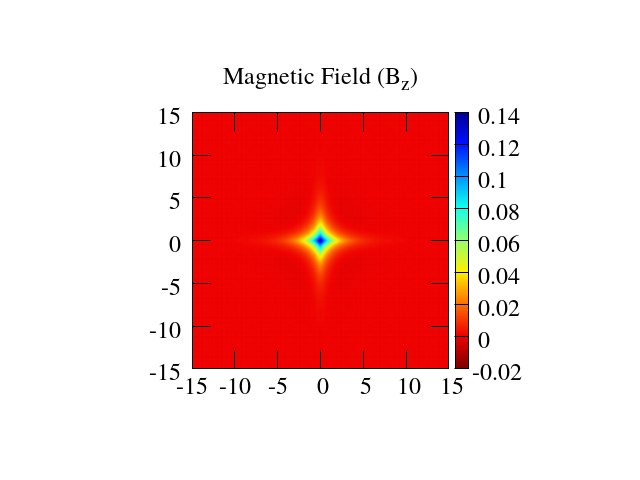}
  \includegraphics[width=0.35\linewidth,trim={5.1cm 2.5cm 3.5cm 1.6cm},clip]{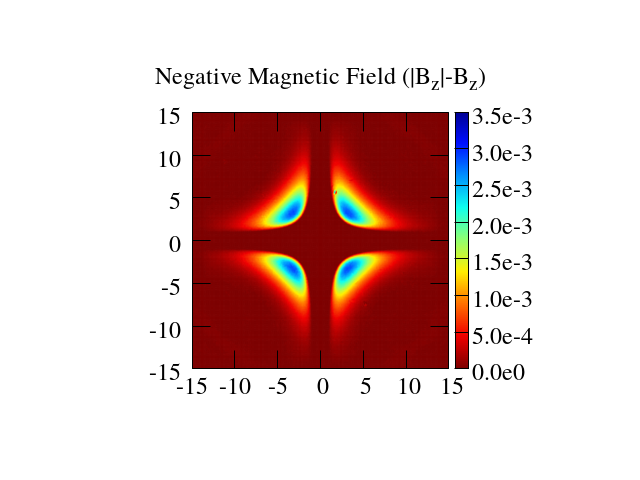}
  \includegraphics[width=0.35\linewidth,trim={5.1cm 2.5cm 3.5cm 1.6cm},clip]{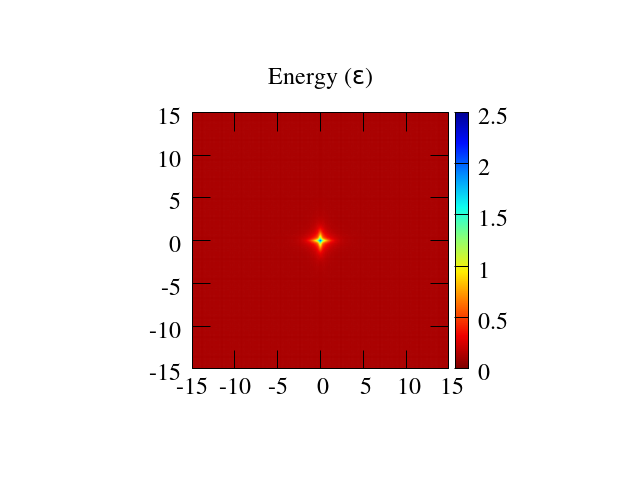}}
  \centerline{
  \includegraphics[width=0.35\linewidth,trim={5.1cm 2.5cm 3.5cm 1.6cm},clip]{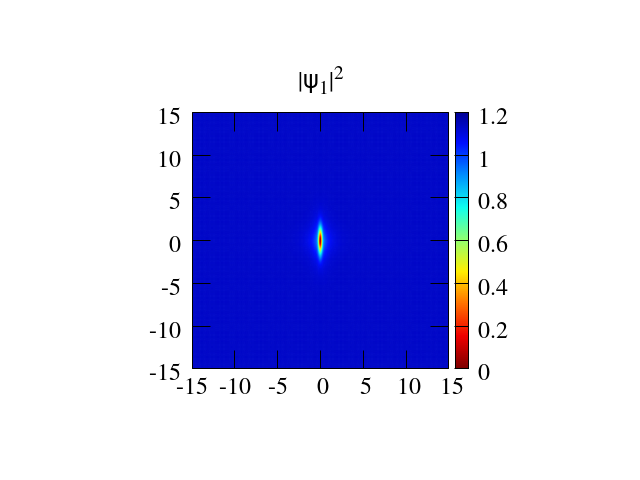}
  \includegraphics[width=0.35\linewidth,trim={5.1cm 2.5cm 3.5cm 1.6cm},clip]{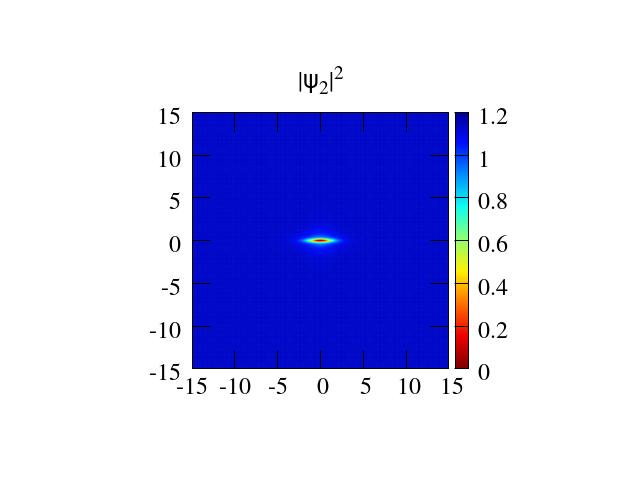}
  \includegraphics[width=0.35\linewidth,trim={5.1cm 2.5cm 3.5cm 1.6cm},clip]{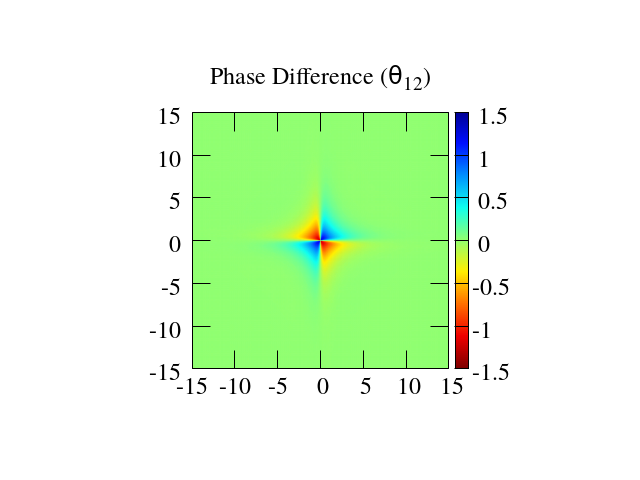}  
  }
 \caption{\label{Fig:charge1opp} (Colour online)
 $N=1$ single quanta numerical solution for strong anisotropy in opposite directions 
 $\lambda_{x1}=\lambda_{y2}=1$, $\lambda^{-1}_{y1}=\lambda^{-1}_{x2} = 0.1$, $\eta_{12} = 0.5$ and 
 $\gamma_{1}=\gamma_{2}=2$. The contour plots are (a) $B_z$ magnetic field 
 (b) $\left|B_z\right| - B_z$ negative magnetic field (c) $\mathcal{E}$ energy density 
 (d) $\left|\phi_1\right|^2$ (e)$\left|\phi_2\right|^2$ (f)$\theta_{12}$ phase difference. }
 \end{figure}

For anisotropy in different directions however, we see a more pronounced effect by the anisotropy, as can be seen in Fig. \ref{Fig:charge1opp} for parameters $\lambda_{x1}=\lambda_{y2}=\lambda$, 
$\lambda_{y1}=\lambda_{x2} = 10\lambda$, 
$\eta_{12} = 0.5$. When compared to a similar solution with anisotropy in a single band in Fig. \ref{Fig:charge1one} the magnetic field inversion is more notable and the deformation of the shape while retaining the $D_4$ symmetry of the free energy is more deformed.

 \subsection{Field inversion beyond the London limit}
We now consider the effect of altering the Josephson coupling strength. If we increase the coupling to become very strong it leads to a diminished disparity of the magnetic field penetration lengths. Thus the magnetic field becomes more localized and the inversion less pronounced. If we take the Josephson coupling to be of a similar scale as the covariant derivative pre-factors, then the anisotropic effects are maximal, as predicted in the London model. Finally taking the Josephson strength to be small we observe the magnetic field becoming continually longer range and the field inversion less pronounced. 

\begin{figure}[tb!]
  \centerline{
  \includegraphics[width=0.35\linewidth,trim={5.1cm 2.5cm 3.5cm 1.6cm},clip]{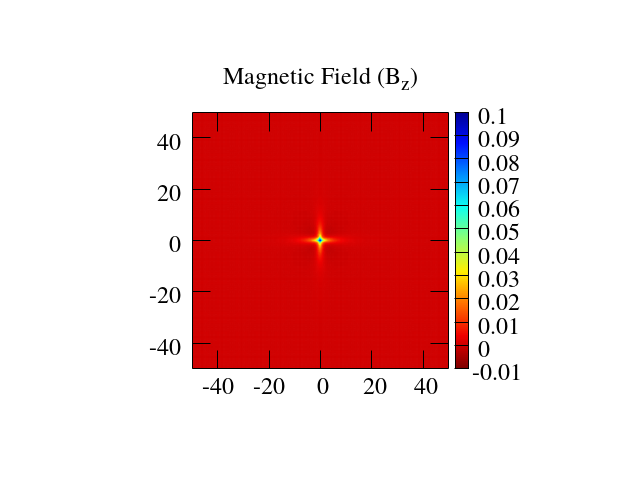}
  \includegraphics[width=0.35\linewidth,trim={5.1cm 2.5cm 3.5cm 1.6cm},clip]{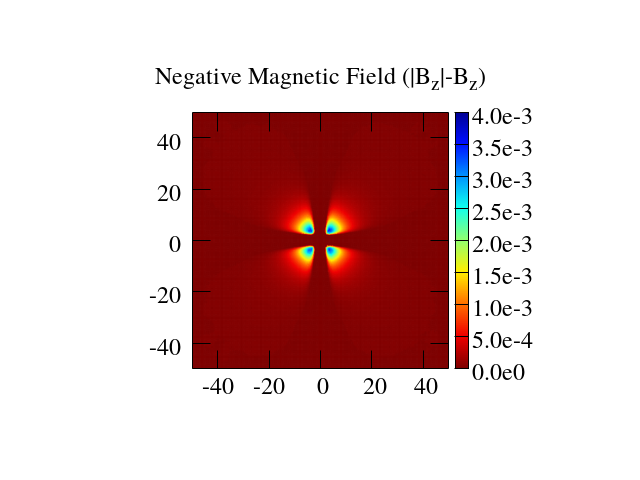}
  \includegraphics[width=0.35\linewidth,trim={5.1cm 2.5cm 3.5cm 1.6cm},clip]{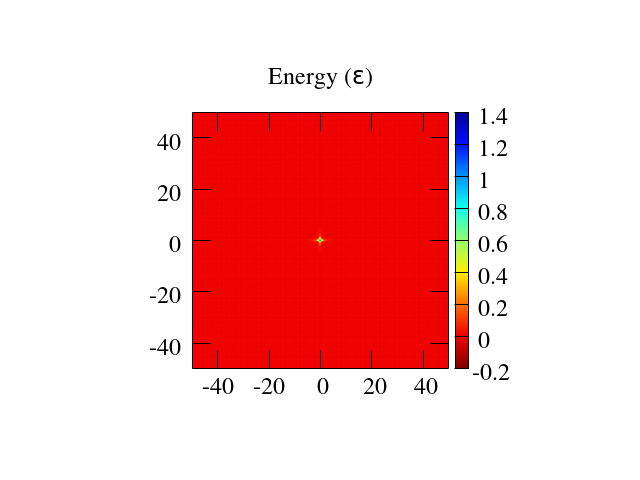}}
  \centerline{
  \includegraphics[width=0.35\linewidth,trim={5.1cm 2.5cm 3.5cm 1.6cm},clip]{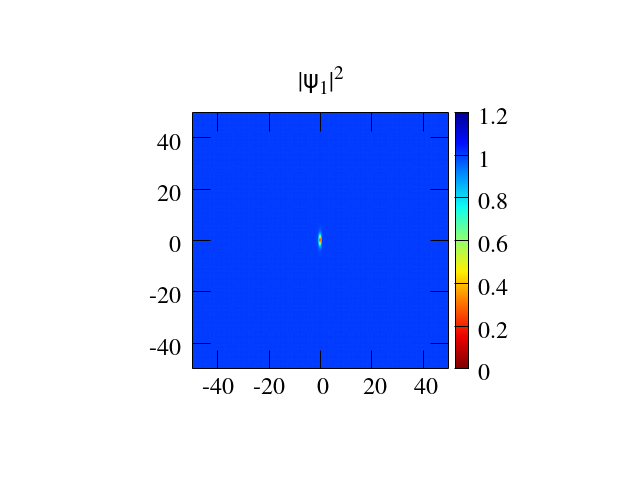}
  \includegraphics[width=0.35\linewidth,trim={5.1cm 2.5cm 3.5cm 1.6cm},clip]{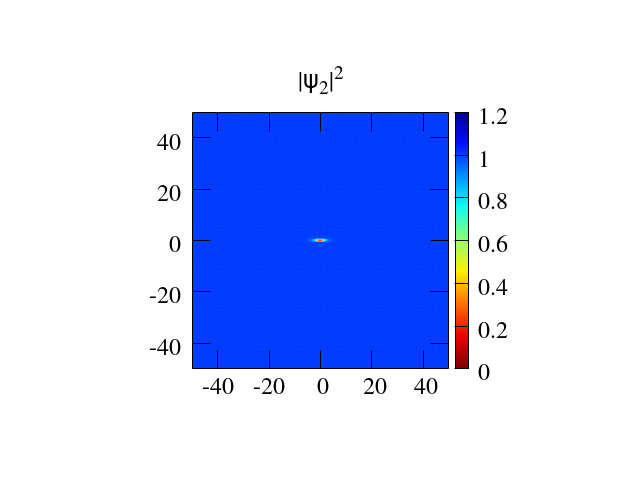}
  \includegraphics[width=0.35\linewidth,trim={5.1cm 2.5cm 3.5cm 1.6cm},clip]{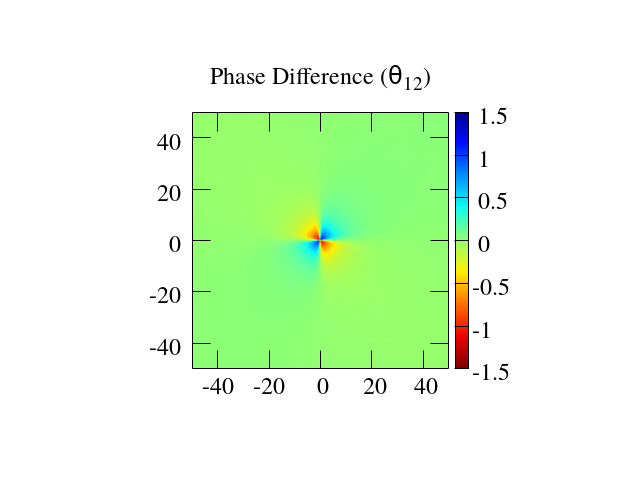}}
  \caption{\label{Fig:nok} (Colour online)
 $N=1$ single quanta numerical solution for strong anisotropy in opposite directions but with no Josephson coupling 
 $\gamma_1=\gamma_2=2$, $\eta_{12}=0.0$, $\lambda_{x1}=\lambda_{y2}=1$ and 
 $\lambda^{-1}_{y1}=\lambda^{-1}_{x2} = 0.1$. Note that the grid size of $50\times 50$ is due to the long range nature of the negative magnetic field. (a) $B_z$ magnetic field (b) $\left|B_z\right| - B_z$ negative magnetic field 
 (c) $\mathcal{E}$ energy density (d) $\left|\phi_1\right|^2$ (e)$\left|\phi_2\right|^2$ (f)$\theta_{12}$ phase difference. }
 \end{figure}

In the limit of zero Josephson coupling the London model predicts that the Leggett mode decouples from the magnetic field. That is, as $\eta_{12}\rightarrow 0$ one of the modes becomes zero and the magnetic field decay is described  by a single exponential, hence the London model predicts absence of magnetic field inversion. However this is not what is found in the full model as can be seen in Fig. \ref{Fig:nok} for the parameters $\gamma_1=\gamma_2=2$, $\eta_{12}=0.0$, $\lambda_{x1}=\lambda_{y2}=\lambda$ and $\lambda_{y1}=\lambda_{x2} = 10\lambda$. 
Here we see that the magnetic field still exhibits inversion, though at very long range. This result refutes the applicability of the London model for this regime.

The origin of this behaviour must be the interplay of anisotropy with an additional effect that appears beyond the London model. It has been proposed that in the isotropic Ginzburg-Landau model, the magnetic response has the form of a massive vector field coupled to Faddeev-Skyrme terms \cite{Babaev2002b,Babaev2009a}.
The Faddeev-Skyrme terms represent magnetic field contribution that is generated by cross-gradients of relative densities and relative phases of components. This leads to magnetic field inversion in fractional vortex solutions \cite{Babaev2009}, 
however in the standard isotropic model, for axially symmetric integer flux vortices, solutions do not have gradients of relative phases and the effect is absent. Hence in the anisotropic case, the self-induced phase differences and relative density gradients lead to magnetic field inversion even in the case of zero Josephson coupling, beyond the London limit. For finite Josephson coupling this effect coexists with the contributions discussed above in the London model.
  

 \section{Vortex bound states} 
 \label{Sec:VortexBoundStates}
In this section we consider inter-vortex interactions, that are likely to lead to non-trivial multi-solitons, due to the non-monotonic nature of the magnetic field and the property of field inversion. If we return to the London-Leggett model energy formulated in Eq. (\ref{Eq:LondonEnergy}), we can expand with respect to the key terms 
$\theta_{12}$, $\boldsymbol{p}_s$ and $\boldsymbol{B}$,
\begin{eqnarray}
\varepsilon &=& \left(\frac{\Phi_0}{2\pi}\right)^2\left[\frac{1}{2}\lambda_L^{-2}\boldsymbol{\nabla}\theta_{12}\cdot \boldsymbol{\nabla}\theta_{12} + 2\lambda_L^{-2}\boldsymbol{p}_s \cdot \boldsymbol{p}_s \nonumber \right.\\ & & +  \left. 2\left(\lambda_1^{-2} - \lambda_2^{-2}\right)\boldsymbol{\nabla}\theta_{12}\right] + E_J \theta_{12}^2 + \bm B^2. \label{Eq:SepEnergy}
\end{eqnarray}

We now want to find the interaction energy of two composite vortices with winding in both components in the London model. We assume that they are well separated, such that we can write the various terms as the sums of the tail interactions of the two solitons, $\boldsymbol{B} = \boldsymbol{B}^{(1)} + \boldsymbol{B}^{(2)}$, $\theta_{12} = \theta_{12}^{(1)} + \theta_{12}^{(2)}$ and $\boldsymbol{p}_s = \boldsymbol{p}_s^{(1)} + \boldsymbol{p}_s^{(2)}$. If we then separate and integrate equation \ref{Eq:SepEnergy} by parts, we can use the equations of motion to reduce the interaction energy to the simple form,
\begin{equation}
\varepsilon_{int} = 2\Phi_0\left[\boldsymbol{B}^{(1)}(\boldsymbol{x}_2) \cdot \boldsymbol{n}_{v2} + \boldsymbol{B}^{(2)}(\boldsymbol{x}_1) \cdot \boldsymbol{n}_{v1}\right]
\label{Eq:IntEnergy}
\end{equation}
Hence, by substituting in the form of the parallel field around a single composite vortex given in Eq. (\ref{Eq:BZFin}), we can calculate the total interaction energy of a given configuration. For the system in question, the only required data to represent a given configuration is then the positions of each individual composite vortex or a collection of $2n$ parameters, where $n$ is the total number of quanta or winding number of the system. We can now minimise the interaction energy for a given winding number over the $2n$-dim space of positions, to find the optimal configurations for type 2 composite vortices in the London model. Eq.(\ref{Eq:IntEnergy}) was minimised using a simulated annealing method. 

\begin{figure}[tb!]
  \centerline{
  \includegraphics[width=0.2\linewidth]{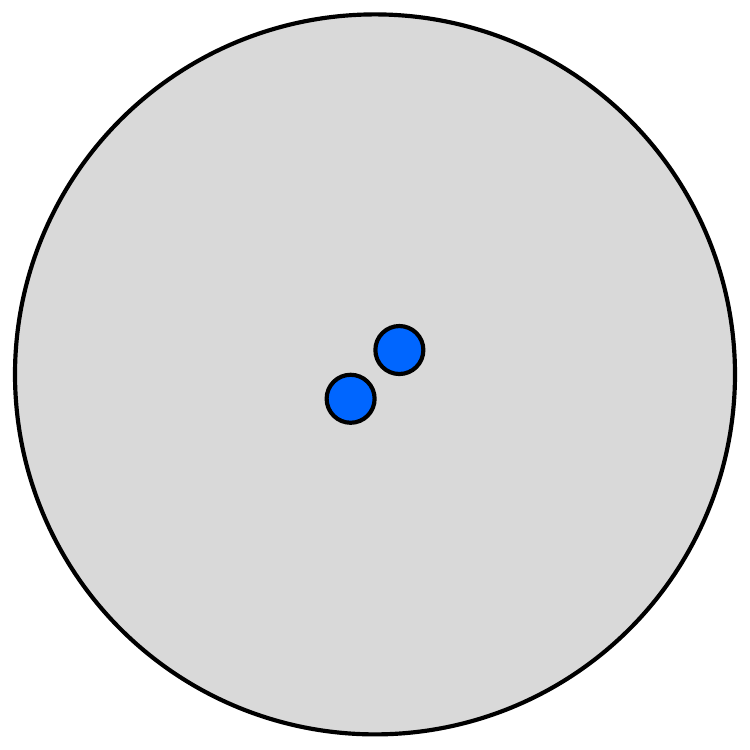}  
  \includegraphics[width=0.2\linewidth]{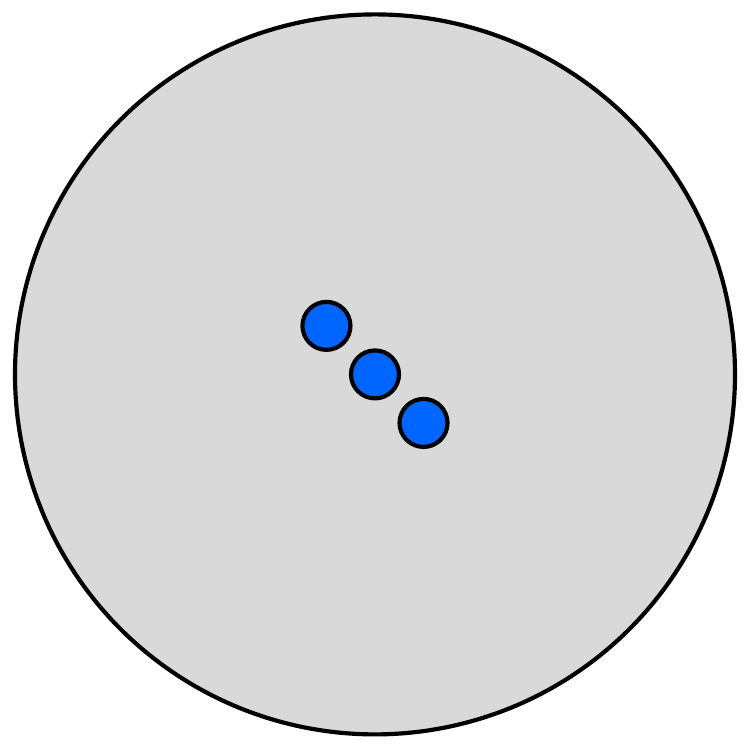}
  \includegraphics[width=0.2\linewidth]{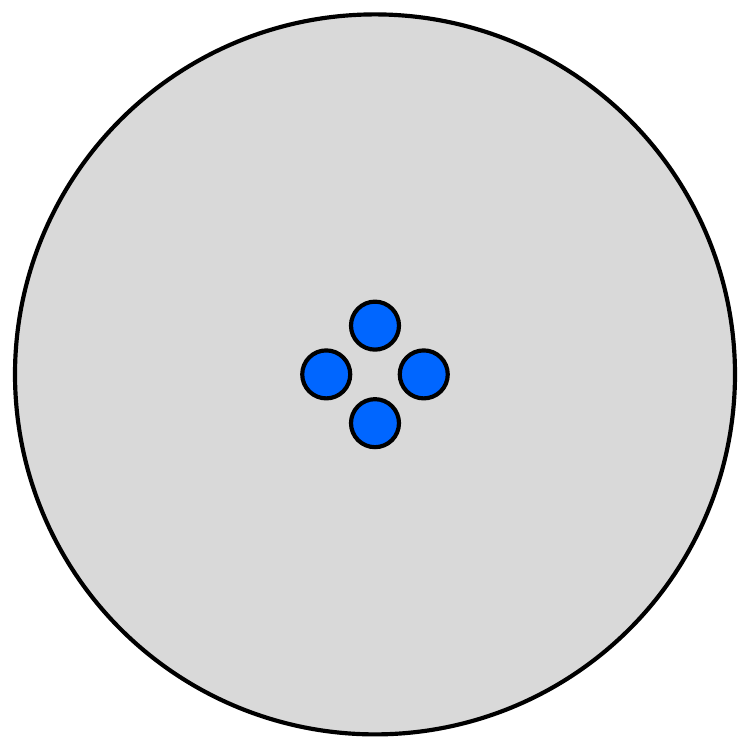}
  \includegraphics[width=0.2\linewidth]{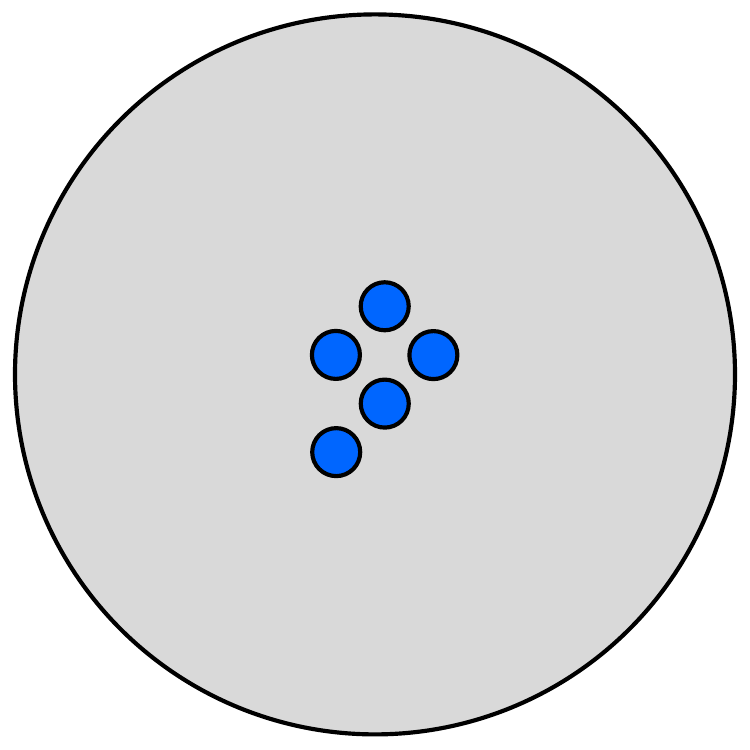}
  \includegraphics[width=0.2\linewidth]{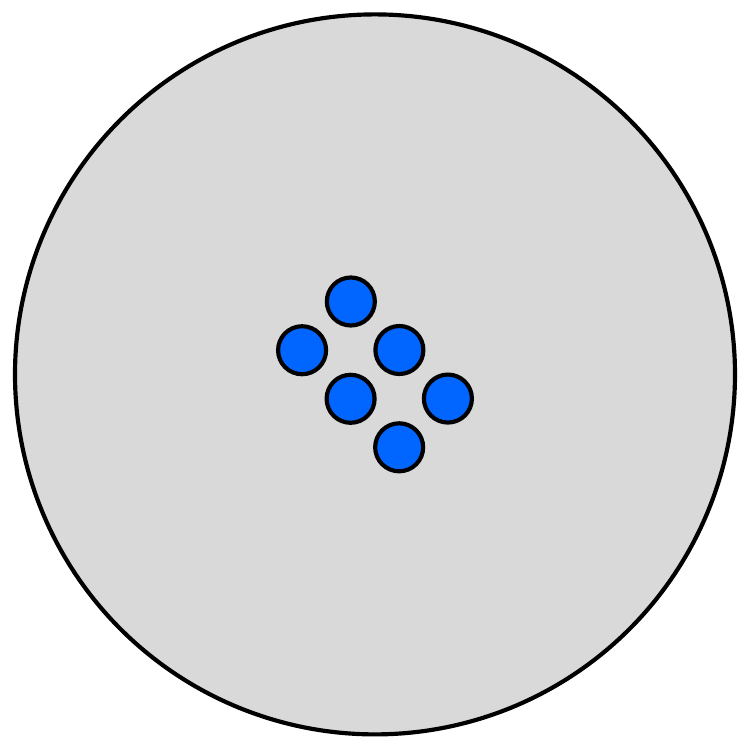}}  
  \centerline{
  \includegraphics[width=0.2\linewidth]{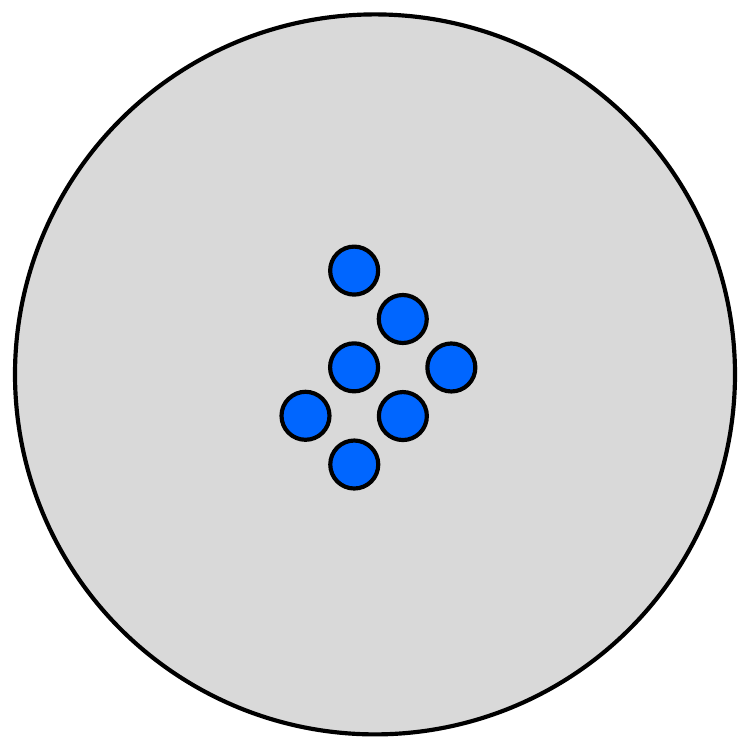}
  \includegraphics[width=0.2\linewidth]{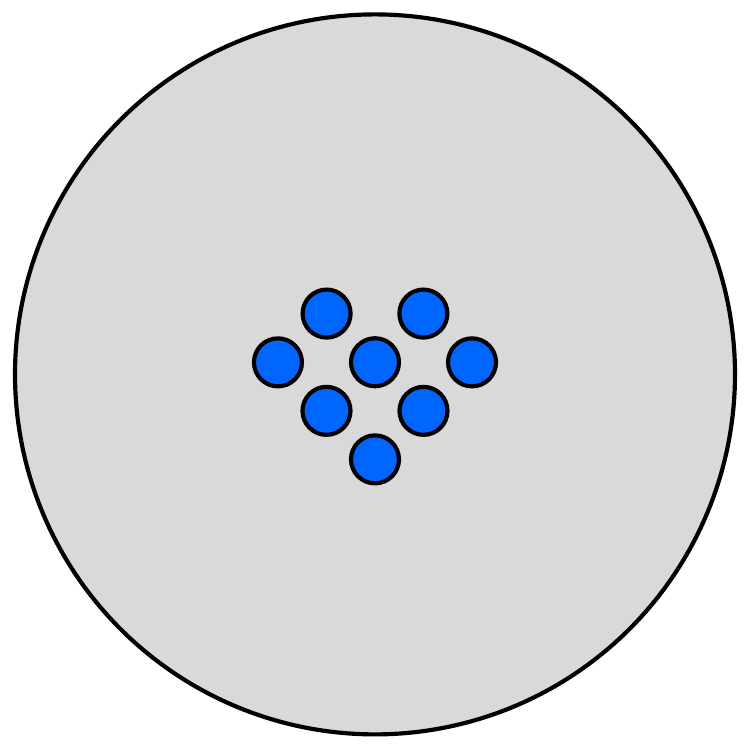}
  \includegraphics[width=0.2\linewidth]{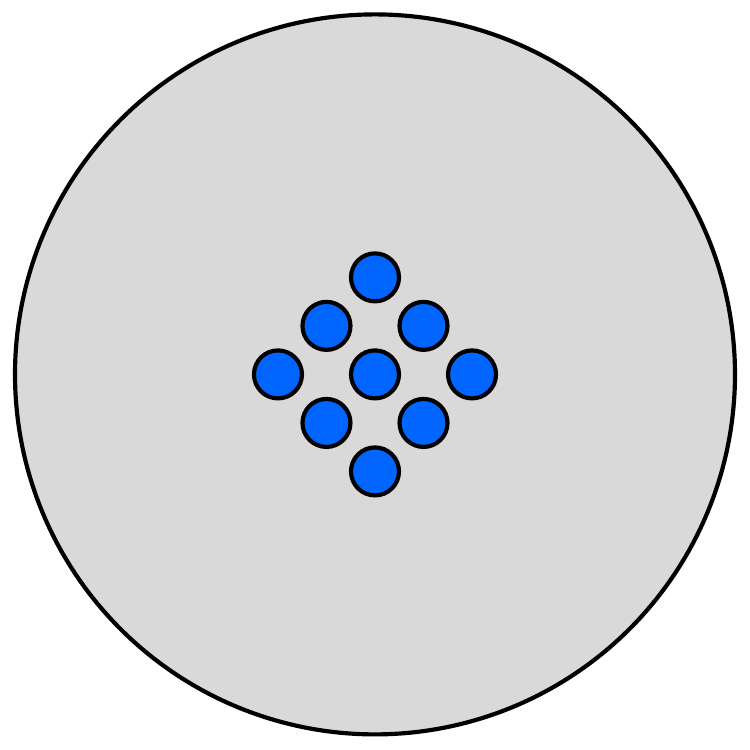}
  \includegraphics[width=0.2\linewidth]{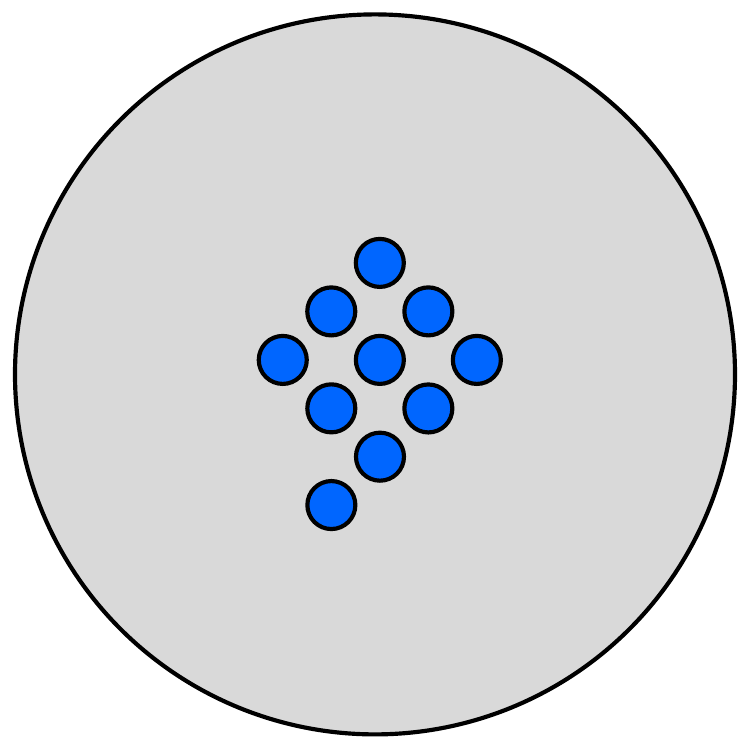}}
 \caption{\label{Fig:approxoppweak} Minimal energy multi-quanta solution to the London model with weak anisotropy in both bands, in opposite directions, found using simulated annealing for composite vortices. The parameters are $\lambda_{1x}=\lambda_{2y}=\lambda$, 
 $\lambda_{2x}=\lambda_{1y}=0.5\lambda$ and $k_0 = 0.7/\lambda$. } 
 \end{figure}

\begin{figure}[tb!]
\centerline{
\includegraphics[width=0.2\linewidth]{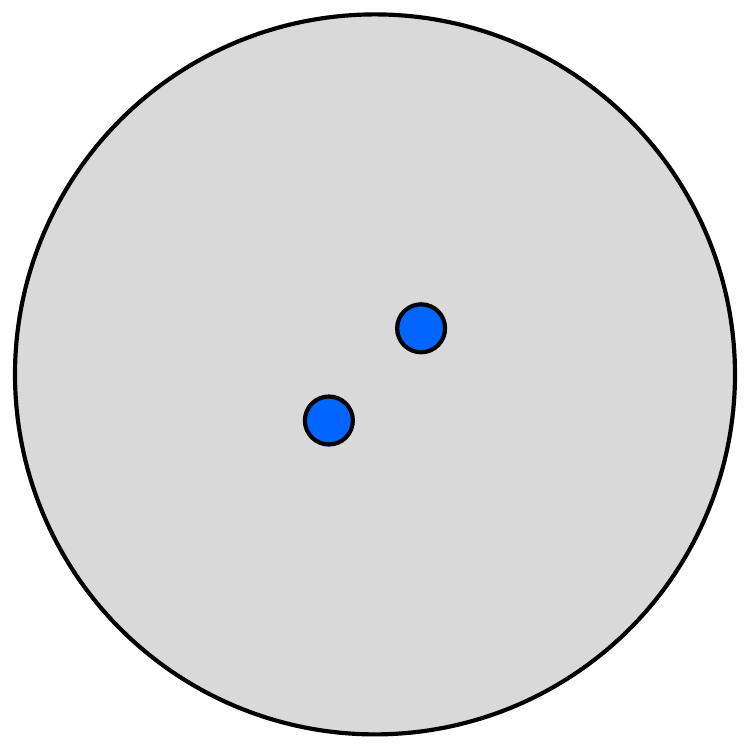}
\includegraphics[width=0.2\linewidth]{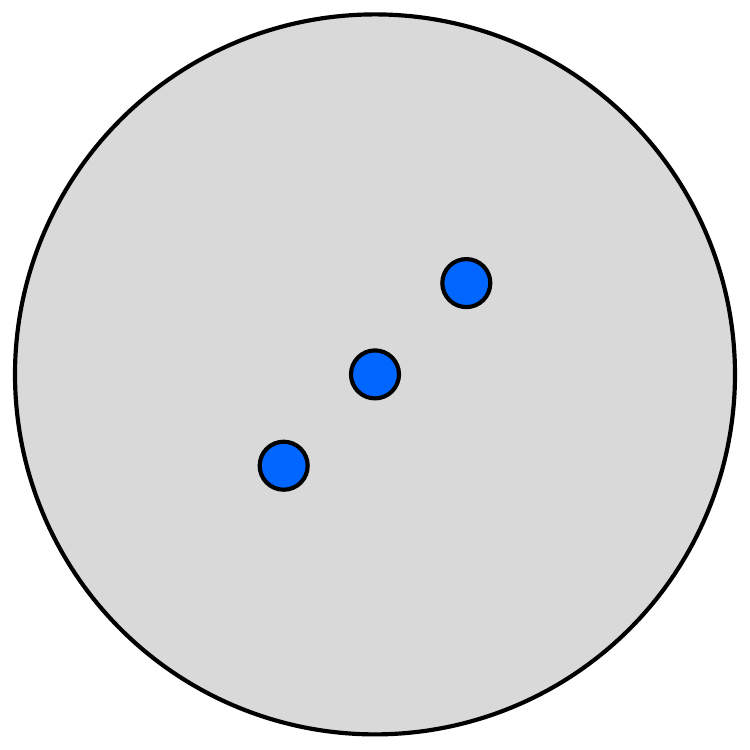}
\includegraphics[width=0.2\linewidth]{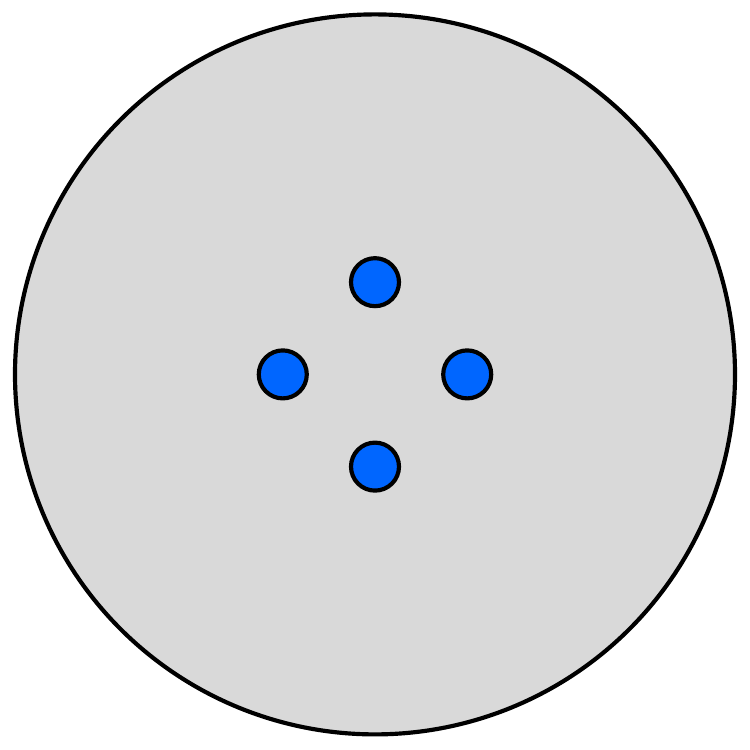}
\includegraphics[width=0.2\linewidth]{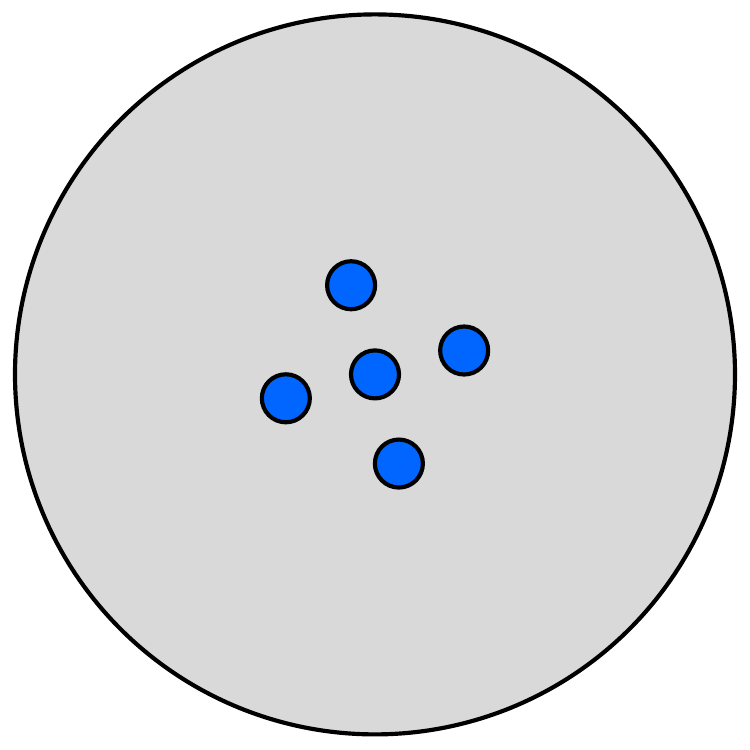}
\includegraphics[width=0.2\linewidth]{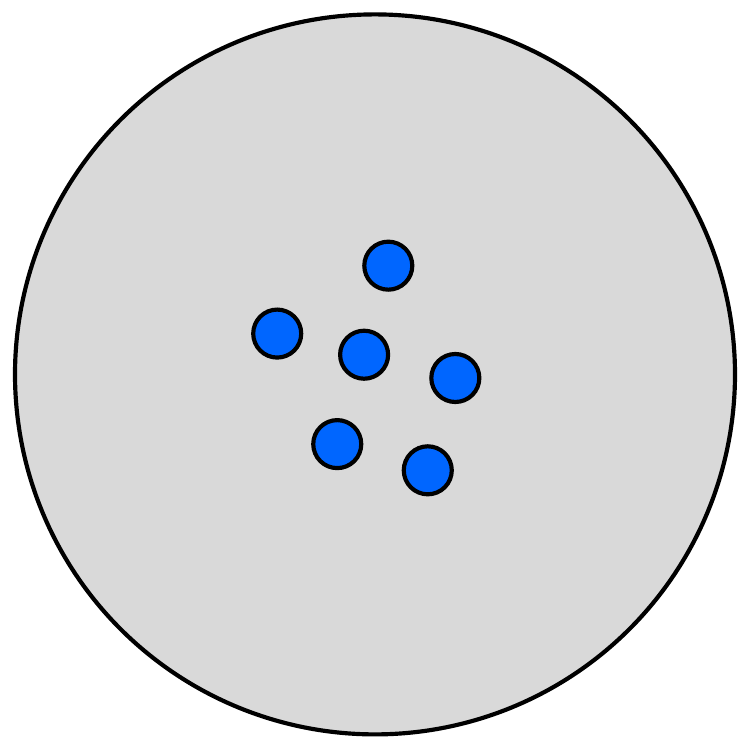}} 
 \centerline{
 \includegraphics[width=0.2\linewidth]{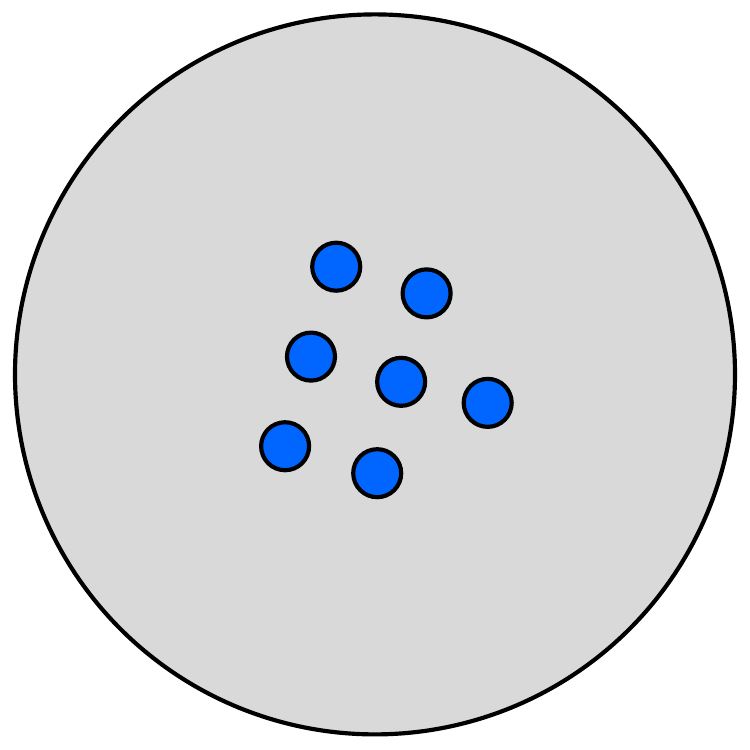} 
 \includegraphics[width=0.2\linewidth]{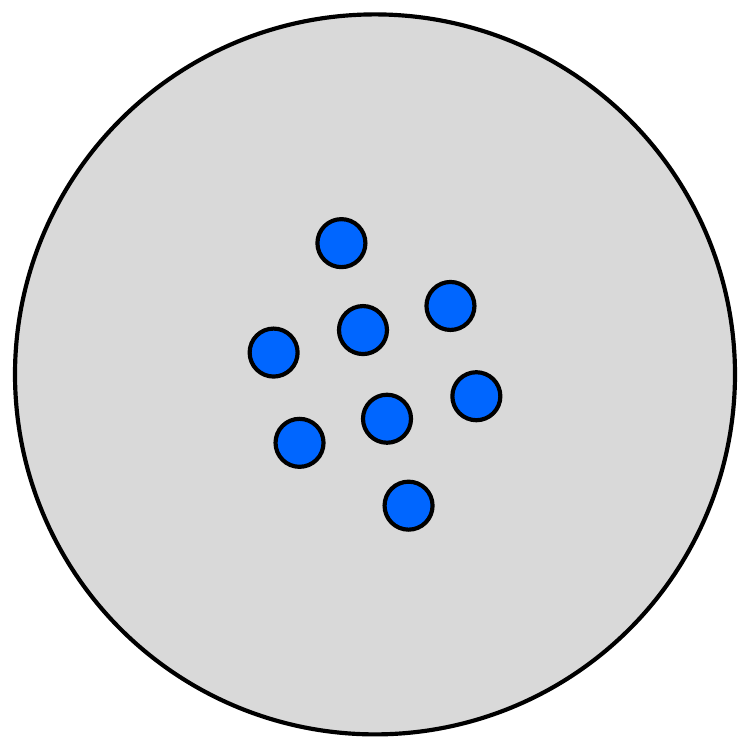}
 \includegraphics[width=0.2\linewidth]{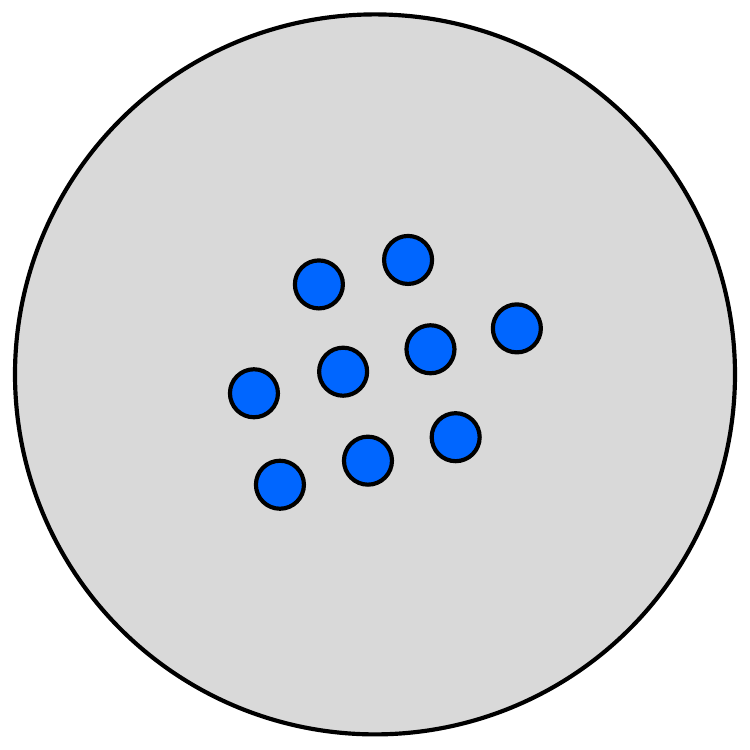}
 \includegraphics[width=0.2\linewidth]{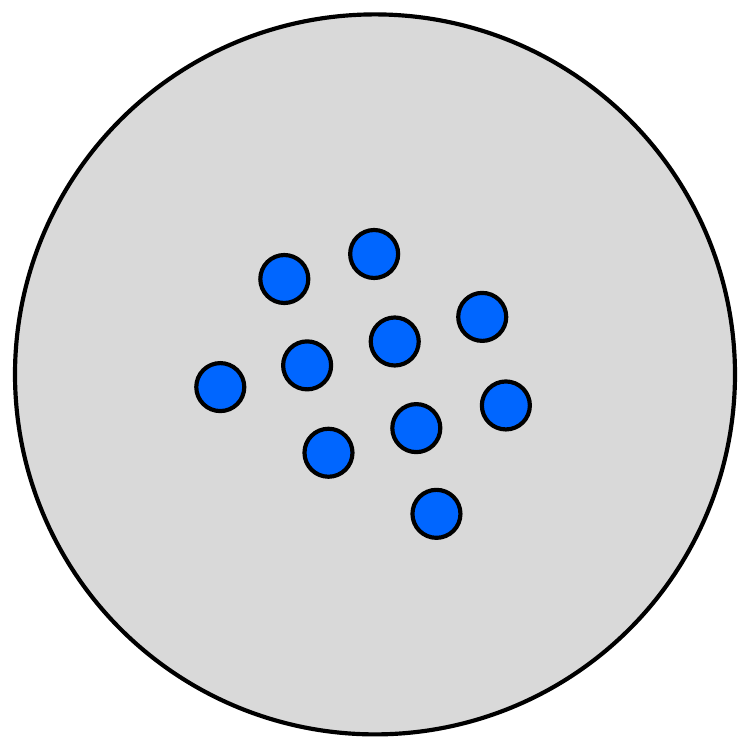}}
 \caption{\label{Fig:approxoppnormal} Minimal energy multi-quanta solution to the London model with stronger anisotropy in both bands, in opposite directions, found using simulated annealing for composite vortices. The parameters are 
 $\lambda_{1x}=\lambda_{2y}=\lambda$, $\lambda_{2x}=\lambda_{1y}=0.3\lambda$ and $k_0 = 0.7/\lambda$. } 
\end{figure}

\begin{figure}[tb!]
\centerline{\includegraphics[width=0.2\linewidth]{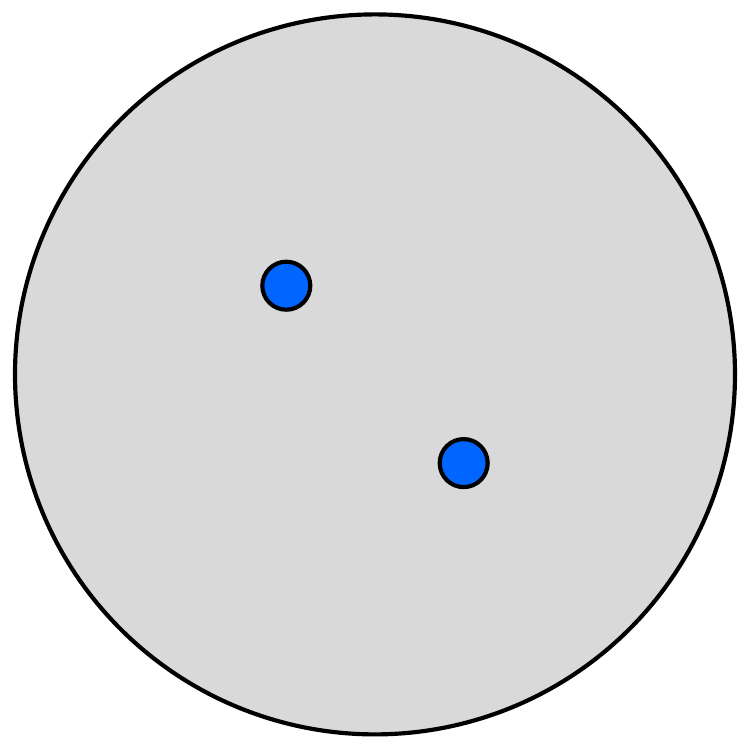}\includegraphics[width=0.2\linewidth]{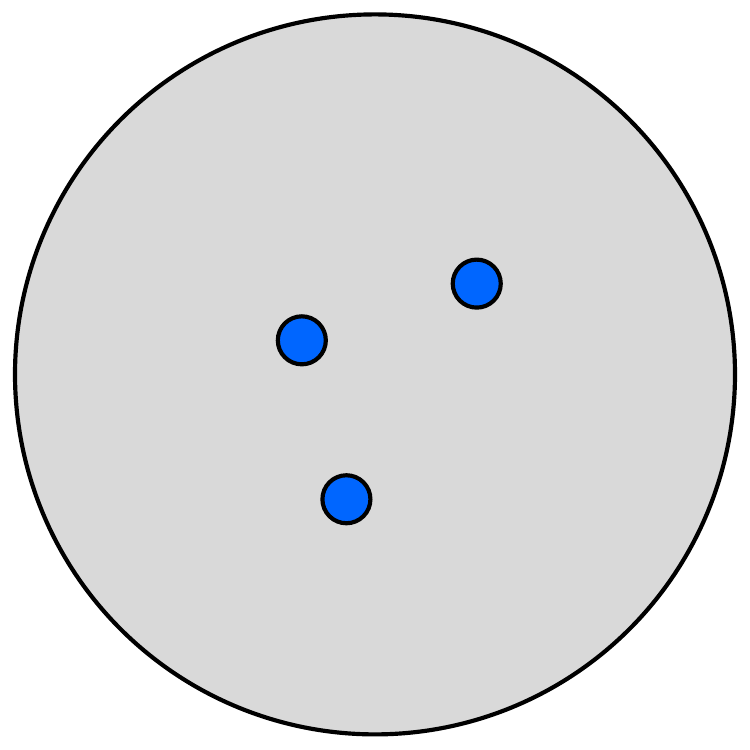}\includegraphics[width=0.2\linewidth]{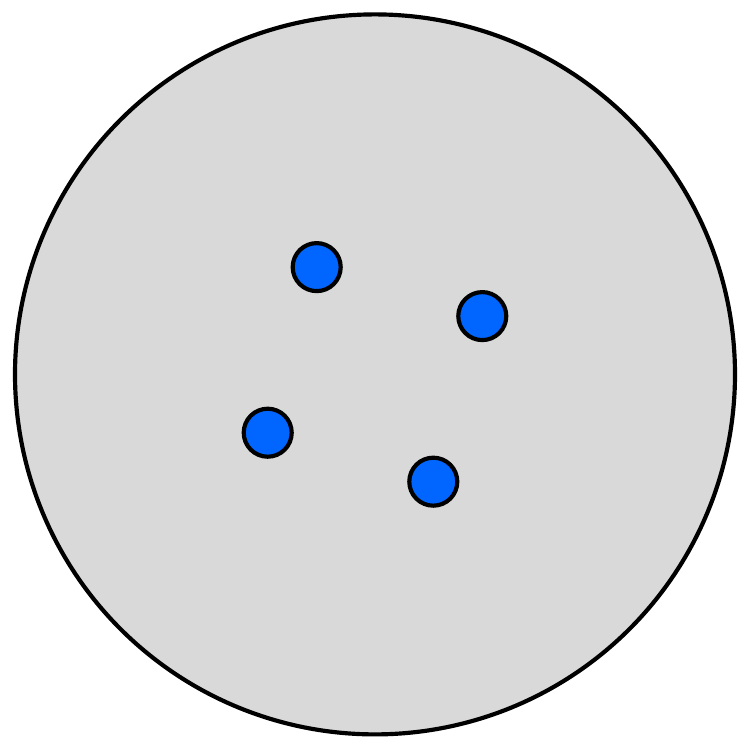}\includegraphics[width=0.2\linewidth]{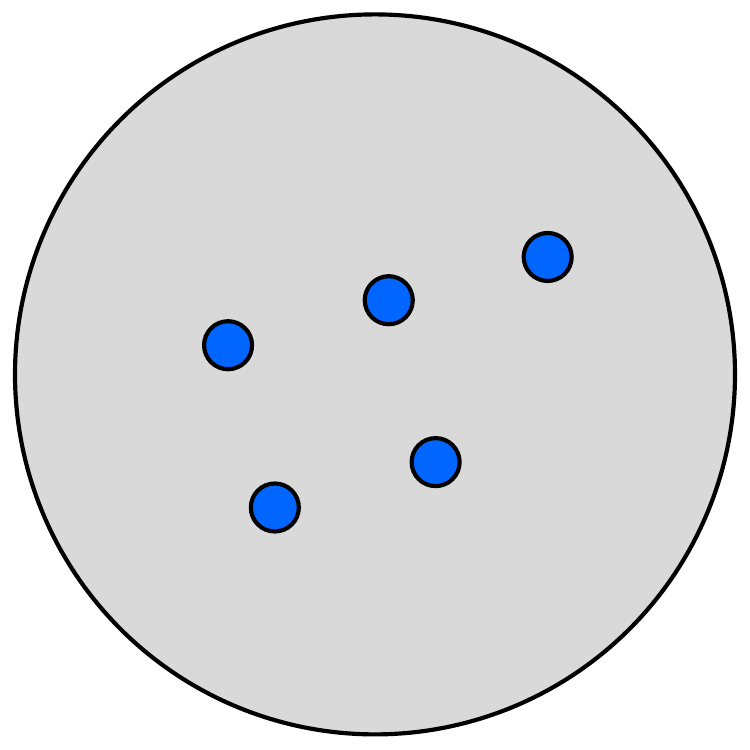}\includegraphics[width=0.2\linewidth]{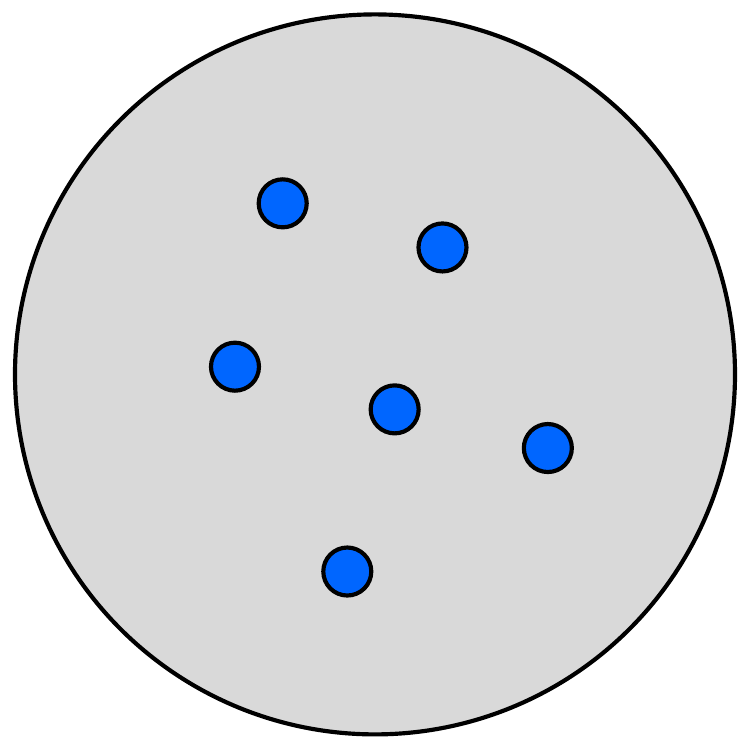}}  \centerline{\includegraphics[width=0.2\linewidth]{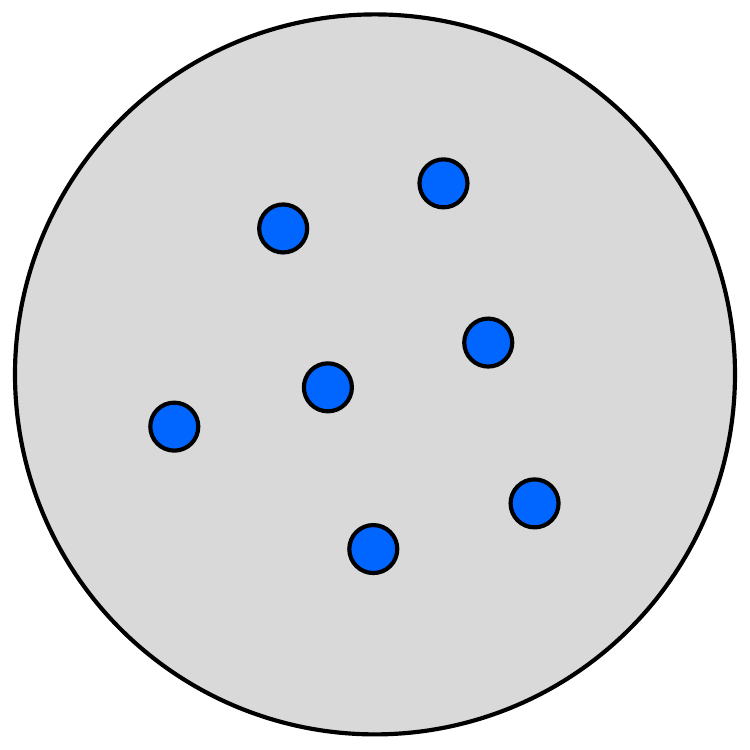}\includegraphics[width=0.2\linewidth]{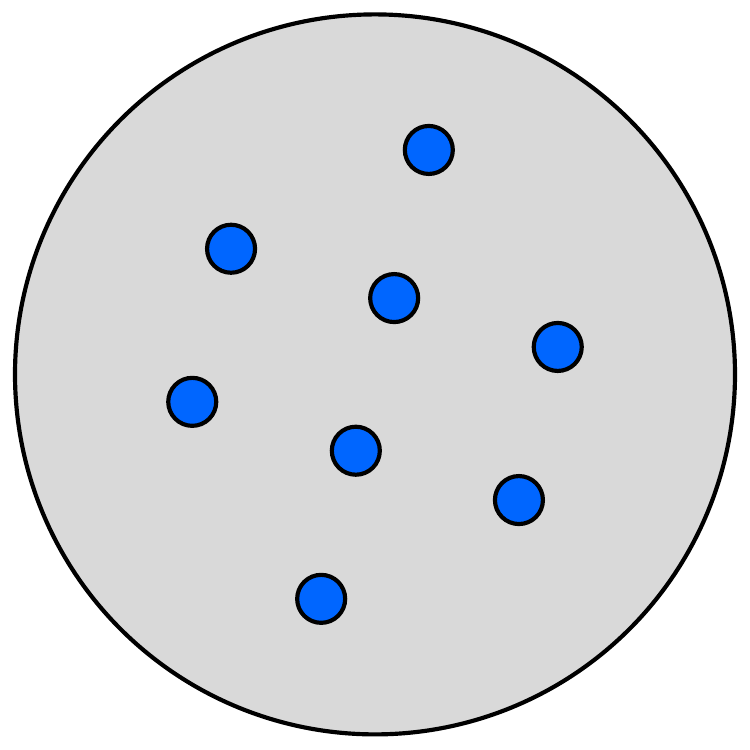}\includegraphics[width=0.2\linewidth]{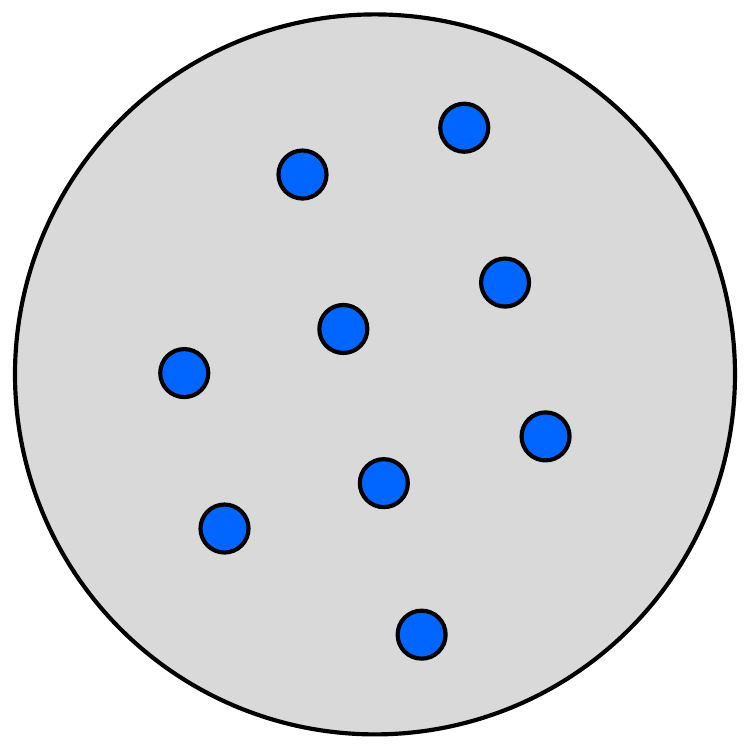}\includegraphics[width=0.2\linewidth]{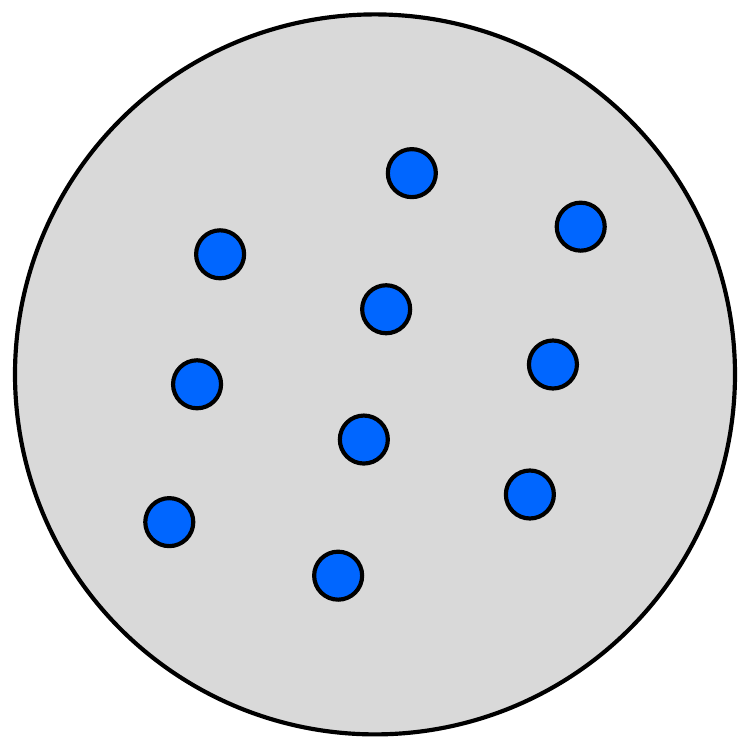}}
\caption{
\label{Fig:approxoppstrong} Minimal energy multi-quanta solution to the London model with stronger anisotropy in both bands, in opposite directions, found using simulated annealing for composite vortices. The parameters are 
$\lambda_{1x}=\lambda_{2y}=\lambda$, $\lambda_{2x}=\lambda_{1y}=0.1\lambda$ and $k_0 = 0.7/\lambda$. } 
\end{figure}

We first consider the solutions for equal and opposite anisotropies for various strengths of anisotropy, some of which are plotted in Fig.s \ref{Fig:approxoppweak}, \ref{Fig:approxoppnormal} and \ref{Fig:approxoppstrong}. If we consider the solution for weaker anisotropies, presented in Fig. \ref{Fig:approxoppweak} for $\lambda_{1x}=\lambda_{2y}=0.5\lambda$, 
$\lambda_{2x}=\lambda_{1y}=\lambda$ and $k_0 = 0.84/\lambda$, we see that the form of the minimal energy solutions is that of polyominoes (geometric plane Fig.s formed by connecting $n$ sqaures along their edges, each square representing a $D_4$ symmetric composite vortex). The rule for the minimal energy polyominoe is then the one that maximises the total number of neighbours for all the vortices. This is no surprise, as the chosen anisotropy leads to a $D_4$ dihedral symmetry to the magnetic field density of the vortex. To minimise the form of the interaction energy \ref{Eq:IntEnergy}, a good candidate is placing the vortex positions into the maximal negative magnetic field locations of the other composite vortices. This would suggest that local minima should appear for each of the various polyominoes in the London model. Similar types of interaction and corresponding minimal soliton configurations have been studied before in the baby Skyrme model \cite{Jennings2014}, 
 where it appears for different reasons.

Increasing the strength of the anisotropy to $\lambda_{1x}=\lambda_{2y}=0.3\lambda$, 
$\lambda_{2x}=\lambda_{1y}=\lambda$ and $k_0 = 0.84/\lambda$ the polyominoes pattern no longer applies and the form of the minimal energy solutions are more complex, as can be seen in Fig. \ref{Fig:approxoppnormal}. Up to $n=4$ we see similar solutions to before, but for $n=5$ we see a rotated $N=4$ solutions with an additional vortex in the centre of the configuration. As the winding number increases a pattern emerges, that of chains that are just off the $\frac{\pi}{4}$ diagonal, interlaced with each other, such that the chains are staggered.

Finally if we increase the strength of anisotropy to extremely strong values, we get a continuation of the above to more extreme behaviours. This can be seen in Fig. \ref{Fig:approxoppstrong} for $\lambda_{1x}=\lambda_{2y}=0.1\lambda$,
$\lambda_{2x}=\lambda_{1y}=\lambda$ and $k_0 = 0.84/\lambda$. We now see the above shifting away from the 
polyominoe form for $n\geq 3$.

\begin{figure}[tb!]
\centerline{\includegraphics[width=0.2\linewidth]{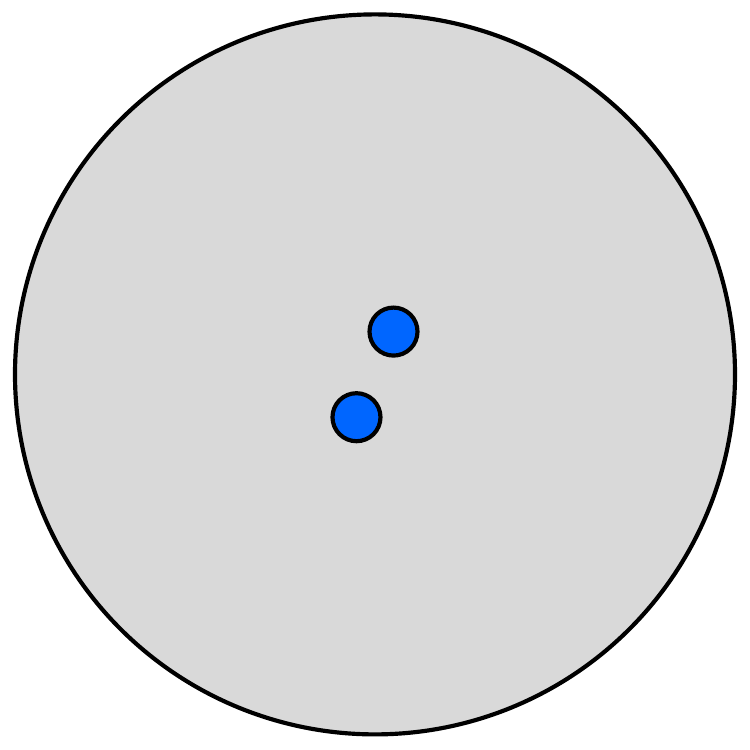}\includegraphics[width=0.2\linewidth]{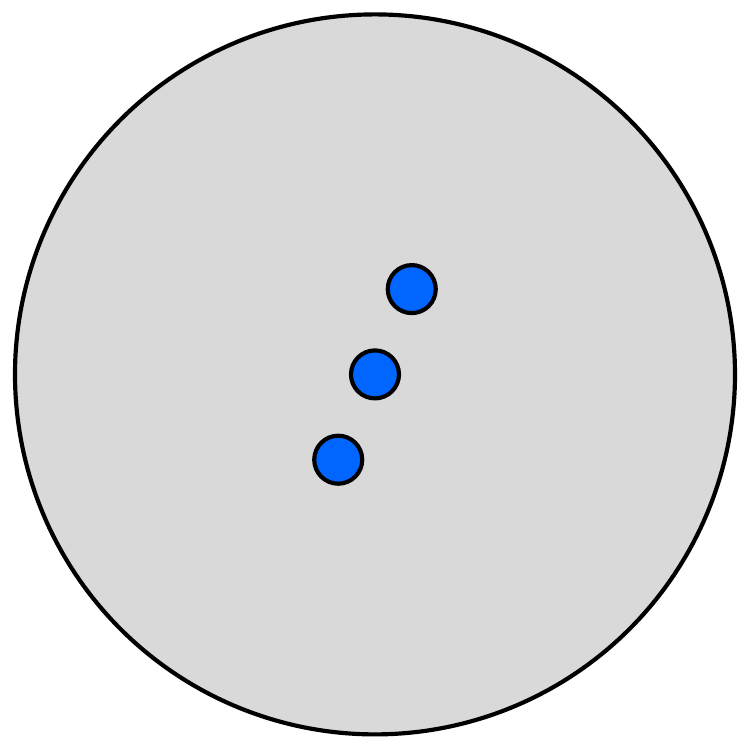}\includegraphics[width=0.2\linewidth]{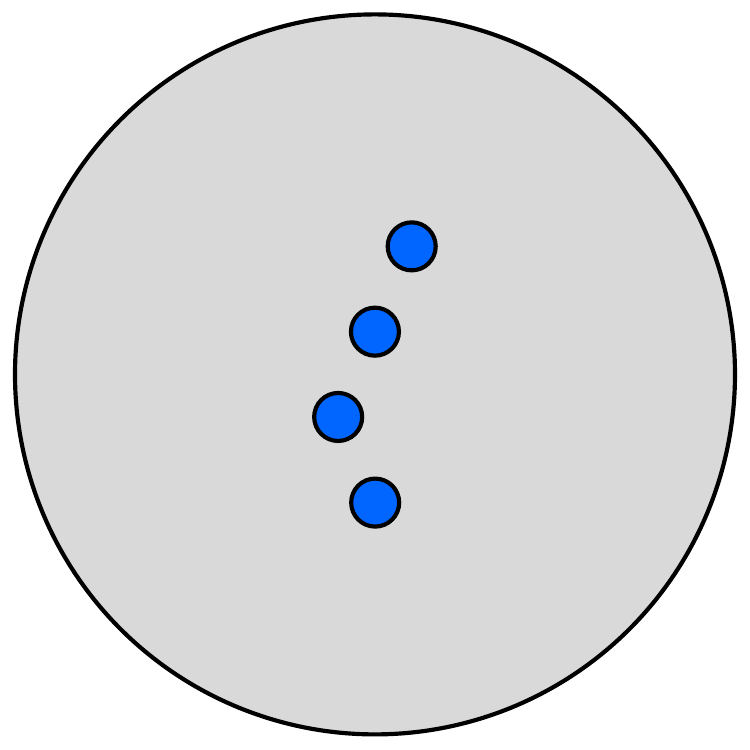}\includegraphics[width=0.2\linewidth]{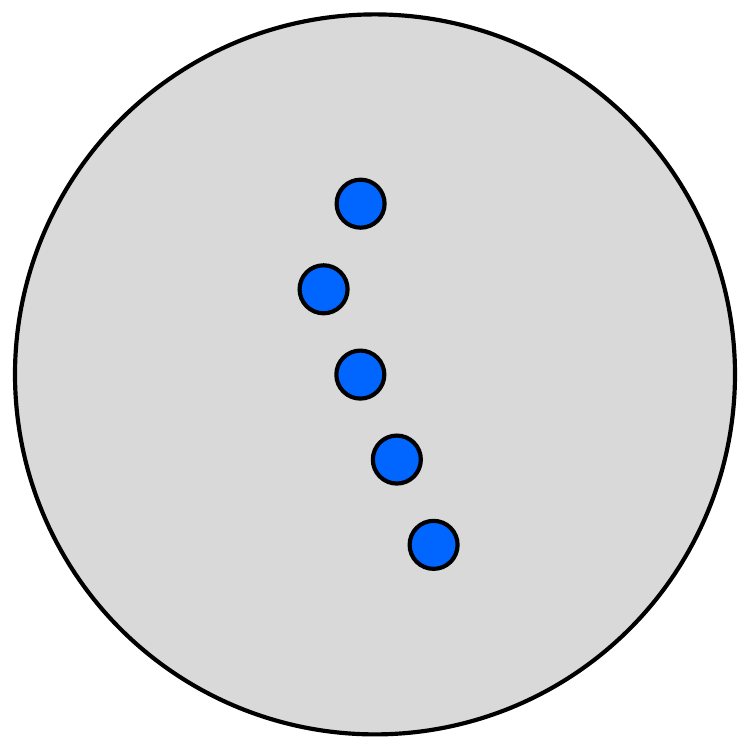}\includegraphics[width=0.2\linewidth]{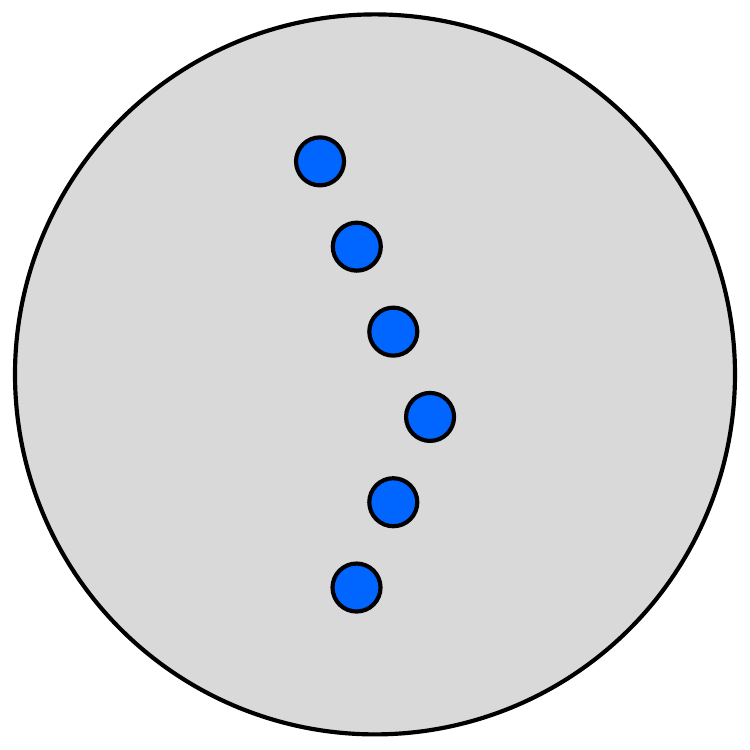}}  \centerline{\includegraphics[width=0.2\linewidth]{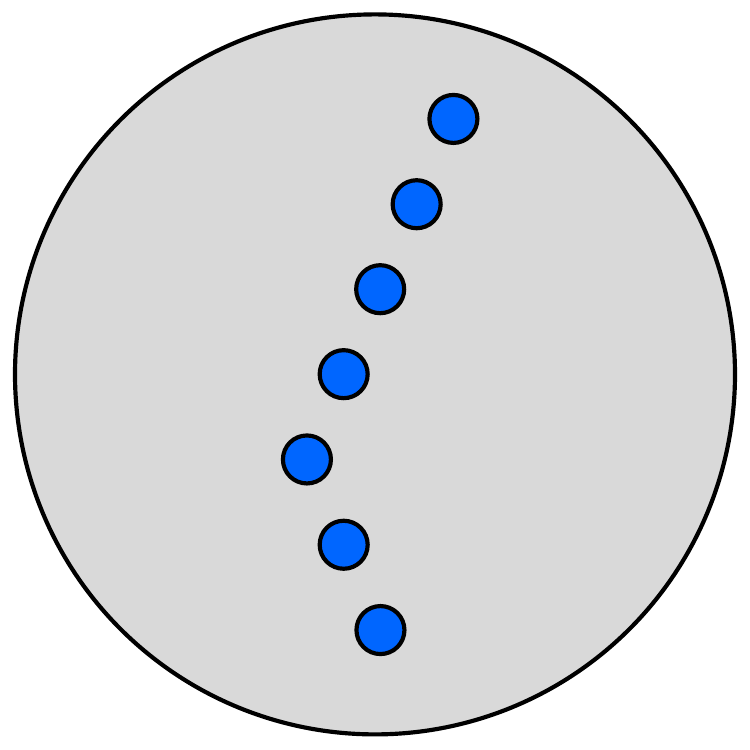}\includegraphics[width=0.2\linewidth]{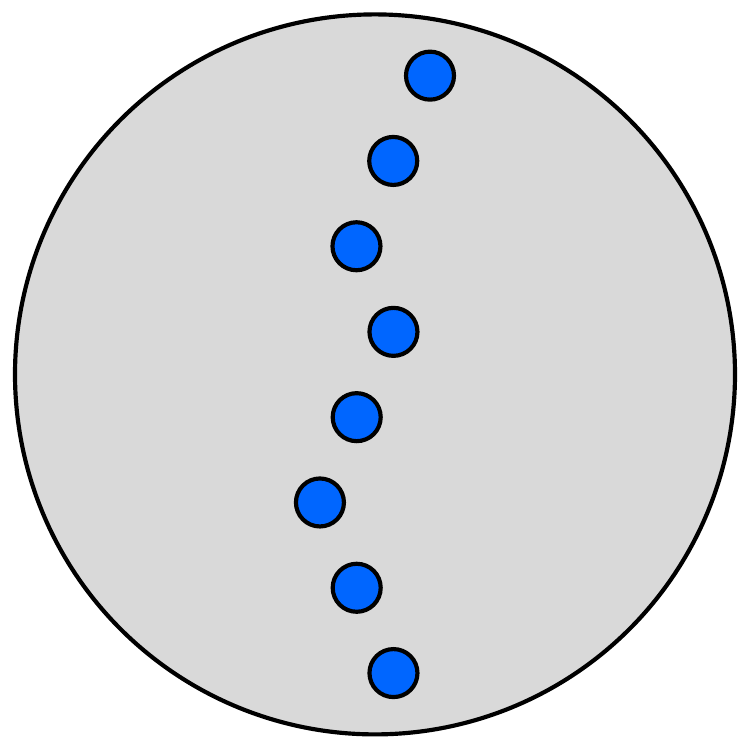}\includegraphics[width=0.2\linewidth]{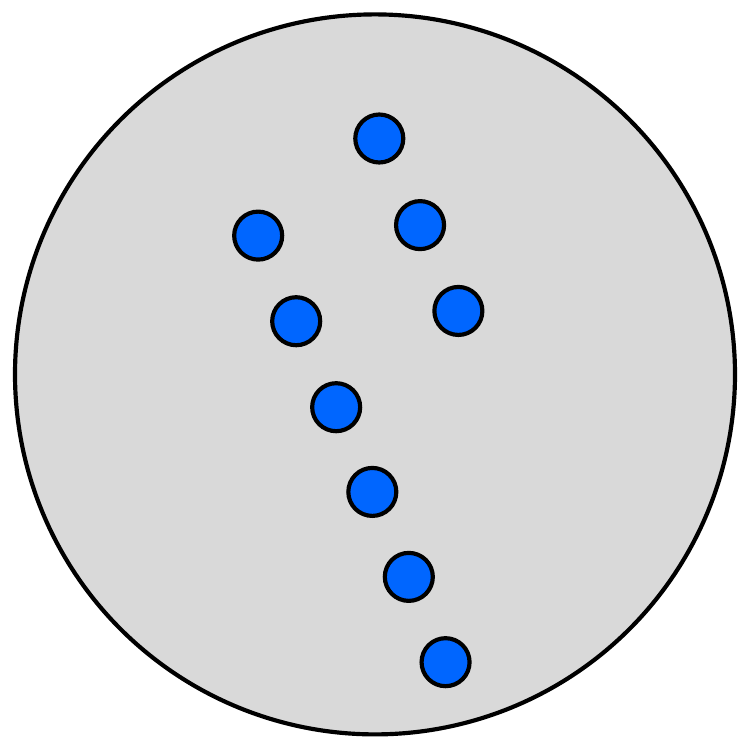}\includegraphics[width=0.2\linewidth]{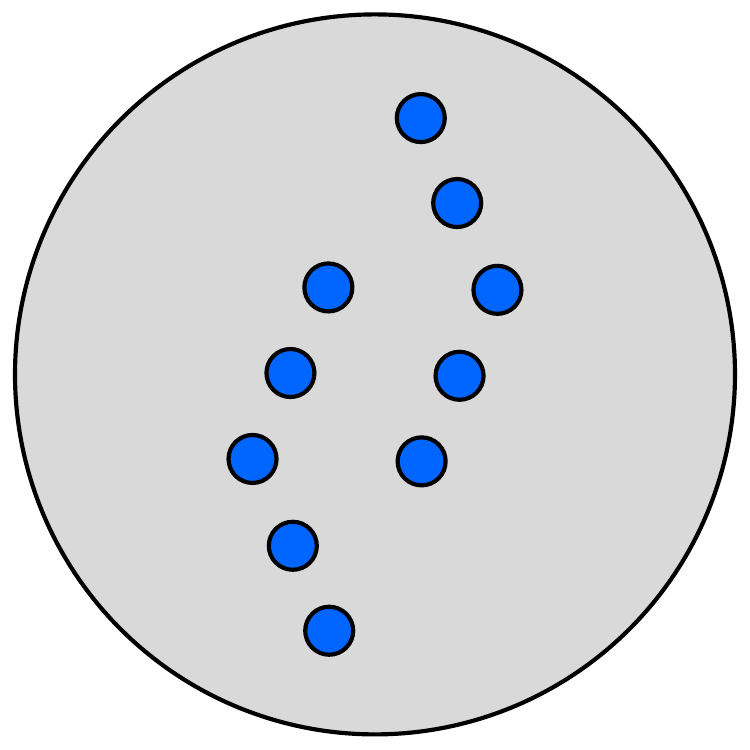}}
\caption{\label{Fig:approxonestrong} Minimal energy multi-quanta solution to the London model with stronger anisotropy in one band, found using simulated annealing for composite vortices. The parameters are $\lambda_{1x}=\lambda_{2y}=\lambda_{1y}=\lambda$, 
$\lambda_{2x}=0.3\lambda$ and $k_0 = 0.7/\lambda$.} 
\end{figure}

If we now consider anisotropy in a single direction we get a very different result, as shown in Fig. \ref{Fig:approxonestrong} 
for $\lambda_{1x}=0.3\lambda$, $\lambda_{1y}=\lambda_{2x}=\lambda_{2y}=\lambda$ and $k_0 = 0.84/\lambda$. Here we see that the form of the solutions is now of chains. As the winding number increases the chains develop kinks, looking at the solution for $n=4$ this is due to the additional vortex being far enough away from the one directly above it to interact weakly, but close enough to the one at the tip of the chain to be affected by the negative magnetic field which is longer range. The chains form on a line with an angle to the x-axis determined by the form of the anisotropy. As the winding number increases further, the interlacing effect of the chains appears as before.

If we were to consider more complicated forms of anisotropy it is likely the minimal energy solution will take some hybrid of the two presented above based upon the how warped the symmetry is away from the maximal $D_4$.

\begin{figure}[tb!]
\centerline{
 \includegraphics[width=0.36\linewidth,trim={5.1cm 2.5cm 3.5cm 1.6cm},clip]{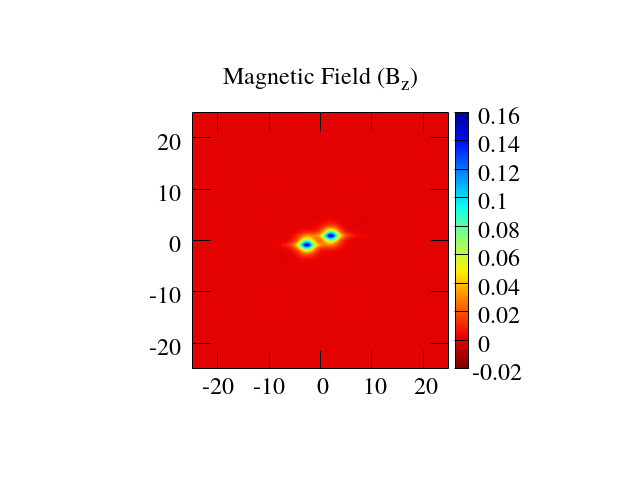}  
 \includegraphics[width=0.36\linewidth,trim={5.1cm 2.5cm 3.5cm 1.6cm},clip]{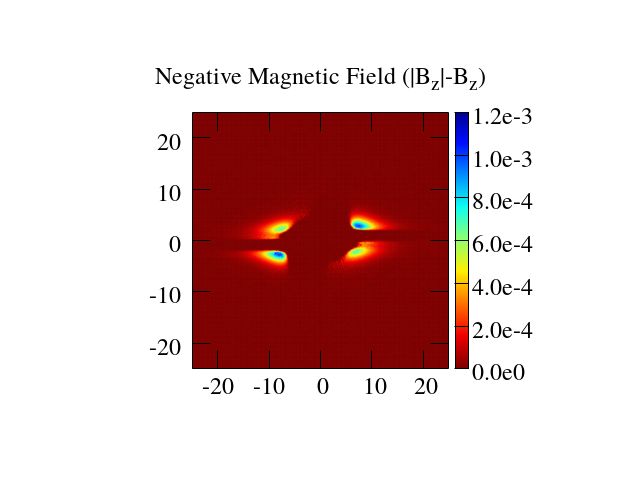} 
 \includegraphics[width=0.36\linewidth,trim={5.1cm 2.5cm 3.5cm 1.6cm},clip]{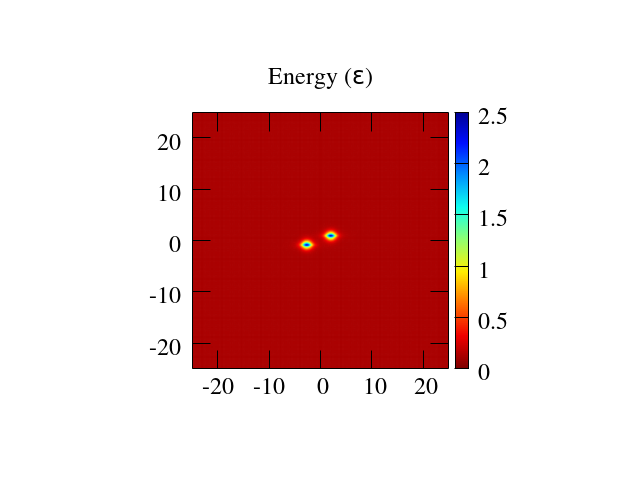}}
 \centerline{
 \includegraphics[width=0.36\linewidth,trim={5.1cm 2.5cm 3.5cm 1.6cm},clip]{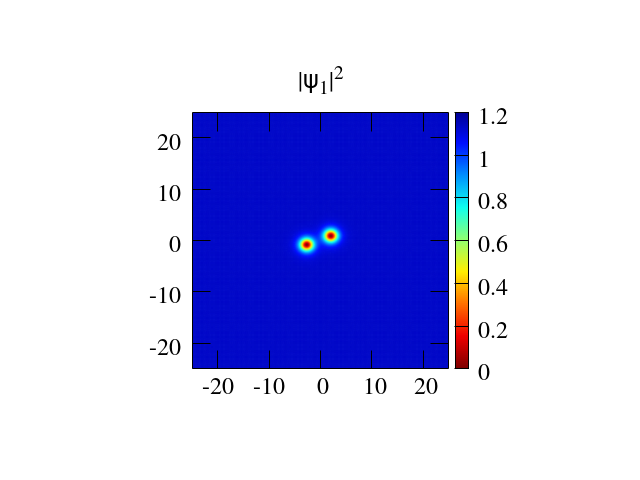}
 \includegraphics[width=0.36\linewidth,trim={5.1cm 2.5cm 3.5cm 1.6cm},clip]{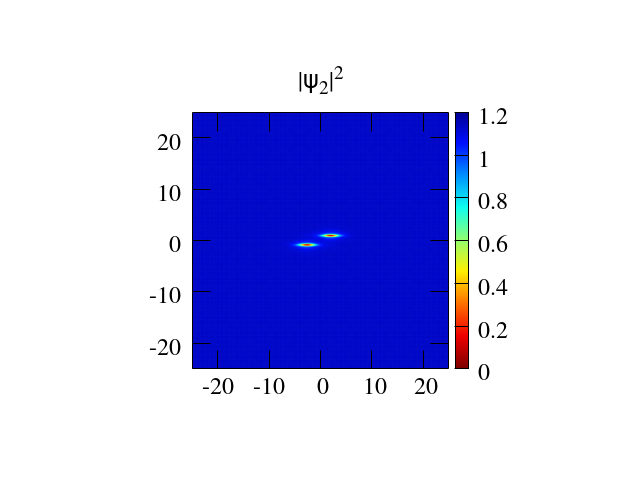}
 \includegraphics[width=0.36\linewidth,trim={5.1cm 2.5cm 3.5cm 1.6cm},clip]{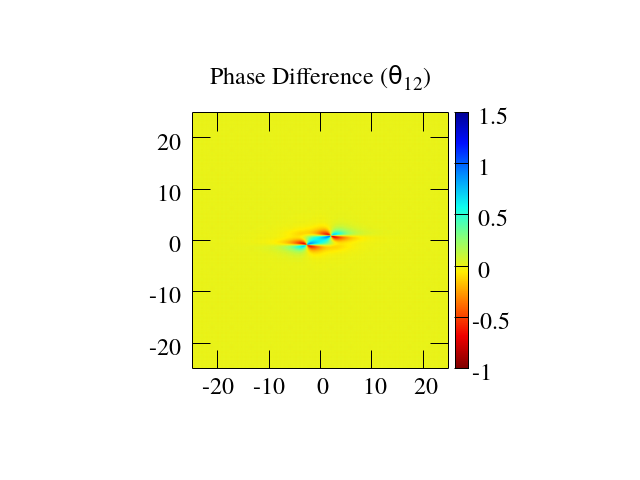}}
 \caption{\label{Fig:twoquantaone} (Colour online) 
 $N=2$ two quanta numerical solution for strong anisotropy in a single component 
 $\gamma_1=\gamma_2=2$, $\eta_{12}=0.5$, $\lambda_{x1}=\lambda_{y1}=\lambda_{x2}=1$ and 
 $\lambda^{-1}_{y2} = 0.1$ (a) $B_z$ magnetic field (b) $\left|B_z\right| - B_z$ negative magnetic field 
 (c) $\mathcal{E}$ energy density (d) $\left|\phi_1\right|^2$ (e)$\left|\phi_2\right|^2$ (f)$\theta_{12}$ phase difference.}
 \end{figure}

\begin{figure}[tb!]
  \centerline{
  \includegraphics[width=0.36\linewidth,trim={5.1cm 2.5cm 3.5cm 1.6cm},clip]{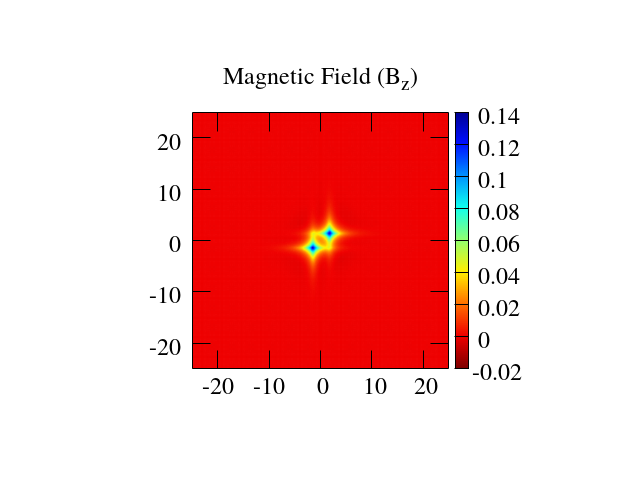}  
  \includegraphics[width=0.36\linewidth,trim={5.1cm 2.5cm 3.5cm 1.6cm},clip]{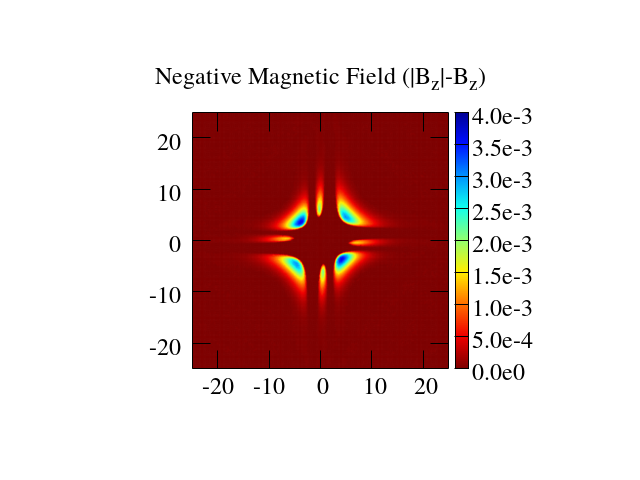}
  \includegraphics[width=0.36\linewidth,trim={5.1cm 2.5cm 3.5cm 1.6cm},clip]{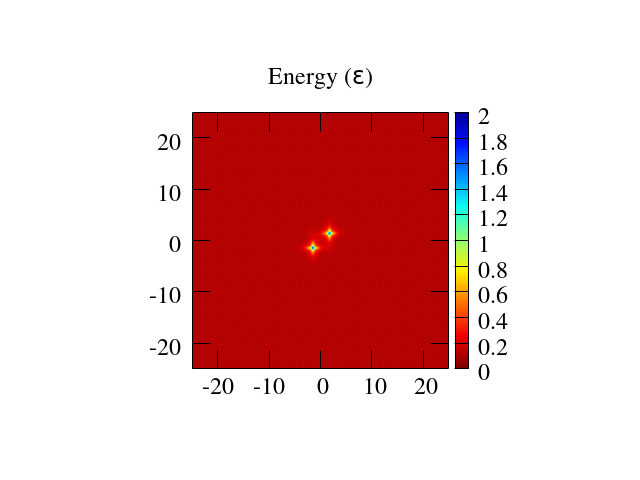}}
  \centerline{
  \includegraphics[width=0.36\linewidth,trim={5.1cm 2.5cm 3.5cm 1.6cm},clip]{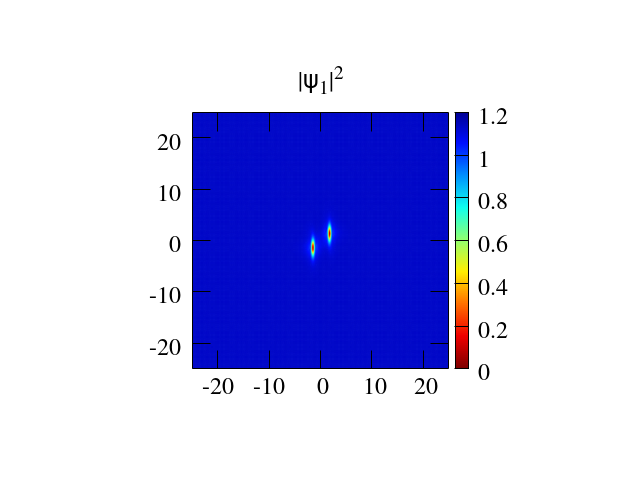}
  \includegraphics[width=0.36\linewidth,trim={5.1cm 2.5cm 3.5cm 1.6cm},clip]{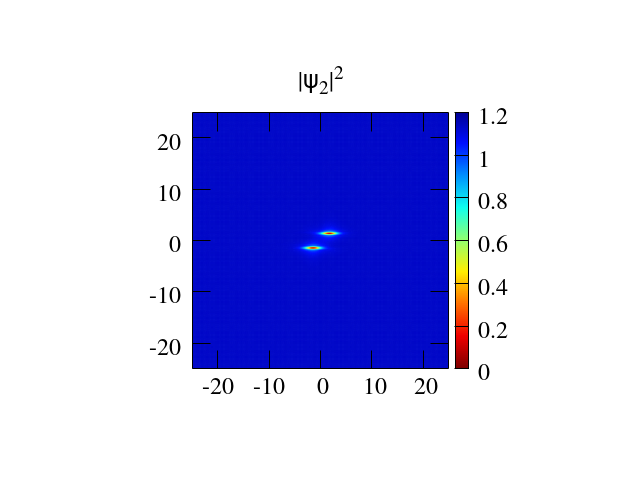}
  \includegraphics[width=0.36\linewidth,trim={5.1cm 2.5cm 3.5cm 1.6cm},clip]{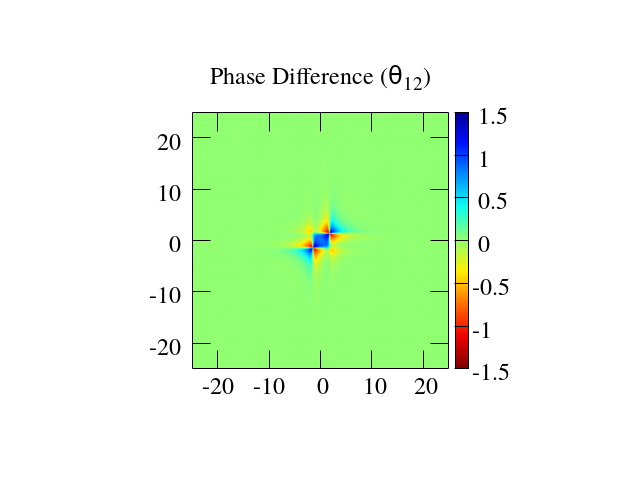}}
 \caption{\label{Fig:twoquantaopp} (Colour online) 
 $N=2$ two quanta numerical solution for strong anisotropy in both bands in opposite directions $\gamma_1=\gamma_2=2$, 
 $\eta_{12}=0.5$, $\lambda_{y1}=\lambda_{x2}=1$ and $\lambda^{-1}_{x1}=\lambda^{-1}_{y2} = 0.1$ 
 (a) $B_z$ magnetic field (b) $\left|B_z\right| - B_z$ negative magnetic field 
 (c) $\mathcal{E}$ energy density (d) $\left|\phi_1\right|^2$ (e)$\left|\phi_2\right|^2$ (f)$\theta_{12}$ phase difference.}
 \end{figure}

We can now compare this to the results of the Ginzburg-Landau field theory. We naturally start with the 2 quanta $N=2$ solutions, shown in Fig. \ref{Fig:twoquantaone} for anisotropy in one band and Fig. \ref{Fig:twoquantaopp} for anisotropy in both bands in opposite directions, demonstrating that the bound states do exist in the full model also. The direction of separation also matches that predicted by the above London model and is based on the parameters of the model. We can find the energies of this formation by simulating the 2 quanta, 1 quanta and vacuum solution on the same grid to compare energies. For the parameters $\gamma_1=\gamma_2=2$, $\eta_{12}=0.5$, 
$\lambda_{x1}=\lambda_{y1}=\lambda_{x2}=\lambda$ and $\lambda_{y2} = 10\lambda$ shown in Fig. \ref{Fig:twoquantaone} we get $E'_1=E_1 - E_0 = 1.6416$ and $E'_2=E_2-E_0=3.2773$, where each $E_i$ is the minimal energy solution for the $i$ quanta system. Which means the 2 quanta normalised energy is lower than the single vortex normalised energy per vortex $E'_2<2E'_1$ and a bound state has been formed. The binding energy of this bound state is small which is no surprise, due to the small levels of the magnetic field inversion compared to other terms in the free energy.

\begin{figure*}[tb!]
  \centerline{\includegraphics[width=0.2\linewidth,trim={4.8cm 1.3cm 3.2cm 1.6cm},clip]{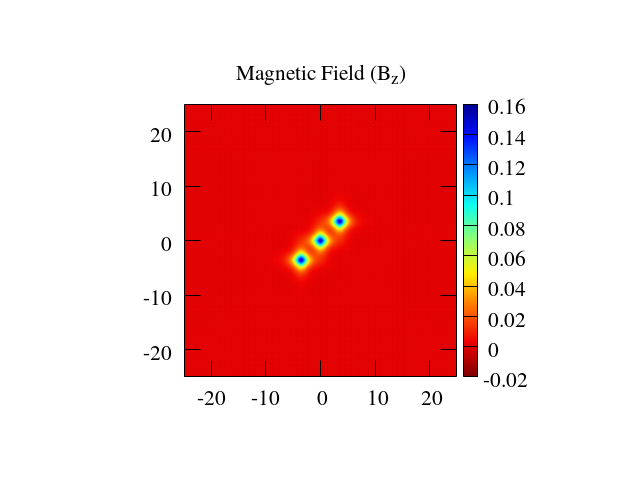}\includegraphics[width=0.2\linewidth,trim={4.8cm 1.3cm 3.2cm 1.6cm},clip]{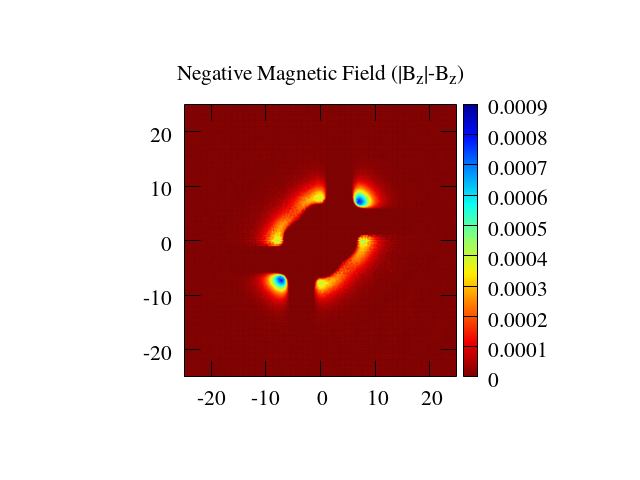}\includegraphics[width=0.2\linewidth,trim={4.8cm 1.3cm 3.2cm 1.6cm},clip]{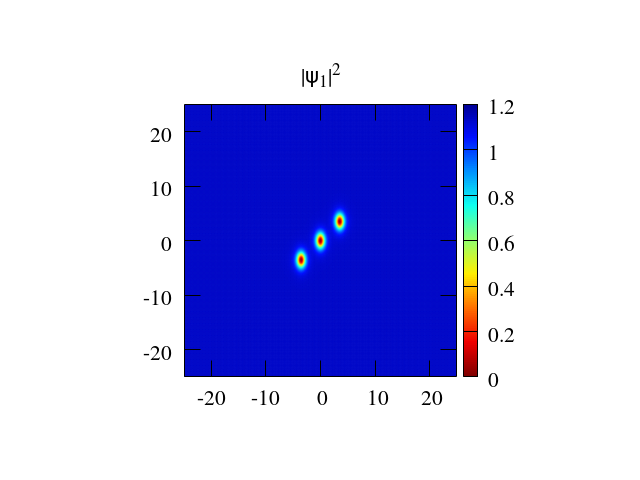}\includegraphics[width=0.2\linewidth,trim={4.8cm 1.3cm 3.2cm 1.6cm},clip]{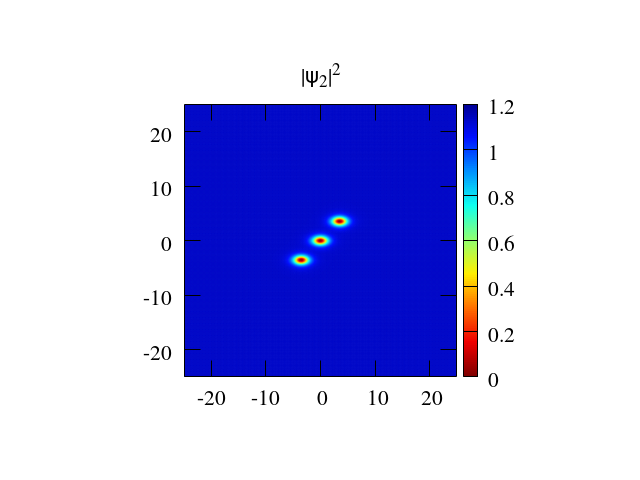}\includegraphics[width=0.2\linewidth,trim={4.8cm 1.3cm 3.2cm 1.6cm},clip]{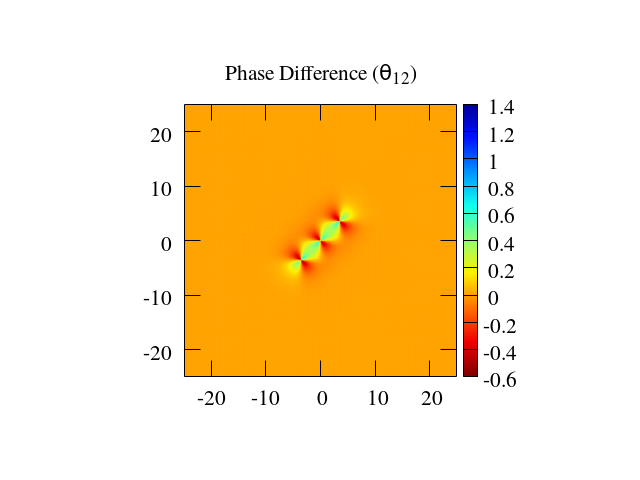}}
    \centerline{\includegraphics[width=0.2\linewidth,trim={4.8cm 1.3cm 3.2cm 1.6cm},clip]{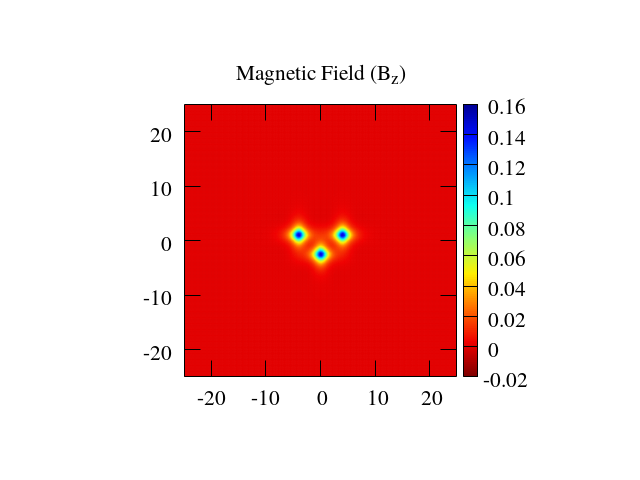}\includegraphics[width=0.2\linewidth,trim={4.8cm 1.3cm 3.2cm 1.6cm},clip]{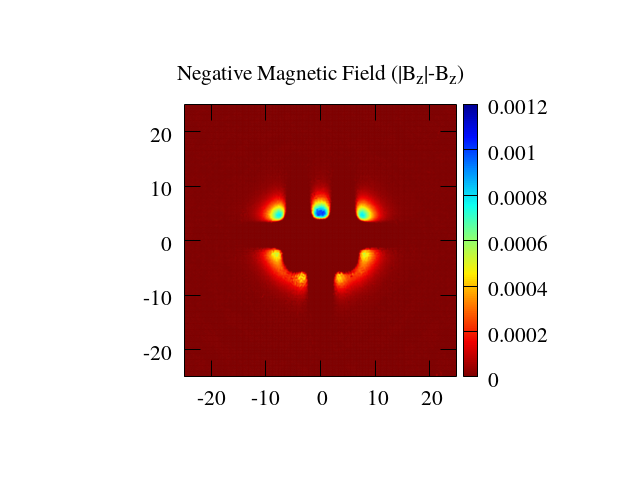}\includegraphics[width=0.2\linewidth,trim={4.8cm 1.3cm 3.2cm 1.6cm},clip]{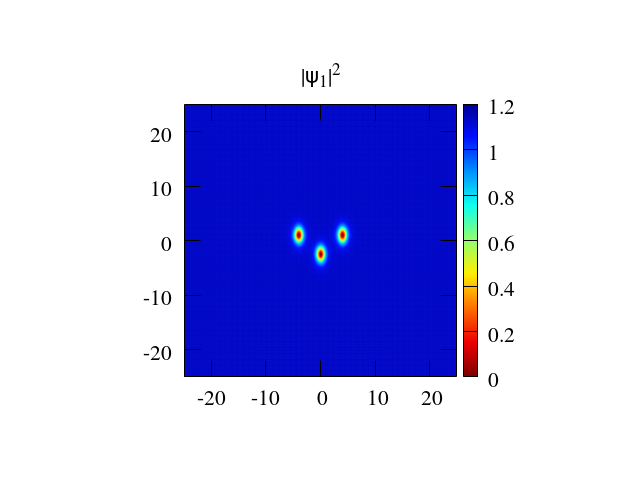}\includegraphics[width=0.2\linewidth,trim={4.8cm 1.3cm 3.2cm 1.6cm},clip]{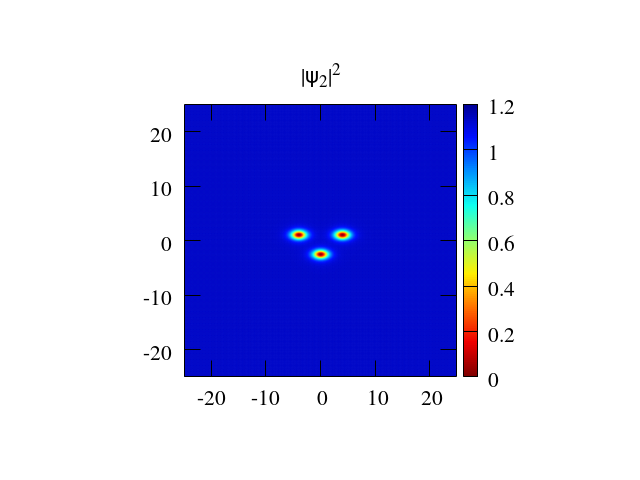}\includegraphics[width=0.2\linewidth,trim={4.8cm 1.3cm 3.2cm 1.6cm},clip]{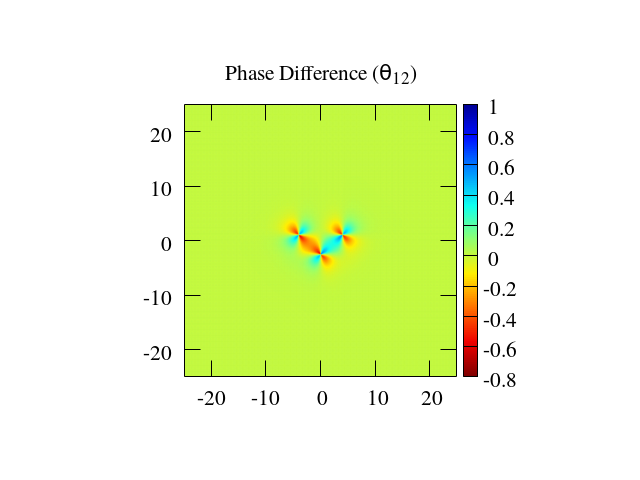}}
      \centerline{\includegraphics[width=0.2\linewidth,trim={4.8cm 1.3cm 3.2cm 1.6cm},clip]{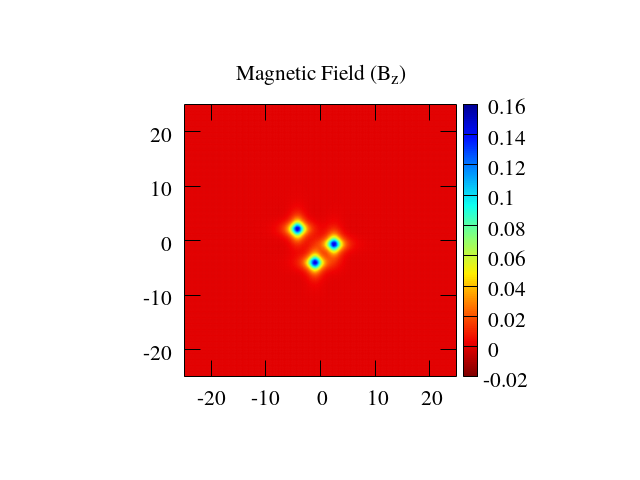}\includegraphics[width=0.2\linewidth,trim={4.8cm 1.3cm 3.2cm 1.6cm},clip]{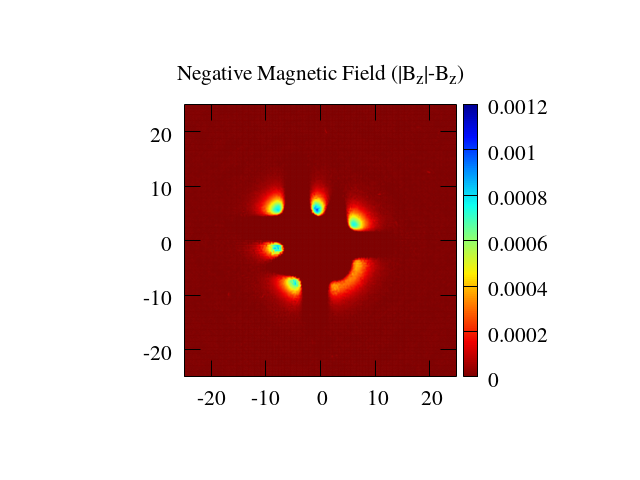}\includegraphics[width=0.2\linewidth,trim={4.8cm 1.3cm 3.2cm 1.6cm},clip]{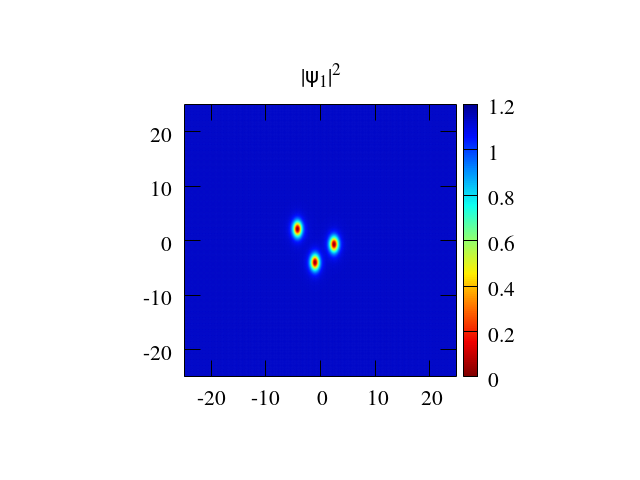}\includegraphics[width=0.2\linewidth,trim={4.8cm 1.3cm 3.2cm 1.6cm},clip]{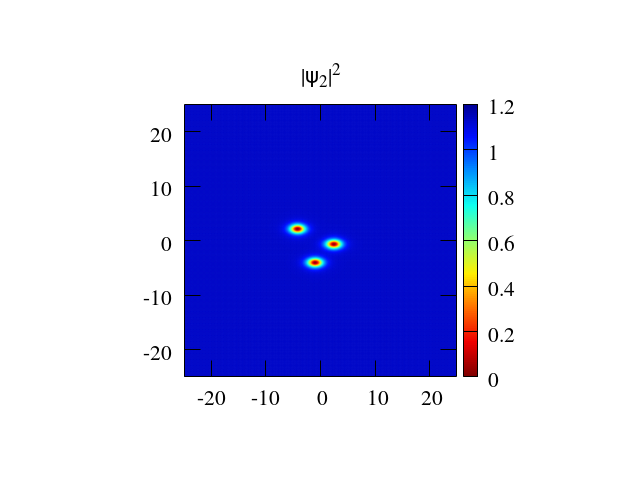}\includegraphics[width=0.2\linewidth,trim={4.8cm 1.3cm 3.2cm 1.6cm},clip]{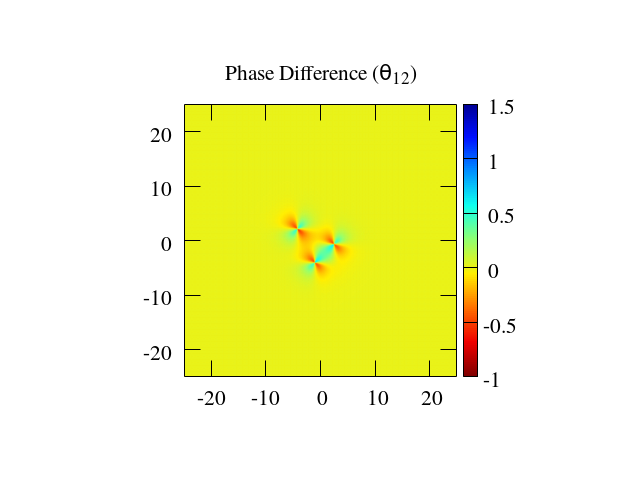}}
 \caption{\label{Fig:type2opp5B3} $N=3$ three quanta local solutions for anisotropy in both bands in opposite directions $\gamma_1=\gamma_2=2$, $\eta_{12}=0.5$, $\lambda_{x2}=\lambda_{y1}=1$ and $\lambda^{-1}_{y2}=\lambda^{-1}_{x1}= 0.5$ (a) $B_z$ magnetic field (b) $\left|B_z\right| - B_z$ negative magnetic field (c) $\left|\phi_1\right|^2$ (d)$\left|\phi_2\right|^2$ (e)$\theta_{12}$ phase difference. } 
 \end{figure*}

We now move onto the $n=3$ quanta solutions, for equal anisotropies in opposite directions we are interested in stronger and weaker anisotropies and the effect on the minimal energy solutions. For weaker anisotropy, shown in Fig. \ref{Fig:type2opp5B3}, we observe 3 key local solutions in the form of a line, an L shape and a T shape. The first two of these are polyominoes, the second is not. If we compare the energies we get the following, $E'_{line} = 4.7652$, $E'_{L} = 4.7678$  and $E'_{T} = 4.7689$, which gives the line solution as the global minima, as predicted by the London model. It is instructive however to compare the energies of the other two local minima, as the London model predicts polyominoes being more favoured over interlaced solutions, thus one would expect the L solution to have lower energy than the T solution, which is the case.

\begin{figure*}[tb!]
  \centerline{\includegraphics[width=0.2\linewidth,trim={4.8cm 1.3cm 3.2cm 1.6cm},clip]{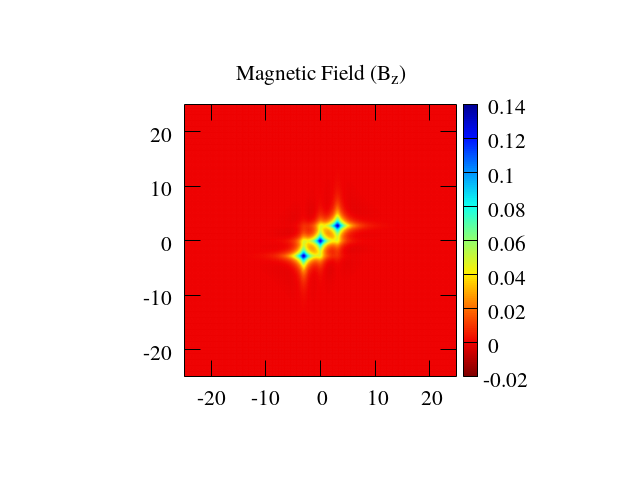}\includegraphics[width=0.2\linewidth,trim={4.8cm 1.3cm 3.2cm 1.6cm},clip]{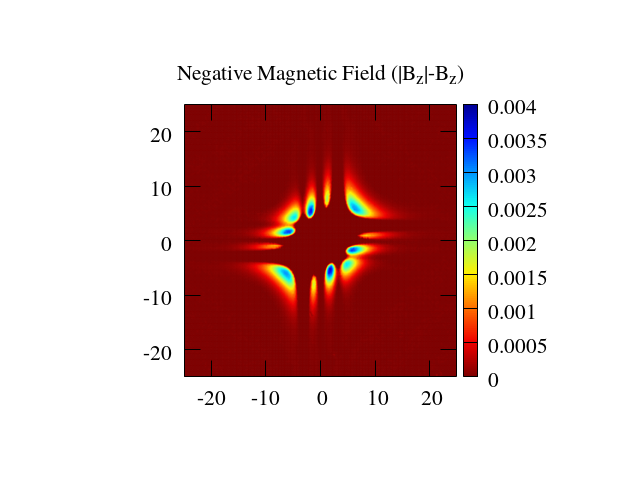}\includegraphics[width=0.2\linewidth,trim={4.8cm 1.3cm 3.2cm 1.6cm},clip]{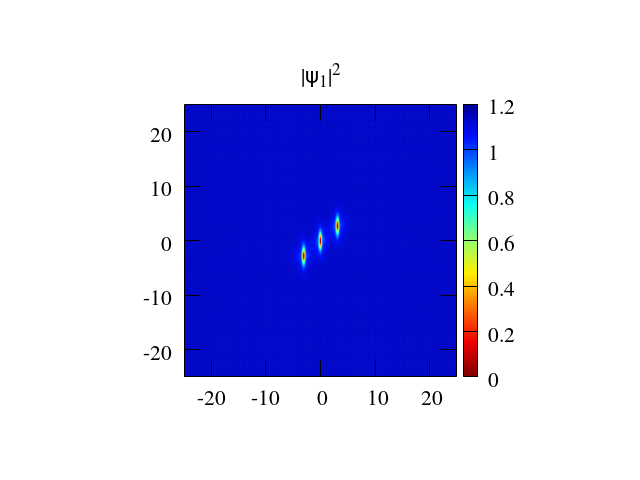}\includegraphics[width=0.2\linewidth,trim={4.8cm 1.3cm 3.2cm 1.6cm},clip]{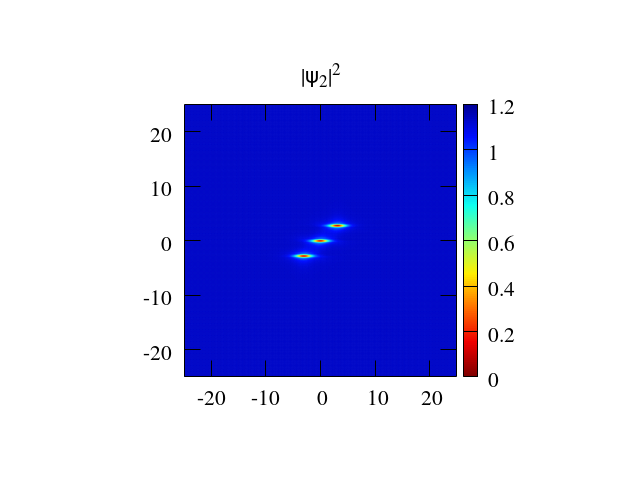}\includegraphics[width=0.2\linewidth,trim={4.8cm 1.3cm 3.2cm 1.6cm},clip]{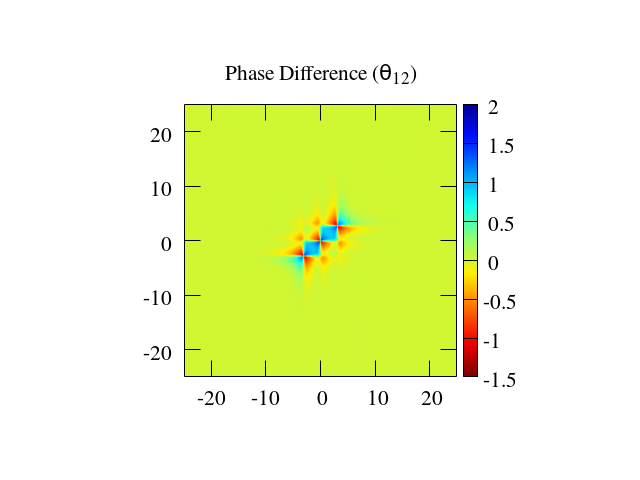}}
    \centerline{\includegraphics[width=0.2\linewidth,trim={4.8cm 1.3cm 3.2cm 1.6cm},clip]{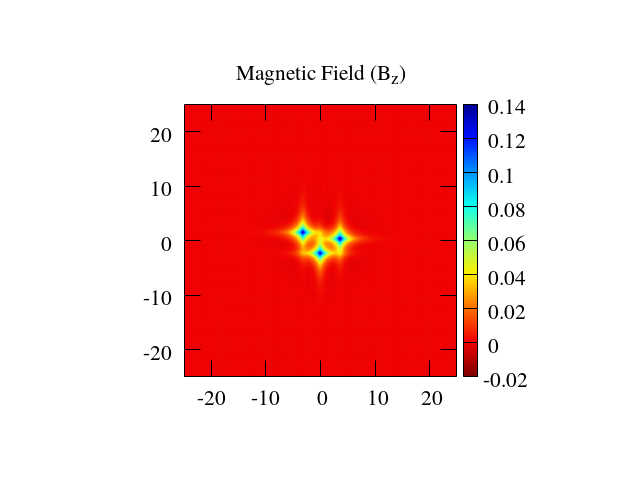}\includegraphics[width=0.2\linewidth,trim={4.8cm 1.3cm 3.2cm 1.6cm},clip]{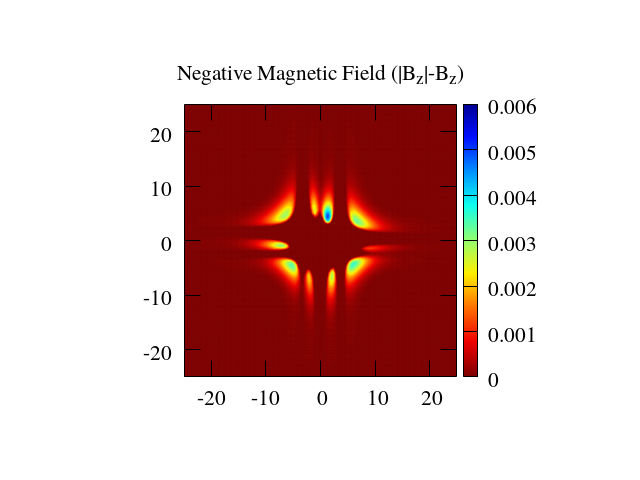}\includegraphics[width=0.2\linewidth,trim={4.8cm 1.3cm 3.2cm 1.6cm},clip]{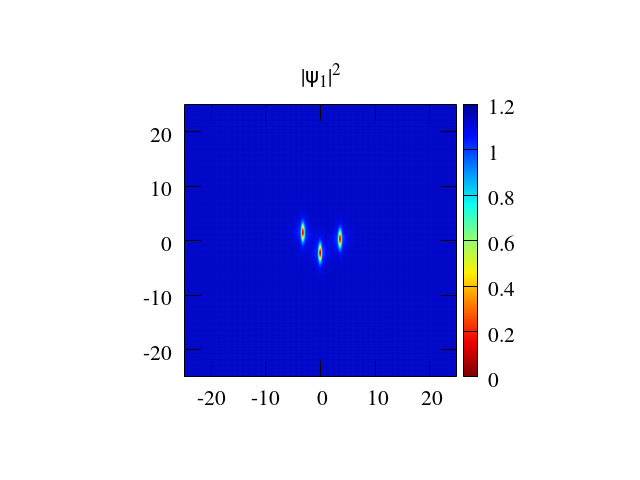}\includegraphics[width=0.2\linewidth,trim={4.8cm 1.3cm 3.2cm 1.6cm},clip]{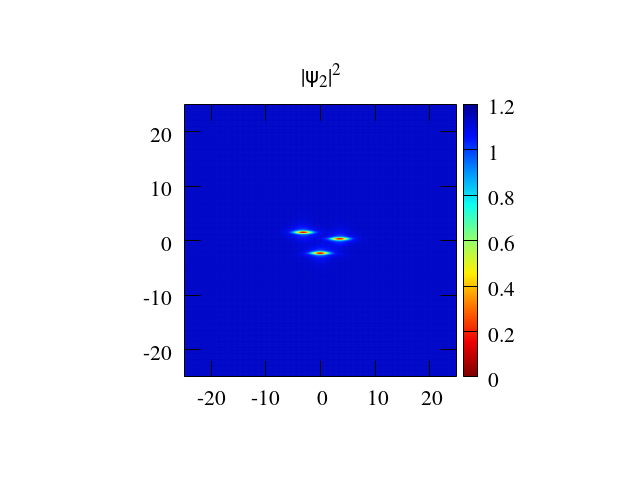}\includegraphics[width=0.2\linewidth,trim={4.8cm 1.3cm 3.2cm 1.6cm},clip]{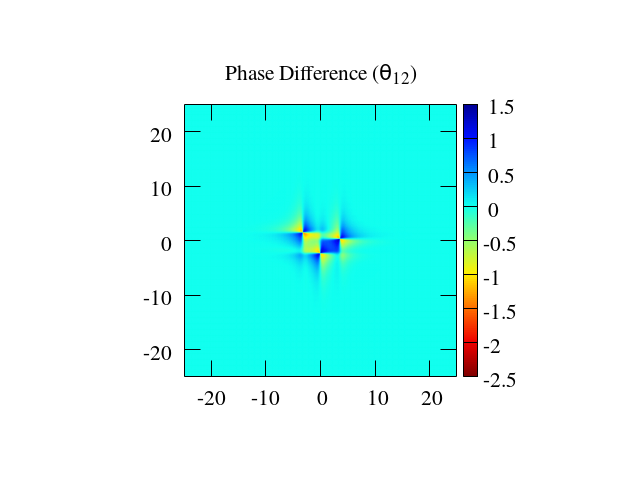}}
      \centerline{\includegraphics[width=0.2\linewidth,trim={4.8cm 1.3cm 3.2cm 1.6cm},clip]{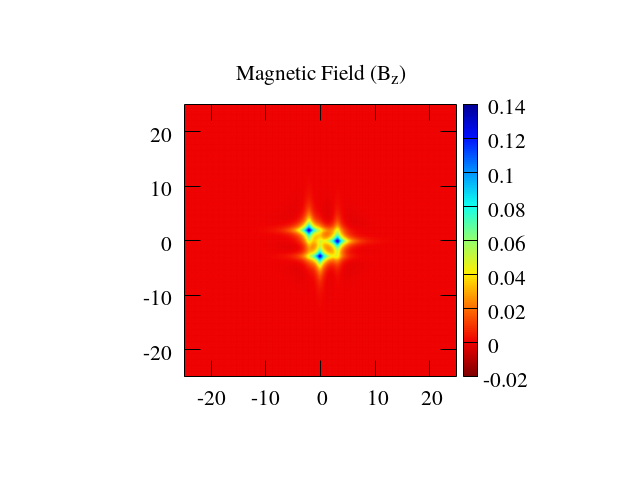}\includegraphics[width=0.2\linewidth,trim={4.8cm 1.3cm 3.2cm 1.6cm},clip]{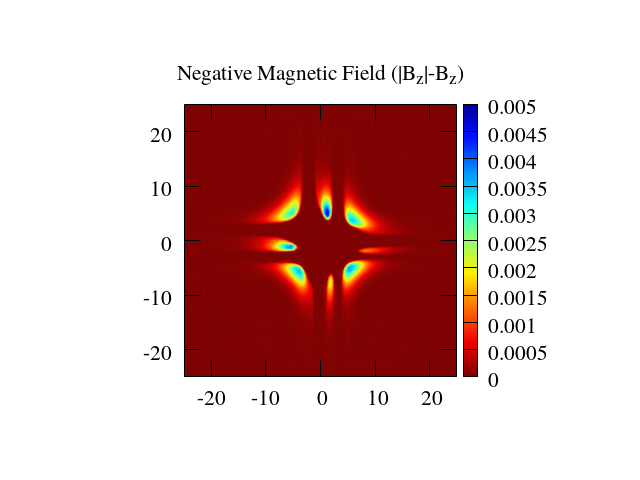}\includegraphics[width=0.2\linewidth,trim={4.8cm 1.3cm 3.2cm 1.6cm},clip]{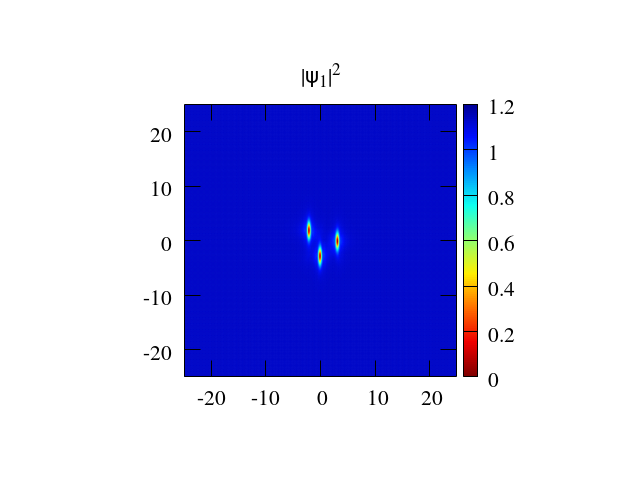}\includegraphics[width=0.2\linewidth,trim={4.8cm 1.3cm 3.2cm 1.6cm},clip]{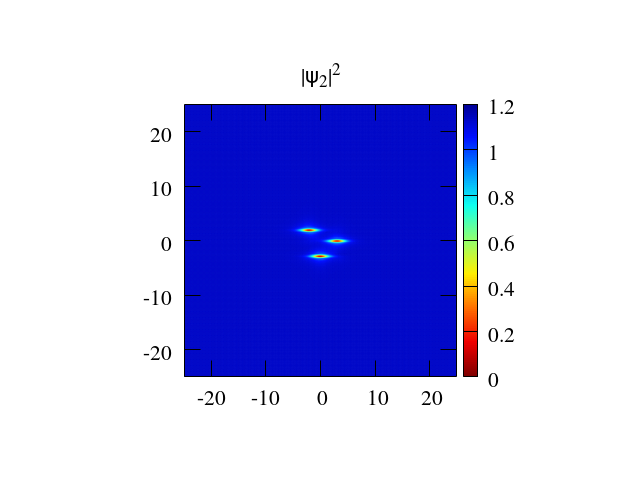}\includegraphics[width=0.2\linewidth,trim={4.8cm 1.3cm 3.2cm 1.6cm},clip]{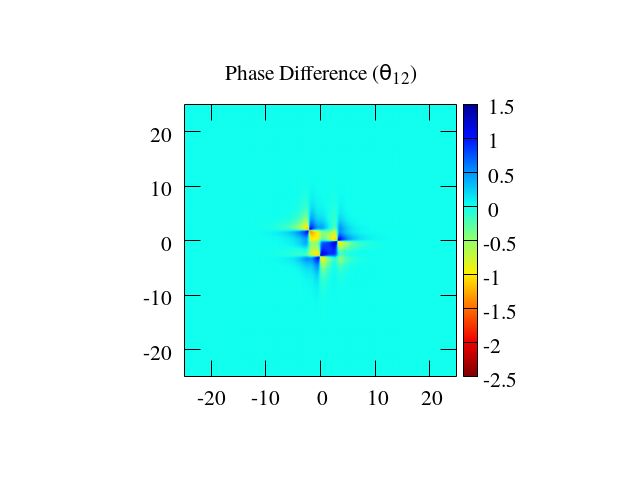}}
 \caption{\label{Fig:type2opp1B3} 
 $N=3$ three quanta local solutions for strong anisotropy in both bands in opposite directions $\gamma_1=\gamma_2=2$, $\eta_{12}=0.5$, $\lambda_{x2}=\lambda_{y1}=1$ and $\lambda^{-1}_{y2}=\lambda^{-1}_{x1}= 0.1$ (a) $B_z$ magnetic field (b) $\left|B_z\right| - B_z$ negative magnetic field (c) $\left|\phi_1\right|^2$ (d)$\left|\phi_2\right|^2$ (e)$\theta_{12}$ phase difference. } 
 \end{figure*}

For stronger anisotropy, shown in Fig. \ref{Fig:type2opp1B3}, we can see the same local solutions, however the energies are now given as $E'_{line} = 3.1098$, $E'_{L} = 3.1199$  and $E'_{T} = 3.1110$, such that the line solution is still the global minima. Here we see that Ginzburg-Landau solutions do not agree with the London model. However the energy difference has decreased between the various solutions so the trend is the same. This is likely due to the London model not allowing the form of the individual vortices to deform. Note that the minimal value when comparing the L and T solutions has now switched to the T solution. This means that the interlaced solutions are more favoured now, as predicted by the London model above.

\begin{figure*}[tb!]
      \centerline{\includegraphics[width=0.2\linewidth,trim={4.8cm 1.3cm 3.2cm 1.6cm},clip]{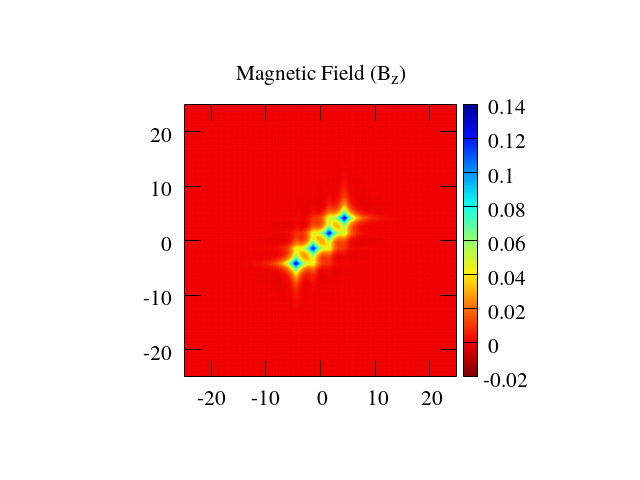}\includegraphics[width=0.2\linewidth,trim={4.8cm 1.3cm 3.2cm 1.6cm},clip]{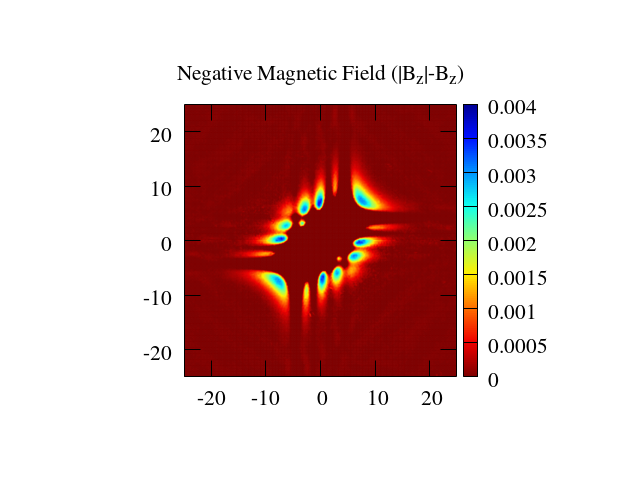}\includegraphics[width=0.2\linewidth,trim={4.8cm 1.3cm 3.2cm 1.6cm},clip]{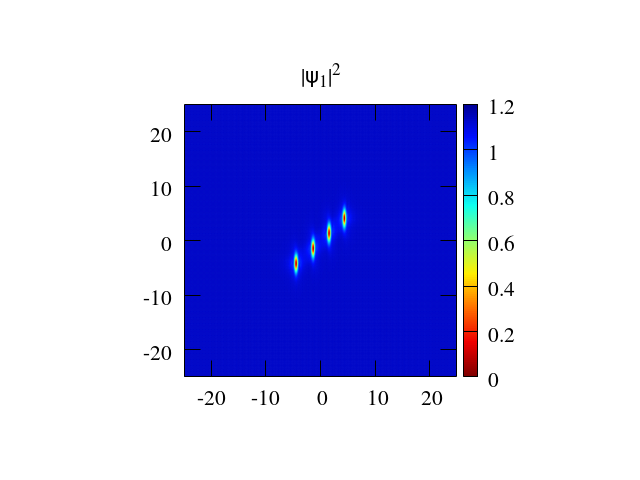}\includegraphics[width=0.2\linewidth,trim={4.8cm 1.3cm 3.2cm 1.6cm},clip]{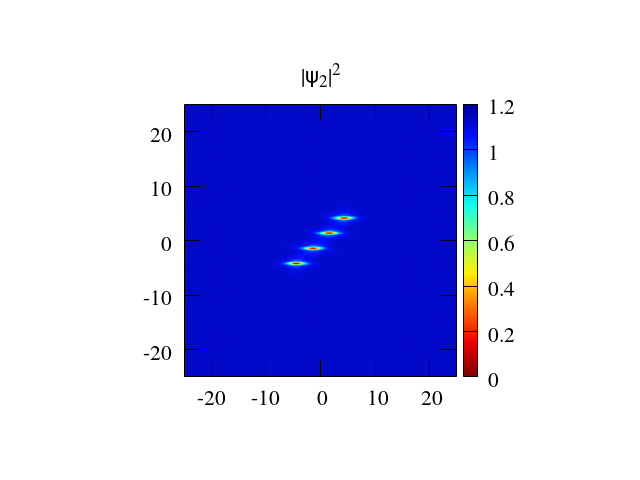}\includegraphics[width=0.2\linewidth,trim={4.8cm 1.3cm 3.2cm 1.6cm},clip]{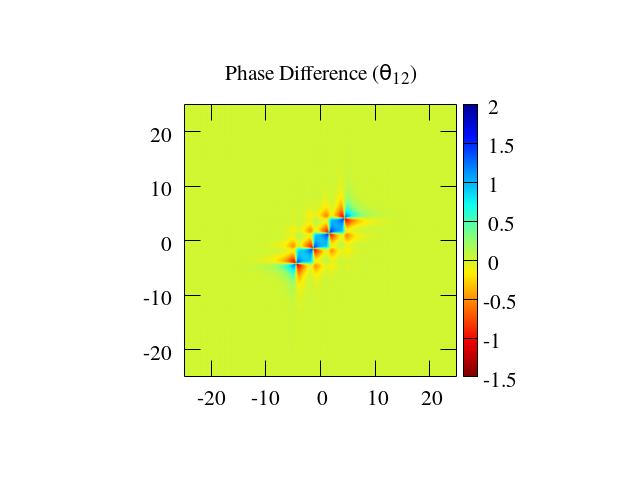}}
      \centerline{\includegraphics[width=0.2\linewidth,trim={4.8cm 1.3cm 3.2cm 1.6cm},clip]{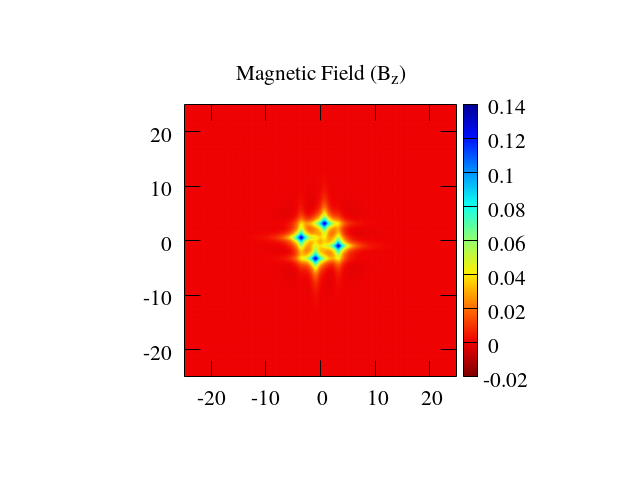}\includegraphics[width=0.2\linewidth,trim={4.8cm 1.3cm 3.2cm 1.6cm},clip]{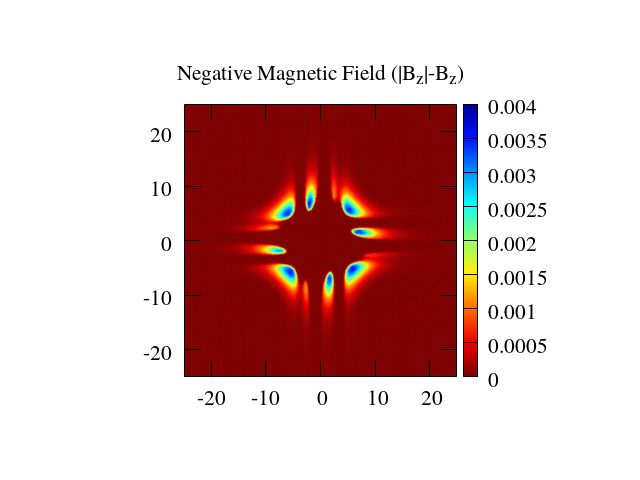}\includegraphics[width=0.2\linewidth,trim={4.8cm 1.3cm 3.2cm 1.6cm},clip]{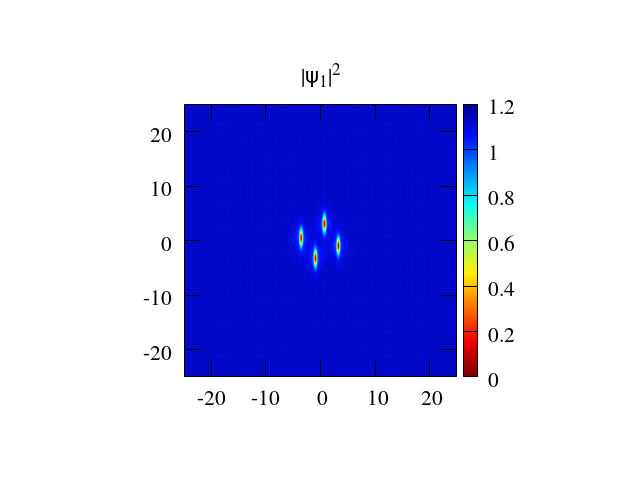}\includegraphics[width=0.2\linewidth,trim={4.8cm 1.3cm 3.2cm 1.6cm},clip]{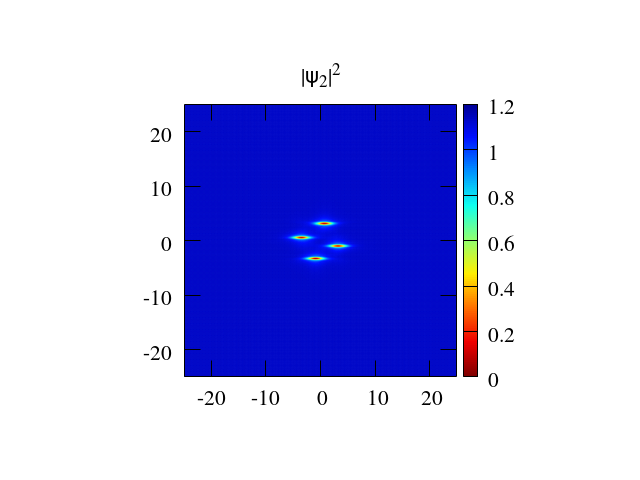}\includegraphics[width=0.2\linewidth,trim={4.8cm 1.3cm 3.2cm 1.6cm},clip]{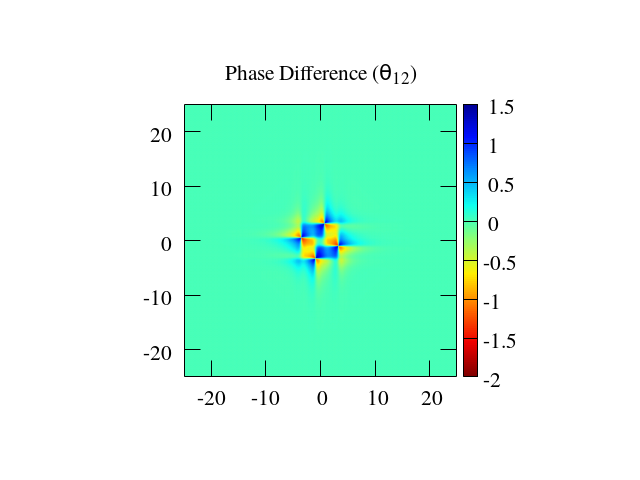}}
      \centerline{\includegraphics[width=0.2\linewidth,trim={4.8cm 1.3cm 3.2cm 1.6cm},clip]{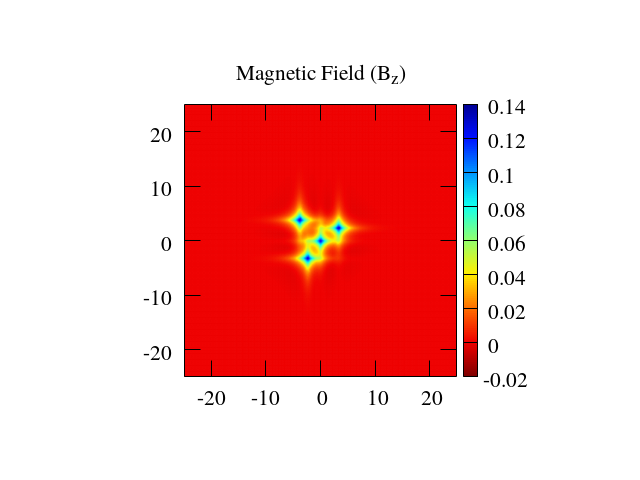}\includegraphics[width=0.2\linewidth,trim={4.8cm 1.3cm 3.2cm 1.6cm},clip]{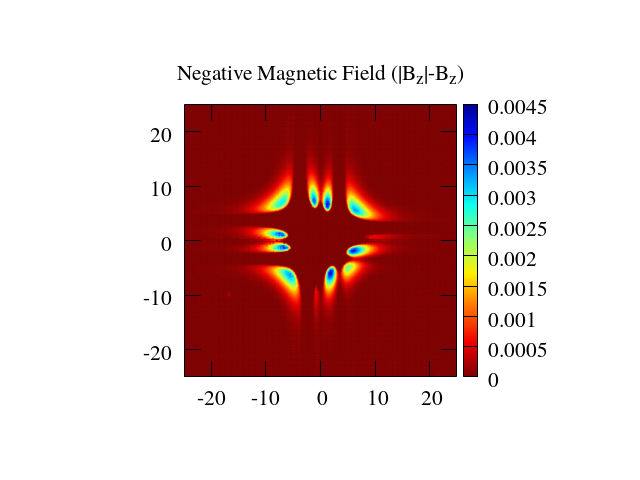}\includegraphics[width=0.2\linewidth,trim={4.8cm 1.3cm 3.2cm 1.6cm},clip]{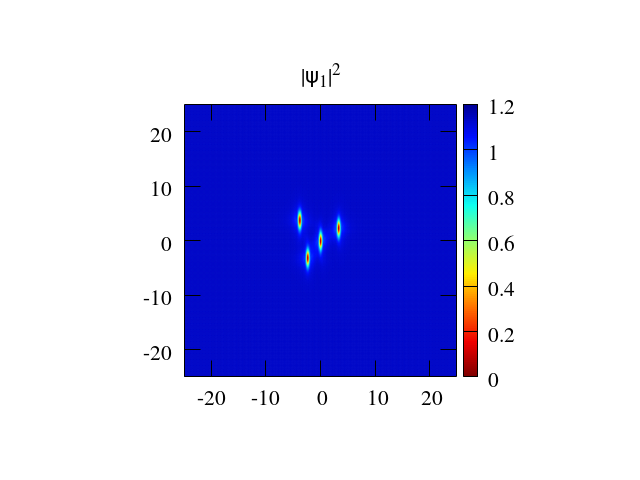}\includegraphics[width=0.2\linewidth,trim={4.8cm 1.3cm 3.2cm 1.6cm},clip]{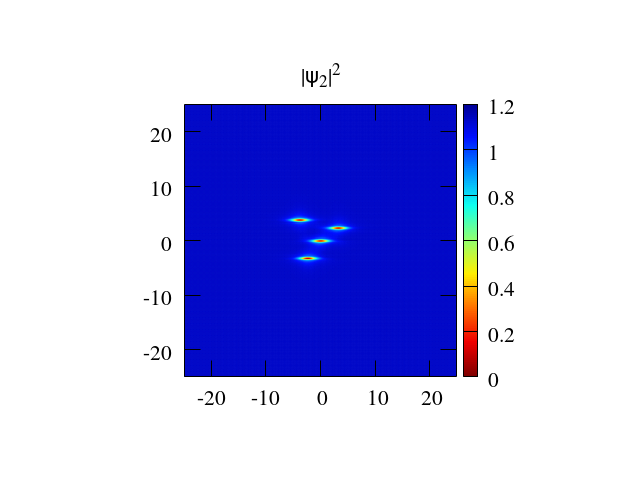}\includegraphics[width=0.2\linewidth,trim={4.8cm 1.3cm 3.2cm 1.6cm},clip]{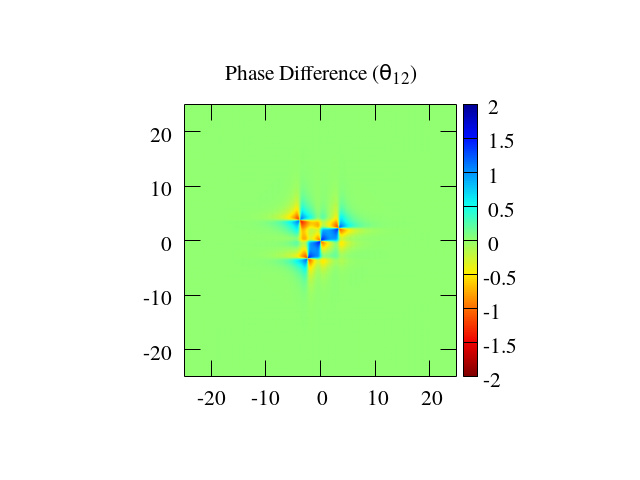}}
       \centerline{\includegraphics[width=0.2\linewidth,trim={4.8cm 1.3cm 3.2cm 1.6cm},clip]{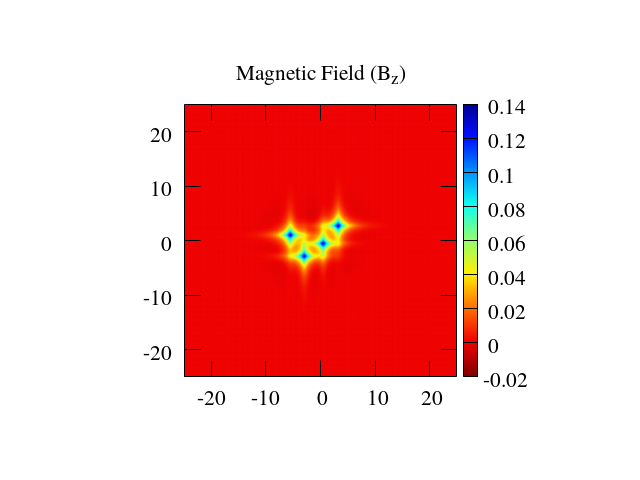}\includegraphics[width=0.2\linewidth,trim={4.8cm 1.3cm 3.2cm 1.6cm},clip]{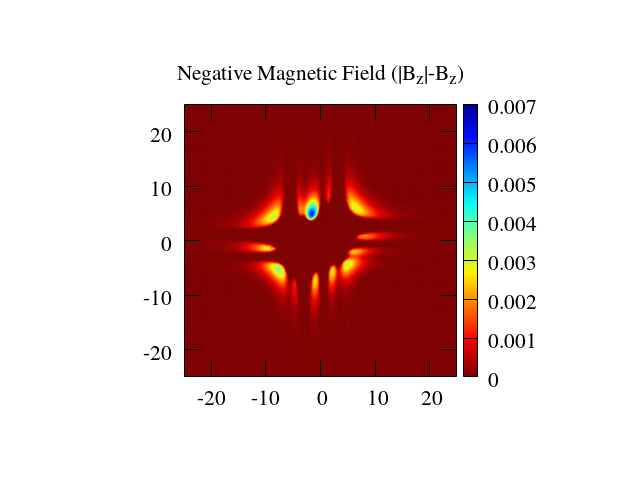}\includegraphics[width=0.2\linewidth,trim={4.8cm 1.3cm 3.2cm 1.6cm},clip]{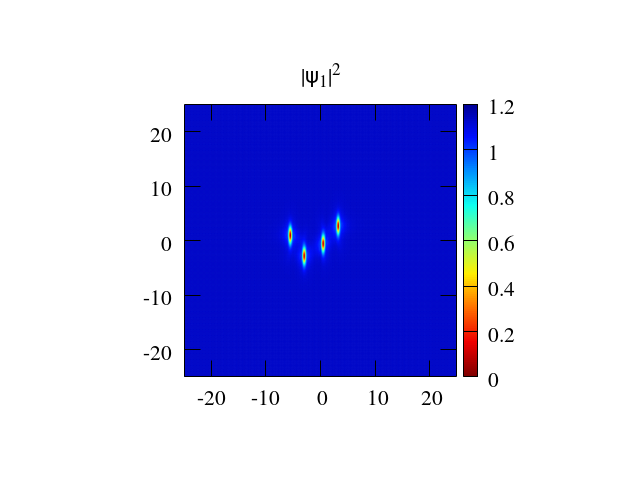}\includegraphics[width=0.2\linewidth,trim={4.8cm 1.3cm 3.2cm 1.6cm},clip]{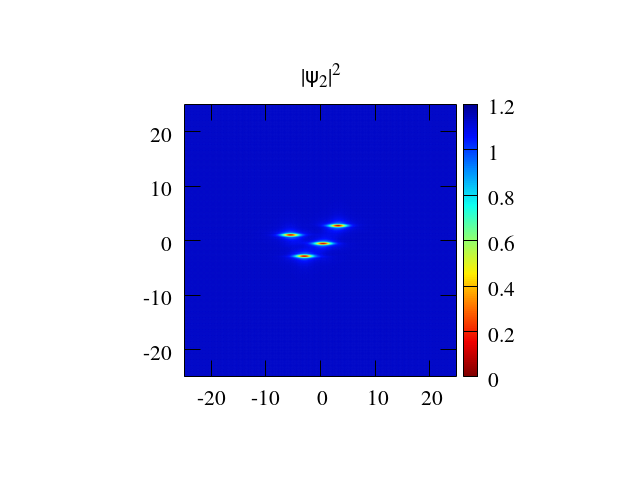}\includegraphics[width=0.2\linewidth,trim={4.8cm 1.3cm 3.2cm 1.6cm},clip]{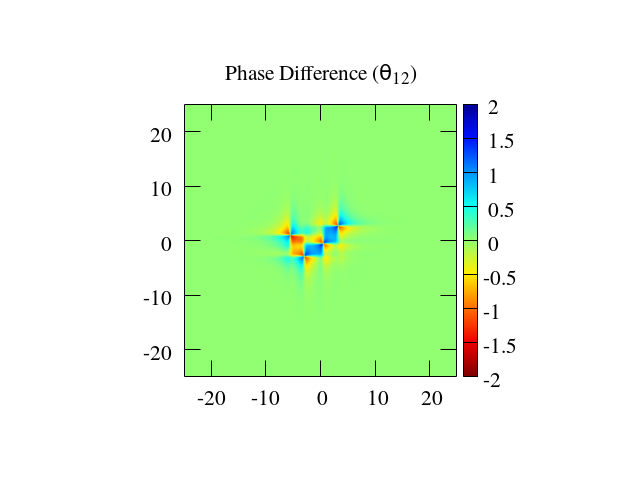}}
        \centerline{\includegraphics[width=0.2\linewidth,trim={4.8cm 1.3cm 3.2cm 1.6cm},clip]{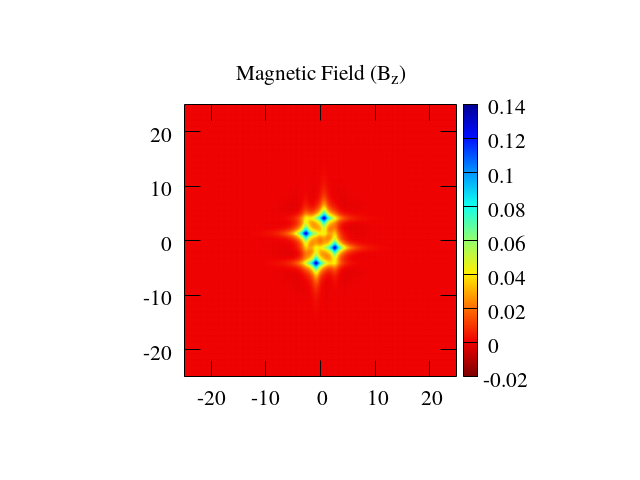}\includegraphics[width=0.2\linewidth,trim={4.8cm 1.3cm 3.2cm 1.6cm},clip]{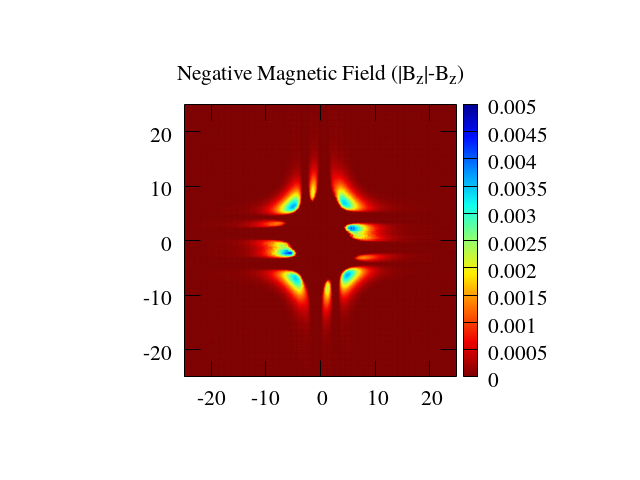}\includegraphics[width=0.2\linewidth,trim={4.8cm 1.3cm 3.2cm 1.6cm},clip]{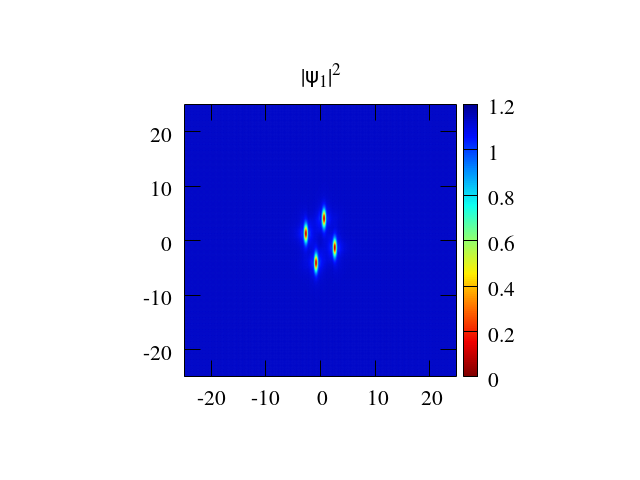}\includegraphics[width=0.2\linewidth,trim={4.8cm 1.3cm 3.2cm 1.6cm},clip]{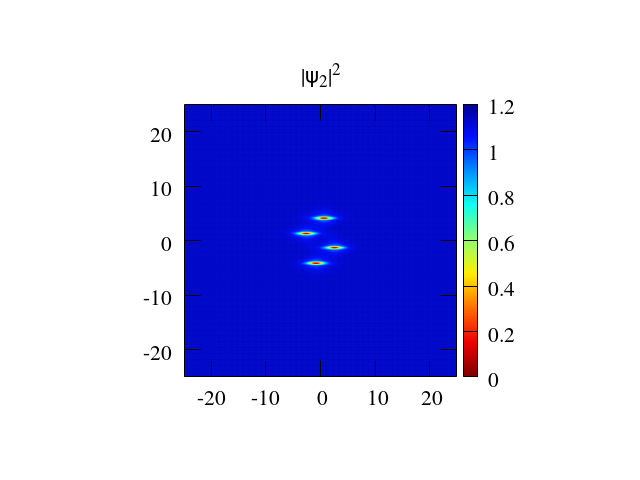}\includegraphics[width=0.2\linewidth,trim={4.8cm 1.3cm 3.2cm 1.6cm},clip]{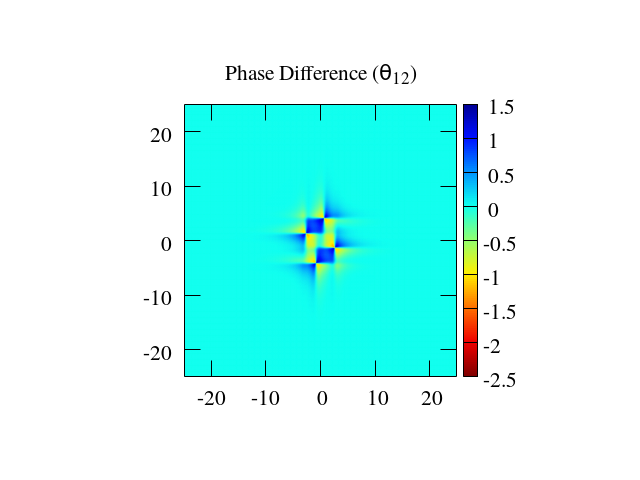}}
 \caption{\label{Fig:type2oppB4p1} 
 $N=4$ four quanta local solutions for strong anisotropy in both bands in opposite directions $\gamma_1=\gamma_2=2$, $\eta_{12}=0.5$, $\lambda_{x2}=\lambda_{y1}=1$ and $\lambda^{-1}_{y2}=\lambda^{-1}_{x1}= 0.1$ (a) $B_z$ magnetic field (b) $\left|B_z\right| - B_z$ negative magnetic field (c) $\left|\phi_1\right|^2$ (d)$\left|\phi_2\right|^2$ (e)$\theta_{12}$ phase difference. } 
 \end{figure*}

\begin{figure*}[tb!]
      \centerline{\includegraphics[width=0.2\linewidth,trim={4.8cm 1.3cm 3.2cm 1.6cm},clip]{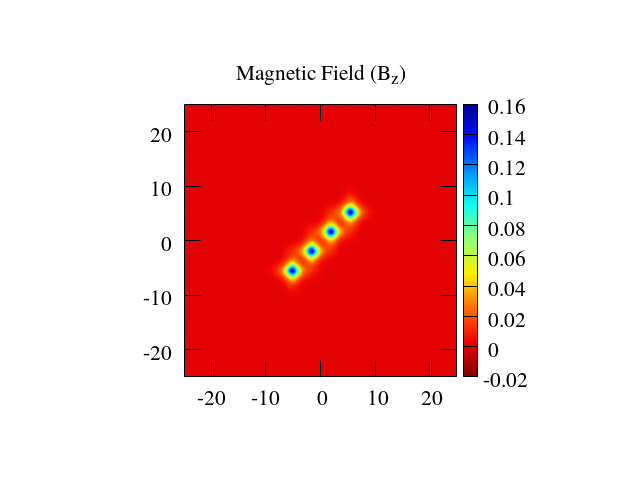}\includegraphics[width=0.2\linewidth,trim={4.8cm 1.3cm 3.2cm 1.6cm},clip]{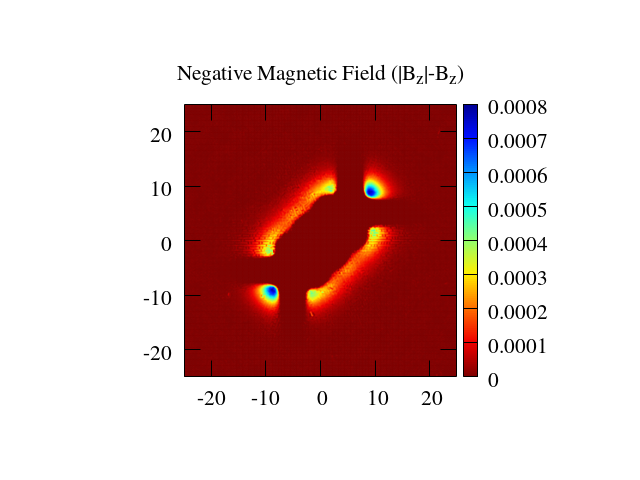}\includegraphics[width=0.2\linewidth,trim={4.8cm 1.3cm 3.2cm 1.6cm},clip]{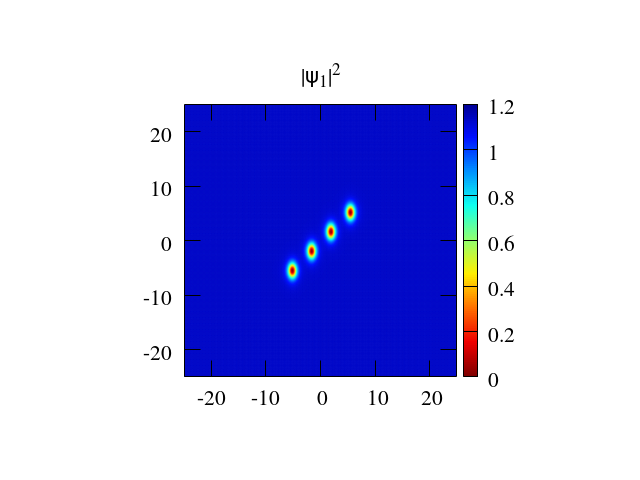}\includegraphics[width=0.2\linewidth,trim={4.8cm 1.3cm 3.2cm 1.6cm},clip]{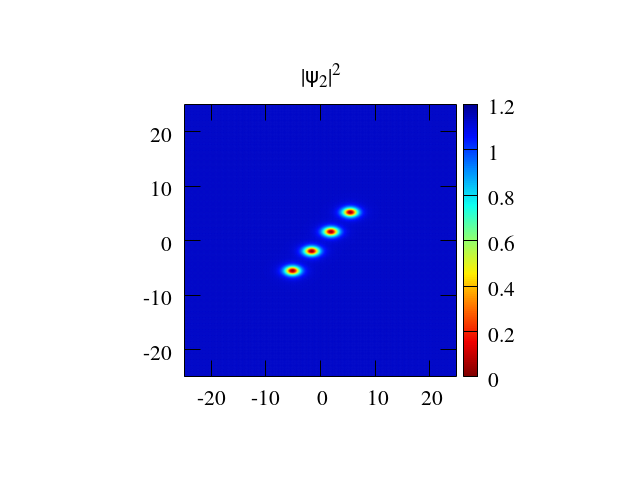}\includegraphics[width=0.2\linewidth,trim={4.8cm 1.3cm 3.2cm 1.6cm},clip]{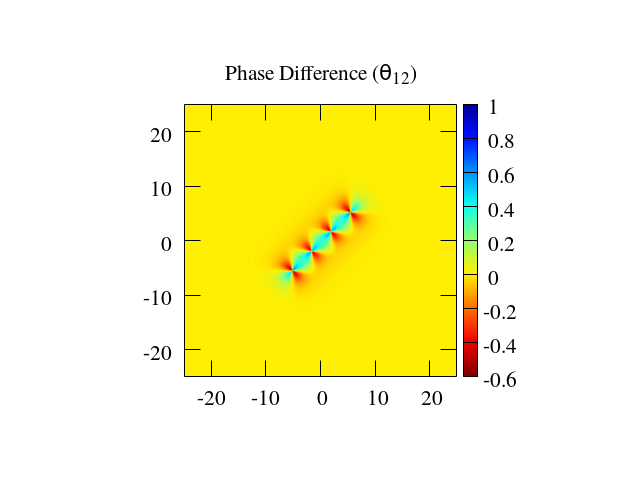}}
      \centerline{\includegraphics[width=0.2\linewidth,trim={4.8cm 1.3cm 3.2cm 1.6cm},clip]{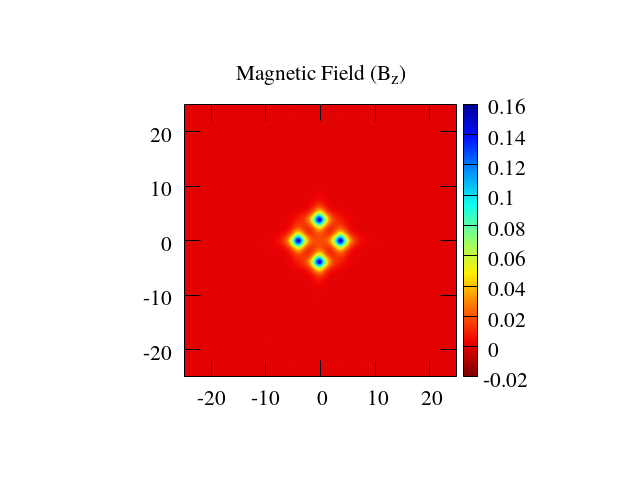}\includegraphics[width=0.2\linewidth,trim={4.8cm 1.3cm 3.2cm 1.6cm},clip]{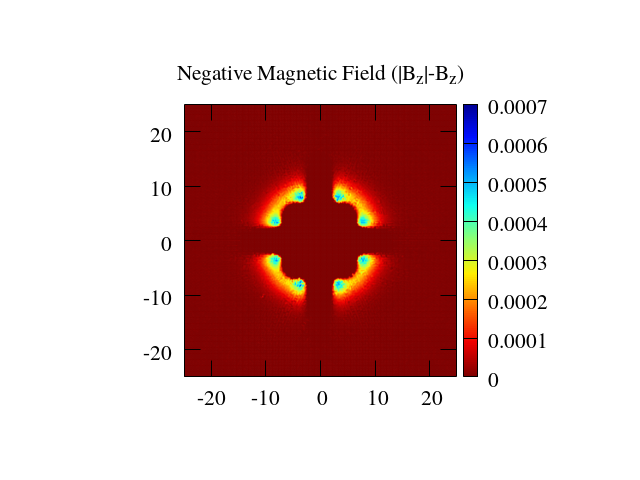}\includegraphics[width=0.2\linewidth,trim={4.8cm 1.3cm 3.2cm 1.6cm},clip]{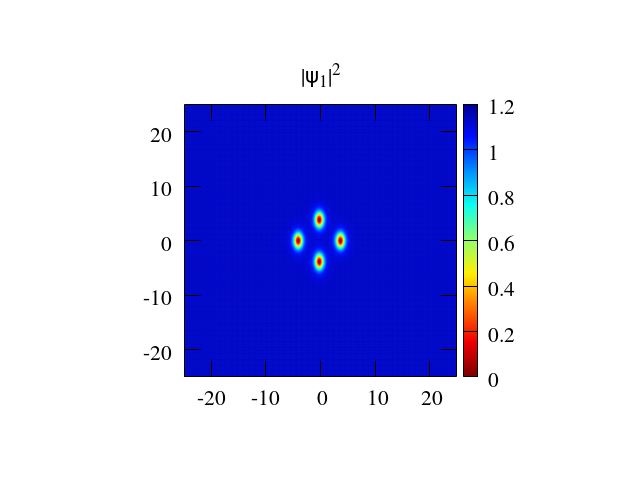}\includegraphics[width=0.2\linewidth,trim={4.8cm 1.3cm 3.2cm 1.6cm},clip]{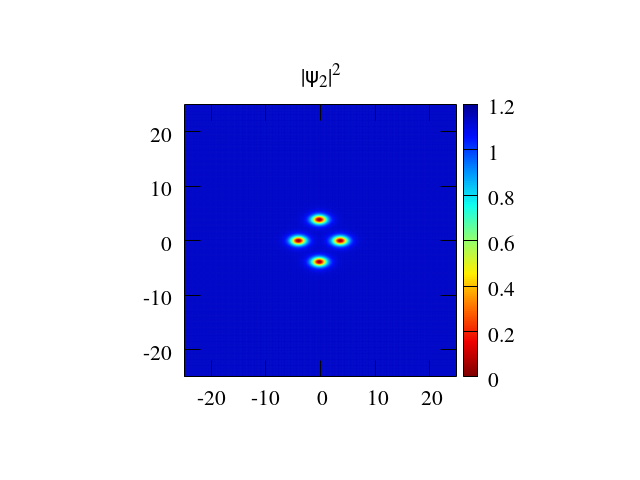}\includegraphics[width=0.2\linewidth,trim={4.8cm 1.3cm 3.2cm 1.6cm},clip]{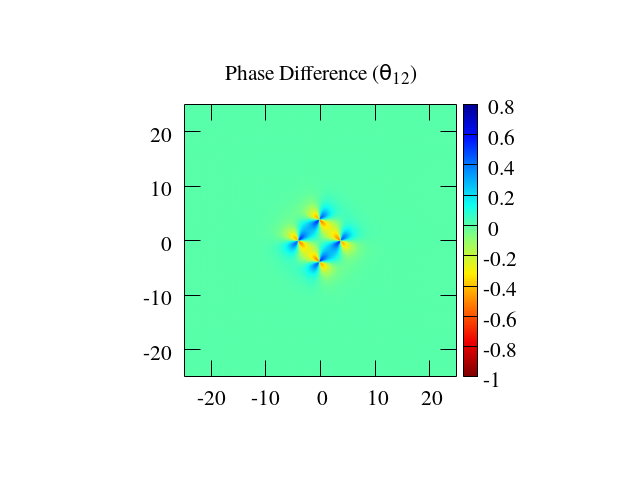}}
      \centerline{\includegraphics[width=0.2\linewidth,trim={4.8cm 1.3cm 3.2cm 1.6cm},clip]{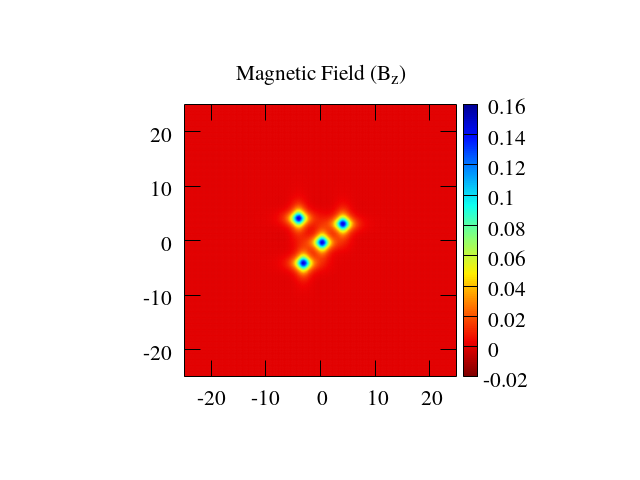}\includegraphics[width=0.2\linewidth,trim={4.8cm 1.3cm 3.2cm 1.6cm},clip]{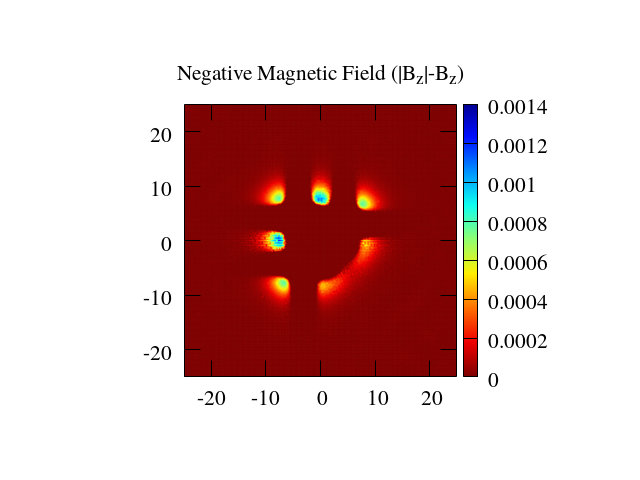}\includegraphics[width=0.2\linewidth,trim={4.8cm 1.3cm 3.2cm 1.6cm},clip]{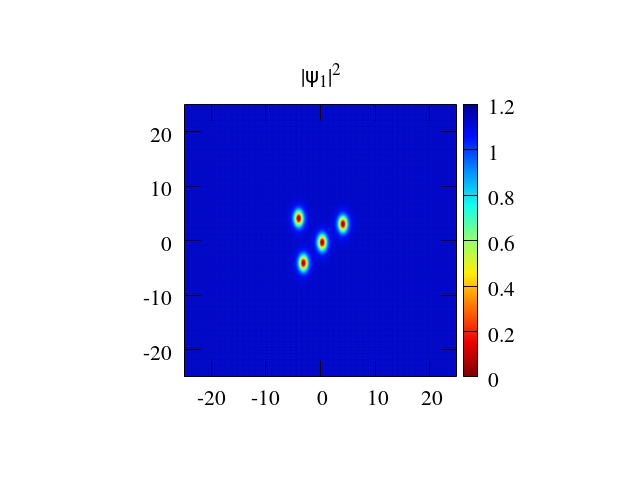}\includegraphics[width=0.2\linewidth,trim={4.8cm 1.3cm 3.2cm 1.6cm},clip]{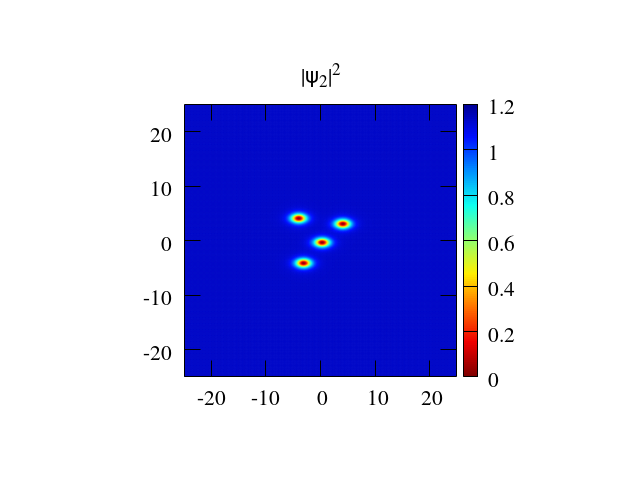}\includegraphics[width=0.2\linewidth,trim={4.8cm 1.3cm 3.2cm 1.6cm},clip]{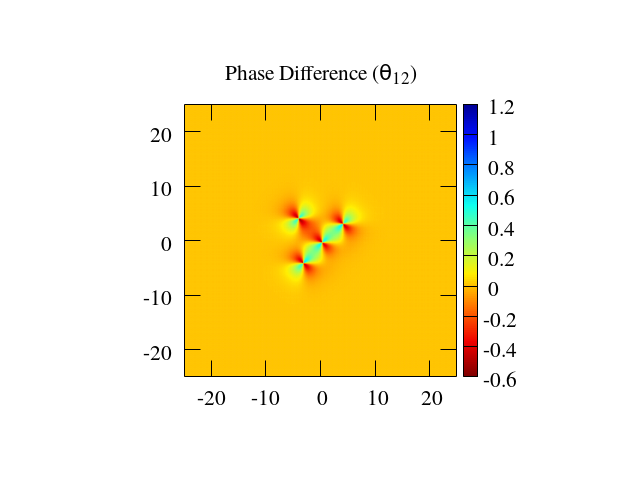}}
       \centerline{\includegraphics[width=0.2\linewidth,trim={4.8cm 1.3cm 3.2cm 1.6cm},clip]{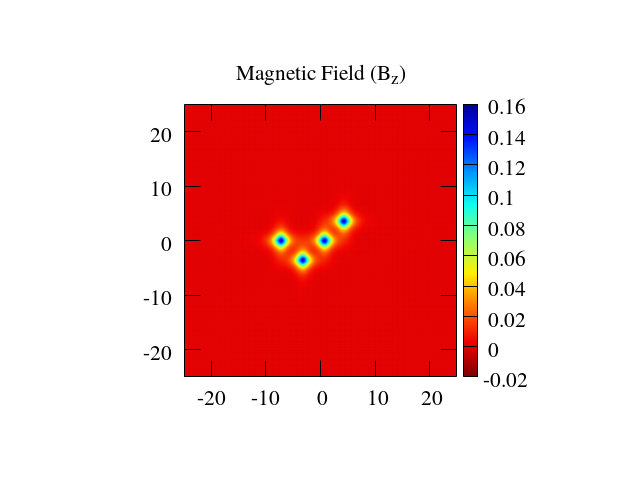}\includegraphics[width=0.2\linewidth,trim={4.8cm 1.3cm 3.2cm 1.6cm},clip]{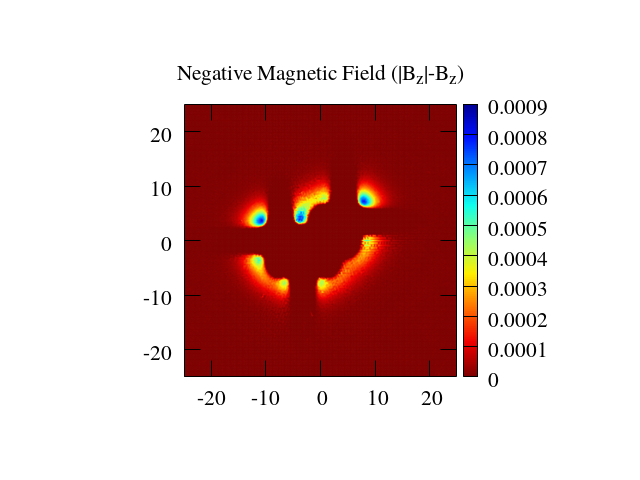}\includegraphics[width=0.2\linewidth,trim={4.8cm 1.3cm 3.2cm 1.6cm},clip]{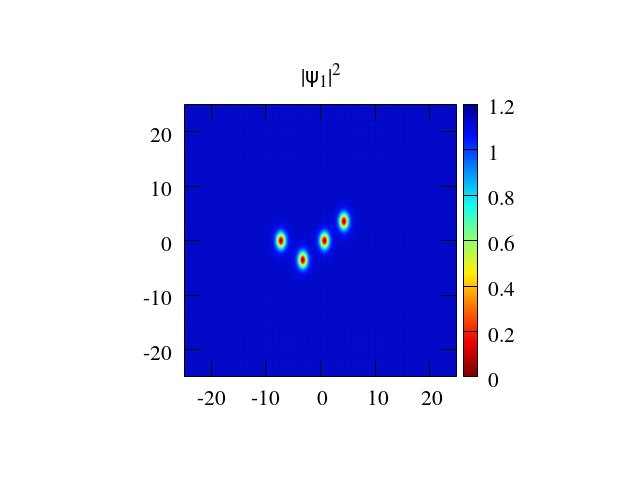}\includegraphics[width=0.2\linewidth,trim={4.8cm 1.3cm 3.2cm 1.6cm},clip]{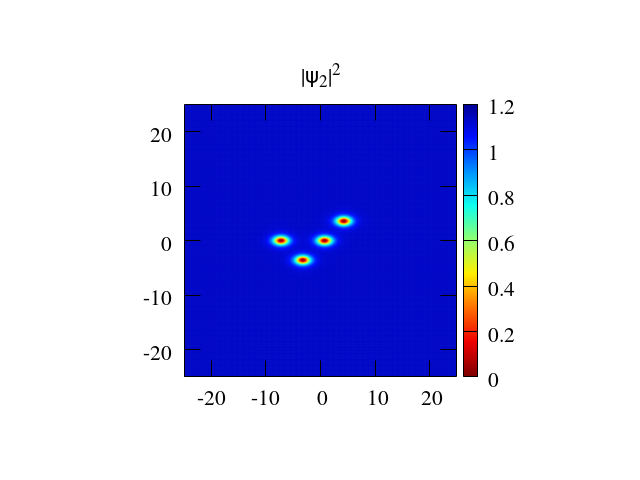}\includegraphics[width=0.2\linewidth,trim={4.8cm 1.3cm 3.2cm 1.6cm},clip]{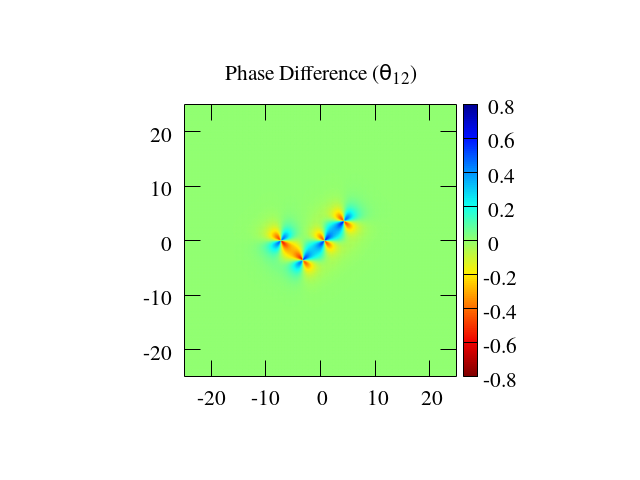}}
        \centerline{\includegraphics[width=0.2\linewidth,trim={4.8cm 1.3cm 3.2cm 1.6cm},clip]{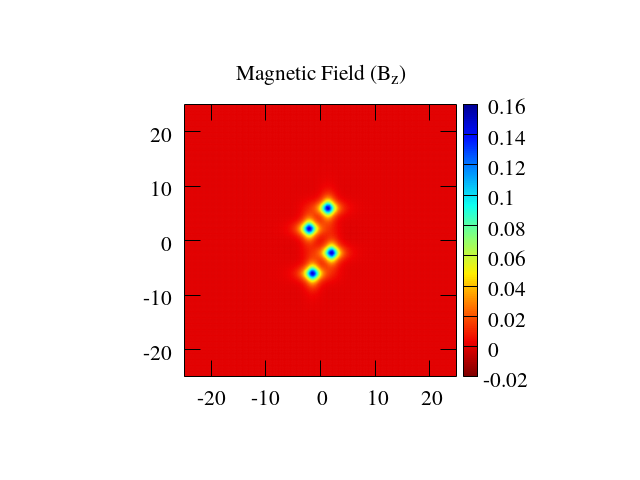}\includegraphics[width=0.2\linewidth,trim={4.8cm 1.3cm 3.2cm 1.6cm},clip]{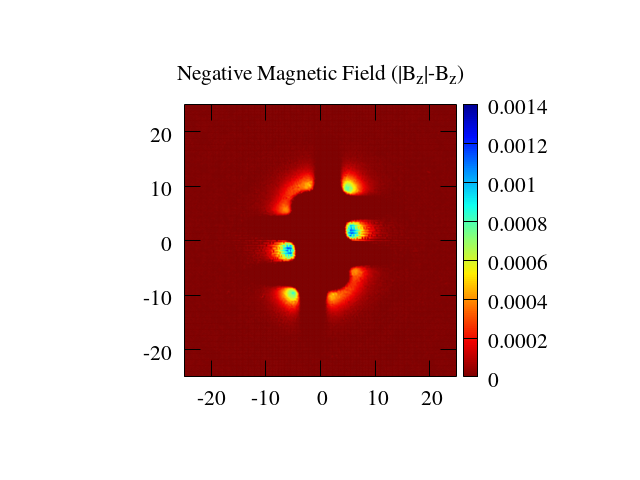}\includegraphics[width=0.2\linewidth,trim={4.8cm 1.3cm 3.2cm 1.6cm},clip]{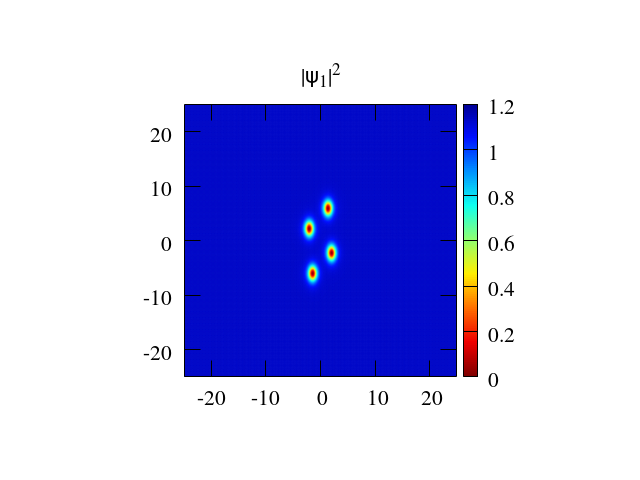}\includegraphics[width=0.2\linewidth,trim={4.8cm 1.3cm 3.2cm 1.6cm},clip]{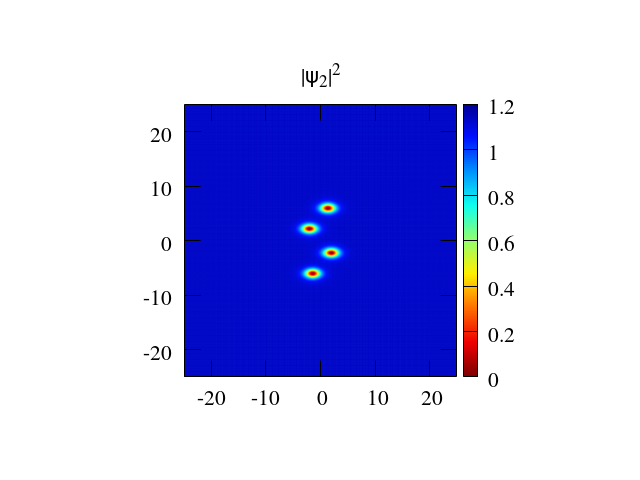}\includegraphics[width=0.2\linewidth,trim={4.8cm 1.3cm 3.2cm 1.6cm},clip]{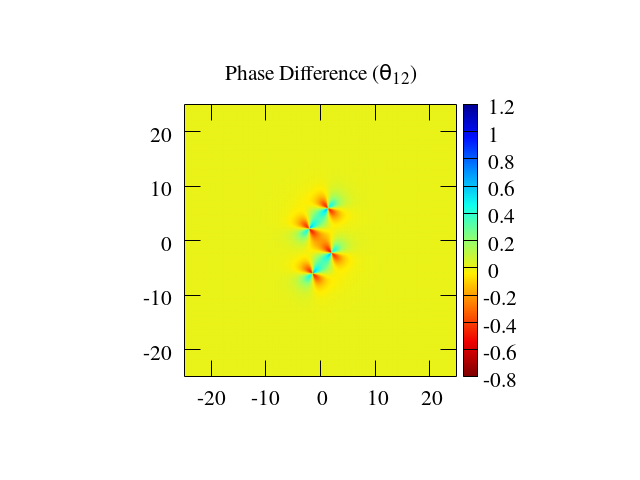}}
 \caption{\label{Fig:type2oppB4p5} 
 $N=4$ our quanta local solutions for anisotropy in both bands in opposite directions $\gamma_1=\gamma_2=2$, $\eta_{12}=0.5$, $\lambda^{-1}_{x2}=\lambda^{-1}_{y1}=1$ and $\lambda^{-1}_{y2}=\lambda^{-1}_{x1}= 0.5$ (a) $B_z$ magnetic field (b) $\left|B_z\right| - B_z$ negative magnetic field (c) $\left|\phi_1\right|^2$ (d)$\left|\phi_2\right|^2$ (e)$\theta_{12}$ phase difference. } 
 \end{figure*}

We finally show the local solutions for $N=4$ for equal anisotropies in the bands in Figs. \ref{Fig:type2oppB4p5} and \ref{Fig:type2oppB4p1}. The global minima for weak anisotropy is the line however the square solution is extremely close in energy, as the London model predicted. For stronger anisotropy we observe the square being rotated and deformed, as predicted by the London model. It is now the Z solution that is close to the line solution, which is again as the London model predicted.

\begin{figure*}[tb!]
  \centerline{\includegraphics[width=0.2\linewidth,trim={4.8cm 1.3cm 3.2cm 1.6cm},clip]{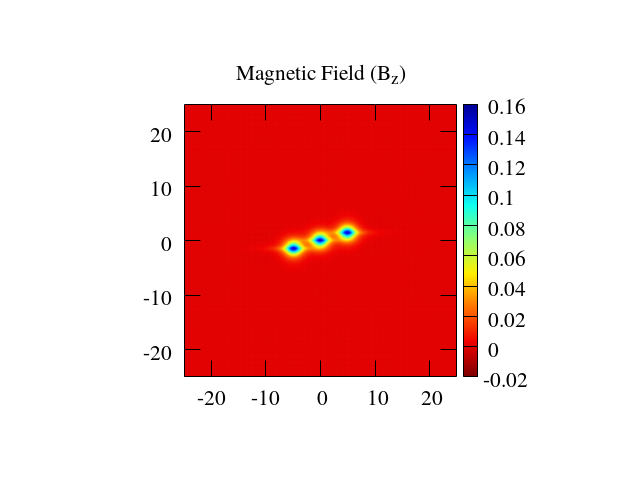}\includegraphics[width=0.2\linewidth,trim={4.8cm 1.3cm 3.2cm 1.6cm},clip]{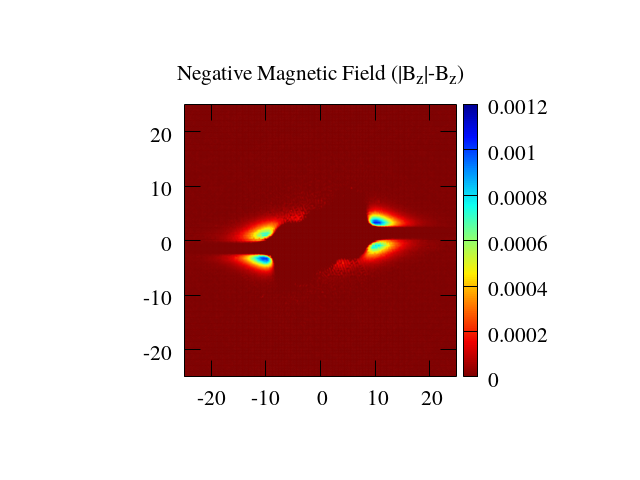}\includegraphics[width=0.2\linewidth,trim={4.8cm 1.3cm 3.2cm 1.6cm},clip]{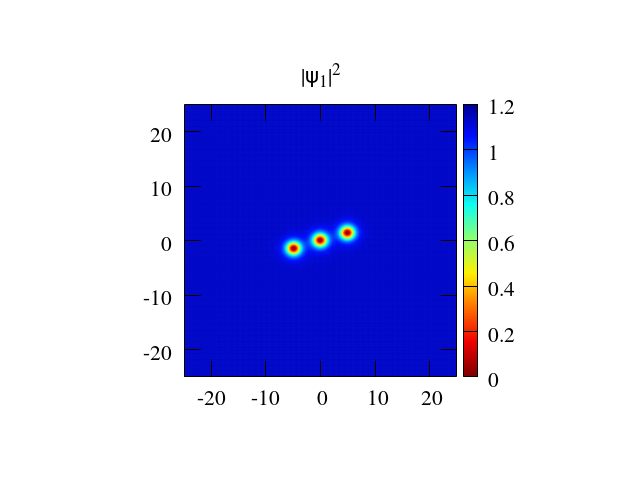}\includegraphics[width=0.2\linewidth,trim={4.8cm 1.3cm 3.2cm 1.6cm},clip]{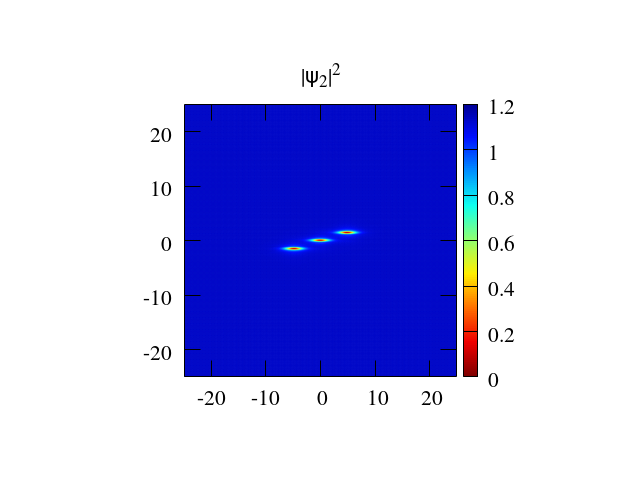}\includegraphics[width=0.2\linewidth,trim={4.8cm 1.3cm 3.2cm 1.6cm},clip]{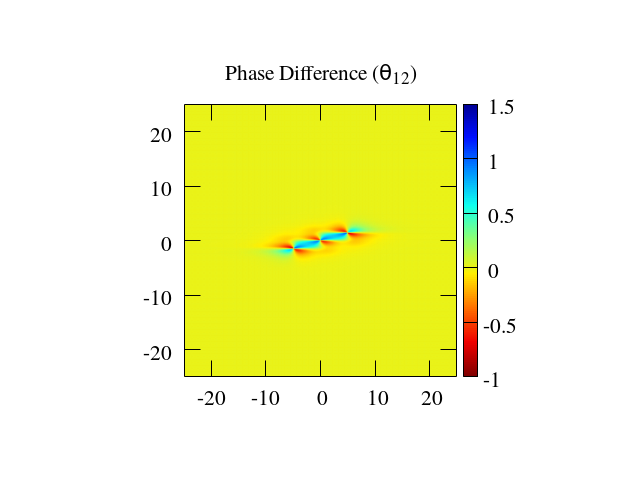}}
    \centerline{\includegraphics[width=0.2\linewidth,trim={4.8cm 1.3cm 3.2cm 1.6cm},clip]{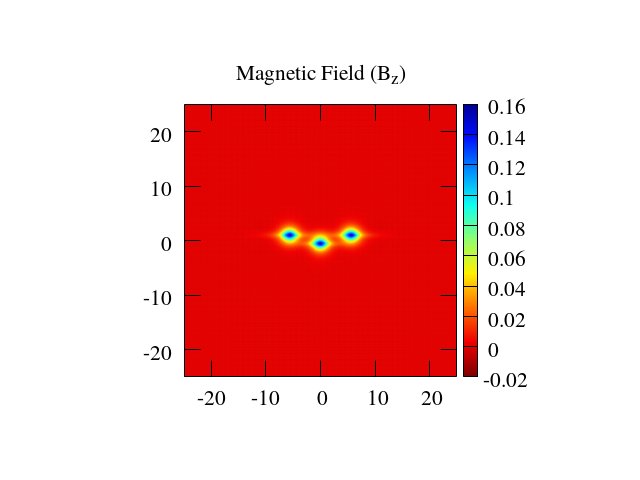}\includegraphics[width=0.2\linewidth,trim={4.8cm 1.3cm 3.2cm 1.6cm},clip]{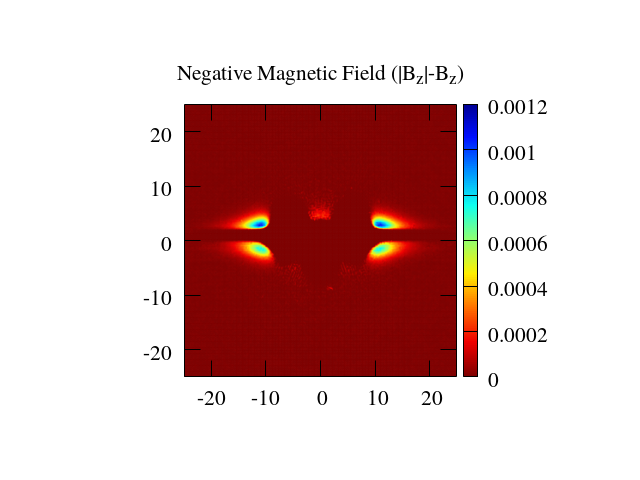}\includegraphics[width=0.2\linewidth,trim={4.8cm 1.3cm 3.2cm 1.6cm},clip]{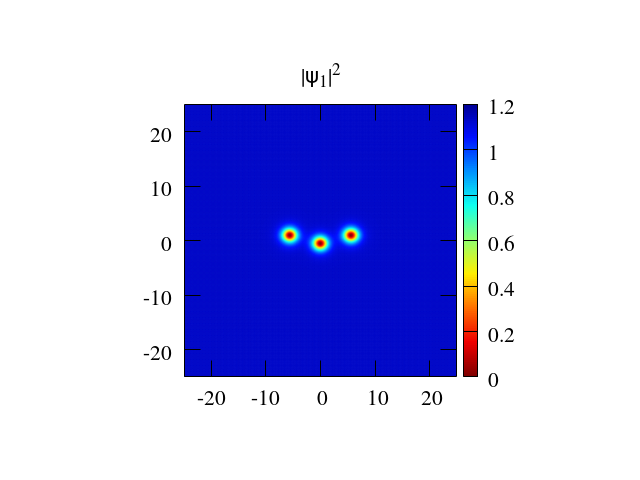}\includegraphics[width=0.2\linewidth,trim={4.8cm 1.3cm 3.2cm 1.6cm},clip]{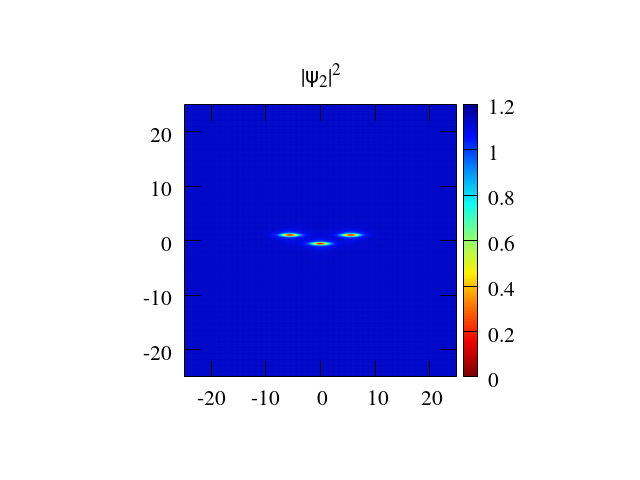}\includegraphics[width=0.2\linewidth,trim={4.8cm 1.3cm 3.2cm 1.6cm},clip]{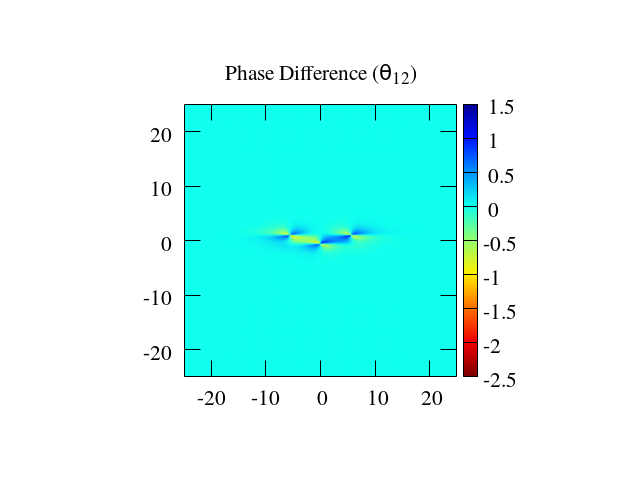}}
      \centerline{\includegraphics[width=0.2\linewidth,trim={4.8cm 1.3cm 3.2cm 1.6cm},clip]{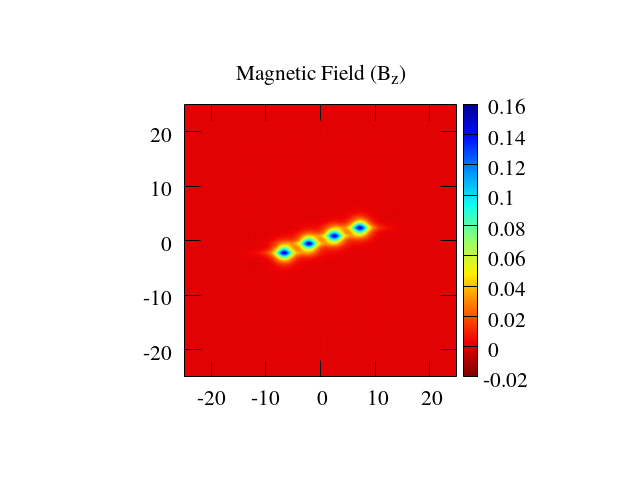}\includegraphics[width=0.2\linewidth,trim={4.8cm 1.3cm 3.2cm 1.6cm},clip]{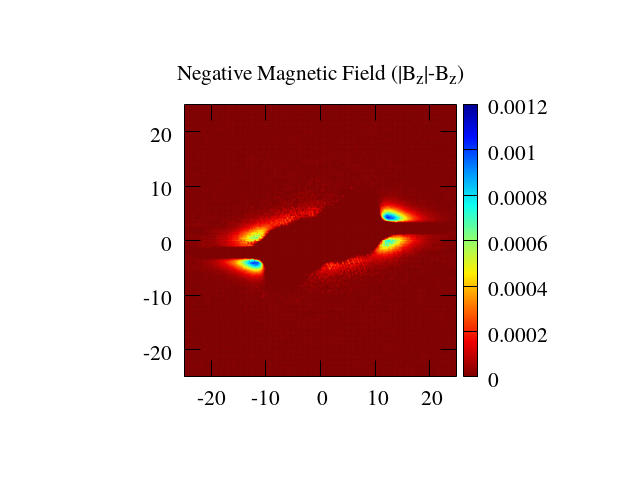}\includegraphics[width=0.2\linewidth,trim={4.8cm 1.3cm 3.2cm 1.6cm},clip]{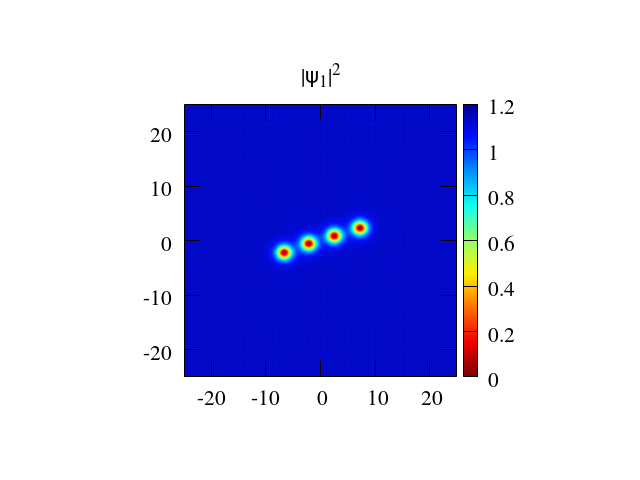}\includegraphics[width=0.2\linewidth,trim={4.8cm 1.3cm 3.2cm 1.6cm},clip]{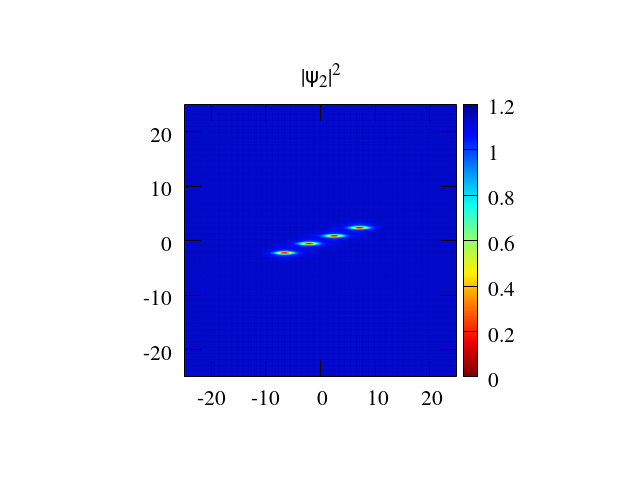}\includegraphics[width=0.2\linewidth,trim={4.8cm 1.3cm 3.2cm 1.6cm},clip]{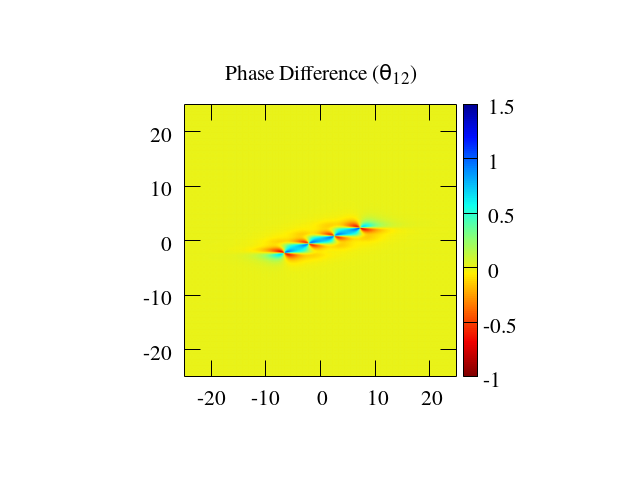}}
      \centerline{\includegraphics[width=0.2\linewidth,trim={4.8cm 1.3cm 3.2cm 1.6cm},clip]{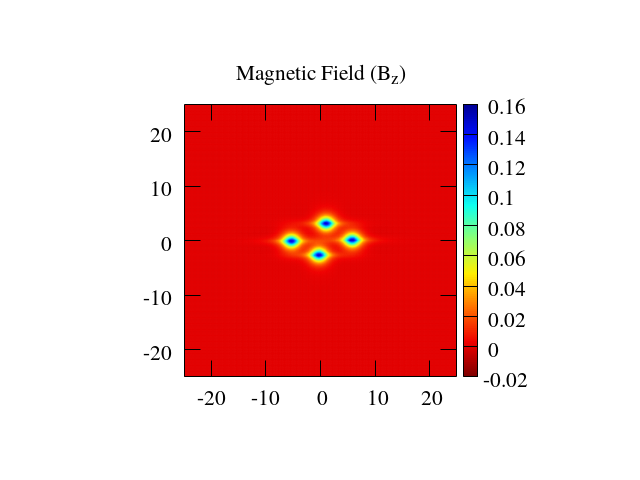}\includegraphics[width=0.2\linewidth,trim={4.8cm 1.3cm 3.2cm 1.6cm},clip]{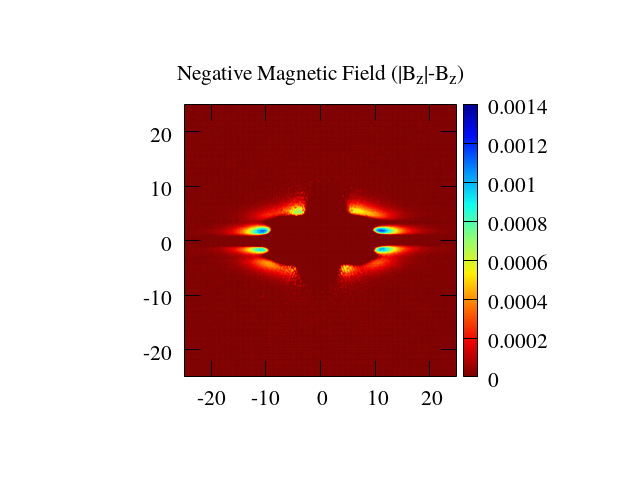}\includegraphics[width=0.2\linewidth,trim={4.8cm 1.3cm 3.2cm 1.6cm},clip]{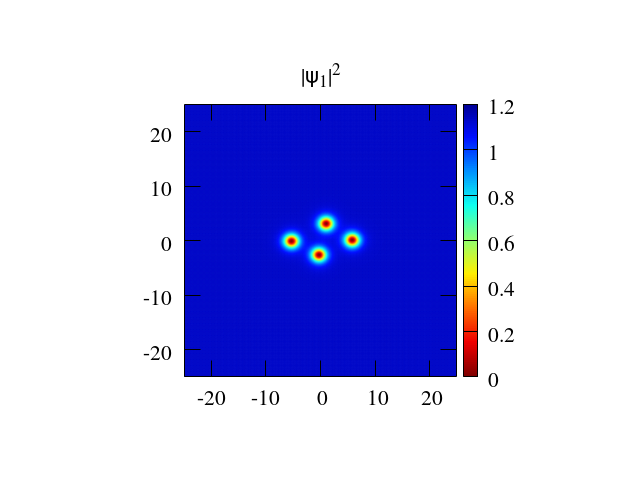}\includegraphics[width=0.2\linewidth,trim={4.8cm 1.3cm 3.2cm 1.6cm},clip]{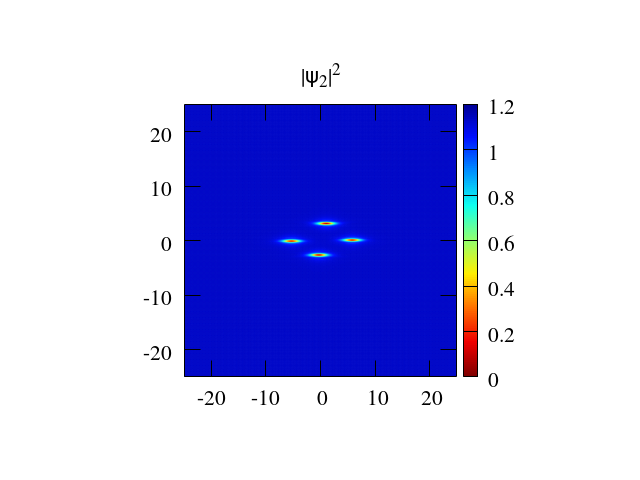}\includegraphics[width=0.2\linewidth,trim={4.8cm 1.3cm 3.2cm 1.6cm},clip]{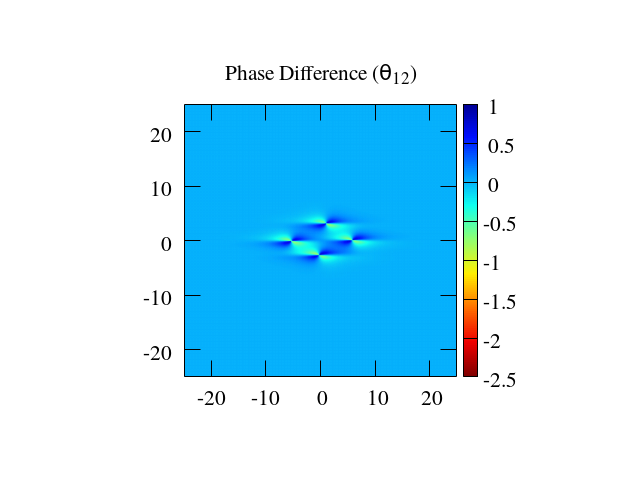}}
      \centerline{\includegraphics[width=0.2\linewidth,trim={4.8cm 1.3cm 3.2cm 1.6cm},clip]{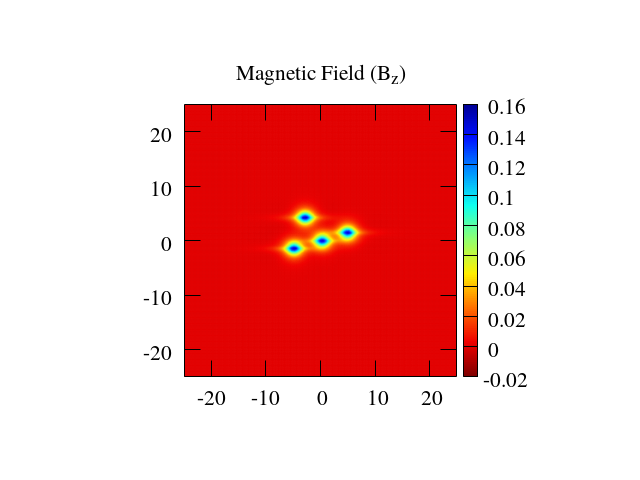}\includegraphics[width=0.2\linewidth,trim={4.8cm 1.3cm 3.2cm 1.6cm},clip]{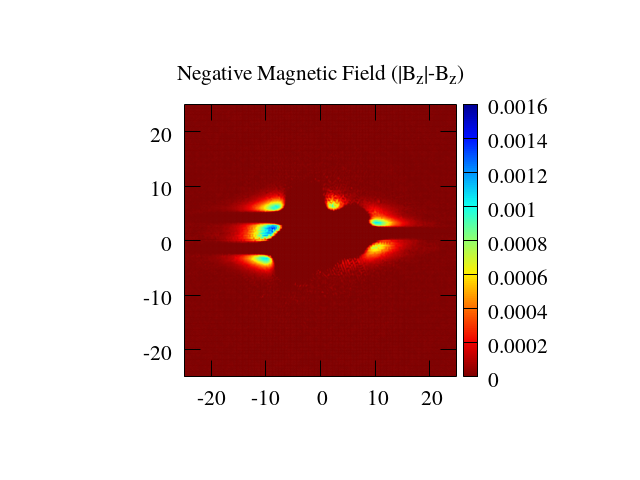}\includegraphics[width=0.2\linewidth,trim={4.8cm 1.3cm 3.2cm 1.6cm},clip]{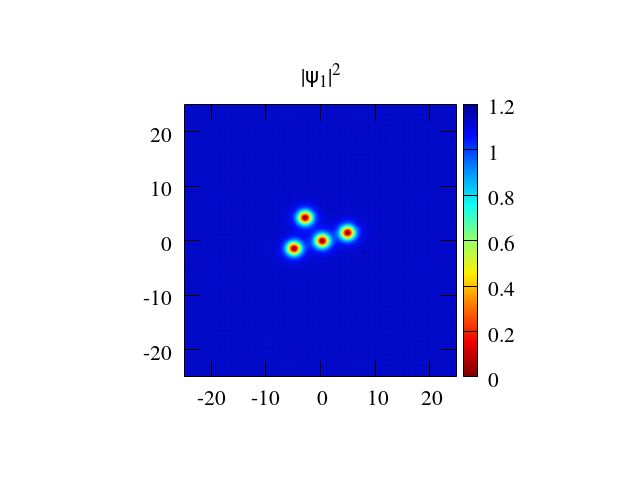}\includegraphics[width=0.2\linewidth,trim={4.8cm 1.3cm 3.2cm 1.6cm},clip]{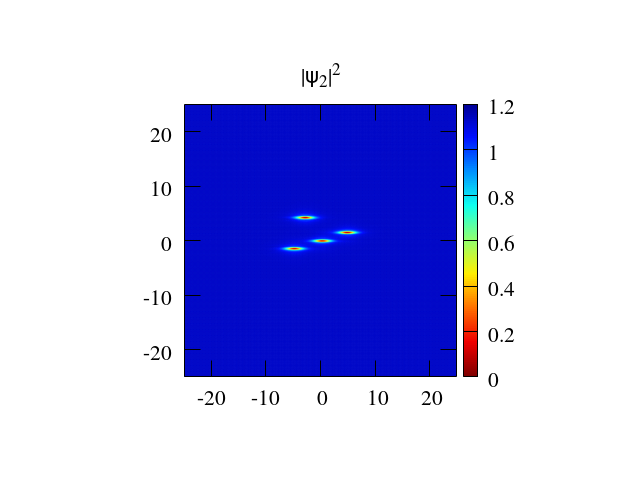}\includegraphics[width=0.2\linewidth,trim={4.8cm 1.3cm 3.2cm 1.6cm},clip]{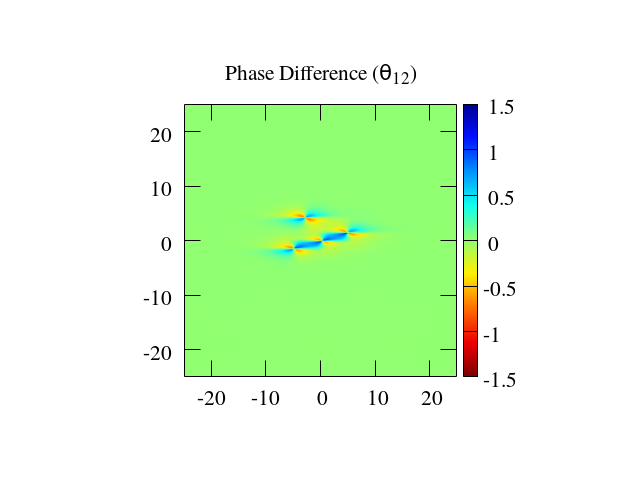}}
       \centerline{\includegraphics[width=0.2\linewidth,trim={4.8cm 1.3cm 3.2cm 1.6cm},clip]{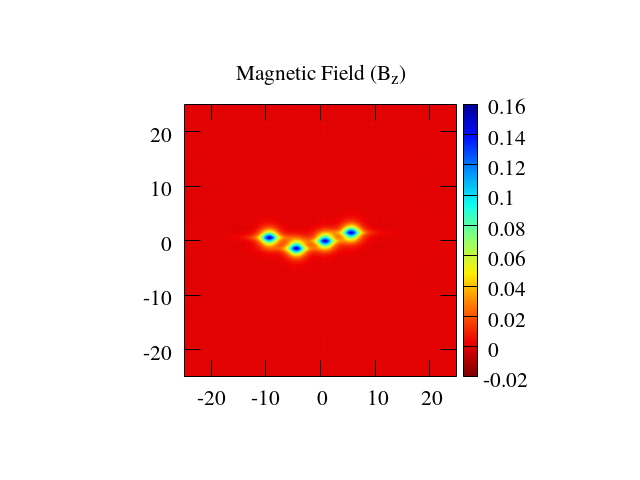}\includegraphics[width=0.2\linewidth,trim={4.8cm 1.3cm 3.2cm 1.6cm},clip]{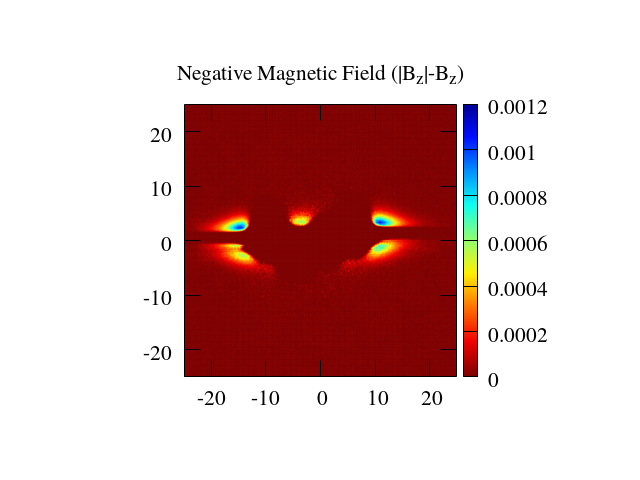}\includegraphics[width=0.2\linewidth,trim={4.8cm 1.3cm 3.2cm 1.6cm},clip]{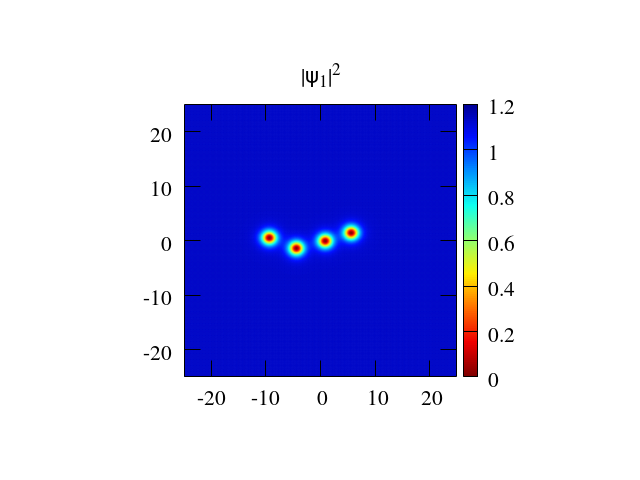}\includegraphics[width=0.2\linewidth,trim={4.8cm 1.3cm 3.2cm 1.6cm},clip]{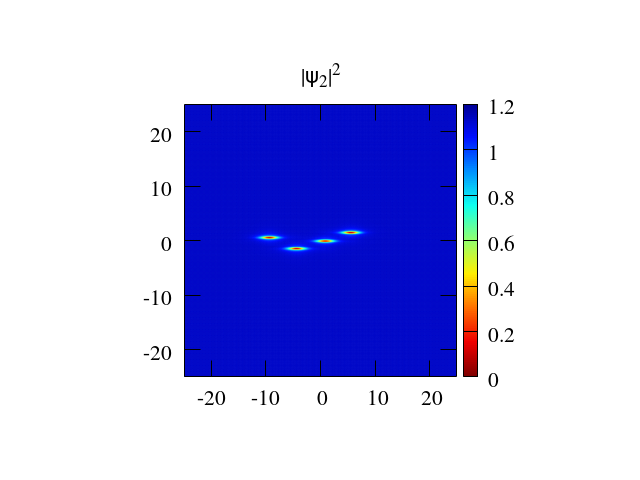}\includegraphics[width=0.2\linewidth,trim={4.8cm 1.3cm 3.2cm 1.6cm},clip]{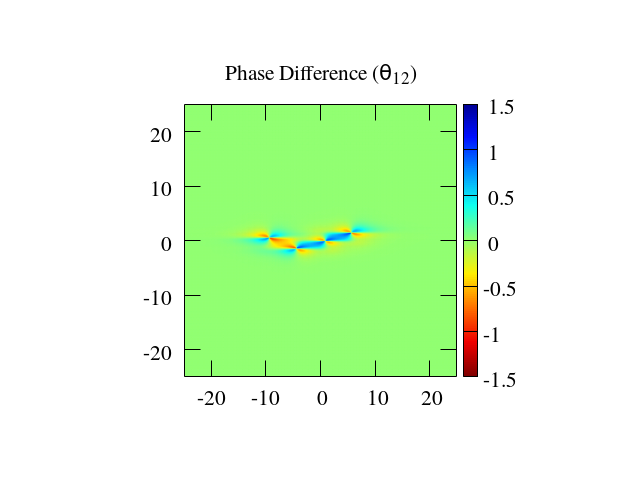}}
        \centerline{\includegraphics[width=0.2\linewidth,trim={4.8cm 1.3cm 3.2cm 1.6cm},clip]{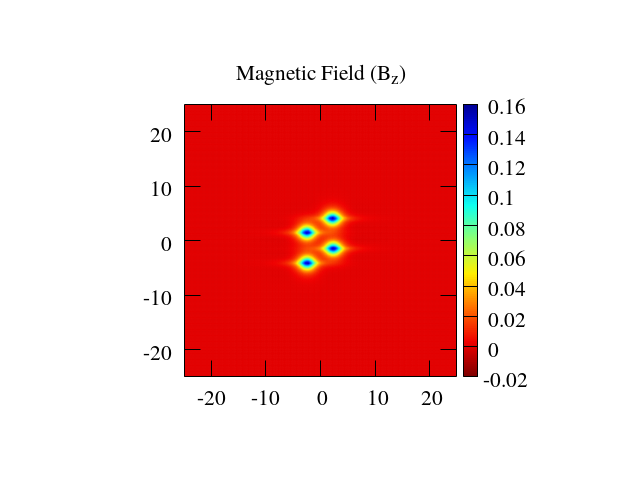}\includegraphics[width=0.2\linewidth,trim={4.8cm 1.3cm 3.2cm 1.6cm},clip]{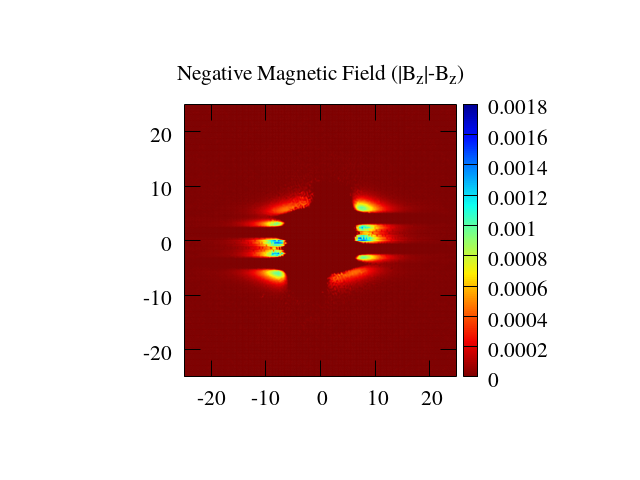}\includegraphics[width=0.2\linewidth,trim={4.8cm 1.3cm 3.2cm 1.6cm},clip]{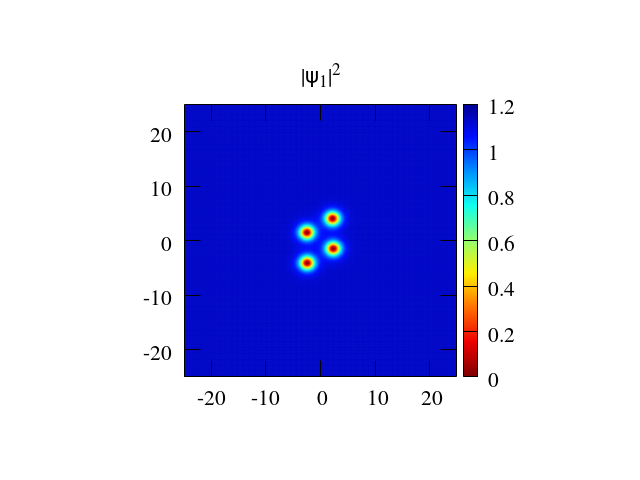}\includegraphics[width=0.2\linewidth,trim={4.8cm 1.3cm 3.2cm 1.6cm},clip]{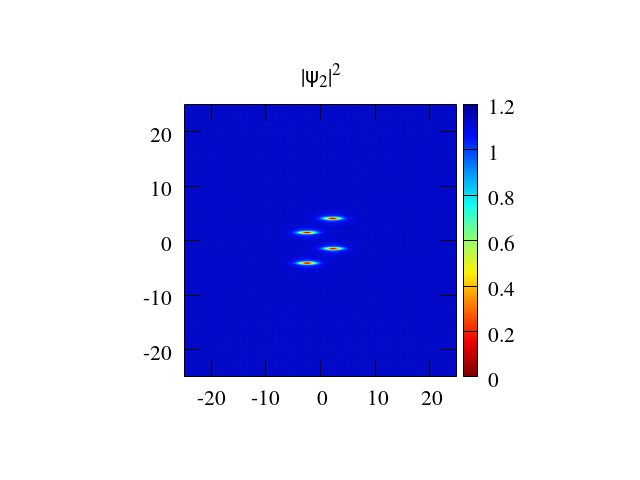}\includegraphics[width=0.2\linewidth,trim={4.8cm 1.3cm 3.2cm 1.6cm},clip]{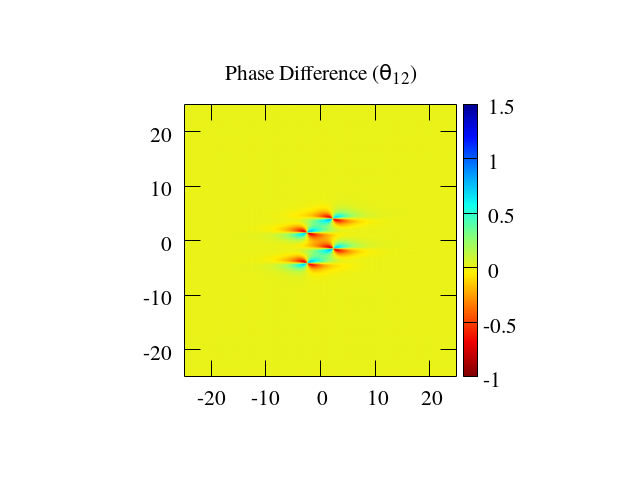}}
 \caption{\label{Fig:type2oneB34} 
 $N=3,4$ three and four quanta local solutions for strong anisotropy in one direction $\gamma_1=\gamma_2=2$, $\eta_{12}=0.5$, $\lambda_{x2}=\lambda_{x1}=\lambda_{y1}=1$ and $\lambda^{-1}_{y2} = 0.1$ (a) $B_z$ magnetic field (b) $\left|B_z\right| - B_z$ negative magnetic field (c) $\left|\phi_1\right|^2$ (d)$\left|\phi_2\right|^2$ (e)$\theta_{12}$ phase difference. } 
 \end{figure*}

Finally, for anisotropy in one direction we observe similar results to the London model, with chain solutions taking the minimal energy solutions as expected. Various local and global minima solutions are plotted for $N=3,4$ in Fig. \ref{Fig:type2oneB34}. Note that the energy of the L solution approaches that of the line solution, due to the deformation from the line discussed in the London model. The other local solutions have much higher energy compared to the L and line solutions.

This means that for the type 2 regime, the London model is good at predicting the qualitative form for the higher quanta solutions. This is despite the fact that when one goes beyond the London limit, there appear additional terms that contribute to field inversion
 \cite{Babaev2009}.
Hence it would be likely to continue to take the various forms predicted for higher quanta above.
A different regime, that appears in the Ginzburg-Landau theory and is not captured
by a London model, is where coherence lengths exceed magnetic field penetration lengths.
In that regime vortex bound states form via a different mechanism (see discussion 
for the isotropic case in \cite{Babaev2005,Carlstroem2011,Silaev2011,BABAEV201720}) 
Addition of anisotropy to these regimes also leads to anisotropic vortex cluster solutions and vortex chains  \cite{winyard2018hierarchies}.
 
\section{Conclusions}
The magnetic response of isotropic single-component superconductors can be cast in the form of a massive vector field theory characterized by the London magnetic field penetration length. However superconducting materials are often multiband and anisotropic. We have demonstrated that this leads to deviation of the magnetic properties of the model from  London's hydromagnetostatics even at the level of the anisotropic multiband London model. We showed that anisotropy leads to hybridization of the Leggett and the London modes, causing the gradients of the phase difference to create transverse charge currents, generating magnetic field. This in turn leads to the existence of several magnetic field penetration lengths (in general $N+1$ magnetic field penetration lengths for an $N$-band London model) and also to a non-local magnetic response in the nominally local London model. For example in the case of the two-band anistropic Meissner state, the magnetic field, directed along one of the crystal axes, decays according to a double-exponential law. In the general case of arbitrary directions of such two-band case, there are three magnetic modes with different penetration lengths. In the limit of vanishingly small mass for the Leggett mode (e.g. near $s+is$ transition, one of the magnetic field penetration lengths diverged, leading to long-range, small-amplitude penetration of magnetic field even far below the superconducting phase transition.

{\sl Under certain conditions the magnetic field is described by a massive vector field
theory with complex mass, which means that magnetic field decay cannot
be entirely characterized by a real length scales but has oscillating behaviour.
Moreover}
 the combination of different magnetic modes gives the overall 
 magnetic field profile a non-monotonic form and even magnetic field inversion. This affects the nature of vortex states: the non-monotonic behaviour and field inversion leads to the formation of vortex bound states. The minimal energy bound states were shown to depend on the symmetry of the system and have forms of polyominoe vortex clusters and chains. 
 
A number of multiband superconductors are currently the subject of detailed experimental research. 
 The examples  studied in this paper give inverted magnetic field up to $10^{-3}$ of applied magnetic field.
 Such field strength makes the effect in principle measurable  either by SQUIDs or in muon spin rotation experiments.
 The interband coupling strength is often difficult to calculate precisely. Measuring the effect that we report for
 samples with different boundaries, cut relative to crystaline axises, can be used as a tool to experimentally assess interband coupling strength and relative anisotropies of bands. It can additionally be used to distinguish the vortex 
 bound states that we report from vortex clusters and chains forming for different reasons.
In anisotropic multiband systems, there are at least two other mechanisms for formation of vortex bound states. Inclusion of density variations in the theory yields ``type-1.5" regimes where coherence lengths are 
 larger than magnetic field penetration lengths and vortex bound states form due to core-core interaction \cite{winyard2018hierarchies}. Compared to the core-core-interaction-driven vortex binding, the mechanism considered on this paper yields much weaker vortex interaction forces. Thus vortex bound states considered here  should be relatively easily destroyed by thermal fluctuations. Note also that one should expect vortices sticking to sample's boundaries due to the fact that those also feature inverted field.
 Different situation  appears for strong anisotropies where integer flux
  vortices split  into bound states of fractional vortices \cite{winyard2018skyrmions}. Those are easily distinguishable, due to the different
  magnetic field profile and coreless nature. 
 
\begin{acknowledgments}
We acknowledge  useful discussions with Julien Garaud,  Vadim Grinenko and Stephen Lee.
The work was supported by the Academy of Finland,
Swedish Research Council Grants
No. 642-2013-7837, VR2016-06122, Goran Gustafsson Foundation for Research in Natural Sciences and Medicine and EPSERC Grant No. EP/P024688/1.
The computations were performed on resources 
provided by the Swedish National Infrastructure for Computing 
(SNIC) at National Supercomputer Center at Link\"oping, Sweden.
\end{acknowledgments}

\appendix

\section{Additional Solutions}

  \begin{figure*}
 \centerline{\includegraphics[width=0.2934\linewidth]{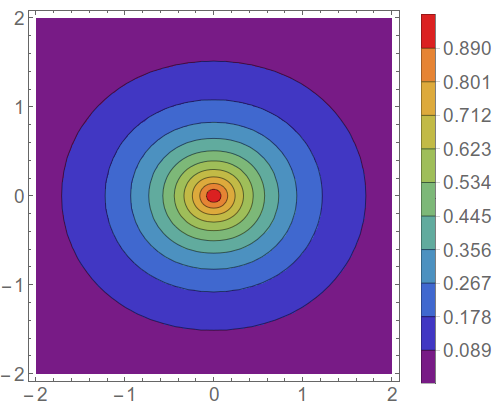}\includegraphics[width=0.3312\linewidth]{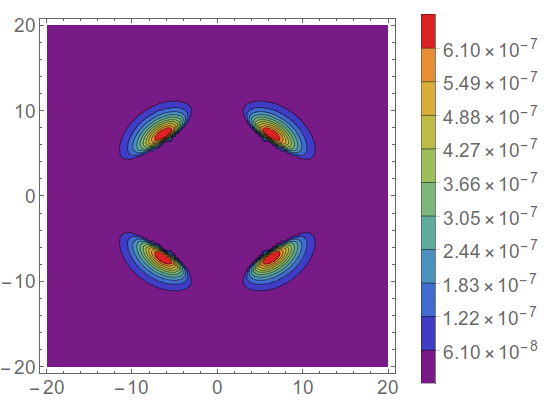}\includegraphics[width=0.2934\linewidth]{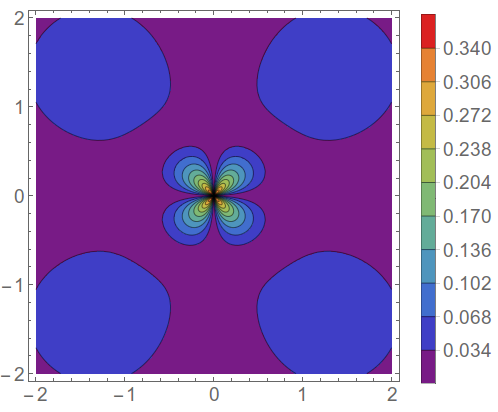}} \centerline{\includegraphics[width=0.2934\linewidth]{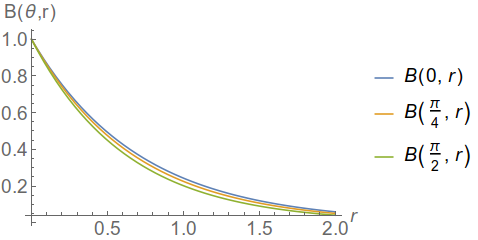}\includegraphics[width=0.3312\linewidth]{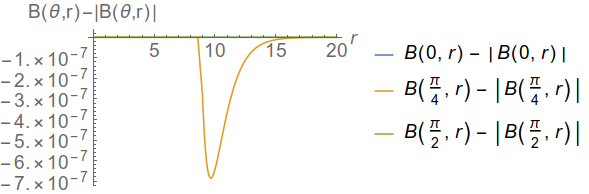}\includegraphics[width=0.2934\linewidth]{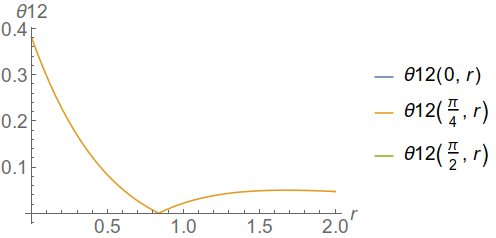}} \centerline{\includegraphics[width=0.2934\linewidth]{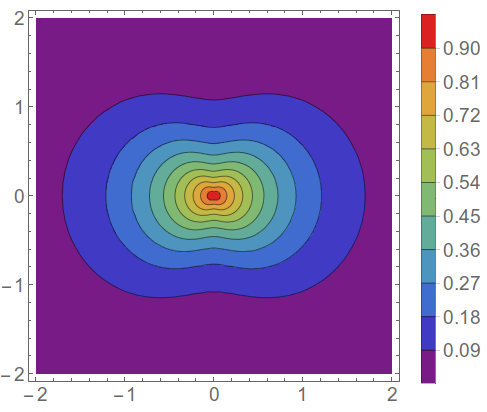}\includegraphics[width=0.3312\linewidth]{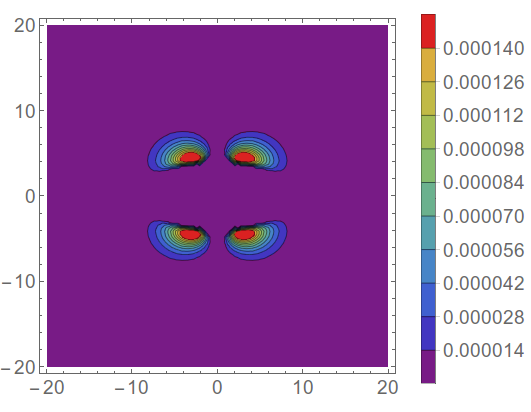}\includegraphics[width=0.2934\linewidth]{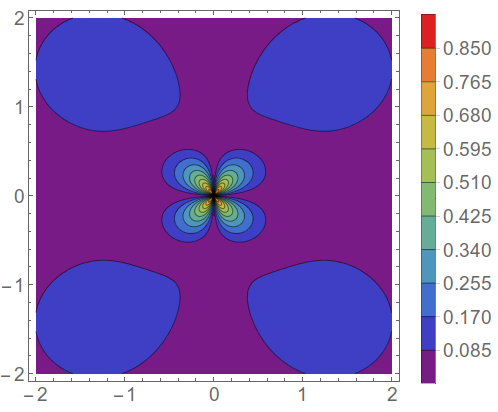}} \centerline{\includegraphics[width=0.2934\linewidth]{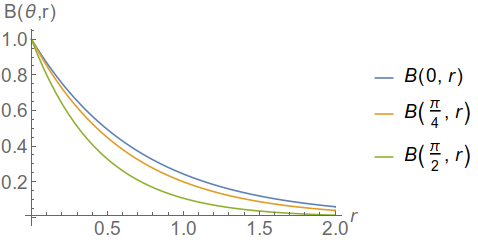}\includegraphics[width=0.3312\linewidth]{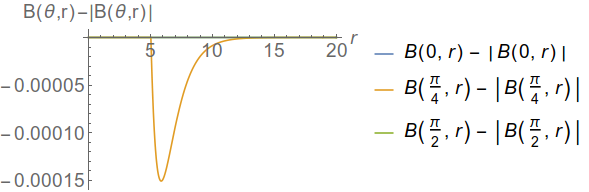}\includegraphics[width=0.2934\linewidth]{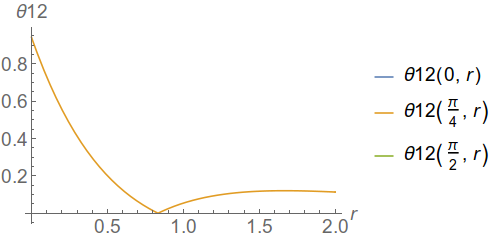}} \centerline{\includegraphics[width=0.2934\linewidth]{boundary_one_p1_p7sq_contour.png}\includegraphics[width=0.3312\linewidth]{boundary_one_p1_p7sq_negcontour.png}\includegraphics[width=0.2934\linewidth]{boundary_one_p1_p7sq_thetacontour.png}} \centerline{\includegraphics[width=0.2934\linewidth]{boundary_one_p1_p7sq_line.png}\includegraphics[width=0.3312\linewidth]{boundary_one_p1_p7sq_negline.png}\includegraphics[width=0.2934\linewidth]{boundary_one_p1_p7sq_thetaline.png}}
 \caption{A contour plot and a radial slice at various angles of the magnetic field, negative magnetic field and phase difference for the 1d boundary problem solution with anisotropy in a single component ($\lambda_{2x}=\lambda_{2y}=\lambda_{1y}=1$). A radial curve from the centre of the plot represents the field orthogonal to a 1d boundary crossing the origin in the x-y plane. This way every possible direction (or $\theta$) is plotted for equations \ref{Eq:LondonsolB} and \ref{Eq:Londonsoltheta} with the radial distance representing $r$. The plot quantities are Magnetic field $B_z$ (left), negative magnetic field $|B_z|-B_z$ (centre) and phase difference $\theta_{12}$ (right) for various strengths of anisotropy from weak to strong (a) $\lambda_{1x}=0.8\lambda_{1y}$ (b) $\lambda_{1x}=0.5\lambda_{1y}$ (c) $\lambda_{1x}=0.1\lambda_{1y}$.}
 \label{Fig:LondonBoundarySameAll}
   \end{figure*}
   
\begin{figure*}
 \centerline{\includegraphics[width=0.2934\linewidth]{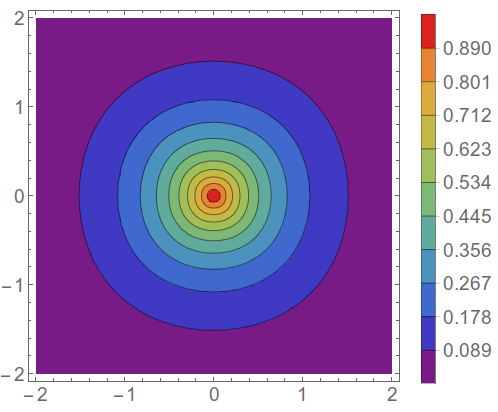}\includegraphics[width=0.3312\linewidth]{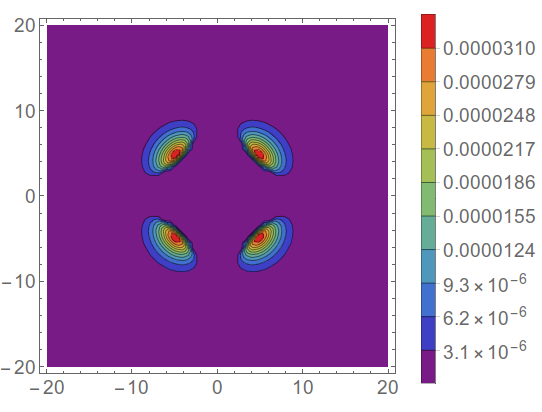}\includegraphics[width=0.2934\linewidth]{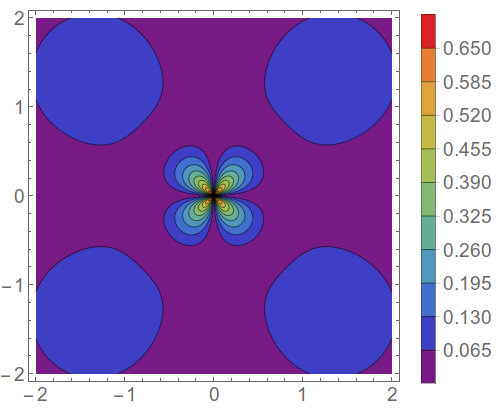}} \centerline{\includegraphics[width=0.2934\linewidth]{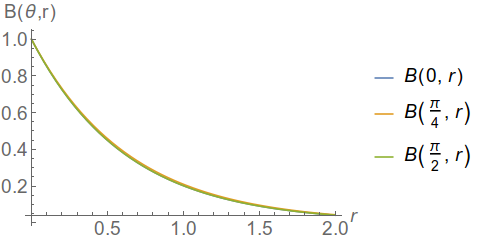}\includegraphics[width=0.3312\linewidth]{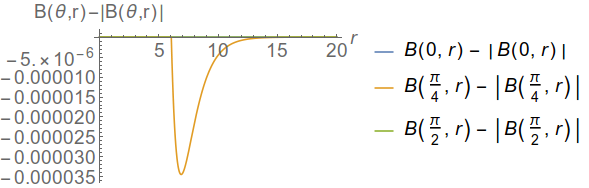}\includegraphics[width=0.2934\linewidth]{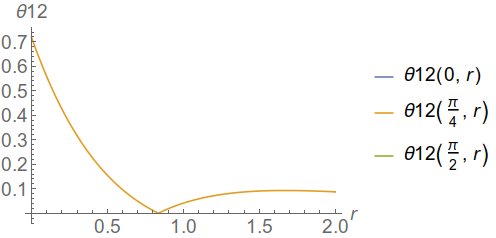}} \centerline{\includegraphics[width=0.2934\linewidth]{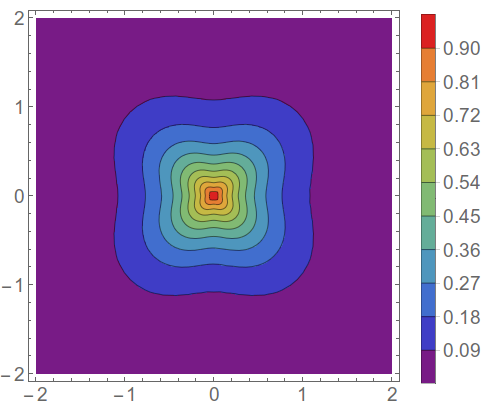}\includegraphics[width=0.3312\linewidth]{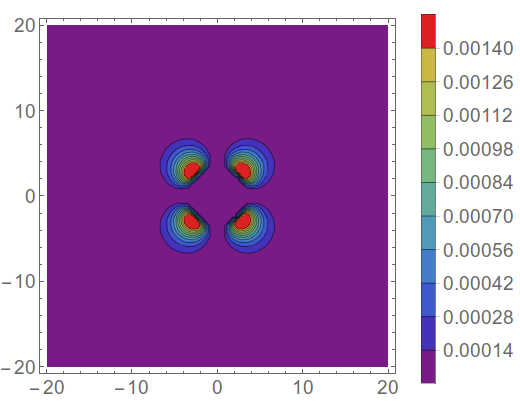}\includegraphics[width=0.2934\linewidth]{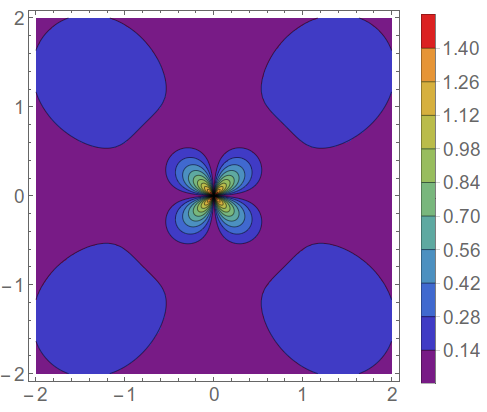}} \centerline{\includegraphics[width=0.2934\linewidth]{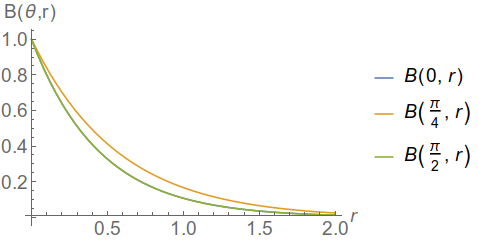}\includegraphics[width=0.3312\linewidth]{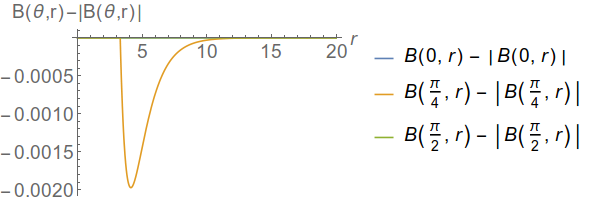}\includegraphics[width=0.2934\linewidth]{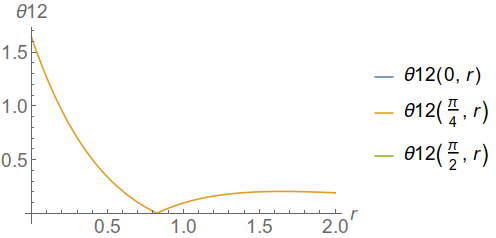}} \centerline{\includegraphics[width=0.2934\linewidth]{boundary_opp_p1_p7sq_contour.png}\includegraphics[width=0.3312\linewidth]{boundary_opp_p1_p7sq_negcontour.png}\includegraphics[width=0.2934\linewidth]{boundary_opp_p1_p7sq_thetacontour.png}} \centerline{\includegraphics[width=0.2934\linewidth]{boundary_opp_p1_p7sq_line.png}\includegraphics[width=0.3312\linewidth]{boundary_opp_p1_p7sq_negline.png}\includegraphics[width=0.2934\linewidth]{boundary_opp_p1_p7sq_thetaline.png}}
 \caption{A contour plot and a radial slice at various angles for the 1d boundary problem solution with anisotropy in opposite directions in each component ($\lambda_{2x}=\lambda_{1y}=1$ and $\lambda_{1x}=\lambda_{2y}$). A radial curve from the centre of the plot represents the field orthogonal to a 1d boundary crossing the origin in the x-y plane. This way every possible direction (or $\theta$) is plotted for equations \ref{Eq:LondonsolB} and \ref{Eq:Londonsoltheta} with the radial distance representing $r$. The plot quantities are Magnetic field $B_z$ (left), negative magnetic field $|B_z|-B_z$ (centre) and phase difference $\theta_{12}$ (right) for various strengths of anisotropy from weak to strong (a) $\lambda_{1x}=\lambda_{2y}=0.8\lambda_{1y}$ (b) $\lambda_{1x}=\lambda_{2y}=0.5\lambda_{1y}$ (c) $\lambda_{1x}=\lambda_{2y}=0.1\lambda_{1y}$.}
 \label{Fig:LondonBoundaryOppAll}
   \end{figure*}
   
   \bibliography{LiteratureAnisotropic}

\begin{thebibliography}{44}
\expandafter\ifx\csname natexlab\endcsname\relax\def\natexlab#1{#1}\fi
\expandafter\ifx\csname bibnamefont\endcsname\relax
  \def\bibnamefont#1{#1}\fi
\expandafter\ifx\csname bibfnamefont\endcsname\relax
  \def\bibfnamefont#1{#1}\fi
\expandafter\ifx\csname citenamefont\endcsname\relax
  \def\citenamefont#1{#1}\fi
\expandafter\ifx\csname url\endcsname\relax
  \def\url#1{\texttt{#1}}\fi
\expandafter\ifx\csname urlprefix\endcsname\relax\def\urlprefix{URL }\fi
\providecommand{\bibinfo}[2]{#2}
\providecommand{\eprint}[2][]{\url{#2}}

\bibitem[{\citenamefont{London and London}(1935)}]{London1935}
\bibinfo{author}{\bibfnamefont{F.}~\bibnamefont{London}} \bibnamefont{and}
  \bibinfo{author}{\bibfnamefont{H.}~\bibnamefont{London}},
  \bibinfo{journal}{Proc R Soc Lond A Math Phys Sci}
  \textbf{\bibinfo{volume}{149}}, \bibinfo{pages}{71} (\bibinfo{year}{1935}),
  \urlprefix\url{http://rspa.royalsocietypublishing.org/content/149/866/71.abstract}.

\bibitem[{\citenamefont{Anderson}(1963)}]{Anderson1963}
\bibinfo{author}{\bibfnamefont{P.~W.} \bibnamefont{Anderson}},
  \bibinfo{journal}{Phys. Rev.} \textbf{\bibinfo{volume}{130}},
  \bibinfo{pages}{439} (\bibinfo{year}{1963}),
  \urlprefix\url{https://link.aps.org/doi/10.1103/PhysRev.130.439}.

\bibitem[{\citenamefont{Mazin and Antropov}(2003)}]{Mazin2003}
\bibinfo{author}{\bibfnamefont{I.~I.} \bibnamefont{Mazin}} \bibnamefont{and}
  \bibinfo{author}{\bibfnamefont{V.~P.} \bibnamefont{Antropov}},
  \bibinfo{journal}{Physica C: Superconductivity}
  \textbf{\bibinfo{volume}{385}}, \bibinfo{pages}{49} (\bibinfo{year}{2003}),
  ISSN \bibinfo{issn}{0921-4534},
  \urlprefix\url{http://www.sciencedirect.com/science/article/pii/S0921453402022992}.

\bibitem[{\citenamefont{Damascelli et~al.}(2000)\citenamefont{Damascelli, Lu,
  Shen, Armitage, Ronning, Feng, Kim, Shen, Kimura, Tokura
  et~al.}}]{Damascelli2000}
\bibinfo{author}{\bibfnamefont{A.}~\bibnamefont{Damascelli}},
  \bibinfo{author}{\bibfnamefont{D.~H.} \bibnamefont{Lu}},
  \bibinfo{author}{\bibfnamefont{K.~M.} \bibnamefont{Shen}},
  \bibinfo{author}{\bibfnamefont{N.~P.} \bibnamefont{Armitage}},
  \bibinfo{author}{\bibfnamefont{F.}~\bibnamefont{Ronning}},
  \bibinfo{author}{\bibfnamefont{D.~L.} \bibnamefont{Feng}},
  \bibinfo{author}{\bibfnamefont{C.}~\bibnamefont{Kim}},
  \bibinfo{author}{\bibfnamefont{Z.-X.} \bibnamefont{Shen}},
  \bibinfo{author}{\bibfnamefont{T.}~\bibnamefont{Kimura}},
  \bibinfo{author}{\bibfnamefont{Y.}~\bibnamefont{Tokura}},
  \bibnamefont{et~al.}, \bibinfo{journal}{Phys. Rev. Lett.}
  \textbf{\bibinfo{volume}{85}}, \bibinfo{pages}{5194} (\bibinfo{year}{2000}),
  \urlprefix\url{https://link.aps.org/doi/10.1103/PhysRevLett.85.5194}.

\bibitem[{\citenamefont{Kamihara et~al.}(2008)\citenamefont{Kamihara, Watanabe,
  Hirano, and Hosono}}]{Kamihara2008}
\bibinfo{author}{\bibfnamefont{Y.}~\bibnamefont{Kamihara}},
  \bibinfo{author}{\bibfnamefont{T.}~\bibnamefont{Watanabe}},
  \bibinfo{author}{\bibfnamefont{M.}~\bibnamefont{Hirano}}, \bibnamefont{and}
  \bibinfo{author}{\bibfnamefont{H.}~\bibnamefont{Hosono}},
  \bibinfo{journal}{Journal of the American Chemical Society}
  \textbf{\bibinfo{volume}{130}}, \bibinfo{pages}{3296} (\bibinfo{year}{2008}),
  \urlprefix\url{http://pubs.acs.org/doi/abs/10.1021/ja800073m}.

\bibitem[{\citenamefont{Mazin et~al.}(2008)\citenamefont{Mazin, Singh,
  Johannes, and Du}}]{Mazin2008}
\bibinfo{author}{\bibfnamefont{I.~I.} \bibnamefont{Mazin}},
  \bibinfo{author}{\bibfnamefont{D.~J.} \bibnamefont{Singh}},
  \bibinfo{author}{\bibfnamefont{M.~D.} \bibnamefont{Johannes}},
  \bibnamefont{and} \bibinfo{author}{\bibfnamefont{M.~H.} \bibnamefont{Du}},
  \bibinfo{journal}{Phys. Rev. Lett.} \textbf{\bibinfo{volume}{101}},
  \bibinfo{pages}{057003} (\bibinfo{year}{2008}),
  \urlprefix\url{http://link.aps.org/doi/10.1103/PhysRevLett.101.057003}.

\bibitem[{\citenamefont{Kuroki et~al.}(2008)\citenamefont{Kuroki, Onari, Arita,
  Usui, Tanaka, Kontani, and Aoki}}]{Kuroki2008}
\bibinfo{author}{\bibfnamefont{K.}~\bibnamefont{Kuroki}},
  \bibinfo{author}{\bibfnamefont{S.}~\bibnamefont{Onari}},
  \bibinfo{author}{\bibfnamefont{R.}~\bibnamefont{Arita}},
  \bibinfo{author}{\bibfnamefont{H.}~\bibnamefont{Usui}},
  \bibinfo{author}{\bibfnamefont{Y.}~\bibnamefont{Tanaka}},
  \bibinfo{author}{\bibfnamefont{H.}~\bibnamefont{Kontani}}, \bibnamefont{and}
  \bibinfo{author}{\bibfnamefont{H.}~\bibnamefont{Aoki}},
  \bibinfo{journal}{Phys. Rev. Lett.} \textbf{\bibinfo{volume}{101}},
  \bibinfo{pages}{087004} (\bibinfo{year}{2008}),
  \urlprefix\url{https://link.aps.org/doi/10.1103/PhysRevLett.101.087004}.

\bibitem[{\citenamefont{Chubukov et~al.}(2008)\citenamefont{Chubukov, Efremov,
  and Eremin}}]{Chubukov2008}
\bibinfo{author}{\bibfnamefont{A.~V.} \bibnamefont{Chubukov}},
  \bibinfo{author}{\bibfnamefont{D.~V.} \bibnamefont{Efremov}},
  \bibnamefont{and} \bibinfo{author}{\bibfnamefont{I.}~\bibnamefont{Eremin}},
  \bibinfo{journal}{Phys. Rev. B} \textbf{\bibinfo{volume}{78}},
  \bibinfo{pages}{134512} (\bibinfo{year}{2008}),
  \urlprefix\url{http://link.aps.org/doi/10.1103/PhysRevB.78.134512}.

\bibitem[{\citenamefont{B\"oker et~al.}(2017)\citenamefont{B\"oker, Volkov,
  Efetov, and Eremin}}]{Boeker2017}
\bibinfo{author}{\bibfnamefont{J.}~\bibnamefont{B\"oker}},
  \bibinfo{author}{\bibfnamefont{P.~A.} \bibnamefont{Volkov}},
  \bibinfo{author}{\bibfnamefont{K.~B.} \bibnamefont{Efetov}},
  \bibnamefont{and} \bibinfo{author}{\bibfnamefont{I.}~\bibnamefont{Eremin}},
  \bibinfo{journal}{Phys. Rev. B} \textbf{\bibinfo{volume}{96}},
  \bibinfo{pages}{014517} (\bibinfo{year}{2017}),
  \urlprefix\url{https://link.aps.org/doi/10.1103/PhysRevB.96.014517}.

\bibitem[{\citenamefont{Bardeen
  et~al.}(1957{\natexlab{a}})\citenamefont{Bardeen, Cooper, and
  Schrieffer}}]{Bardeen1957}
\bibinfo{author}{\bibfnamefont{J.}~\bibnamefont{Bardeen}},
  \bibinfo{author}{\bibfnamefont{L.~N.} \bibnamefont{Cooper}},
  \bibnamefont{and} \bibinfo{author}{\bibfnamefont{J.~R.}
  \bibnamefont{Schrieffer}}, \bibinfo{journal}{Phys. Rev.}
  \textbf{\bibinfo{volume}{106}}, \bibinfo{pages}{162}
  (\bibinfo{year}{1957}{\natexlab{a}}),
  \urlprefix\url{https://link.aps.org/doi/10.1103/PhysRev.106.162}.

\bibitem[{\citenamefont{Bardeen
  et~al.}(1957{\natexlab{b}})\citenamefont{Bardeen, Cooper, and
  Schrieffer}}]{Bardeen1957a}
\bibinfo{author}{\bibfnamefont{J.}~\bibnamefont{Bardeen}},
  \bibinfo{author}{\bibfnamefont{L.~N.} \bibnamefont{Cooper}},
  \bibnamefont{and} \bibinfo{author}{\bibfnamefont{J.~R.}
  \bibnamefont{Schrieffer}}, \bibinfo{journal}{Phys. Rev.}
  \textbf{\bibinfo{volume}{108}}, \bibinfo{pages}{1175}
  (\bibinfo{year}{1957}{\natexlab{b}}),
  \urlprefix\url{https://link.aps.org/doi/10.1103/PhysRev.108.1175}.

\bibitem[{\citenamefont{Pippard}(1953)}]{Pippard1953}
\bibinfo{author}{\bibfnamefont{A.}~\bibnamefont{Pippard}},
  \bibinfo{journal}{Proc R Soc Lond A Math Phys Sci}
  \textbf{\bibinfo{volume}{216}}, \bibinfo{pages}{547} (\bibinfo{year}{1953}),
  \urlprefix\url{http://rspa.royalsocietypublishing.org/content/216/1127/547.abstract}.

\bibitem[{\citenamefont{Kogan}(1981)}]{Kogan1981}
\bibinfo{author}{\bibfnamefont{V.~G.} \bibnamefont{Kogan}},
  \bibinfo{journal}{Phys. Rev. B} \textbf{\bibinfo{volume}{24}},
  \bibinfo{pages}{1572} (\bibinfo{year}{1981}),
  \urlprefix\url{https://link.aps.org/doi/10.1103/PhysRevB.24.1572}.

\bibitem[{\citenamefont{Balatskii et~al.}(1986)\citenamefont{Balatskii,
  Burlachkov, and Gor'kov}}]{Balatskii}
\bibinfo{author}{\bibfnamefont{A.}~\bibnamefont{Balatskii}},
  \bibinfo{author}{\bibfnamefont{L.}~\bibnamefont{Burlachkov}},
  \bibnamefont{and} \bibinfo{author}{\bibfnamefont{L.}~\bibnamefont{Gor'kov}},
  \bibinfo{journal}{Sov. Phys. JETP.} \textbf{\bibinfo{volume}{63}},
  \bibinfo{pages}{866} (\bibinfo{year}{1986}).

\bibitem[{\citenamefont{Buzdin and Simonov}(1990)}]{Buzdin}
\bibinfo{author}{\bibfnamefont{A.}~\bibnamefont{Buzdin}} \bibnamefont{and}
  \bibinfo{author}{\bibfnamefont{A.}~\bibnamefont{Simonov}},
  \bibinfo{journal}{Sov. Phys. JETP.} \textbf{\bibinfo{volume}{71}},
  \bibinfo{pages}{1165} (\bibinfo{year}{1990}).

\bibitem[{\citenamefont{Leggett}(1966)}]{Leggett1966}
\bibinfo{author}{\bibfnamefont{A.~J.} \bibnamefont{Leggett}},
  \bibinfo{journal}{Progress of Theoretical Physics}
  \textbf{\bibinfo{volume}{36}}, \bibinfo{pages}{901} (\bibinfo{year}{1966}),
  \urlprefix\url{http://ptp.ipap.jp/link?PTP/36/901/}.

\bibitem[{\citenamefont{Smiseth et~al.}(2005)\citenamefont{Smiseth,
  Sm{\o}rgrav, Babaev, and Sudb{\o}}}]{smiseth2005field}
\bibinfo{author}{\bibfnamefont{J.}~\bibnamefont{Smiseth}},
  \bibinfo{author}{\bibfnamefont{E.}~\bibnamefont{Sm{\o}rgrav}},
  \bibinfo{author}{\bibfnamefont{E.}~\bibnamefont{Babaev}}, \bibnamefont{and}
  \bibinfo{author}{\bibfnamefont{A.}~\bibnamefont{Sudb{\o}}},
  \bibinfo{journal}{Physical Review B} \textbf{\bibinfo{volume}{71}},
  \bibinfo{pages}{214509} (\bibinfo{year}{2005}).

\bibitem[{\citenamefont{Carlstr\"om
  et~al.}(2011{\natexlab{a}})\citenamefont{Carlstr\"om, Garaud, and
  Babaev}}]{Carlstroem2011a}
\bibinfo{author}{\bibfnamefont{J.}~\bibnamefont{Carlstr\"om}},
  \bibinfo{author}{\bibfnamefont{J.}~\bibnamefont{Garaud}}, \bibnamefont{and}
  \bibinfo{author}{\bibfnamefont{E.}~\bibnamefont{Babaev}},
  \bibinfo{journal}{Phys. Rev. B} \textbf{\bibinfo{volume}{84}},
  \bibinfo{pages}{134518} (\bibinfo{year}{2011}{\natexlab{a}}),
  \urlprefix\url{http://link.aps.org/doi/10.1103/PhysRevB.84.134518}.

\bibitem[{\citenamefont{Weston and Babaev}(2013)}]{Weston2013}
\bibinfo{author}{\bibfnamefont{D.}~\bibnamefont{Weston}} \bibnamefont{and}
  \bibinfo{author}{\bibfnamefont{E.}~\bibnamefont{Babaev}},
  \bibinfo{journal}{Phys. Rev. B} \textbf{\bibinfo{volume}{88}},
  \bibinfo{pages}{214507} (\bibinfo{year}{2013}),
  \urlprefix\url{https://link.aps.org/doi/10.1103/PhysRevB.88.214507}.

\bibitem[{\citenamefont{Silaev et~al.}(2015)\citenamefont{Silaev, Garaud, and
  Babaev}}]{Silaev2015c}
\bibinfo{author}{\bibfnamefont{M.}~\bibnamefont{Silaev}},
  \bibinfo{author}{\bibfnamefont{J.}~\bibnamefont{Garaud}}, \bibnamefont{and}
  \bibinfo{author}{\bibfnamefont{E.}~\bibnamefont{Babaev}},
  \bibinfo{journal}{Phys. Rev. B} \textbf{\bibinfo{volume}{92}},
  \bibinfo{pages}{174510} (\bibinfo{year}{2015}),
  \urlprefix\url{http://link.aps.org/doi/10.1103/PhysRevB.92.174510}.

\bibitem[{\citenamefont{Garaud et~al.}(2016)\citenamefont{Garaud, Silaev, and
  Babaev}}]{Garaud2016}
\bibinfo{author}{\bibfnamefont{J.}~\bibnamefont{Garaud}},
  \bibinfo{author}{\bibfnamefont{M.}~\bibnamefont{Silaev}}, \bibnamefont{and}
  \bibinfo{author}{\bibfnamefont{E.}~\bibnamefont{Babaev}},
  \bibinfo{journal}{Phys. Rev. Lett.} \textbf{\bibinfo{volume}{116}},
  \bibinfo{pages}{097002} (\bibinfo{year}{2016}),
  \urlprefix\url{https://link.aps.org/doi/10.1103/PhysRevLett.116.097002}.

\bibitem[{\citenamefont{Mazin et~al.}(2002)\citenamefont{Mazin, Andersen,
  Jepsen, Dolgov, Kortus, Golubov, Kuz’menko, and van~der Marel}}]{Mazin2002}
\bibinfo{author}{\bibfnamefont{I.~I.} \bibnamefont{Mazin}},
  \bibinfo{author}{\bibfnamefont{O.~K.} \bibnamefont{Andersen}},
  \bibinfo{author}{\bibfnamefont{O.}~\bibnamefont{Jepsen}},
  \bibinfo{author}{\bibfnamefont{O.~V.} \bibnamefont{Dolgov}},
  \bibinfo{author}{\bibfnamefont{J.}~\bibnamefont{Kortus}},
  \bibinfo{author}{\bibfnamefont{A.~A.} \bibnamefont{Golubov}},
  \bibinfo{author}{\bibfnamefont{A.~B.} \bibnamefont{Kuz’menko}},
  \bibnamefont{and} \bibinfo{author}{\bibfnamefont{D.}~\bibnamefont{van~der
  Marel}}, \bibinfo{journal}{Phys. Rev. Lett.} \textbf{\bibinfo{volume}{89}},
  \bibinfo{pages}{107002} (\bibinfo{year}{2002}),
  \urlprefix\url{https://link.aps.org/doi/10.1103/PhysRevLett.89.107002}.

\bibitem[{\citenamefont{Moshchalkov et~al.}(2009)\citenamefont{Moshchalkov,
  Menghini, Nishio, Chen, Silhanek, Dao, Chibotaru, Zhigadlo, and
  Karpinski}}]{Moshchalkov2009}
\bibinfo{author}{\bibfnamefont{V.}~\bibnamefont{Moshchalkov}},
  \bibinfo{author}{\bibfnamefont{M.}~\bibnamefont{Menghini}},
  \bibinfo{author}{\bibfnamefont{T.}~\bibnamefont{Nishio}},
  \bibinfo{author}{\bibfnamefont{Q.~H.} \bibnamefont{Chen}},
  \bibinfo{author}{\bibfnamefont{A.~V.} \bibnamefont{Silhanek}},
  \bibinfo{author}{\bibfnamefont{V.~H.} \bibnamefont{Dao}},
  \bibinfo{author}{\bibfnamefont{L.~F.} \bibnamefont{Chibotaru}},
  \bibinfo{author}{\bibfnamefont{N.~D.} \bibnamefont{Zhigadlo}},
  \bibnamefont{and}
  \bibinfo{author}{\bibfnamefont{J.}~\bibnamefont{Karpinski}},
  \bibinfo{journal}{Phys. Rev. Lett.} \textbf{\bibinfo{volume}{102}},
  \bibinfo{pages}{117001} (\bibinfo{year}{2009}),
  \urlprefix\url{https://link.aps.org/doi/10.1103/PhysRevLett.102.117001}.

\bibitem[{\citenamefont{Pal et~al.}(2006)\citenamefont{Pal, DeBeer-Schmitt,
  Bera, Cubitt, Dewhurst, Jun, Zhigadlo, Karpinski, Kogan, and
  Eskildsen}}]{Pal2006}
\bibinfo{author}{\bibfnamefont{D.}~\bibnamefont{Pal}},
  \bibinfo{author}{\bibfnamefont{L.}~\bibnamefont{DeBeer-Schmitt}},
  \bibinfo{author}{\bibfnamefont{T.}~\bibnamefont{Bera}},
  \bibinfo{author}{\bibfnamefont{R.}~\bibnamefont{Cubitt}},
  \bibinfo{author}{\bibfnamefont{C.~D.} \bibnamefont{Dewhurst}},
  \bibinfo{author}{\bibfnamefont{J.}~\bibnamefont{Jun}},
  \bibinfo{author}{\bibfnamefont{N.~D.} \bibnamefont{Zhigadlo}},
  \bibinfo{author}{\bibfnamefont{J.}~\bibnamefont{Karpinski}},
  \bibinfo{author}{\bibfnamefont{V.~G.} \bibnamefont{Kogan}}, \bibnamefont{and}
  \bibinfo{author}{\bibfnamefont{M.~R.} \bibnamefont{Eskildsen}},
  \bibinfo{journal}{Phys. Rev. B} \textbf{\bibinfo{volume}{73}},
  \bibinfo{pages}{012513} (\bibinfo{year}{2006}),
  \urlprefix\url{https://link.aps.org/doi/10.1103/PhysRevB.73.012513}.

\bibitem[{\citenamefont{Brinkman et~al.}(2002)\citenamefont{Brinkman, Golubov,
  Rogalla, Dolgov, Kortus, Kong, Jepsen, and Andersen}}]{Brinkman2002}
\bibinfo{author}{\bibfnamefont{A.}~\bibnamefont{Brinkman}},
  \bibinfo{author}{\bibfnamefont{A.~A.} \bibnamefont{Golubov}},
  \bibinfo{author}{\bibfnamefont{H.}~\bibnamefont{Rogalla}},
  \bibinfo{author}{\bibfnamefont{O.~V.} \bibnamefont{Dolgov}},
  \bibinfo{author}{\bibfnamefont{J.}~\bibnamefont{Kortus}},
  \bibinfo{author}{\bibfnamefont{Y.}~\bibnamefont{Kong}},
  \bibinfo{author}{\bibfnamefont{O.}~\bibnamefont{Jepsen}}, \bibnamefont{and}
  \bibinfo{author}{\bibfnamefont{O.~K.} \bibnamefont{Andersen}},
  \bibinfo{journal}{Phys. Rev. B} \textbf{\bibinfo{volume}{65}},
  \bibinfo{pages}{180517} (\bibinfo{year}{2002}),
  \urlprefix\url{https://link.aps.org/doi/10.1103/PhysRevB.65.180517}.

\bibitem[{\citenamefont{Huang et~al.}(2016)\citenamefont{Huang, Scaffidi,
  Sigrist, and Kallin}}]{Huang2016}
\bibinfo{author}{\bibfnamefont{W.}~\bibnamefont{Huang}},
  \bibinfo{author}{\bibfnamefont{T.}~\bibnamefont{Scaffidi}},
  \bibinfo{author}{\bibfnamefont{M.}~\bibnamefont{Sigrist}}, \bibnamefont{and}
  \bibinfo{author}{\bibfnamefont{C.}~\bibnamefont{Kallin}},
  \bibinfo{journal}{Phys. Rev. B} \textbf{\bibinfo{volume}{94}},
  \bibinfo{pages}{064508} (\bibinfo{year}{2016}),
  \urlprefix\url{http://link.aps.org/doi/10.1103/PhysRevB.94.064508}.

\bibitem[{\citenamefont{Mackenzie and Maeno}(2003)}]{Mackenzie2003}
\bibinfo{author}{\bibfnamefont{A.~P.} \bibnamefont{Mackenzie}}
  \bibnamefont{and} \bibinfo{author}{\bibfnamefont{Y.}~\bibnamefont{Maeno}},
  \bibinfo{journal}{Rev. Mod. Phys.} \textbf{\bibinfo{volume}{75}},
  \bibinfo{pages}{657} (\bibinfo{year}{2003}),
  \urlprefix\url{http://link.aps.org/doi/10.1103/RevModPhys.75.657}.

\bibitem[{\citenamefont{Kortus et~al.}(2001)\citenamefont{Kortus, Mazin,
  Belashchenko, Antropov, and Boyer}}]{Kortus2001}
\bibinfo{author}{\bibfnamefont{J.}~\bibnamefont{Kortus}},
  \bibinfo{author}{\bibfnamefont{I.~I.} \bibnamefont{Mazin}},
  \bibinfo{author}{\bibfnamefont{K.~D.} \bibnamefont{Belashchenko}},
  \bibinfo{author}{\bibfnamefont{V.~P.} \bibnamefont{Antropov}},
  \bibnamefont{and} \bibinfo{author}{\bibfnamefont{L.~L.} \bibnamefont{Boyer}},
  \bibinfo{journal}{Phys. Rev. Lett.} \textbf{\bibinfo{volume}{86}},
  \bibinfo{pages}{4656} (\bibinfo{year}{2001}),
  \urlprefix\url{https://link.aps.org/doi/10.1103/PhysRevLett.86.4656}.

\bibitem[{\citenamefont{Lin and Hu}(2012)}]{Lin2012}
\bibinfo{author}{\bibfnamefont{S.-Z.} \bibnamefont{Lin}} \bibnamefont{and}
  \bibinfo{author}{\bibfnamefont{X.}~\bibnamefont{Hu}}, \bibinfo{journal}{Phys.
  Rev. Lett.} \textbf{\bibinfo{volume}{108}}, \bibinfo{pages}{177005}
  (\bibinfo{year}{2012}),
  \urlprefix\url{http://link.aps.org/doi/10.1103/PhysRevLett.108.177005}.

\bibitem[{\citenamefont{Hecht}(2012)}]{MR3043640}
\bibinfo{author}{\bibfnamefont{F.}~\bibnamefont{Hecht}}, \bibinfo{journal}{J.
  Numer. Math.} \textbf{\bibinfo{volume}{20}}, \bibinfo{pages}{251}
  (\bibinfo{year}{2012}), ISSN \bibinfo{issn}{1570-2820}.

\bibitem[{\citenamefont{Hecht et~al.}(2007)\citenamefont{Hecht, Pironneau,
  Le~Hyaric, and Ohtsuka}}]{Hecht2007}
\bibinfo{author}{\bibfnamefont{F.}~\bibnamefont{Hecht}},
  \bibinfo{author}{\bibfnamefont{O.}~\bibnamefont{Pironneau}},
  \bibinfo{author}{\bibfnamefont{A.}~\bibnamefont{Le~Hyaric}},
  \bibnamefont{and} \bibinfo{author}{\bibfnamefont{K.}~\bibnamefont{Ohtsuka}},
  \emph{\bibinfo{title}{{Freefem++ (manual). \emph{www.freefem.org}}}}
  (\bibinfo{publisher}{web}, \bibinfo{address}{www.freefem.org},
  \bibinfo{year}{2007}), \urlprefix\url{www.freefem.org}.

\bibitem[{\citenamefont{Babaev}(2002)}]{Babaev2002a}
\bibinfo{author}{\bibfnamefont{E.}~\bibnamefont{Babaev}},
  \bibinfo{journal}{Phys. Rev. Lett.} \textbf{\bibinfo{volume}{89}},
  \bibinfo{pages}{067001} (\bibinfo{year}{2002}),
  \urlprefix\url{https://link.aps.org/doi/10.1103/PhysRevLett.89.067001}.

\bibitem[{\citenamefont{Silaev}(2011)}]{Silaev2011b}
\bibinfo{author}{\bibfnamefont{M.~A.} \bibnamefont{Silaev}},
  \bibinfo{journal}{Phys. Rev. B} \textbf{\bibinfo{volume}{83}},
  \bibinfo{pages}{144519} (\bibinfo{year}{2011}),
  \urlprefix\url{https://link.aps.org/doi/10.1103/PhysRevB.83.144519}.

\bibitem[{\citenamefont{Babaev et~al.}(2002)\citenamefont{Babaev, Faddeev, and
  Niemi}}]{Babaev2002b}
\bibinfo{author}{\bibfnamefont{E.}~\bibnamefont{Babaev}},
  \bibinfo{author}{\bibfnamefont{L.~D.} \bibnamefont{Faddeev}},
  \bibnamefont{and} \bibinfo{author}{\bibfnamefont{A.~J.} \bibnamefont{Niemi}},
  \bibinfo{journal}{Phys. Rev. B} \textbf{\bibinfo{volume}{65}},
  \bibinfo{pages}{100512} (\bibinfo{year}{2002}),
  \urlprefix\url{https://link.aps.org/doi/10.1103/PhysRevB.65.100512}.

\bibitem[{\citenamefont{Babaev}(2009)}]{Babaev2009a}
\bibinfo{author}{\bibfnamefont{E.}~\bibnamefont{Babaev}},
  \bibinfo{journal}{Phys. Rev. B} \textbf{\bibinfo{volume}{79}},
  \bibinfo{pages}{104506} (\bibinfo{year}{2009}),
  \urlprefix\url{https://link.aps.org/doi/10.1103/PhysRevB.79.104506}.

\bibitem[{\citenamefont{Babaev et~al.}(2009)\citenamefont{Babaev,
  J{\"a}ykk{\"a}, and Speight}}]{Babaev2009}
\bibinfo{author}{\bibfnamefont{E.}~\bibnamefont{Babaev}},
  \bibinfo{author}{\bibfnamefont{J.}~\bibnamefont{J{\"a}ykk{\"a}}},
  \bibnamefont{and} \bibinfo{author}{\bibfnamefont{M.}~\bibnamefont{Speight}},
  \bibinfo{journal}{Phys. Rev. Lett.} \textbf{\bibinfo{volume}{103}},
  \bibinfo{pages}{237002} (\bibinfo{year}{2009}),
  \urlprefix\url{https://link.aps.org/doi/10.1103/PhysRevLett.103.237002}.

\bibitem[{\citenamefont{Jennings and Winyard}(2014)}]{Jennings2014}
\bibinfo{author}{\bibfnamefont{P.}~\bibnamefont{Jennings}} \bibnamefont{and}
  \bibinfo{author}{\bibfnamefont{T.}~\bibnamefont{Winyard}},
  \bibinfo{journal}{Journal of High Energy Physics}
  \textbf{\bibinfo{volume}{2014}}, \bibinfo{pages}{122} (\bibinfo{year}{2014}),
  ISSN \bibinfo{issn}{1029-8479},
  \urlprefix\url{https://doi.org/10.1007/JHEP01(2014)122}.

\bibitem[{\citenamefont{Babaev and Speight}(2005)}]{Babaev2005}
\bibinfo{author}{\bibfnamefont{E.}~\bibnamefont{Babaev}} \bibnamefont{and}
  \bibinfo{author}{\bibfnamefont{M.}~\bibnamefont{Speight}},
  \bibinfo{journal}{Phys. Rev. B} \textbf{\bibinfo{volume}{72}},
  \bibinfo{pages}{180502} (\bibinfo{year}{2005}),
  \urlprefix\url{http://link.aps.org/doi/10.1103/PhysRevB.72.180502}.

\bibitem[{\citenamefont{Carlstr\"om
  et~al.}(2011{\natexlab{b}})\citenamefont{Carlstr\"om, Babaev, and
  Speight}}]{Carlstroem2011}
\bibinfo{author}{\bibfnamefont{J.}~\bibnamefont{Carlstr\"om}},
  \bibinfo{author}{\bibfnamefont{E.}~\bibnamefont{Babaev}}, \bibnamefont{and}
  \bibinfo{author}{\bibfnamefont{M.}~\bibnamefont{Speight}},
  \bibinfo{journal}{Phys. Rev. B} \textbf{\bibinfo{volume}{83}},
  \bibinfo{pages}{174509} (\bibinfo{year}{2011}{\natexlab{b}}),
  \urlprefix\url{http://prb.aps.org/abstract/PRB/v83/i17/e174509}.

\bibitem[{\citenamefont{Silaev and Babaev}(2011)}]{Silaev2011}
\bibinfo{author}{\bibfnamefont{M.}~\bibnamefont{Silaev}} \bibnamefont{and}
  \bibinfo{author}{\bibfnamefont{E.}~\bibnamefont{Babaev}},
  \bibinfo{journal}{Phys. Rev. B} \textbf{\bibinfo{volume}{84}},
  \bibinfo{pages}{094515} (\bibinfo{year}{2011}),
  \urlprefix\url{https://link.aps.org/doi/10.1103/PhysRevB.84.094515}.

\bibitem[{\citenamefont{Babaev et~al.}(2017)\citenamefont{Babaev,
  Carlstr{\"o}m, Silaev, and Speight}}]{BABAEV201720}
\bibinfo{author}{\bibfnamefont{E.}~\bibnamefont{Babaev}},
  \bibinfo{author}{\bibfnamefont{J.}~\bibnamefont{Carlstr{\"o}m}},
  \bibinfo{author}{\bibfnamefont{M.}~\bibnamefont{Silaev}}, \bibnamefont{and}
  \bibinfo{author}{\bibfnamefont{J.}~\bibnamefont{Speight}},
  \bibinfo{journal}{Physica C: Superconductivity and its Applications}
  \textbf{\bibinfo{volume}{533}}, \bibinfo{pages}{20 } (\bibinfo{year}{2017}),
  ISSN \bibinfo{issn}{0921-4534}, \bibinfo{note}{ninth international conference
  on Vortex Matter in nanostructured Superdonductors},
  \urlprefix\url{http://www.sciencedirect.com/science/article/pii/S0921453416301198}.

\bibitem[{\citenamefont{Winyard
  et~al.}(2018{\natexlab{a}})\citenamefont{Winyard, Silaev, and
  Babaev}}]{winyard2018hierarchies}
\bibinfo{author}{\bibfnamefont{T.}~\bibnamefont{Winyard}},
  \bibinfo{author}{\bibfnamefont{M.}~\bibnamefont{Silaev}}, \bibnamefont{and}
  \bibinfo{author}{\bibfnamefont{E.}~\bibnamefont{Babaev}},
  \bibinfo{journal}{arXiv preprint arXiv:1801.09274}
  (\bibinfo{year}{2018}{\natexlab{a}}).

\bibitem[{\citenamefont{Winyard
  et~al.}(2018{\natexlab{b}})\citenamefont{Winyard, Silaev, and
  Babaev}}]{winyard2018skyrmions}
\bibinfo{author}{\bibfnamefont{T.}~\bibnamefont{Winyard}},
  \bibinfo{author}{\bibfnamefont{M.}~\bibnamefont{Silaev}}, \bibnamefont{and}
  \bibinfo{author}{\bibfnamefont{E.}~\bibnamefont{Babaev}},
  \bibinfo{journal}{arXiv preprint arXiv:1801.10463}
  (\bibinfo{year}{2018}{\natexlab{b}}).

\bibitem[{\citenamefont{Garaud et~al.}(2017)\citenamefont{Garaud, Silaev, and
  Babaev}}]{Garaud2017a}
\bibinfo{author}{\bibfnamefont{J.}~\bibnamefont{Garaud}},
  \bibinfo{author}{\bibfnamefont{M.}~\bibnamefont{Silaev}}, \bibnamefont{and}
  \bibinfo{author}{\bibfnamefont{E.}~\bibnamefont{Babaev}},
  \bibinfo{journal}{Physica C: Superconductivity and its Applications}
  \textbf{\bibinfo{volume}{533}}, \bibinfo{pages}{63 } (\bibinfo{year}{2017}),
  ISSN \bibinfo{issn}{0921-4534},
  \urlprefix\url{http://www.sciencedirect.com/science/article/pii/S0921453416300983}.

\end{thebibliography}

 \end{document}